\documentclass[runningheads]{Harvardderived}
\usepackage[T1]{fontenc}
\usepackage{fix-cm}
\usepackage{graphicx}
\usepackage{hyperref}
\usepackage{bm}
\usepackage{amsfonts}
\usepackage{float}
\usepackage{xcolor}
\usepackage{booktabs}
\usepackage{orcidlink}

\newcommand{\ntest}{4,087}
\newcommand{\ibm}{ibm\_kingston}
\newcommand{\qgan}{QGAN}
\newcommand{\qgans}{QGANs}

\usepackage{blindtext}
\usepackage{multirow}
\usepackage{placeins}

\def\keyFont{\fontsize{8}{11}\helveticabold }
\def\firstAuthorLast{Baglio {et~al.}}
\def\Authors{Julien Baglio\,$^{1,2,*}$\orcidlink{0000-0001-6691-6108}, Yacine Haddad\,$^{3,4,*}$\orcidlink{0000-0003-4916-7752} and Richard Polifka\,$^{2}$\orcidlink{0000-0003-1036-3844}}
\def\Address{$^{1}$Center for Quantum Computing and Quantum Coherence (QC2), Department of Physics, University of Basel, Klingelbergstrasse 82, CH-4056 Basel, Switzerland\\
$^{2}$QuantumBasel, Schorenweg 44b, CH-4144 Arlesheim, Switzerland\\
$^{3}$Laboratory for High Energy Physics (LHEP), Physics Institute, Faculty of Sciences, University of Bern, Sidlerstrasse 5, CH-3012 Bern, Switzerland\\
$^{4}$ European Organization for Nuclear Research (CERN), Espl. des Particules, CH-1211 Geneva 23, Switzerland}

\def\corrAuthor{julien.baglio@unibas.ch}
\def\corrEmail{yacine.haddad@unibe.ch}
\usepackage[backend=biber,
            style=chem-acs,
            natbib=true, 
            doi=true, 
            hyperref=true,
            giveninits=true,
            pageranges = true,
            articletitle = true,
            autocite=superscript,
            autopunct = false
]{biblatex}
\begin{document}
%
%\linenumbers

\onecolumn
\firstpage{1}
%%%%%

\title[Latent Style-based Quantum WGAN for Drug Design]{Latent Style-based Quantum Wasserstein GAN for Drug Design}
%%%%%

\author[\firstAuthorLast ]{\Authors}
\address{}
\correspondance{}

\extraAuth{}
%%%%%
\maketitle
\begin{abstract}

\section{}
%%%%%%
The development of new drugs is a tedious, time-consuming, and expensive process, for which the average costs are estimated to be up to around \$2.5 billion. The first step in this long process is the design of the new drug, for which de novo drug design, assisted by artificial intelligence, has blossomed in recent years and has revolutionized the field. In particular, generative artificial intelligence has delivered promising results in drug discovery and development, reducing costs and time to solution. However, classical generative models, such as generative adversarial networks (GANs), can be difficult to train and are prone to mode collapse. Quantum computing may be an avenue to overcome some of these issues, providing compact models with far fewer trainable parameters and thereby enhancing the generalizability of GANs.
We propose a new latent style-based quantum Wasserstein GAN (\qgan) for drug design, combining a variational autoencoder that maps molecular structures into a latent space with a hybrid quantum GAN. Our pipeline is benchmarked against the classical models of the industry-relevant MOSES suite. Style-based noise reuploading at every rotational gate and a Wasserstein gradient penalty are used to improve trainability and to mitigate mode collapse. Training and large-scale sampling with circuits of up to 15 qubits are performed on noiseless quantum simulators to validate our architecture.
We present an inference-only demonstration on a 156-qubit IBM Heron quantum computer using a five-qubit setup, sampling 2,500 new molecules, to investigate the effect of quantum noise. Our default \qgan, using 110 trainable generator parameters, reaches statistically comparable overall performance and statistically significant gains on selected drug-relevant metrics relative to both the MOSES classical latent GAN baseline, which contains about 700,000 trainable parameters for the generator, and a classical latent GAN with matched generator capacity.

\tiny
 \keyFont{ \section{Keywords:} drug design, smiles, artificial intelligence, latent space, variational autoencoder, quantum machine learning, quantum generative adversarial networks, superconducting-qubit quantum computer} 
%%%%%%
\end{abstract}
%
%%%%%%%%%%%%%%%%%%%%%%%%%%%%%%%%%%%%%%%%%%%%
%

\section{Introduction}
\label{sec:intro}

The development of new pharmaceutical drugs is a lengthy, tedious, and expensive process. It usually takes 15 years, and the lower estimate of mean expense is around \$1.3 billion~\autocite{Wouters2020,Mullard2020,Wouters2022,Sertkaya2024}, and can rise to around \$2.5 billion~\autocite{DIMASI201620,Deloitte2025}.
The development cycle consists of the following main stages: target discovery, molecular design, preclinical studies, clinical trials, and finally the regulatory approval to enter the market. In the first two stages, there exist two key strategies for identifying viable drug candidates: either by modifying already existing commercial drugs, for example by performing a virtual screening, or by identifying new relationships between already known molecular structures and the desired activity; or by generating entirely new molecules, also called de novo drug design. In both methods, there are multiple challenges, such as the safety of the candidate molecule, the efficiency, and eventually the ability to manufacture the drug candidate.

In this context, the emergence of artificial intelligence (AI) has revolutionized the drug design approach, particularly advances in generative AI, which have been used for computer-assisted drug design in the past decade~\autocite{GANGWAL2024103992,Tang2024,Gangwal2024}. Generative adversarial networks (GAN)~\autocite{goodfellow2014generativeadversarialnetworks} are prominent AI models applied for drug design~\autocite{TRIPATHI2022100045,kotkondawar2025generative}. A GAN is composed of two neural networks: 1) a generator, which is trained to create new molecular structures from a random noise input distribution; 2) a discriminator, which is trained to classify between the generator (fake) output and the (real) molecular structures contained in the training input dataset. The training scheme is an adversarial game, where the generator aims to fool the discriminator as much as possible while the discriminator aims to distinguish between real data and newly generated fake data, which in turn will help the generator to perform better. While GANs (and more generally AI models for drug design) have blossomed in the past decade, their training can be computationally expensive when their scale is very large, and the associated massive parameter space requires careful optimization~\autocite{TRIPATHI2022100045}. GANs can suffer from nonconvergence of the adversarial training procedure, mode collapse when the generator only generates a subsample of the entire diversity of the molecules present in the training set, or falling into a nonrobust training pattern with great sensitivity to model hyperparameters and to the initialization of the model weights.

Quantum machine learning aims to provide new avenues to overcome some of these challenges. Quantum generative models~\autocite{amin2018quantum,liu2018differentiable,barthe2025parameterized,demidik2025expressive,demidik2025sample} are generative models relying on the rules of quantum mechanics: Starting from quantum states encoding the problem, subject to the rules of superposition and entanglement, they build up the final probabilistic output distribution using quantum measurements. Using spectral analysis, it has been shown that quantum neural networks probe different frequency modes compared to their classical counterparts, a feature which seems to be linked to higher expressivity of quantum machine learning~\autocite{mhiri2025constrained,jaderberg2024let, xu2024frequency, duffy2025spectral}. Quantum GANs (\qgans)~\autocite{dallaire2018quantum,
lloyd2018quantum,hu2019quantum,Zoufal:2019rxf,niu2022entangling,chaudhary2023towards} thus offer new avenues for powerful generative neural networks and have been applied to drug design~\autocite{Li2021,Diptanshu2023,Kao2023,Anoshin2024}. In \autocite{Kao2023,Anoshin2024}, the authors adopted the Wasserstein distance as the backbone of the loss function (WGAN)~\autocite{Arjovsky2017}, leading to an improvement over early \qgan\ investigations for drug discovery~\autocite{Li2021,Diptanshu2023}. In \autocite{Anoshin2024}, this was also supplemented by gradient penalty (WGAN-GP)~\autocite{gulrajani2017improvedtrainingwassersteingans} to improve trainability. In \autocite{Kao2023,Anoshin2024}, the quantum architecture was a modification of the MolGAN architecture~\autocite{decao2022}, the standard in computer-assisted drug design, and the two studies applied their quantum models on datasets of small molecules, the QM9 dataset~\autocite{Ramakrishnan2014} and the PC9 dataset~\autocite{Glavatskikh2019}.

In this work, we propose two improvements: 1) going beyond small-molecule generation and 2) investigating several architectural modifications over the previous \qgan\ studies for computer-assisted drug design. First, we use the MOSES dataset and benchmark suite~\autocite{moses}, which is an industry-relevant dataset of SMILES molecules~\autocite{Weininger1988} not restricted to small molecules. In addition, to exploit potential quantum advantages for sampling efficiency and model expressiveness and to address the challenge of mode collapse, we introduce a latent-space approach in the quantum computing pipeline, following classical GAN studies for drug design~\autocite{Prykhodko2019}. The input SMILES molecules of the training dataset are projected into an abstract vector space on which training is performed, before decoding the output back into the practical, physical space. Our last improvement over previous quantum studies is combining this latent-space approach with a hybrid style-based quantum GAN~\autocite{BravoPrieto2022,baglio2024}, which uses data reuploading, where the input noise distribution for the generator is inserted in all layers and not only in the first initial layer of the neural network. The style architecture was introduced for classical GANs in the context of image generation~\autocite{Karras2021}, and the latent style-based \qgan\ using WGAN-GP has been successfully tested for image generation~\autocite{chang2024} but has never been investigated for drug design. It should also be noted that the barren plateaus, where vanishing gradients render the training exponentially harder in the exploration of the loss function landscape, are less problematic when using the latent style-based approach for \qgan~\autocite{chang2024}. In addition, a recent study hinted toward a potential quantum exponential advantage with style-based \qgan\ in the number of trainable parameters~\autocite{liepelt2026}: an exponentially smaller number of trainable parameters for \qgans\ compared to their classical counterpart for nearly comparable results in the given metrics, a feature which is of high interest for a complex computer-assisted drug design pipeline. Further studies are still needed to verify this observation and check whether it is a genuine effect of the quantum pipeline, or rather an artifact of the classical baseline which was not pushed to its limits.

The molecular structure is encoded following the SMILES approach. The SMILES strings are embedded into vectors of real numbers that are fed into the autoencoder projecting these vectors into an abstract latent vector space. The discriminator remains classical, and we only use a parametrized quantum circuit for the generator. We benchmark our new quantum drug design architecture against a classical latent GAN approach based on the widely used MOSES benchmark suite and show that our proposed latent style-based \qgan\ can be very efficient for drug design with far fewer trainable parameters than the nominal MOSES latent generator. Our pipeline uses RDKit~\autocite{RDKit} to check that the newly generated molecular structures from our pipeline do correspond to valid (bio)chemical compounds and to calculate relevant metrics. The training of our quantum computing pipeline is performed on a noiseless quantum simulator, including the sampling of 30,000 new molecules to validate the pipeline. We also perform an illustrative, inference-only sampling of 2,500 new, viable molecular structures on \ibm, a 156-qubit IBM Heron quantum computer, using the model trained on quantum simulator.
While Li~\textit{et al.}~\autocite{Li2021} previously reported \qgan\ inference on IBM hardware for QM9 graph molecules, the recent WGAN-based MolGAN extensions~\autocite{Kao2023,Anoshin2024} assess device noise only on quantum simulators. To our knowledge, the inference-only sampling we perform in this work is the first quantum execution on the industry-relevant MOSES benchmark, the largest molecular \qgan\ hardware inference study published to date with 2,500 generated structures, compared with 1,000 in~\autocite{Li2021}, and the first on an IBM Heron-class processor.
Using the MOSES benchmark suite, we also quantify the quality of our output with metrics such as internal diversity, molecular weight, LogP, or quantitative estimation of drug-likeness. Our proposed quantum pipeline, tested on a real-world dataset not restricted to small molecules, enhances the explainability of computer-assisted drug design by using far fewer trainable parameters than the nominal, conventional classical AI pipeline proposed in the MOSES benchmark suite. The statistical overall score of our pipeline demonstrates that it is competitive with a classical AI pipeline, with even statistically significant improvements over classical GANs for some of the individual metrics we have investigated, even when we design a latent classical GAN matching the capacity of our proposed \qgan.

%%%%%%%%%%%%%%%%%%%%%%%%%%%%%%%%%%%%%%%%%%%%
%
\section{Materials and methods}
\label{sec:setup}

\subsection{Dataset construction}
\label{sec:setup:dataset}

We use the MOSES dataset~\autocite{moses} as our main analysis dataset. This dataset is built out of the ZINC Clean Leads collection~\autocite{Irwin2012} and contains over 1.9M molecules of interest for biochemistry applications. We have further applied a random sampling of the dataset to obtain 12,000 molecules for training and \ntest~molecules for validation.

The reason for this downsizing of the original MOSES dataset is twofold: 1) we use a reduced training set to study the low-data regime, in which only a limited number of training molecules is available. This setting is of high practical relevance because labeled or curated chemical data are often scarce and expensive to obtain, and because models with fewer trainable parameters are expected to generalize better when data are limited; 2) in the proof-of-concept phase of our new quantum architecture, it helps to validate the concept within a short runtime. This downsizing is not meant to simplify the generation task but to define a low-data benchmark in which the quantum and classical generators are compared under the same pipeline. The MOSES dataset is a high-dimensional dataset that is currently impractical to be used directly on a quantum computer, due to limited availability of resources of current quantum hardware. We thus preprocess the data by transforming the SMILES strings into high-dimensional molecular descriptors using RDKit features~\autocite{RDKit}, which will be used to calculate the various metrics. The SMILES strings are then fed to an autoencoder, see below, to compress the (high-dimensional) molecular alphabet into a latent representation (low-dimensional features), which is compact enough to be used for the quantum part of our pipeline. The decoder part of the autoencoder allows mapping the generated latent features back into valid SMILES molecules. In order to check the chemical properties of both the randomly selected samples for our final training and testing dataset as well as the generated samples, we have used RDKit filters~\autocite{RDKit}. Our procedure balances chemical validity, through the autoencoder and the RDKit validation, with quantum feasibility allowed by the low-dimensional latent-feature encoding.

When generating new molecules, we have sampled 30,000 molecules unless stated otherwise, following the default value used in the MOSES benchmark suite.

\subsection{Description of the pipeline}
\label{sec:setup:description}

The structure of the molecules contained in our dataset is encoded using SMILES strings~\autocite{Weininger1988}. A standard method in classical generative approaches for drug design uses a pretrained autoencoder to project the SMILES strings into a compact latent space, which is much more tractable than the original strings for the generative models. Note that the very first step before the autoencoder is the extraction of chemical features from the SMILES strings using RDKit~\autocite{RDKit} in order to get the relevant metrics. The autoencoder captures salient chemical features while reducing dimensionality, allowing us to operate in a manageable latent representation. Following this approach, also used in the MOSES benchmark suite, our pipeline starts with a pretrained variational autoencoder (VAE) to encode the drug-like SMILES molecules into the latent space. 

To perform the generation of new molecules, we use a generative adversarial network (GAN)~\autocite{goodfellow2014generativeadversarialnetworks}. A GAN is composed of two networks competing against each other: a generator, sampling new candidates from a random number input distribution; and a discriminator, acting as a binary classifier trained to distinguish between real input data and generated samples from the generator. The training is an adversarial procedure, a two-player minimax game between the two networks, until ideally a Nash equilibrium is reached when the generator has become so good that the discriminator can no longer distinguish between real input data and generated samples. We will compare the classical GAN implementation as provided in the MOSES benchmark suite to our proposed quantum GAN architecture.

Quantum computing is a different paradigm for performing computation. Instead of classical bits, which have binary outcomes (0 or 1), quantum computers use qubits (quantum bits), which are quantum states of a controlled two-level quantum system. Following the laws of quantum mechanics, the qubits are subject to superposition and entanglement, which in turn means that they live in a much more complex space than traditional bits and are not restricted to binary outputs, allowing for a more diverse exploration of the solution space for generative modeling applications. A class of such quantum algorithms is parametrized quantum circuits (PQC)~\autocite{Benedetti2019}, where the qubits are subjected to quantum operations (gates) which are parametrized by real numbers (rotation angles) and arranged in a circuit of consecutive operations. The final quantum state is then measured, and the output measurements are collected to statistically create the final outcome. A classical training procedure allows for optimization of these rotation angles, which are the trainable parameters of the quantum network formed by this PQC. In our pipeline, the generator will be a quantum network that will be trained on an ideal (noiseless) quantum simulator, while the discriminator remains a classical network. The generator outputs are latent vectors in the same latent space as that of the VAE. To obtain the final generated molecules, the generator output is fed back to the decoder part of the VAE.

The same pretrained VAE, including both encoder and decoder, is used for the classical and quantum pipelines. This is a fixed component of the comparison and any difference in the molecular metrics between the classical GAN and the \qgan, including the aggregated $\langle Z_0\rangle$ score defined in Section~\ref{sec:setup:metric}, therefore reflects the classical-versus quantum nature of the generator and not the VAE. Our results should be understood as characterizing a compact quantum latent-space generator within a VAE-based molecular generation pipeline. The absolute metric values also inherit a contribution from the pretrained decoder, which we report separately through the VAE-only baseline (the ``Tuned VAE'' column of Table~\ref{tab:vae_gan_metrics}) and the decoder study in Supporting Information Section~\ref{sec:app:vae_dec}.

Our analysis consists of the following steps:
\begin{enumerate}
    \item Train the VAE and optimize the VAE hyperparameters;
    \item Use the encoder part of the VAE to transform the input dataset into input latent vectors, and train the GAN with these input latent vectors;
    \item Optimize the latent classical GAN hyperparameters relevant for the discriminator; the classical generator is already optimized and follows the MOSES benchmark suite; 
    \item Replace the classical generator by a quantum generator, train, and optimize;
    \item Use the decoder and an RDKit validation to produce final generated molecule samples; Perform a comparison between classical and quantum GAN performances with a set of metrics (see below), perform inference on a real quantum hardware.
\end{enumerate}

\begin{figure}[t]
    \centering
    \includegraphics[width=1.02\linewidth]{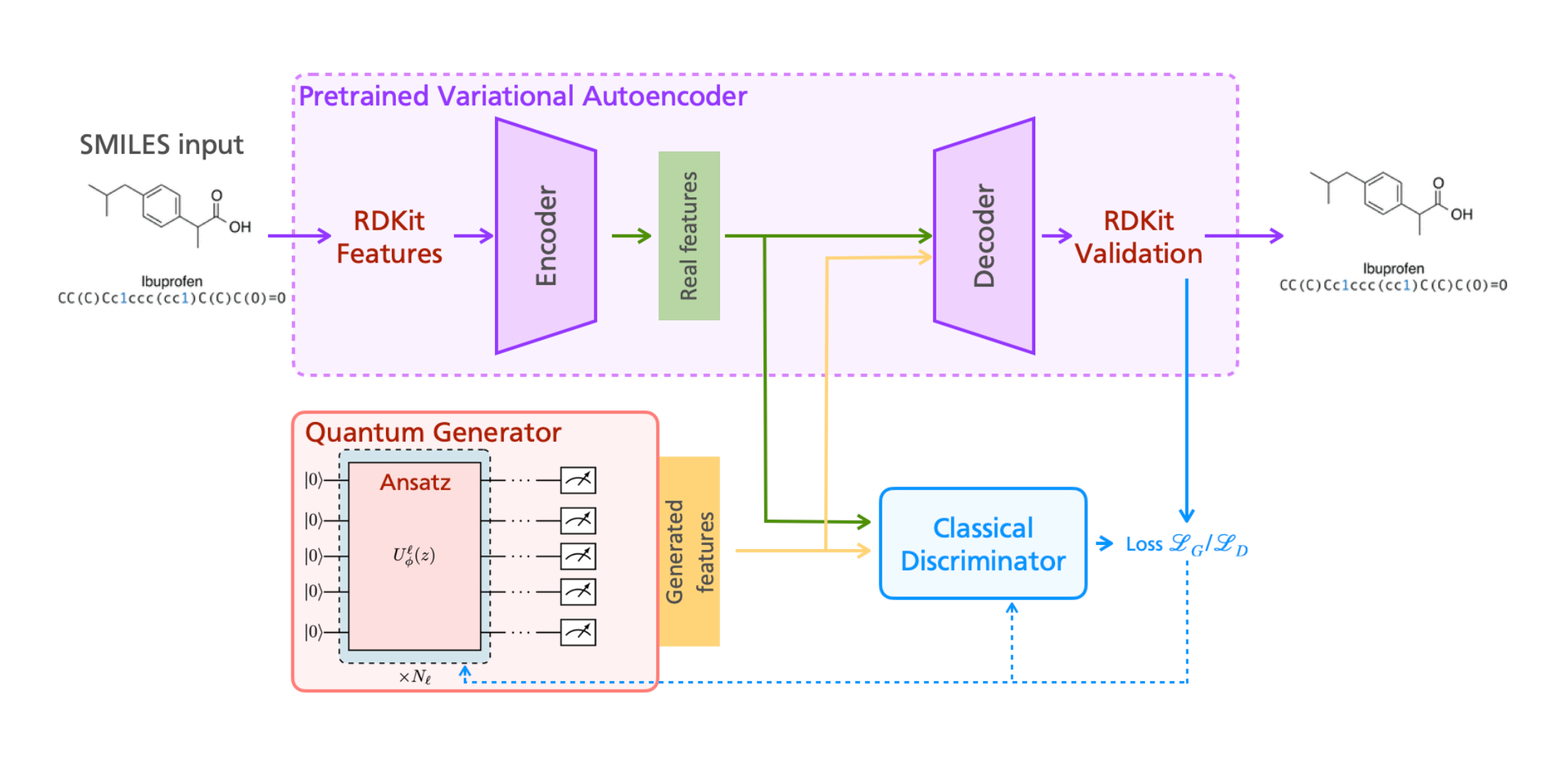}
\caption{Schematic representation of the quantum-augmented drug-design pipeline. The model comprises a pretrained VAE on the MOSES dataset using SMILES input (in purple), that embeds the original molecules into a low-dimensional latent space, and a quantum GAN, with a quantum generator (in red) and a classical discriminator (in blue). The features extracted by the VAE from the input dataset are used to train the GAN. The final generated SMILES molecules are reconstructed by inversely transforming the features generated by the quantum generator using the pretrained VAE decoder, and finally validated using RDKit.}
    \label{fig:pipeline}
\end{figure}

We present a compact representation of the proposed drug-design pipeline in Figure~\ref{fig:pipeline}. In our benchmark against a classical drug-design pipeline, the quantum generator depicted in red is replaced by a classical generator with a structure following that of the latent GAN from the MOSES benchmark suite. We have used {\tt Torch v2.8.0}~\autocite{2019arXiv191201703P} and {\tt Pennylane v0.42.3}~\autocite{Bergholm2018} to implement our pipeline ran on an NVIDIA A100 GPU with 40GB of memory.

\subsubsection{VAE and latent dimension}
\label{sec:setup:lat_dim}

The VAE architecture, taken from the MOSES benchmark suite with default parameter settings, uses gated recurrent units (GRUs) as building blocks. The encoder is made of one bidirectional GRU ($n_{l}=1$) with a hidden dimension $D_h = 256$, followed by two linear output layers that transform the output vector of the GRU to a vector of the latent space. The decoder is made of three layers of GRU ($n_{l}=3$) with hidden dimensions $D_h = 512$, with each layer (except the last one) followed by a dropout layer. The default dropout probability out of the MOSES benchmark suite is zero (hence no dropout layer in practice), but we will later explore the impact of nonzero dropout probabilities.

The SMILES string is tokenized, such that each character of the string is associated with an integer ID, which in turn is embedded as a vector (initially a one-hot vector, but the embedding is also part of the training of the VAE to increase efficiency) according to the SMILES vocabulary table, the dimension of that vector corresponding to the number of characters in the vocabulary table. As such, the GRU processes the input as sequential data $x_t$ where each element of the sequence is a vector representing the embedded token ID. The variable $t$ indexes the character ID positioning in the SMILES string from left to right, equivalent to a temporal index. At a given step $t$, a GRU uses two gates to update the internal hidden state $h_t$: the reset gate $r_t$ and the update gate $z_t$. The hidden state $h_t$ is thus computed from the hidden state $h_{t-1}$ and the new candidate hidden state $n_t$ as given in Equation~\ref{eq:gru}:
\begin{equation}
\begin{aligned}
r_t &= \sigma\left( W_{r} x_t + U_{r} h_{t-1} + b_{r} \right),\\
z_t &= \sigma\left( W_{z} x_t + U_{z} h_{t-1} + b_{z} \right),\\
n_t &= \tanh\Big( W_{n} x_t  + b_{n} + r_t \odot \left( U_{n} h_{t-1} + b_{n,u} \right) \Big),\\
h_t &= (1-z_t) \odot n_t + z_t \odot h_{t-1},
\end{aligned}
\label{eq:gru}
\end{equation}
where the function $\sigma$ is the sigmoid function controlling the gating mechanism and $\odot$ stands for Hadamard product (or element-wise product). As a slight abuse of notation, both $\tanh$ and $\sigma$ functions are meant to be applied element-wise on each element of the vector argument of these functions. The matrices $W_{r/t/n}$ and $U_{r/t/n}$ are the input trainable weights and recurrent trainable weights for the reset gate, the update gate, and the candidate hidden state, respectively. The vectors $b_{r/z/n}$ and $b_{n,u}$ are the associated trainable biases for each gate. Note that for the first letter (equivalent to $t=0$), $h_{t-1}$ corresponds to the initial hidden state at time $t=0$.

For a bidirectional GRU, the sequence $(x_t)$ is processed in both forward and backward directions, producing a final state which has doubled dimension, $h_{GRU}^{bi} = (h^{\rightarrow},h^{\leftarrow})$. At each step, the output is a concatenation of the forward direction (processing the SMILES string from left to right) and the backward direction (processing the SMILES string from right to left). This is important for the VAE encoder, as it captures information from both past and future contexts in the SMILES string, allowing for a richer and more accurate encoding of the SMILES sequence. When there is more than one layer in the GRU, the input of layer $l$ is the final hidden state (output) of layer $(l-1)$, $h^{(l-1)}_t$. When a dropout layer is included, the actual input of layer $l$ is a modification of $h^{(l-1)}_t$ into $\delta_t^{(l-1)} h^{(l-1)}_t$, where $\delta_t^{(l-1)}$ is a Bernoulli random variable equal to zero with a probability equal to dropout.

The dimensionality of the latent space is a key parameter for the pipeline. The output of the generator, for both the classical and quantum GANs, has to match the dimension of the latent space. For the quantum generator, there is a linear correspondence between the number of qubits used in the quantum circuit and the latent dimension $D_l$: either a one-to-one correspondence or a one-to-two correspondence, depending on how the measurement outcomes from the quantum network are collected; see the next sections.

As current quantum computers are still limited in the number of qubits and as simulating quantum states with classical resources is exponentially costly, our default setup for latent space is a latent dimension of 10, hence using at most 10 qubits. We have studied a latent dimension up to 30 and validated our pipeline by performing training on noiseless quantum simulators using classical resources. We have also performed inference on superconducting-qubit hardware, the \ibm\ computer with 156 qubits on an IBM Heron chip.

\subsubsection{Classical and hybrid quantum GAN architecture}
\label{sec:setup:sqGAN}

The heart of our pipeline is the GAN architecture. Let us define the generator as $G$ and the discriminator as $D$. The training of the adversarial minimax game of the GAN uses the Earth-Mover (or Wasserstein) distance, together with gradient penalty (WGAN-GP), to ensure an efficient mitigation of the effect of mode collapse~\autocite{gulrajani2017improvedtrainingwassersteingans}. The training of the WGAN-GP follows from
\begin{equation}
\min_G \max_D \;   \mathcal{L}_{WGAN}(G,D) + 
\lambda\, \mathbb{E}_{\boldsymbol{\hat{x}} \sim p(\boldsymbol{\hat{x}})}
\Big[\left(||\nabla_{\boldsymbol{\hat{x}}} D(\boldsymbol{\hat{x}})||_2 -1\right)^2 \Big],
\label{eq:lganloss}
\end{equation}
where $\boldsymbol{\hat{x}}$ are samples drawn randomly from a uniform distribution $p(\boldsymbol{\hat{x}})$, along straight lines between
pairs of points sampled from the input data distribution and the distribution of the generated output from the generator $G$. Explicitly, the interpolated samples are constructed as $\boldsymbol{\hat{x}} = \varepsilon\, \mathbf{x} + (1-\varepsilon)\, G(\boldsymbol{\xi})$, where $\mathbf{x} \sim p_{\mathrm{data}}$, $G(\boldsymbol{\xi})$ is a generated sample, and the interpolation coefficient $\varepsilon$ is sampled from a uniform distribution $U[0,1]$. We denote the input data distribution as $p_{\mathrm{data}}$ and $p(\boldsymbol{\xi})$ the noise distribution which serves as input for the generator $G$. Throughout this work, the generator input noise $\boldsymbol{\xi}$ is sampled from a standard Gaussian (normal) distribution $\mathcal{N}(0,1)$ with zero mean and unit variance; for the quantum generator, the noise variables $\xi_q$ directly set the rotation angles of the parametrized quantum circuit through Equation~\ref{eq:angle}. The function $\mathcal{L}_{WGAN}(G,D)$ is the loss function of the original Wasserstein GAN without the gradient penalty, and corresponds to
\begin{equation}
    \mathcal{L}_{WGAN}(G,D) = 
\mathbb{E}_{\boldsymbol{\xi} \sim p(\boldsymbol{\xi})}
\big[D(\mathbf{x} =G(\boldsymbol{\xi})) \big] -
\mathbb{E}_{\mathbf{x} \sim p_{\mathrm{data}}}
\big[D(\mathbf{x}) \big].
\label{eq:lganloss:wasserstein}
\end{equation}
The parameter $\lambda$ controls the impact of the gradient penalty term and is fixed to $\lambda=10$~\autocite{gulrajani2017improvedtrainingwassersteingans}.

A WGAN-GP can still suffer from training instability, difficulties with generalization, or a lack of robustness, which is reflected in a sensitivity to the initialization of the trainable parameters of the GAN. Our pipeline hence replaces the WGAN-GP with a hybrid quantum WGAN-GP (from now on simply called classical GAN and QGAN), where the discriminator is still a classical network but the generator is now a quantum generator described by a PQC. Following previous studies in the context of high-energy physics or high-quality image generation~\autocite{BravoPrieto2022,baglio2024, chang2024}, we adopt a style-based architecture~\autocite{Karras2021} in the quantum generator~\autocite{BravoPrieto2022}. It uses a data reuploading approach where the input random noise for the generator is implemented in all gates of the quantum circuit and not just in the first gates on each qubit. This approach has been shown to allow for more flexibility and richness of the quantum generation mechanism for images or high-energy physics applications but has never been attempted for drug design so far.

The classical GAN architecture in our benchmark follows closely the one from the MOSES benchmark suite~\autocite{moses} introduced in \autocite{Prykhodko2019}. We adopt the established latent-GAN architecture from the MOSES suite as our classical reference: the pretrained VAE, latent space, and WGAN-GP training objective are held fixed, and only the generator is exchanged between a classical network and a quantum circuit. The large difference in the number of trainable parameters between the two generators is therefore not an arbitrary modeling choice, but follows from comparing our quantum generator to the standard MOSES classical baseline. We also include in our study a comparison to a classical latent GAN with matched generator capacity, showing that the standard MOSES classical baseline is overparametrized in the low-data regime we have adopted in this work. We will nonetheless show that this matched classical GAN does not substantially improve over the standard MOSES baseline and that our proposed \qgan\ architecture is competitive against both the standard baseline and the matched-capacity architecture, while delivering substantial improvement over some of the drug-relevant metrics. As such, besides the comparison of the \qgan\ results with the matched-capacity classical GAN, we also quantify the parameter gap of the standard MOSES baseline at comparable performance. The discriminator (same structure for classical and quantum GANs) is made of three linear (dense) layers for which the dimension of each layer follows the sequence $[512,256,1]$, all followed by a LeakyReLU activation function with parameter 0.2, except for the last layer. The classical generator is made of five linear layers with the sequence $[128,256,512,1024,D_l]$, each followed by a LeakyReLU activation function with parameter 0.2, except for the last layer. In addition, the second, third, and fourth layers use batch normalization.

\begin{figure}[h!]
    \centering
    \includegraphics[height=0.2\linewidth]{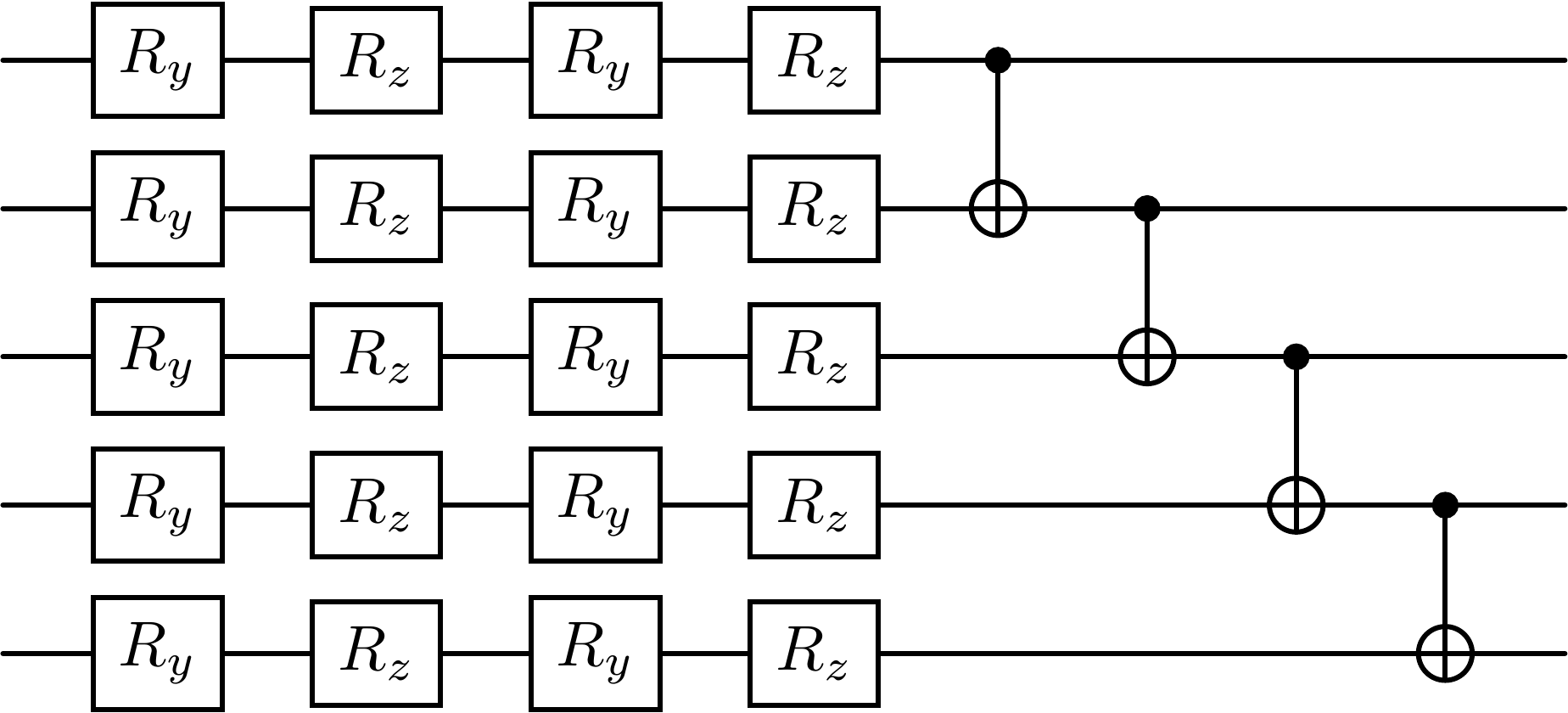}
    \caption{Styled simple circuit with five qubits and one layer.}
    \label{fig:sqGAN:spl}
\end{figure}

\begin{figure}[h!]
    \centering
    \includegraphics[height=0.2\linewidth]{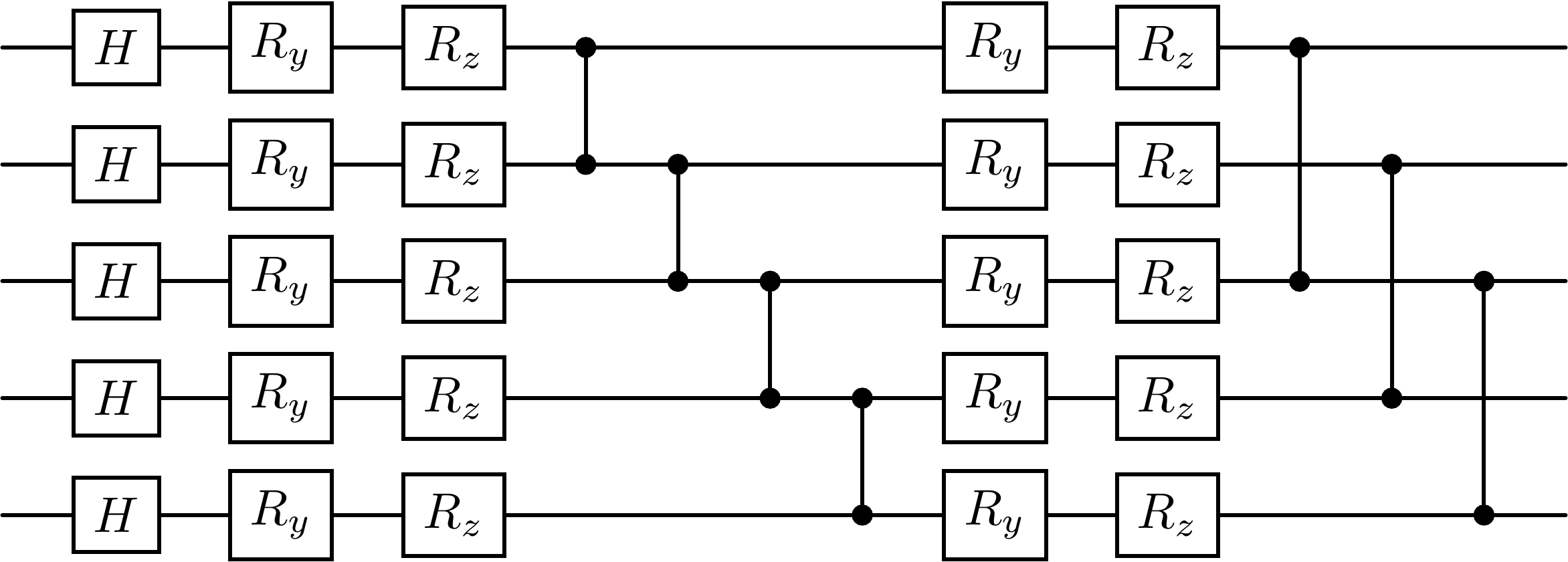}
    \caption{Styled basic entangled layer (BEL) circuit with five qubits and one layer. In addition, there is a final set of one-qubit $R_Y$ gates on each qubit to rotate all qubits in the computational basis before performing the final measurement operation.}
    \label{fig:sqGAN:bel}
\end{figure}

We will study two different ansätze for the quantum circuit of the quantum generator, built from layers that can be repeated $n_{l}$ times to increase the expressive power of the model. Such layered structures follow the hardware-efficient ansätze paradigm widely used in variational quantum algorithms, including the variational quantum eigensolver and various QGAN studies~\autocite{Kandala2017,Peruzzo2014,Benedetti2019,Zoufal:2019rxf}. In particular, the designs are inspired by parametrized circuit templates available in modern quantum machine-learning frameworks such as PennyLane~\autocite{Bergholm2018}. The first architecture corresponds to a simple structure composed of single-qubit rotations and controlled rotations arranged in repeated layers (Fig.~\ref{fig:sqGAN:spl}), while the second architecture introduces a basic entangled layer (BEL) (Fig.~\ref{fig:sqGAN:bel}) that enhances correlations between qubits through additional entangling operations~\autocite{Schuld2020}. We consider representative circuits with five data qubits. All rotation gates, including the controlled rotation gates, are parametrized by angles in the domain $[-\pi,\pi]$, defined as
\begin{equation}
    \theta_{q,\ell,k} = \pi \tanh\!\left( \xi_q W_{q\ell k} + b_{q\ell k}\right),
    \label{eq:angle}
\end{equation}
where $\xi_q$ is the input noise vector element for a given qubit $q$, $W_{q\ell k}$ is the element of the tensor of trainable parameters for a given qubit $q$, a given layer $\ell$, and a given angle label $k$ corresponding to the $k$th rotation gate. The tensor $\mathbf{b}$ contains the biases $b_{q\ell k}$. The weight tensor $\mathbf{W}$ and the bias tensor $\mathbf{b}$ are real-valued and share the same shape: each carries one independent trainable entry per (rotation gate, qubit) pair, indexed jointly by the qubit $q$, the layer $\ell$, and the in-layer gate label $k$. In particular, every parametrized rotation acting on every qubit has its own weight and its own bias, so that biases are not shared across gates, layers, or qubits. For the BEL circuit, which contains five parametrized rotations per layer plus one final rotation per qubit, the resulting number of trainable parameters is $2 \times (5\, n_{l} + 1) \times n_{qb}$ (the factor of two accounting for weights and biases); for instance, $n_{qb}=5$ and $n_{l}=2$ yield $110$ trainable parameters. Using the nonlinear function $\tanh()$ helps to avoid over-rotation during the training phase, further mitigating the (quantum) mode collapse.

We construct output vectors from the quantum circuit by performing quantum measurements of the Pauli operators and computing their expectation values. We can use either single readout or dual readout. In the former case, one Pauli operator (usually chosen as the $Z$ operator) is measured for each qubit so that a quantum generator with $n_{qb}$ qubits produces a latent vector of dimension $n_{qb}$. In the latter case, we exploit the linear independence of the Pauli $X$ and Pauli $Z$ operators to use the same quantum circuit but perform two statistically independent measurements for each qubit. The quantum generator with $n_{qb}$ qubits thus outputs a latent vector of dimension $2 n_{qb}$, at the expense of running twice as many quantum measurements to obtain the same quantum precision of the final outcome compared to that of a single readout.

Because of the exponential cost of simulating quantum circuits, we have restricted ourselves to up to 15 qubits with dual readout in our analysis, corresponding to a maximum latent dimension of 30. We have adopted the latent dimension of 10 as our default setting, with either 10 qubits and single readout or five qubits and dual readout, as used for the quantum inference on \ibm. This choice of default latent dimension is driven by the analysis of the classical latent GAN, where a latent dimension greater than 10 does not change the results significantly; see also Supporting Information Table~\ref{tab:app:scen:cgan_latdim}. We will also present results on a quantum simulator with a latent dimension of 30, with 15 qubits and dual readout.

\subsection{Metrics}
\label{sec:setup:metric}

The definitions of the metrics used in this analysis are taken from the MOSES benchmark suite and are standard for drug design. We list them below:

\begin{itemize}
    \item \textbf{Distinct Fraction ($\epsilon_d$)}: Calculates the fraction of molecules after string deduplication has been applied to the generated SMILES molecules.
    \item \textbf{Fraction of valid molecules ($\epsilon_v$)}: Reports the validity of generated SMILES strings and is defined using the RDKit molecular structure parser that checks atom valency and consistency of bonds in aromatic rings.
    \item \textbf{Fraction of unique molecules ($\epsilon_u$)}: Checks that the model does not collapse to producing only a few typical molecules.
    \item \textbf{Novelty}: Fraction of the generated molecules that are not present in the training set. A low novelty indicates overfitting.
    \item \textbf{Filters}: is the fraction of generated molecules that pass filters applied during dataset construction. While the generated molecules are often chemically valid, they may contain unwanted fragments that were removed from the training set.
    \item \textbf{Internal diversity (IntDiv)}: Assesses the chemical diversity within the generated set of molecules. This metric detects a common failure case of generative models—mode collapse. With mode collapse, the model produces a limited set of samples, ignoring some regions of chemical space.
    \item \textbf{Molecular weight (Weight)}: Sum of atomic weights in a molecule. In general, it is easier to build viable drugs from lighter molecules.
    \item \textbf{LogP}: Octanol-water partition coefficient, a ratio of a chemical concentration in the octanol phase to its concentration in the aqueous phase of a two-phase octanol/water system. It is a measure of lipophilicity and is defined as the ability for a compound to differentially dissolve in a mixture of water and lipids/organic solvents. Using MOSES tools, it is computed with the RDKit Crippen estimator~\autocite{Wildman1999}. According to the Lipinski's rule of 5~\autocite{LIPINSKI20013,LIPINSKI2004337}, a typical good value of LogP for (oral) drugs is between 0 and 5. We have constructed an associated metric, $\epsilon_{LogP}$, which is the fraction of molecules in this range.
    \item \textbf{Synthetic Accessibility Score (SA)}: Heuristic estimate of how hard (10) or how easy (1) it is to synthesize a given molecule.
    \item \textbf{Quantitative Estimation of Drug-likeness (QED)}: Real number in the interval $[0;1]$, estimating how likely a molecule is a viable candidate for a drug.
\end{itemize}

The first six metrics are constrained to the interval $[0;1]$, where the desired value is as close to unity as possible. The last four metrics are reported in two ways: either as the Wasserstein distance (denoted as $\mathcal{W}$) between the test and generated dataset, following the MOSES benchmark suite~\autocite{moses}; or as the mean of the distribution itself, calculated on the generated sample. The latter choice removes any dependency of the results on the test dataset and is the preferred choice in our analysis.

As mentioned in the list above, we have also introduced another metric not present in the MOSES analysis~\autocite{moses}, the fraction $\epsilon_{LogP}$. As the MOSES analysis was not necessarily restricted to (human) drug discovery, this new metric helps to quantify how many drug candidates we obtain in our quantum-computer-assisted drug-design pipeline are compatible with typical LogP values encountered in viable drugs. It is important to avoid a chemical compound with too high lipophilicity (which can lead to accumulation in fat tissues and fatal toxic issues, or difficulties to penetrate certain barriers in the body) or too low lipophilicity (the compound cannot bind to the target molecule at all). Small negative values for LogP are typical of drugs that can be injected, values between 1.3 and 1.8 are ideal oral candidates, values around 2 are good candidates for targeting the central nervous system, for example.

There are three different scenarios where the above-mentioned metrics have been used: 

\begin{enumerate}
    \item \textbf{VAE hyperparameter optimization:} The aim is to ensure a good ability to encode/decode the SMILES information into/from the latent dimension. We thus aim to maximize the internal diversity. Additionally, to suppress over-training and improve generalization, we aim to minimize $\mathcal{W}(\text{LogP})$, $\mathcal{W}(\text{Weight})$, $\mathcal{W}(\text{SA})$, and $\mathcal{W}(\text{QED})$.
    \item \textbf{GAN hyperparameter optimization:} The aim is to improve the stability of the training. For the scope of this analysis, only the ratio of the training frequency between the generator and discriminator within a given epoch has been changed.
    \item  \textbf{Classical and quantum GAN performance comparison:} The aim is to compare performance independently of the training or test samples as well as to maximize the chance for generating truly novel molecules. Therefore, all fractions ($\epsilon_d$, $\epsilon_v$, $\epsilon_u$, $\epsilon_{LogP}$, novelty, filters, and internal diversity) have been maximized as well as the mean of the QED distribution, while the means of the SA and molecular weight distributions have been minimized.
\end{enumerate}

In order to test training robustness (i.e. the sensitivity of the model to the initialization of the training parameters), each model has been trained five times with a different random seed each time, so that the quoted values are the mean and standard deviation over these five runs. 

In order to assess globally the performance of the quantum GAN models against the classical GAN, we have also used as a final metric the $Z_0$ significance defined as:
\begin{equation}
    Z_0 = \sum_m Z_0^m = \sum_m \frac{\Delta^m}{\sqrt{(\sigma_1^m)^2 + (\sigma_2^m)^2}},
    \label{eq:z0def}
\end{equation}
where the sum runs over all the relevant metrics for scenario comparison as listed in the list above (fractions $\epsilon_d$, $\epsilon_v$, $\epsilon_u$, $\epsilon_{LogP}$; novelty; filters; internal diversity; $\langle\text{QED}\rangle$; $\langle\text{SA}\rangle$; $\langle\text{Weight}\rangle$) and $\Delta$ is the difference between the means of the respective metric $m$. The quantities $\sigma_1^m$ and $\sigma_2^m$ in Equation~\ref{eq:z0def} refer to the standard deviations of the metric $m$ for the reference scenario and for the scenario under study, respectively. The construction $\Delta$ depends on whether the metric $m$ is aimed to be maximized or minimized. In the former case, such as for efficiencies, $\Delta^m = \mu_2^m-\mu_1^m$ where $\mu_1^m$ and $\mu_2^m$ refer to the means of the metric $m$ for the reference scenario and for the scenario under study, respectively. In the latter case, such as for the SA or Weight, $\mu_1^m$ and $\mu_2^m$ are swapped, $\Delta^m = \mu_1^m - \mu_2^m$. For global significance, we will take the average of $Z_0$, hence dividing by the number of metrics used in the comparison.
The ground truth in this calculation is the classical GAN result taken as the benchmark reference point. The sign of the significance then indicates improvement or degradation of the quantum GAN designs over the classical GAN. To guide the reader, we remind that $|\langle Z_0 \rangle|<1$ indicates globally comparable performance between classical and quantum pipelines (the variance is less than $1\sigma$), while values above indicate sizable improvement or sizable degradation. More details with an example are presented in Section~\ref{subsec:latentGAN}.

We present more details on the metrics applied to the input distributions in Supporting Information Section~\ref{sec:app:input}. For more details on the sensitivity of the metrics to the number of generated samples, see Supporting Information Section~\ref{sec:app:ngen}.

% \Blindtext 

\subsection{Training details}
\label{sec:setup:preproc}

\subsubsection{Training of the VAE}

The VAE has been trained following the procedure in the MOSES benchmark suite~\autocite{moses}. This means in particular that we have used a batch size of 64 and a gradient clipping threshold of 50. We have used Adam optimizer~\autocite{KingmaB14} with a nominal learning rate equal to $3\times 10^{-4}$. The study of the loss function for training and validation datasets has indicated that $N_{ep}=1,000$ epochs produce an optimal result. 

% Requires: \usepackage{booktabs}
\begin{table}[h]
    \centering
    \setlength{\tabcolsep}{8pt} % default is 6pt
    \begin{tabular}{lcc}
        \toprule
        \textbf{Metrics}  & \textbf{Nominal VAE} & \textbf{Tuned VAE} \\
        \midrule
        Learning rate (lr)                       & $3\times 10^{-4}$ & Scheduled $10^{-3} \rightarrow 10^{-5}$ \\
        \midrule
        Encoder $n_{l}$                    & 1 & 1 \\
        Encoder $D_h$                      & 256 & 512 \\
        \midrule
        Decoder $n_{l}$                    & 3 & 4 \\
        Decoder $D_h$                      & 512 & 512 \\
        Decoder dropout                     & 0 & 0.75 \\
        \bottomrule
    \end{tabular}
    \caption{VAE hyperparameter settings comparison between nominal (MOSES default) and tuned settings.\label{tab:vae_train}}
\end{table}

\begin{figure}[h!]
    \centering
    \includegraphics[width=0.32\linewidth]{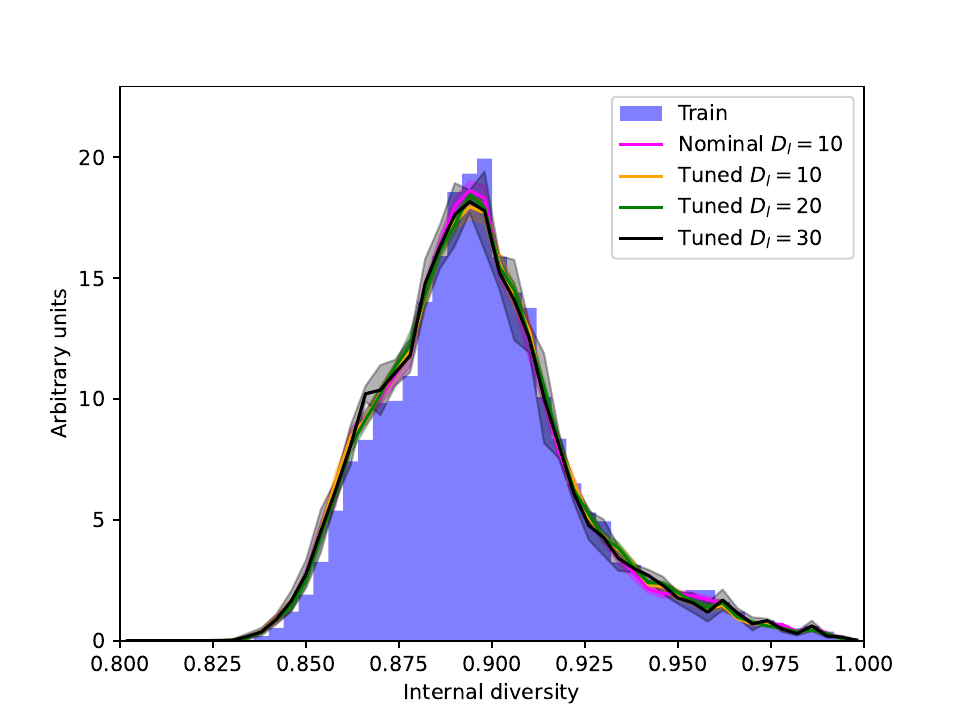}
    \includegraphics[width=0.32\linewidth]{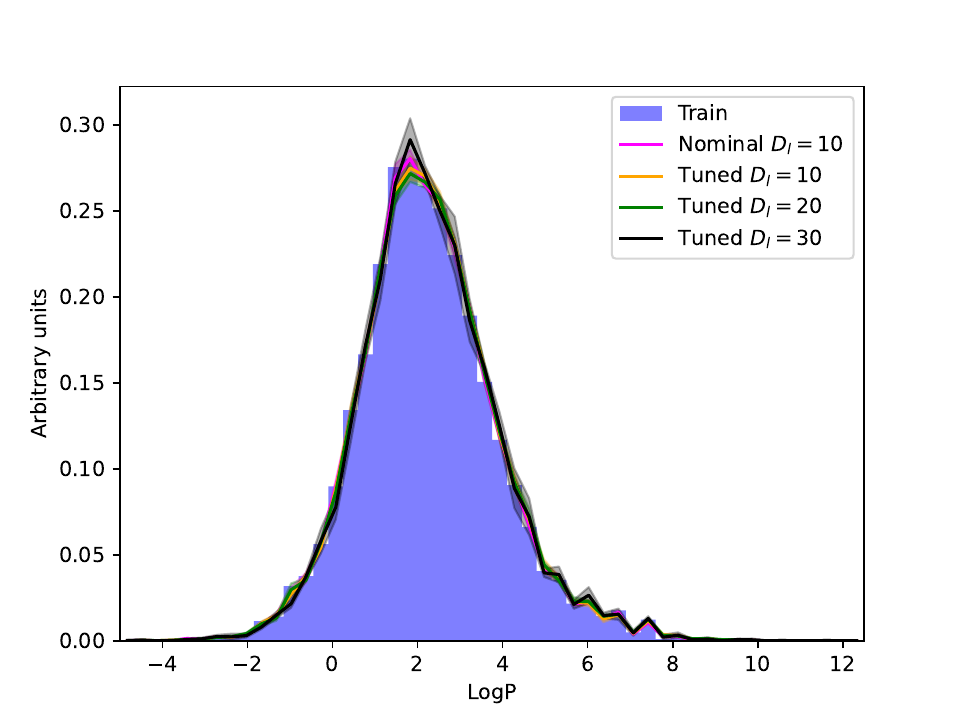}
    \includegraphics[width=0.32\linewidth]{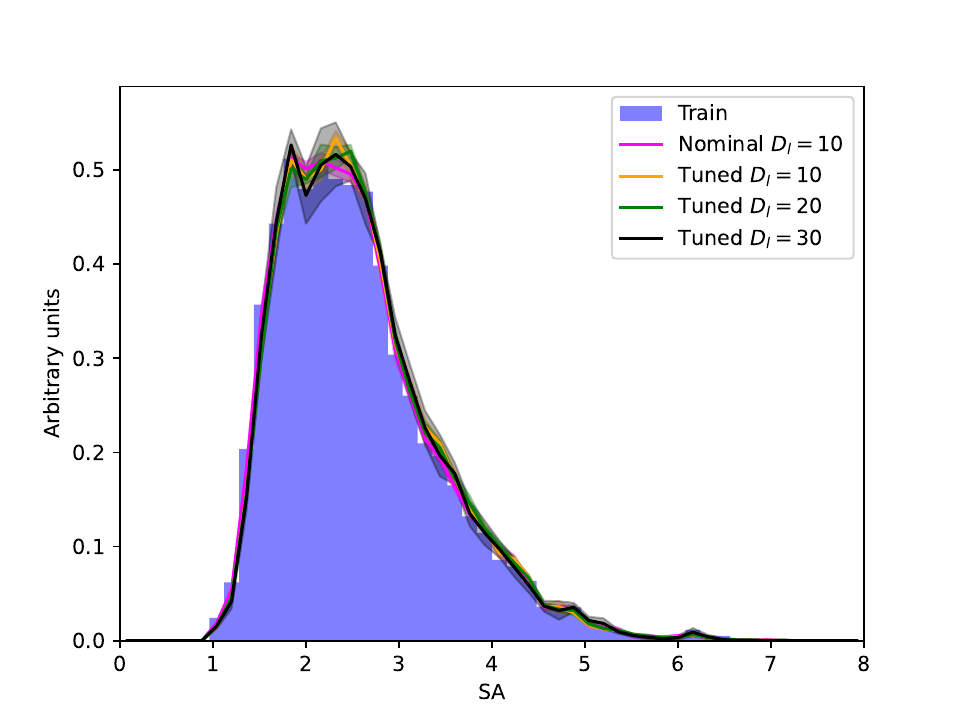}
    \includegraphics[width=0.32\linewidth]{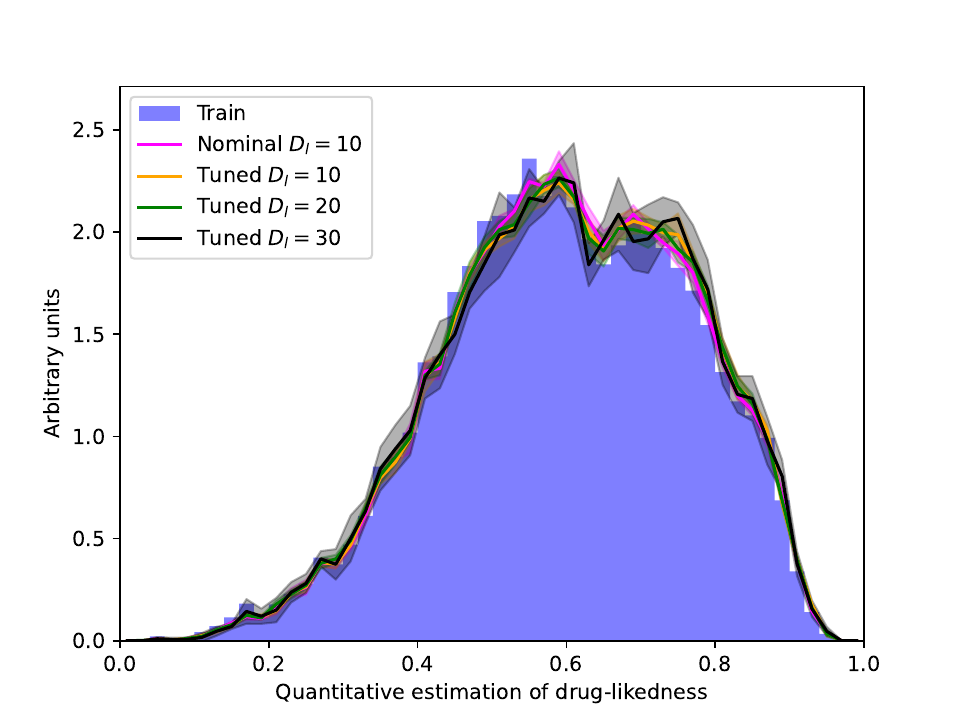}
    \includegraphics[width=0.32\linewidth]{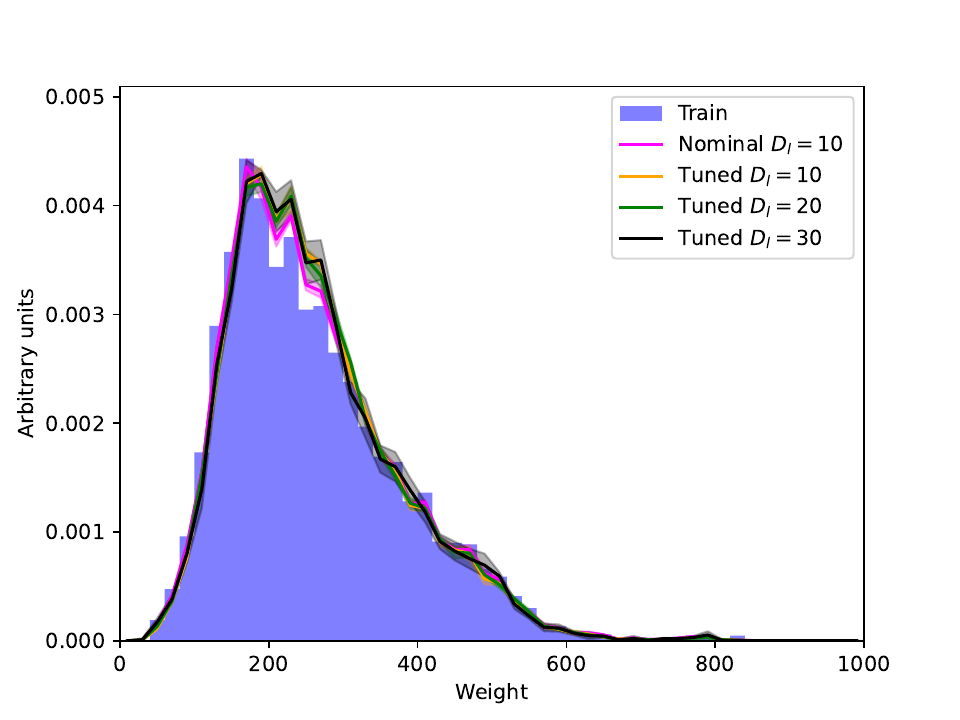}
    \caption{Distributions of metrics derived from the input training set and the generated sets using the nominal and tuned VAE decoder.}
    \label{fig:vae_comp}
\end{figure}

We have performed a small hyperparameter tuning of the VAE by varying the parameters listed in Table~\ref{tab:vae_train}. The learning rate is fixed at $3\times 10^{-4}$ in the nominal setup. We also explore a learning-rate scheduling for which the tuned parameter is a cosine schedule starting from $10^{-3}$ and ending at $10^{-5}$ for the final training epochs. We provide the tuned values of these parameters, leading to the metrics comparison presented in Table~\ref{tab:vae_comp}. The comparison of the various metrics distributions between the nominal and the tuned VAE is presented in Figure~\ref{fig:vae_comp}. We provide more details on the VAE training results in the Supporting Information Section~\ref{sec:app:vae_nep}.

% Requires: \usepackage{booktabs}
\begin{table}[h!]
    \centering
    \setlength{\tabcolsep}{8pt} % default is 6pt
    \begin{tabular}{lccc}
        \toprule
        \textbf{Metrics} & \textbf{Training set } & \textbf{Nominal VAE} & \textbf{Tuned VAE} \\
        \midrule
        IntDiv                          & 0.898 & 0.896 & 0.896 \\
        $\mathcal{W}(\text{LogP})$                   & 0.091 & 0.087 & 0.079 \\
        $\mathcal{W}(\text{SA})$                     & 0.268 & 0.275 & 0.251 \\
        $\mathcal{W}(\text{QED})$                    & 0.027 & 0.025 & 0.022 \\
        $\mathcal{W}(\text{Weight})$                 & 42.8 & 43.9 & 41.8 \\
        $N_{params}$                  & -- & 4,271,250 & 6,472,082 \\
        \bottomrule
    \end{tabular}
    \caption{Metrics derived from the training set and sets generated by the nominal and tuned VAE. Internal diversity is computed as the mean over the respective samples, whereas the remaining four metrics are the average Wasserstein distances between the sample in question and the test sample for the corresponding metric.\label{tab:vae_comp}}
\end{table}

Overall, the performance of all VAE training runs is very good and consistent: The metrics comparison shows an improvement for the tuned VAE, however modest. We have also performed an additional test of the robustness of the VAE decoder to ensure that it can efficiently decode the latent space since this is how we obtain the final generated molecules. This means that we must make sure that the decoder does not collapse the latent space back to a small subset of the molecular space. A shape study of the decoder was performed and demonstrated that it is sensitive to the shape of the latent vector distribution, confirming that the VAE does not collapse to a contrived subspace of the molecular input space. For more details, see Supporting Information Section~\ref{sec:app:vae_dec}.

\subsubsection{Latent GAN} 
\label{subsec:latentGAN}

The latent \qgan\ is trained on a noiseless quantum simulator (state-vector simulation), and the final generated molecules are filtered with RDKIT to check that they are chemically valid. We will also perform an experimental sampling on quantum hardware, the IBM \ibm\ system based on the Heron quantum chip, as an illustrative, inference-only demonstration of the effect of real hardware on inference, see Section~\ref{sec:results:inference}.

\begin{figure}[h!]
    \centering
    \includegraphics[width=0.48\linewidth]{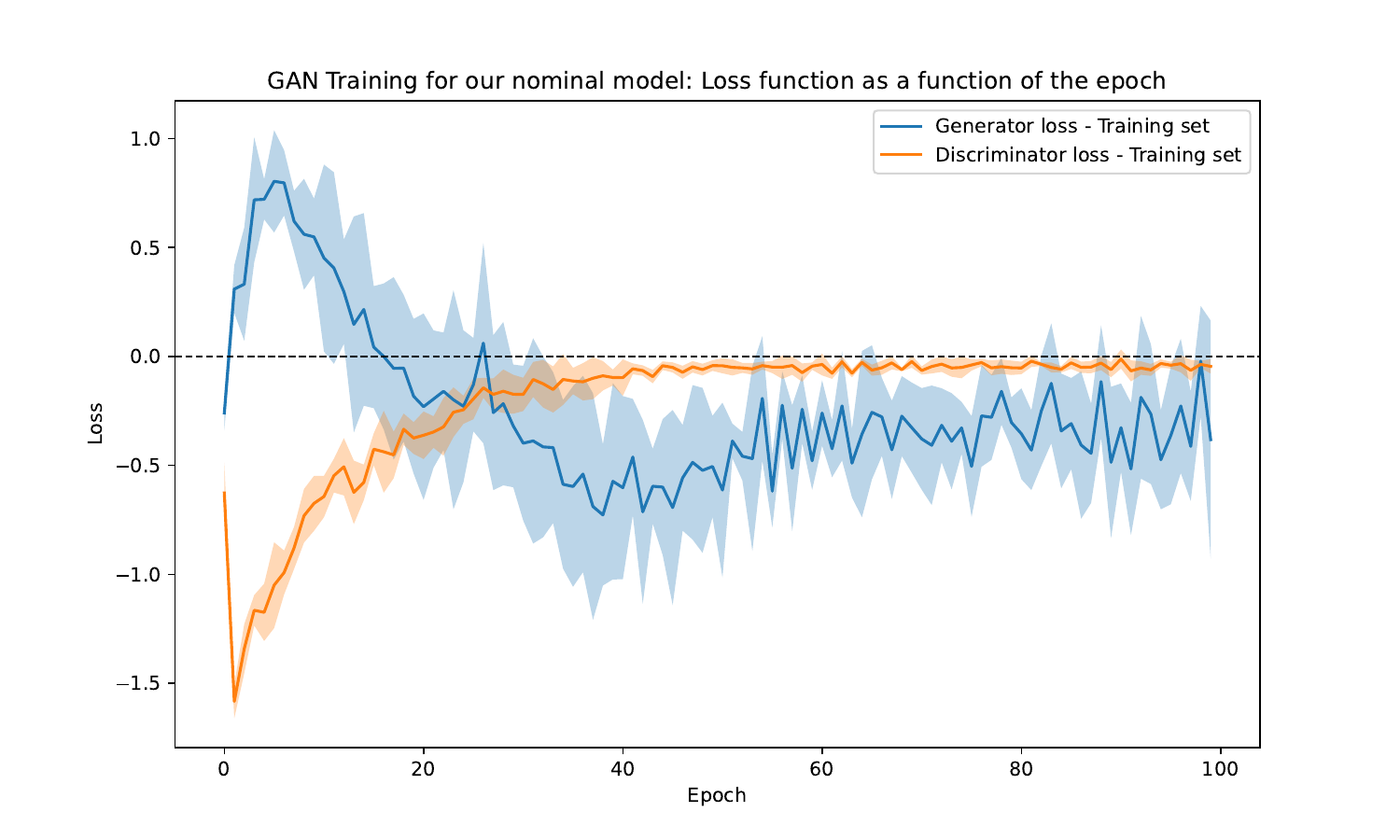}
    \includegraphics[width=0.48\linewidth]{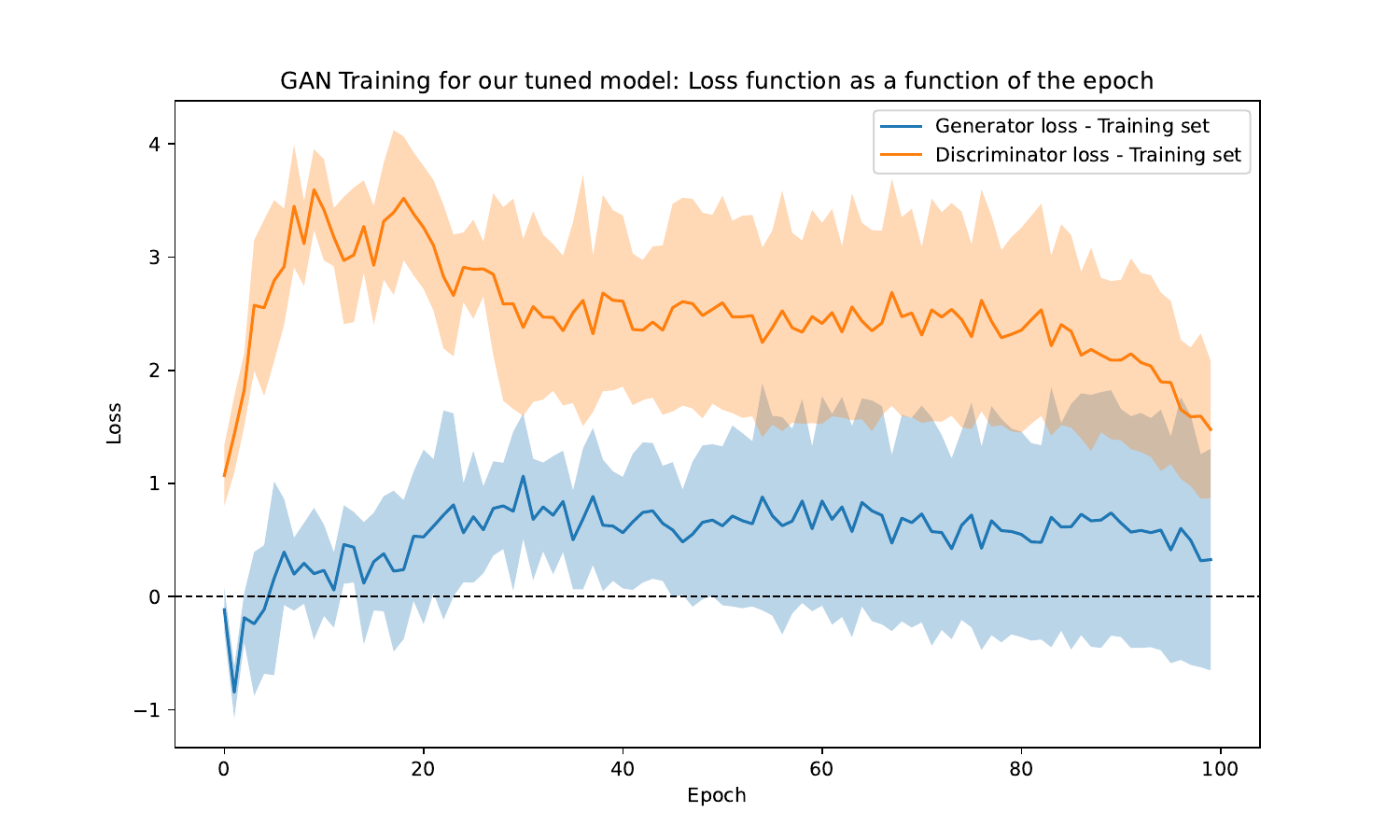}
    \caption{Comparison of the loss functions for the nominal (left panel) and tuned (right panel) GANs. Each model was run with five different seeds, the shaded band denotes a $\pm 1\sigma$ deviation.}
    \label{fig:gan_loss}
\end{figure}

The training of both classical and quantum latent GANs follows from the loss function described in Equations~\ref{eq:lganloss} and \ref{eq:lganloss:wasserstein} using again the Adam optimizer. For the classical GAN, the nominal learning rate is $2 \times 10^{-4}$ with Adam parameters $\beta_1 = 0.5$ and $\beta_2 = 0.9$. The number of epochs has been set to $N_{ep}=100$, and the ratio of the discriminator to generator training frequency was changed from five (default in MOSES benchmark suite) to one (further denoted as ``tuned GAN'').
The loss functions for the nominal and tuned GANs over a number of epochs are presented in Figure~\ref{fig:gan_loss}.

% Requires: \usepackage{booktabs}
\begin{table}[h!]
    \centering
    \setlength{\tabcolsep}{7pt} % default is 6pt
    \begin{tabular}{lcccc}
        \toprule
        \textbf{Metrics} & \textbf{Training set} & \textbf{Tuned VAE} & \textbf{Nominal GAN} & \textbf{Tuned GAN} \\
        \midrule
$\epsilon_d$ & 0.982 & 0.605 $\pm$ 0.017 & 0.510 $\pm$ 0.007 & 0.481 $\pm$ 0.030 \\
$\epsilon_v$ & 1.000 & 0.868 $\pm$ 0.006 & 0.911 $\pm$ 0.006 & 0.917 $\pm$ 0.012 \\
$\epsilon_u$ & 1.000 & 0.993 $\pm$ 0.001 & 0.986 $\pm$ 0.002 & 0.986 $\pm$ 0.003 \\
Novelty & 0.000 & 0.461 $\pm$ 0.026 & 0.501 $\pm$ 0.015 & 0.536 $\pm$ 0.015 \\
Filters & 0.734 & 0.728 $\pm$ 0.002 & 0.723 $\pm$ 0.003 & 0.721 $\pm$ 0.011 \\
IntDiv & 0.898 & 0.896 $\pm$ 0.000 & 0.902 $\pm$ 0.002 & 0.898 $\pm$ 0.004 \\
        \midrule
$\mathcal{W}(\text{LogP})$  & 0.091 & 0.078 $\pm$ 0.004 & 0.291 $\pm$ 0.026 & 0.354 $\pm$ 0.037 \\
$\mathcal{W}(\text{SA})$ & 0.268 & 0.249 $\pm$ 0.004 & 0.375 $\pm$ 0.028 & 0.437 $\pm$ 0.080  \\
$\mathcal{W}(\text{QED})$ & 0.027 & 0.023 $\pm$ 0.001 & 0.039 $\pm$ 0.003 & 0.038 $\pm$ 0.005 \\
$\mathcal{W}(\text{Weight})$  & 42.8 & 42.4 $\pm$ 1.0 & 93.0 $\pm$ 4.6 & 99.4 $\pm$ 10.6 \\
        \midrule
$\epsilon_{LogP}$ & 0.883 & 0.886 $\pm$ 0.003 & 0.895 $\pm$ 0.002 & 0.898 $\pm$ 0.006  \\
$\langle\text{SA}\rangle$ & 2.561 & 2.579 $\pm$ 0.004 & 2.454 $\pm$ 0.028 & 2.391 $\pm$ 0.080  \\
$\langle\text{QED}\rangle$ & 0.596 & 0.604 $\pm$ 0.001 & 0.592 $\pm$ 0.006 & 0.599 $\pm$ 0.007 \\
$\langle\text{Weight}\rangle$ & 261.5 & 260.5 $\pm$ 1.1 & 210.3 $\pm$ 4.6 & 203.9 $\pm$ 10.6 \\
        \midrule
$\langle Z_0\rangle$ & -- & Reference & +1.45 & +0.86 \\
        \midrule
\bottomrule
    \end{tabular}
        \caption{Metrics results for the VAE and classical GANs. All GANs use the tuned VAE for the latent space representation. Unlike the other cases, the nominal GAN uses the nominal GAN hyperparameter settings as discussed in the text. The quoted standard deviation comes from running each scenario five times with different seeds for the initialization of the trainable parameters.  \label{tab:vae_gan_metrics}}
\end{table}

We present in Table~\ref{tab:vae_gan_metrics} the metrics estimated from the samples generated by the respective models. The table is split into four parts:

\begin{enumerate}
    \item The first six metrics are interpretable as efficiencies -- distributions between 0 and 1 estimated from the distributions themselves, where the closer to unity, the better.
    \item The second part, denoted with $\mathcal{W}(X)$, presents the Wasserstein distances between the test sample and the generated sample for a variable $X$ defined by the MOSES analysis.
    \item The third part presents the fraction of the samples with $|LogP|<5$ and mean values of the distributions of the synthetic accessibility score (SA), quantitative estimation of drug-likeness (QED), and molecular weight. 
    \item The last part then presents a performance comparison of scenarios by summing the individual statistical compatibilities of metrics from parts one and three.
\end{enumerate}

As discussed in Section~\ref{sec:setup:metric}, in a situation where the task is to evaluate whether a model is producing novel and unknown molecules with desired properties, maximizing the Wasserstein distance from the test sample should be a good objective. Such an approach poses a subtle yet important problem: if the model produces molecules with metric distributions distant from the test set but similar to the training set, it may incorrectly prefer reproducing the training set. 

\begin{figure}
    \centering
    \includegraphics[width=0.32\linewidth]{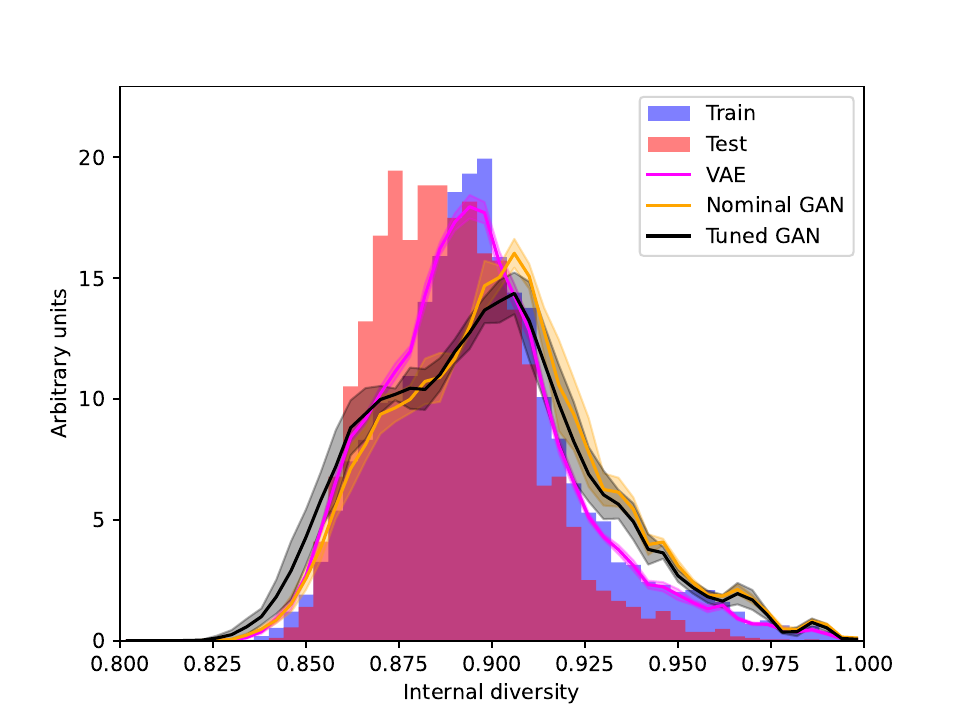}
    \includegraphics[width=0.32\linewidth]{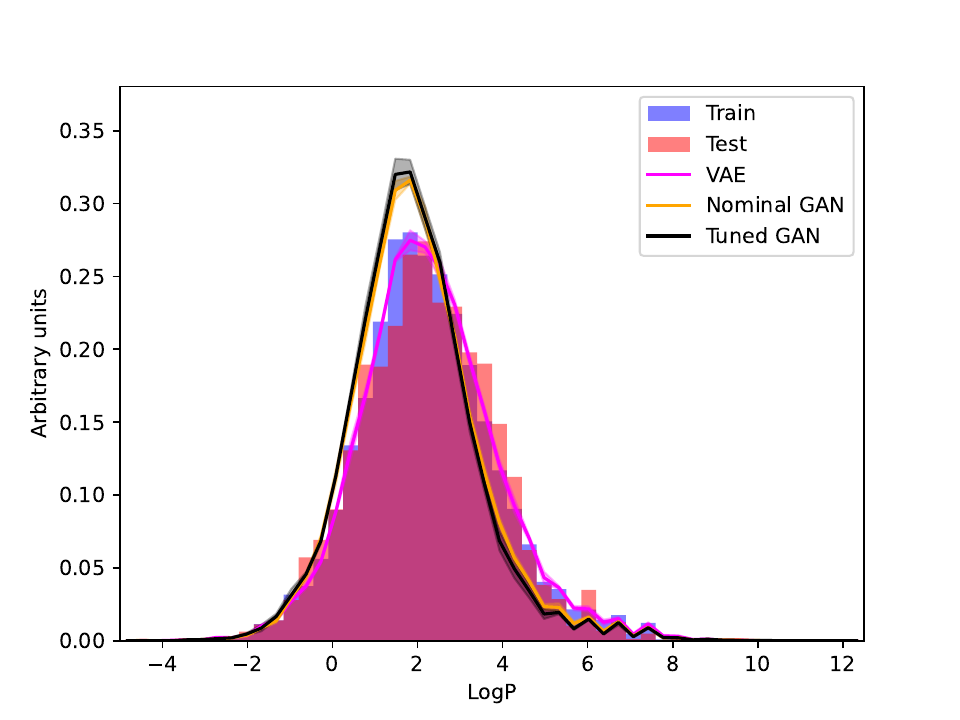}
    \includegraphics[width=0.32\linewidth]{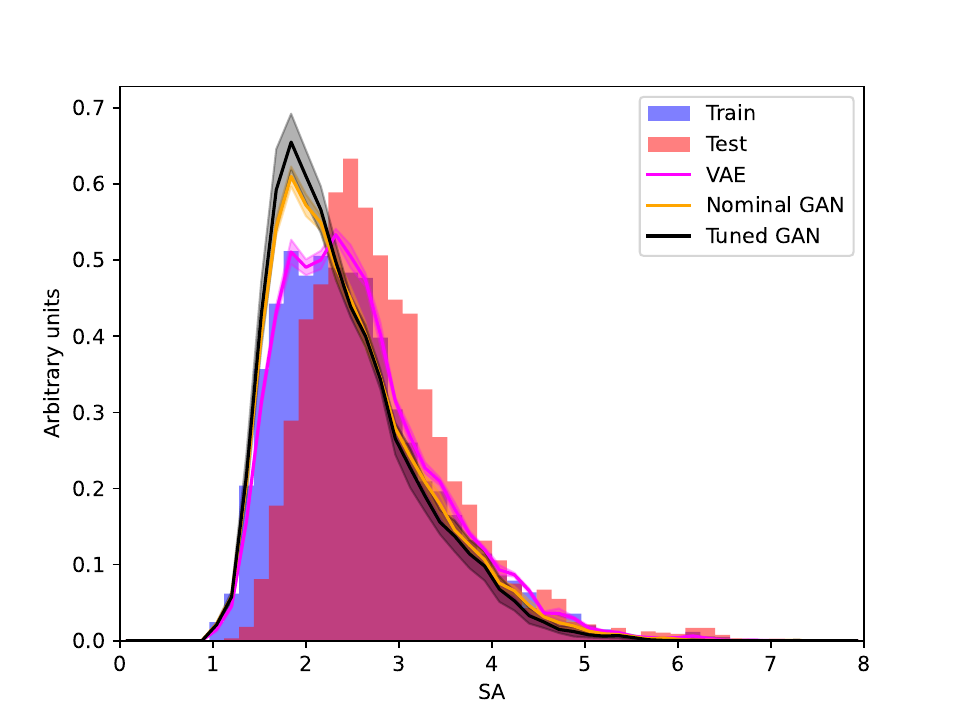}
    \includegraphics[width=0.32\linewidth]{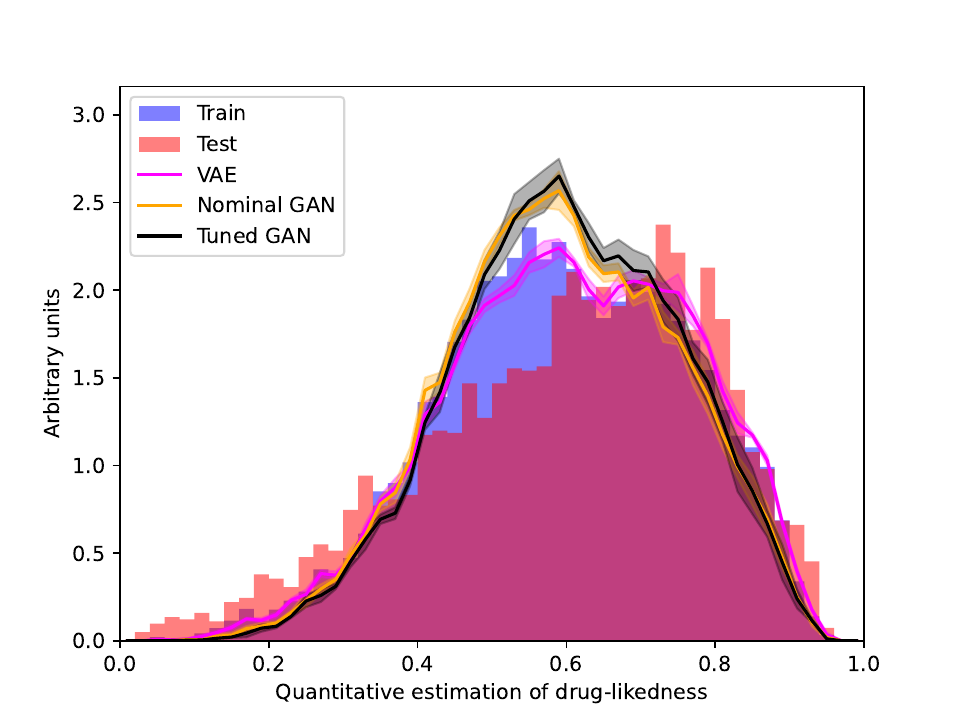}
    \includegraphics[width=0.32\linewidth]{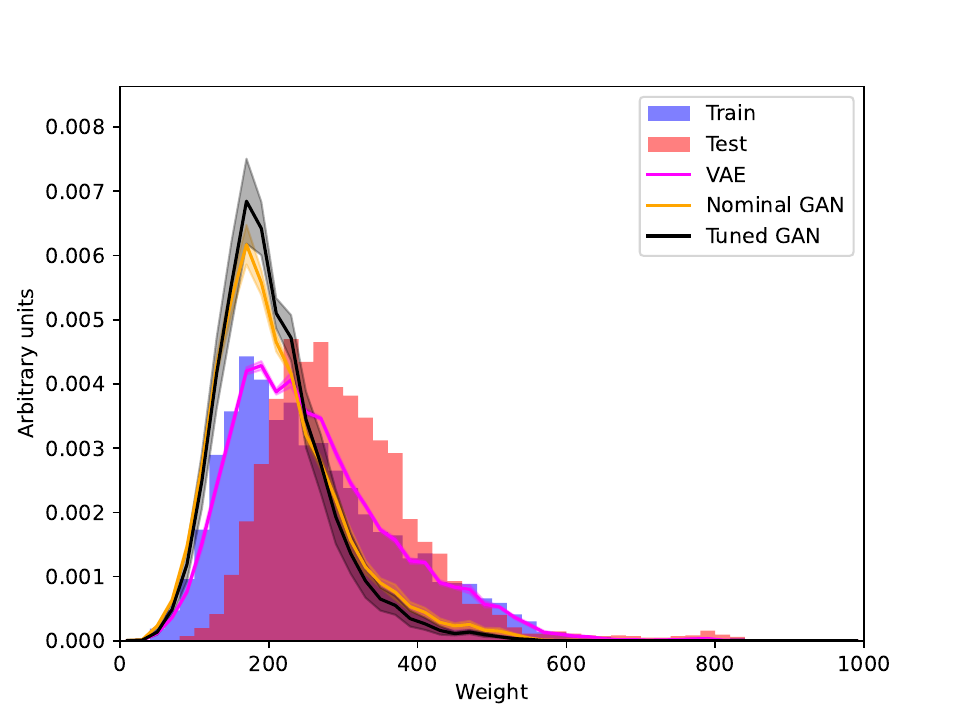}
    \caption{Comparison of the metrics for the models used in the analysis. For the VAE, the tuned version is used. For the GAN, not only the tuned VAE is used, but the GAN is also tuned. Each model was run with five different seeds, the shaded band is defined as $\pm 1\sigma$ deviation.}
    \label{fig:vae_gan_metrics}
\end{figure}

This is exemplified in Figure~\ref{fig:vae_gan_metrics}, comparing training and test input samples with samples generated by the VAE decoder, the nominal GAN, and the tuned GAN models. In all cases, the optimized VAE hyperparameter settings were used. Across all distributions, the GAN exhibits greater capacity to generate samples farther from the training and test datasets, confirming the observation made in \autocite{moses}. However, it is also visible that all distributions tend to be close to the training set. If there were a model positioned in the opposite direction from the test sample relative to the training sample, but closer to the test sample, maximizing the Wasserstein distance would still prefer the model that reproduces the training dataset. The maximization procedure would lead to an incorrect conclusion. For that reason, it has been decided to use only the seven fractions ($\epsilon_d$, $\epsilon_v$, $\epsilon_u$, Novelty, Internal diversity, Filters, and $\epsilon_{LogP}$) and the means of SA, QED, and Weight distributions for the construction of our single scalar average $\langle Z_0\rangle$ significance as a metric for scenario comparison.

Finally, the $\langle Z_0\rangle $ significance difference in Table~\ref{tab:vae_gan_metrics} shows that both GANs are performing better than the VAE, which is in agreement with the intuition we get from the plots.

%%%%%%%%%%%%%%%%%%%%%%%%%%%%%%%%%%%%%%%%%%%%
%
\section{Results}
\label{sec:results}

% \Blindtext 

In this section, we present a comparison of the performance of the classical and quantum GANs. In addition to the metrics discussed in Section~\ref{sec:setup:metric}, we also compare the parameter counts and training stability of the quantum model with those of a classical counterpart.

As explained in Section~\ref{sec:setup:lat_dim}, we present quantum training with a maximum latent dimension of 30, corresponding to quantum circuits with 15 qubits and dual readout. We start with our default scenario with a latent dimension of 10 in Section~\ref{sec:results:nom}, we present a comparison between the different quantum scenarios with up to a latent dimension of 30 in Section~\ref{sec:results:scen}, and we present the results of our quantum inference on the IBM quantum computer \ibm\ in Section~\ref{sec:results:inference}. 

\subsection{Classical vs quantum GANs}
\label{sec:results:nom}
\begin{figure}[h!]
    \centering
    \includegraphics[width=0.48\linewidth]{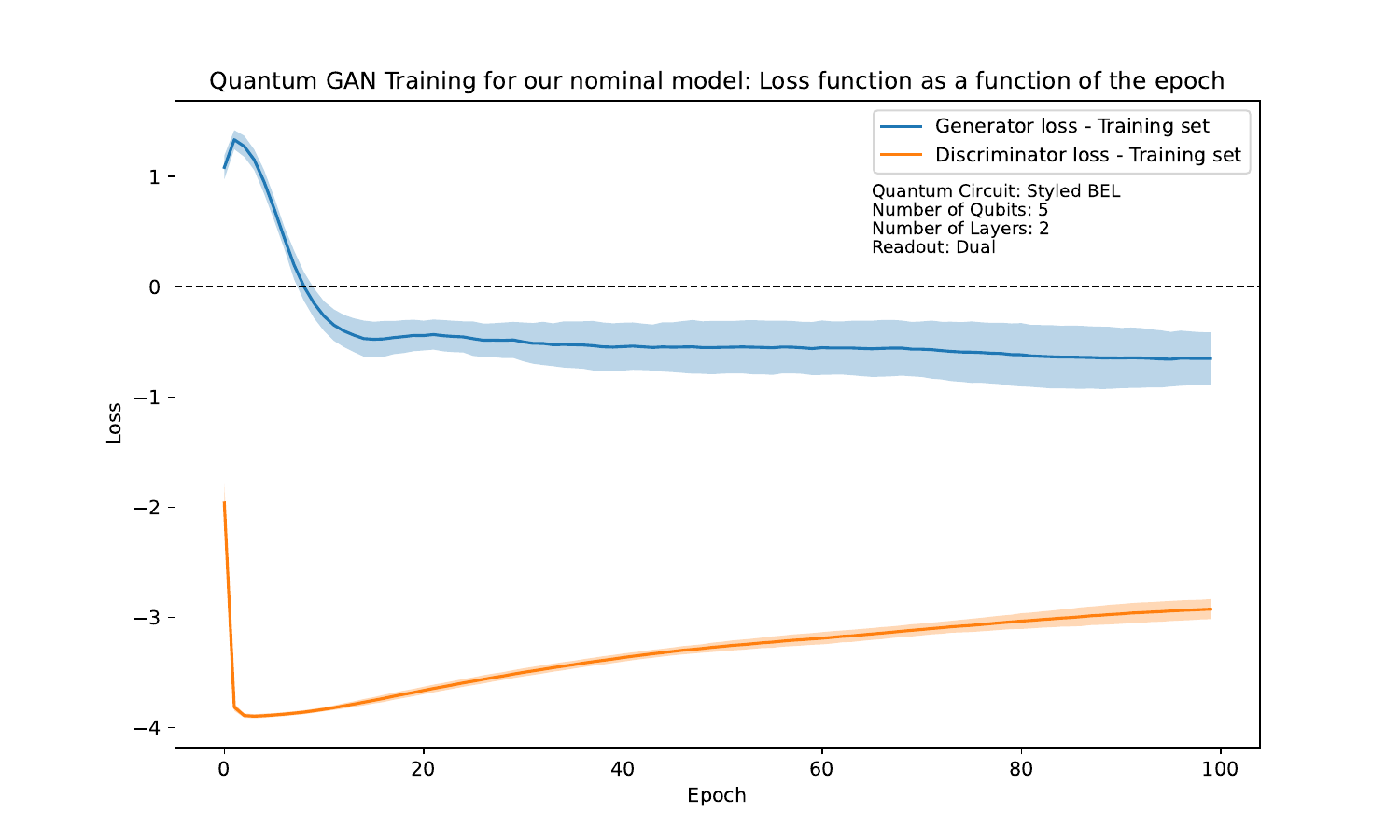}
    \includegraphics[width=0.48\linewidth]{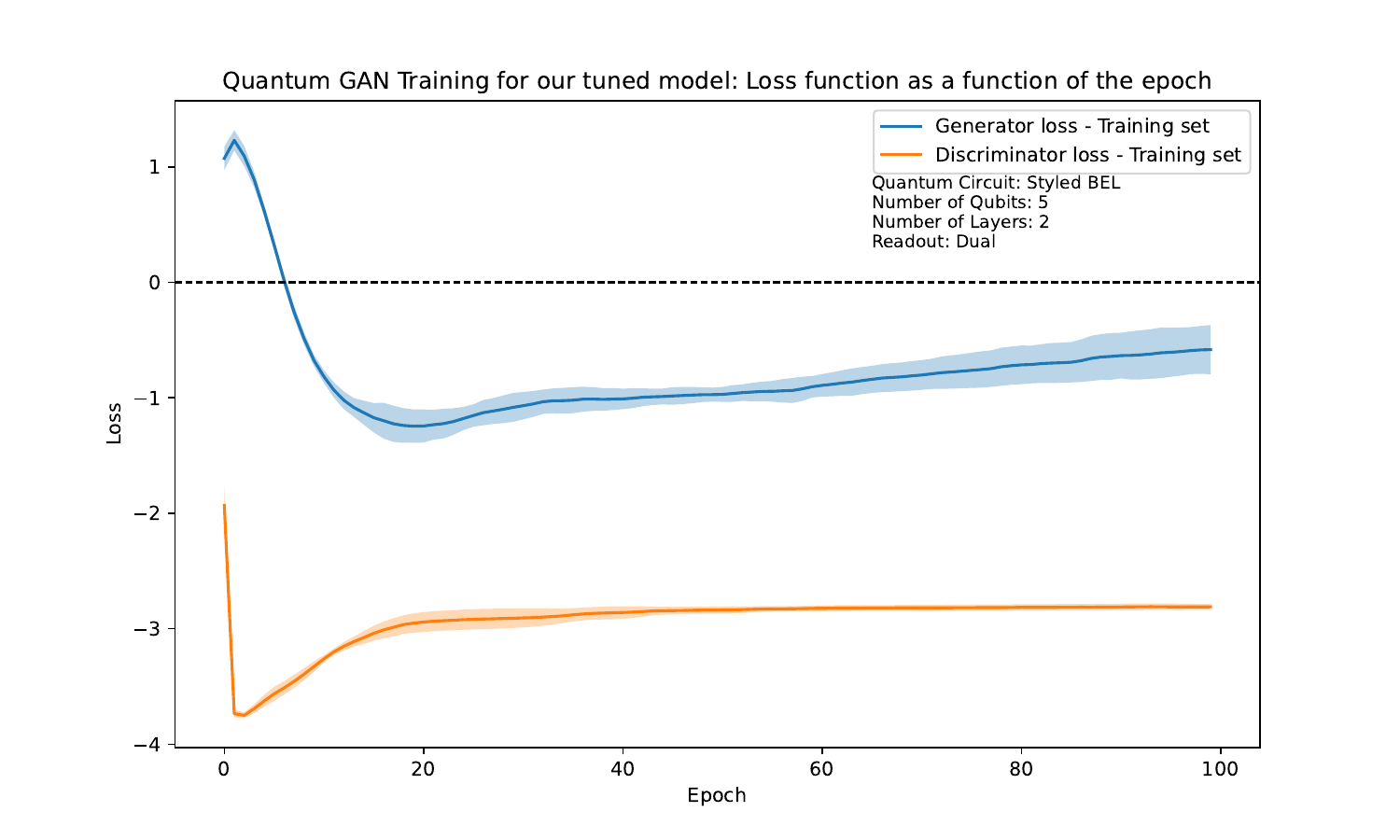}
    \caption{Comparison of the loss functions for the nominal (left panel) and tuned (right panel) quantum GANs with the BEL circuit. Each model was run with five different seeds, the shaded band is defined as $\pm 1\sigma$ deviation.}
    \label{fig:qgan_loss}
\end{figure}

% \Blindtext 
The default scenario we have investigated uses the latent dimension 10, tuned VAE, and classical GAN hyperparameters. The two quantum circuits described in Section~\ref{sec:setup:sqGAN} are compared in the nominal setup of five qubits, two layers, and dual readout. The quantum GANs are trained with 100 epochs. Figure~\ref{fig:qgan_loss} presents the loss function on the training set for the BEL quantum generator for the nominal (left panel) and tuned (right panel) GAN training parameters. Compared to Figure~\ref{fig:gan_loss}, it is clear that the loss function has a much smoother trend and fewer variations. The \qgan\ in our pipeline leads to a more robust pipeline compared to the MOSES baseline with a latent classical GAN. We caution that Figures~\ref{fig:gan_loss} and~\ref{fig:qgan_loss} compare generators of very different capacities (more than $7\times10^5$ trainable parameters for the MOSES classical generator versus 110 for the BEL \qgan). The smoother \qgan\ loss could be primarily attributed to the much smaller parameter count of the quantum architecture, or, conversely, to the high-capacity MOSES classical generator overfitting the data. Supporting Information Section~\ref{sec:app:cgan_lowparam} reports a supplementary control study testing this interpretation with three classical generators of 682, 346, and matched 110 parameters, respectively, the three of them trained using the same VAE, discriminator, and hyperparameters (except for the number of epochs), as in our nominal scenario. Their loss trajectories are not smoother than the reference classical generator. The smoother BEL \qgan\ loss is thus unlikely to follow from parameter count alone down to the exact matched capacity, and we report it as an improved robustness of our pipeline, without necessarily claiming a general quantum optimization advantage. We have also extended this controlled sweep to two other classical GANs with 115 and 94 trainable parameters for the generator. The loss functions behave much better than for the other classical GANs, but still not as smooth as the \qgan\ training curves.

Table~\ref{tab:cgan_qgan_nom} presents the metrics comparison for the nominal circuit setting for the \qgans\ and the classical GAN, including also the classical GAN with matched capacity to the BEL \qgan, obtained with a $[10,D_l]$ single-hidden-layer generator. All reported $\langle Z_0\rangle$ values are computed with respect to the MOSES baseline in the first column. It is worth mentioning that results of the quantum circuits are compatible with the classical GAN within one standard deviation. As expected, the simple circuit performs worse than the BEL circuit. The results of the nominal MOSES classical GAN were obtained with more than 6,400 times more parameters than the results of the BEL \qgan. We note that smaller classical GAN pipelines, presented both in this section for the exact matched capacity and in the Supporting Information Section~\ref{sec:app:cgan_lowparam}, do not generate substantially better results. They are even slightly degraded, except for the classical GAN with 94 trainable parameters for which there is a modest gain compared to the MOSES nominal classical latent GAN, while for some metrics, such as $\langle\text{QED}\rangle$, our \qgan\ pipeline is improving over all the classical baselines tested in this work. Indeed, relative to the MOSES standard reference, the matched $[10,D_l]$ classical generator ($\langle Z_0\rangle=-0.41$) is statistically compatible overall but does not improve $\langle\mathrm{QED}\rangle$ ($0.594\pm 0.001$), whereas the styled BEL \qgan\ ($\langle Z_0\rangle=-0.26$) reaches $0.643\pm 0.001$. We also note that, when comparing the nominal MOSES baseline with the classical latent GAN with matched capacity, there is an overparameterization of the nominal baseline, for a latent dimension of 10.

% Requires: \usepackage{booktabs}
\begin{table}[h!]
    \centering
    \begin{tabular}{lcccc}
        \toprule
        Metrics & Tuned GAN & Simple \qgan &  BEL \qgan & Matched classical GAN \\
        \midrule
        $N_{params}$ & 705,162 & 20 &  110 & 110 \\
        \midrule
        $\epsilon_d$ & $0.481 \pm 0.031$ & $0.529 \pm 0.003$ &  $0.523 \pm 0.003$ & $0.523 \pm 0.001$ \\
        $\epsilon_v$ & $0.917 \pm 0.012$ & $0.831 \pm 0.002$ &  $0.875 \pm 0.003$ & $0.910 \pm 0.001$ \\
        $\epsilon_u$ & $0.986 \pm 0.003$ & $0.994 \pm 0.000$ &  $0.993 \pm 0.001$ & $0.986 \pm 0.001$ \\
        Novelty & $0.536 \pm 0.015$ & $0.572 \pm 0.003$  & $0.548 \pm 0.003$ & $0.495 \pm 0.003$ \\
        IntDiv & $0.898 \pm 0.004$ & $0.882 \pm 0.000$ & $0.883 \pm 0.000$ & $0.902 \pm 0.000$ \\
        Filters & $0.721 \pm 0.011$ & $0.737 \pm 0.003$  & $0.719 \pm 0.002$ & $0.722 \pm 0.003$ \\
        $\epsilon_{LogP}$ & $0.898 \pm 0.007$ & $0.883 \pm 0.001$  & $0.896 \pm 0.002$ & $0.895 \pm 0.002$ \\
        $\langle \mathrm{SA} \rangle$ & $2.391 \pm 0.076$ & $2.575 \pm 0.007$  & $2.448 \pm 0.007$ & $2.472 \pm 0.005$ \\
        $\langle \mathrm{QED} \rangle$ & $0.599 \pm 0.007$ & $0.641 \pm 0.001$  & $0.643 \pm 0.001$ & $0.594 \pm 0.001$ \\
        $\langle \text{Weight} \rangle$ & $203.9 \pm 10.6$ & $294.8 \pm 0.8$ & $261.4 \pm 0.8$ & $215.2 \pm 1.0$ \\
        \midrule
        $\langle Z_0\rangle$ & Reference & $-1.00$ &  $-0.26$ & $-0.41$ \\
        \bottomrule
    \end{tabular}
    \caption{Comparison of the performance of the two types of quantum GANs, as well as a classical latent GAN matched to style-based BEL \qgan\ (single hidden layer structure $[10,D_l]$), with respect to the MOSES baseline classical GAN. All models use tuned VAE and GAN training parameters. The quantum circuits use five qubits, two layers, and dual readout to produce a latent vector of dimension 10.}
    \label{tab:cgan_qgan_nom}
  \end{table}

We summarize here the overall cost of the pipeline in terms of model complexity and runtime. The number of trainable parameters, reflecting directly the model capacity and the cost of the optimization, shows the sharpest contrast: the pretrained VAE uses about $6.5\times10^6$ parameters (Table~\ref{tab:vae_comp}) and the classical latent generator of the nominal MOSES baseline uses $705,162$, whereas our quantum generators use only between $20$ and $620$ trainable parameters, depending on the circuit, the number of layers, and the readout strategy (Table~\ref{tab:scen}). On the NVIDIA A100 40GB GPU used in this work, one-time VAE training over $1,000$ epochs takes about $52$~minutes (Supporting Information Section~\ref{sec:app:vae_nep}) and the generation of $30,000$ molecules takes about $10$~minutes (Supporting Information Section~\ref{sec:app:ngen}); both latent GANs are trained for $100$ epochs. We do not report a wall-clock training-time comparison between the classical and quantum GANs, because the quantum generator is trained on a classical state-vector emulator whose cost grows exponentially with the number of qubits. This restricts the simulation to at most $15$ qubits and motivates the dual- versus single-readout trade-off discussed in Section~\ref{sec:setup:sqGAN}. Such a runtime comparison would measure the cost of emulating a quantum computer on classical hardware rather than the intrinsic cost of the quantum pipeline, and is beyond the scope of this work. Executing the generator natively on quantum hardware, as done for the inference only in Section~\ref{sec:results:inference}, would in principle avoid this exponential emulation cost, although efficient full training on quantum hardware remains an open research challenge~\autocite{jcdk-q3xc}. Investigating such hardware-efficient training schemes is therefore a natural direction for scaling the pipeline to wider latent spaces in future work. Independently of runtime, the much smaller number of trainable parameters of the \qgan\ is already a concrete advantage of the model itself, improving explainability and reducing the optimization burden when compared to the nominal MOSES baseline. Compared to the matched classical GAN, of course, this argument does not apply.

\begin{figure}[h!]
    \centering
    \includegraphics[width=0.32\linewidth]{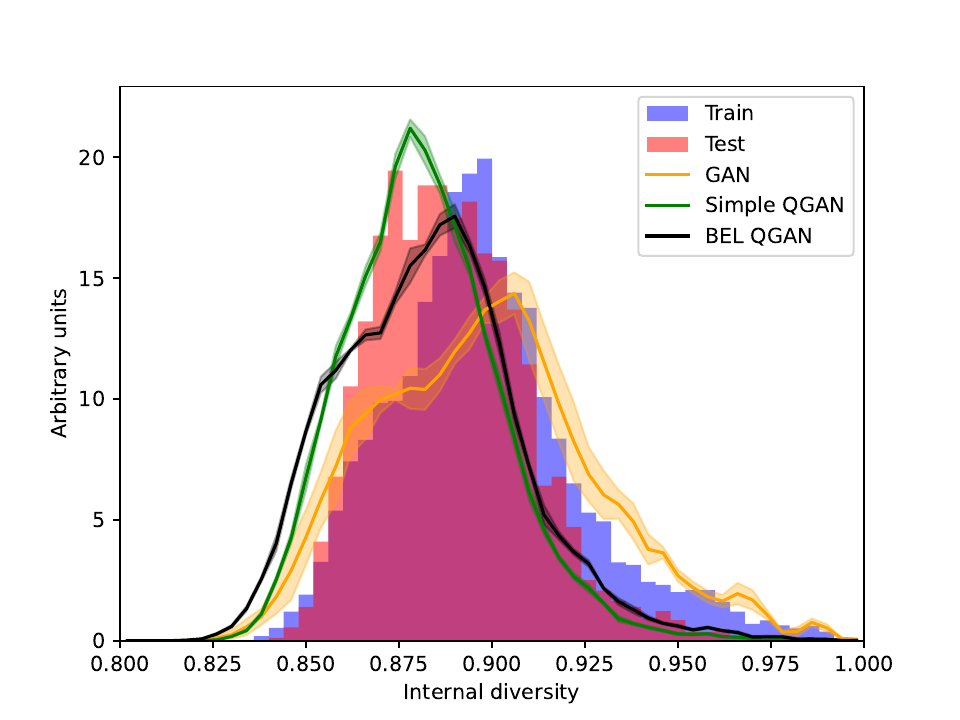}
    \includegraphics[width=0.32\linewidth]{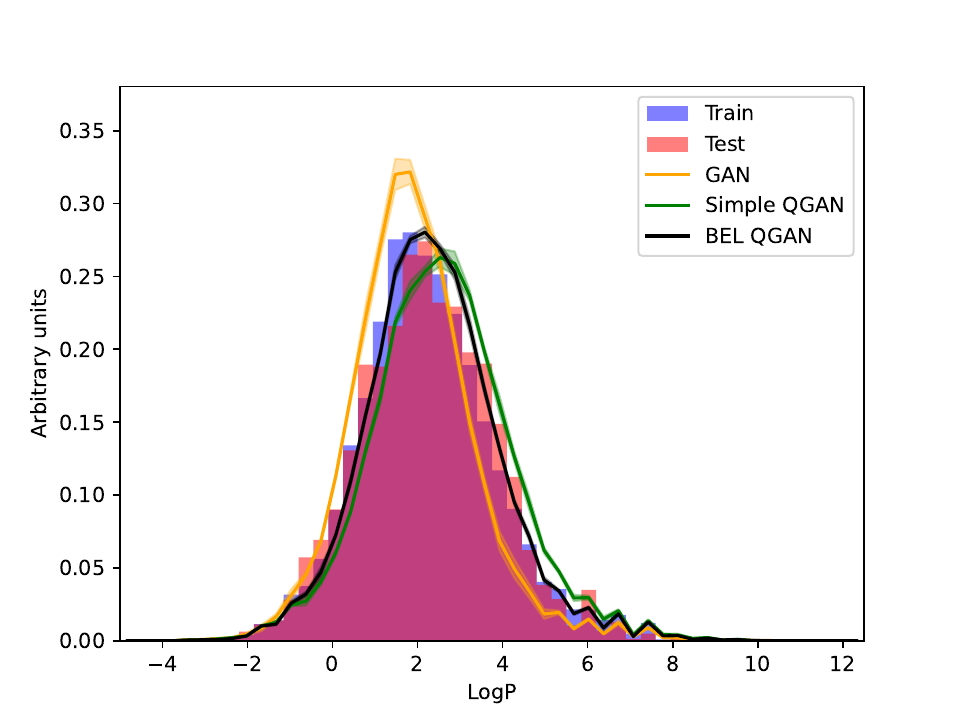}
    \includegraphics[width=0.32\linewidth]{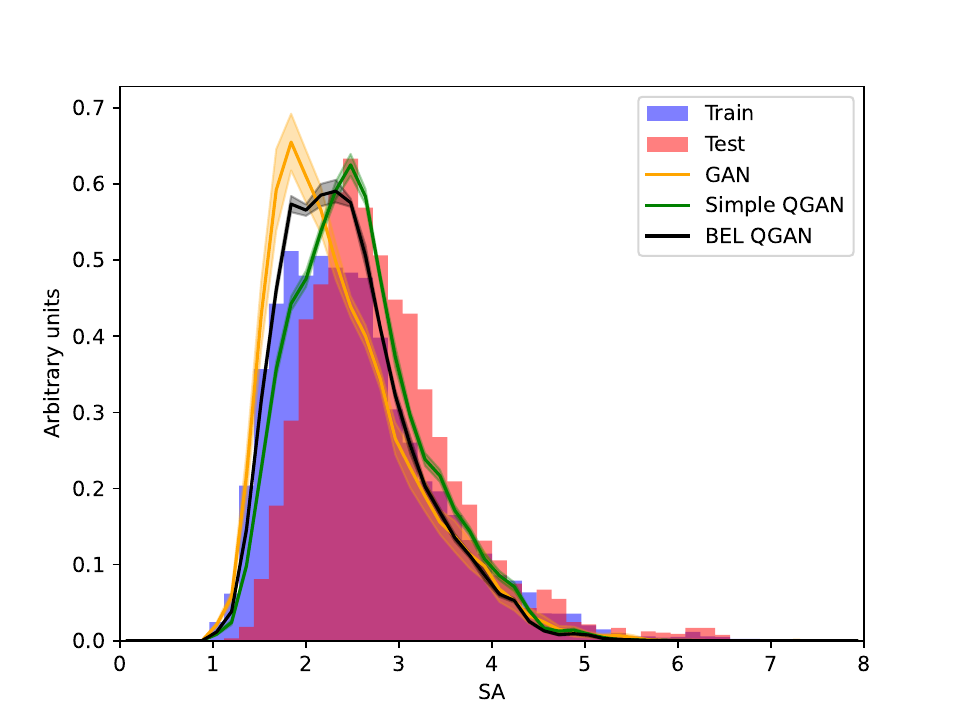}
    \includegraphics[width=0.32\linewidth]{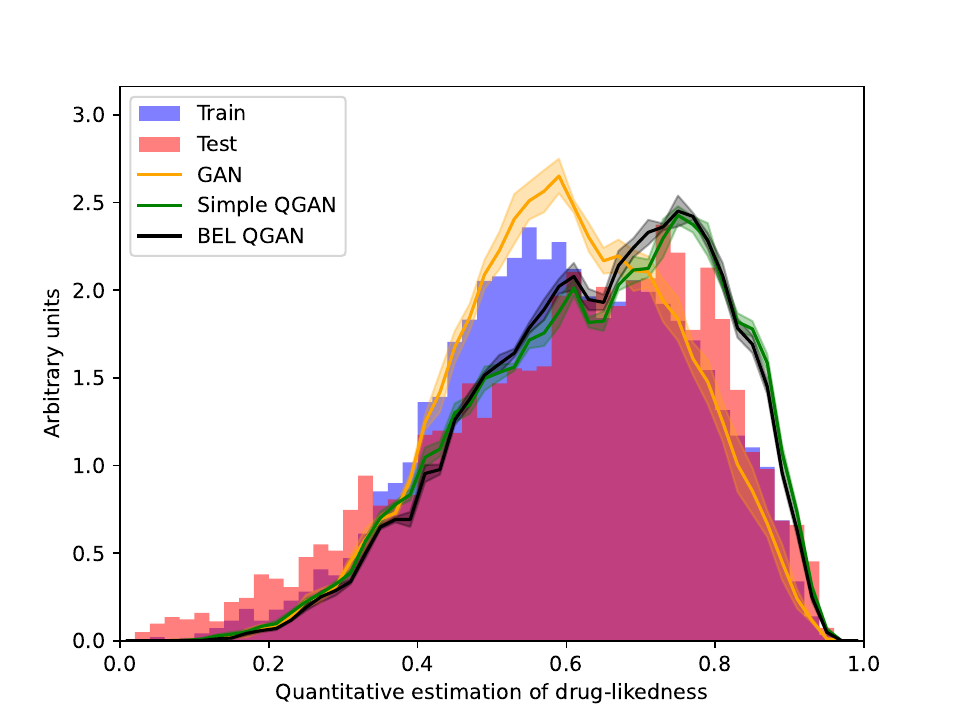}
    \includegraphics[width=0.32\linewidth]{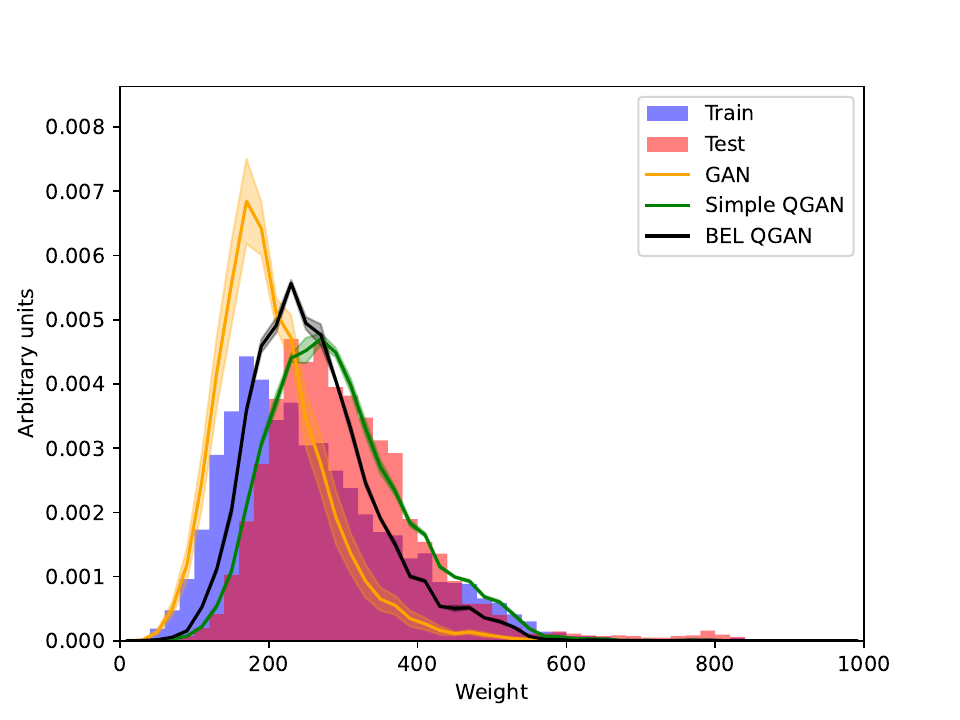}
    \caption{Comparison of the metrics for the two quantum circuits and the nominal MOSES classical GAN. All models use tuned VAE and GAN training parameters. The quantum circuits use five qubits, two layers, and dual readout to produce a latent vector of dimension 10. Each model was run with five different seeds, and the shaded band is defined as $\pm 1\sigma$ deviation.}
    \label{fig:cgan_qgan_nom}
\end{figure}

Figure~\ref{fig:cgan_qgan_nom} presents the comparison of the metric distributions for the nominal classical GAN and two \qgans. The classical and quantum GANs exhibit distinct shapes beyond their respective uncertainty bands. There is also a difference between the simple and BEL circuits. For the distributions of internal diversity and molecular weight, the classical GAN tends to have the distribution skewed toward desirable values -- unity for internal diversity and zero for the weight. However, \qgans\ show a higher likelihood of the molecule being a viable drug (QED).

\begin{figure}[h!]
    \centering
    \includegraphics[width=\linewidth]{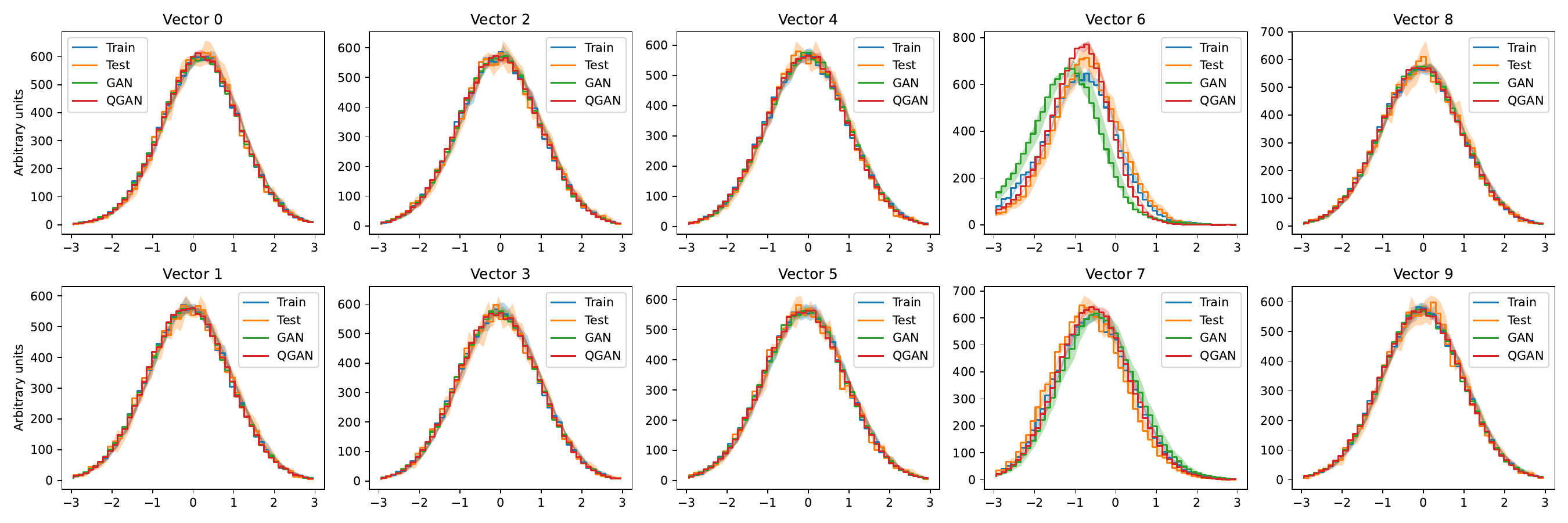}
    \caption{Distribution of latent vectors for the train, test, classical GAN, and BEL \qgan\ samples.}
    \label{fig:cgan_qgan_lat}
\end{figure}

Figure~\ref{fig:cgan_qgan_lat} presents the distributions of the 10 latent vectors for the training, test, classical, and quantum GANs in the VAE encoder coordinate system. The \qgan\ uses the BEL circuit with a nominal setting of five qubits, two layers, and dual readout. When constructing the latent vectors, the VAE encoder adds a random noise term, so even the training and test samples with fixed inputs, when run five times, produce slightly different distributions, and hence the uncertainty band even for those scenarios. All but vectors six and seven show no difference among any of the four samples. For more details on the correlations between the latent vectors, see Supporting Information Section~\ref{app:corr_mat}.
Note that we also present in the Supporting Information a latent vector comparison in the predecoder coordinate system for the quantum-generator readouts rescaled to $[0,1]$, comparing simulator, noisy emulator, and quantum hardware results, at inference time; see Supporting Information Figure~\ref{fig:app:hw:latent}.

\subsection{Scenario comparison}
\label{sec:results:scen}

As mentioned in previous sections, the highest achievable latent dimension is 30, because anything beyond 15 qubits is too computationally complex for the hardware used in this analysis. Taking these limitations into account, a comprehensive study of various scenarios has been performed and presented in Table~\ref{tab:scen}.

The scenario comparison is based on the aggregated $Z_0$ significance estimated for metrics as described in Section~\ref{sec:setup:metric}. As a baseline, the classical GAN, with optimized training hyperparameters and the VAE pipeline with a latent dimension of 10, has been selected. All scenarios are within one standard deviation, hence statistically compatible: any additional difference is therefore statistically negligible. We can make the following additional remarks:

\begin{itemize}
    \item Increasing the size of the latent dimension does not seem to improve the performance of the scenario, even for the classical GANs.
    \item For the simple circuit, the improvement due to single readout (higher number of qubits) is smaller than for the BEL circuit. This is likely connected to circuit complexity.
    \item The number of \qgan\ layers does not seem to have an impact on the result.
    \item For dual readout and a latent dimension of 20, the difference between the \qgan\ and the classical GAN (with the same latent dimension) seems to be much smaller than that for a latent dimension of 10.
    \item For dual readout and a latent dimension of 30, the difference between the \qgan\ and the classical GAN (with the same latent dimension) seems to be much smaller than that for a latent dimension of 10, even if there is a larger difference than for a latent dimension of 20.
    \item The BEL circuit seems to have the same performance for latent dimensions of 20 and 30 (i.e. 10 and 15 qubits).
    \item The only consistent improvement in \qgan\ performance for a given latent dimension is going from dual to single readout. We have restricted our analysis in the single readout case to a latent dimension of 10, as otherwise we would need significantly more computing power for latent dimensions of 20 or 30 and single readout, without a clear hint of improvement with the dataset we have investigated.
\end{itemize}

More details can be found in the Supporting Information Section~\ref{sec:app:scen}.

\begin{table}[h!]
    \centering
    \renewcommand{\arraystretch}{1.5}
    \begin{tabular}{c|c|cc|cc|cccc}
        \toprule
Latent Dimension & \multirow{3}{*}{GAN} & & & \multicolumn{2}{c}{Simple \qgan} & \multicolumn{3}{c}{BEL \qgan} \\
\multirow{2}{*}{($N_{params}$)} & &\multirow{2}{*}{Readout} &\multirow{2}{*}{$n_{qb}$} & $n_{l} = 2$& $n_{l} = 4$& $n_{l} = 2$& $n_{l} = 4$ & $n_{l} = 6$ \\
 & & & & ($n_{par}$) & ($n_{par}$) & ($n_{par}$) & ($n_{par}$) & ($n_{par}$) \\
        \midrule
        \multirow{2}{*}{10} & \multirow{4}{*}{Reference} & \multirow{2}{*}{Dual} & \multirow{2}{*}{5} & -1.00 & -1.01 & -0.26 & -0.27 & \\
        & & & & (20) & (40) & (110) & (210) & \\
        \multirow{2}{*}{(705,162)} & & \multirow{2}{*}{Single} & \multirow{2}{*}{10} & -0.84 & -0.83 & -0.10 & -0.08 & -0.11 \\
        & & & & (40) & (80) & (220) & (420) & (620)\\
        \midrule
        20 & \multirow{2}{*}{-0.25} & \multirow{2}{*}{Dual} & \multirow{2}{*}{10} & -0.28 &  & -0.32 &  & \\
        (716,692) & & & & (40) &  & (220) &  & \\
        \midrule
        30 & \multirow{2}{*}{-0.26} & \multirow{2}{*}{Dual} & \multirow{2}{*}{15} & -0.39 &  & -0.32  &  &  \\
        (728,222) &  &  &  & (60) &  & (330) &  &  \\
        \bottomrule
    \end{tabular}
        \caption{Average $\langle Z_0\rangle$ comparison of different scenarios between classical GANs and \qgans\, using different setup for the \qgans. The numbers in parentheses in the first column are the number of trainable parameters and reflect the model capacity (i.e. complexity).}
    \label{tab:scen}
\end{table}
% \Blindtext 

\subsection{Inference runs on quantum hardware}
\label{sec:results:inference}

To validate the ability of the \qgan\ to generate new molecules using actual quantum hardware, we have performed the sampling of 2,500 molecules on the \ibm\ quantum system provided by IBM using the model of the \qgan\ with five~qubits, two layers, and dual readout trained on a noiseless quantum simulator. This experiment is an inference-only, illustrative demonstration: the model parameters are those obtained from noiseless-simulator training, and the quantum hardware is used to sample from the trained generator. The aim is to test whether a simulator-trained model can be executed on real noisy intermediate-scale quantum hardware and still produce chemically viable samples. End-to-end quantum training is not the subject of this run. We discuss this option, as well as hybrid classical/quantum training schemes, as future work in Section~\ref{sec:discussion}.

To assess the performance of the quantum device and evaluate the effect of quantum noise, we have also generated 2,500 molecules sampled on a quantum simulator, as well as 2,500 samples obtained by the classical generator from the classical GAN. This choice of a lower sample compared to the default choice of 30,000 molecules is driven by cost considerations and time spent on the quantum device. It should be noted that this is already a sufficiently large sample to draw meaningful conclusions; this is in particular larger than the 1,000-sample QM9 hardware inference in~\autocite{Li2021}, which, to our knowledge, was the largest molecular inference on real quantum hardware until our inference in this paper.

In contrast with noiseless quantum simulations using state-vector simulation, which have infinite precision on the quantum state, we have to select the number of measurements performed on the quantum computer to collect the expectation values. We have performed our experiment with 1,000 shots and used the batch mode on IBM Cloud. When running on real quantum hardware, the quantum circuit we have designed needs to be adapted to the actual device topology and to the set of gates available on this device: this is the transpilation step. For our calculation on the \ibm\ device, we have used default Qiskit runtime parameters for the transpilation method (optimization~=~2) and the resilience level (resilience~=~0, no error mitigation applied).
We have not applied IBM Runtime error-mitigation protocols such as twirled readout error extinction (TREX) or zero-noise extrapolation (ZNE). These methods require extra circuit executions at elevated noise levels and therefore increase QPU time and cost~\autocite{Kandala2019Mitigation}. Since this inference run is intended to probe the effect of (native) device noise on a simulator-trained generator, rather than to maximize mitigated output quality, applying them would have exceeded our session budget for sampling 2,500 molecules without a clear benefit for this demonstration. It is also worth noting that hybrid quantum generative pipelines for drug design have reported preservation of chemically relevant metrics under device noise for inference: Anoshin~\textit{et al.} found comparable validity, drug-likeness, and solubility scores from noisy and ideal IBM simulators~\autocite{Anoshin2024}. This differs from variational quantum chemistry applications such as the variational quantum eigensolver, where dedicated error-mitigation protocols are often needed to reach accurate ground-state energies~\autocite{Kandala2019Mitigation}.

The execution settings and job metadata are summarized in the Supporting Information (Tables~\ref{tab:app:hw:config} and~\ref{tab:app:hw:calib}). Transpiled circuit depth, two-qubit-gate count, and physical-qubit layout have been taken from the executed circuits stored in the job records (\texttt{job.inputs}), and device-wide calibration medians in Table~\ref{tab:app:hw:calib} from \ibm\ backend properties on the execution date (2025-10-04 UTC). We also report in Supporting Information Table~\ref{tab:app:hw:noisy} a matched \texttt{FakeKingston} noisy-emulator run on the same trained generator, and in Figure~\ref{fig:app:hw:latent} and Table~\ref{tab:app:hw:latent} a comparison of the latent-vector distributions for the noiseless simulation, the noisy simulation, and the \ibm\ hardware run before VAE decoding. These distributions remain broadly similar to those of the noiseless simulation, in line with the overall robustness of the hardware inference.

Table~\ref{tab:ibm_comp} presents the comparison of metrics between the classical GAN and the \qgan\ with BEL circuit and nominal settings ($n_{qb} = 5$, $n_{l} = 2$, and dual readout). Despite the somewhat limited statistics, the inference run on quantum hardware delivers similar results when compared to those from the noiseless quantum simulator. Running on current noisy intermediate-scale quantum computers does not significantly degrade the overall performance of our pipeline. In fact, four of the seven reported drug-relevant metrics are slightly better on hardware than on the noiseless simulator (see below the detailed bootstrap analysis), but it should be noted that for this comparison we have not performed an inference with five different initializations of the trainable parameters. We do not expect that performing these five different runs would have significantly impacted the observations drawn in this section, however, this would have been quite expensive to run five times our experiment on the quantum computer.
To attach a quantitative uncertainty to this single hardware run without repeating the experiment, we estimate confidence intervals for the metrics of Table~\ref{tab:ibm_comp} by nonparametric bootstrap resampling of the generated molecules. For each of the three runs (classical GAN, \qgan\ on the noiseless simulator, and \qgan\ on \ibm), we draw $B = 10,000$ bootstrap resamples with replacement at the original sample size from the valid canonical molecules of that run, recompute every metric on each resample, and report the mean together with the $95\%$ confidence interval given by the $2.5$th and $97.5$th percentiles of the bootstrap distribution. This quantifies the finite-sample precision of each run; it does not capture run-to-run variability from different parameter initializations, which is the source of the uncertainties quoted for the simulator scenarios elsewhere in this work. The distinctness, validity, and uniqueness fractions ($\epsilon_d$, $\epsilon_v$, $\epsilon_u$) are quoted as point estimates only, since resampling with replacement is known to bias such distinct-count statistics~\autocite{Efron1983,mendelson2016distributionnumberuniqueoriginal} and would artificially lower the distinctness and uniqueness fractions. For comparison with the scenario tables in the Supporting Information, we also evaluate the per-metric significance $Z_0^m$ of Equation~\ref{eq:z0def}, taking the noiseless simulator as reference and the hardware run as the scenario under study and using the bootstrap standard errors of both runs as $\sigma_1^m$ and $\sigma_2^m$ (Supporting Information Table~\ref{tab:app:hw:zscore}). A negative $Z_0^m$ for a maximized metric (e.g.\ $\langle\text{QED}\rangle$) indicates degradation relative to the simulator at this bootstrap precision; $|Z_0^m|<1$ indicates statistical compatibility.

\begin{figure}[h!]
    \centering
    \includegraphics[width=0.32\linewidth]{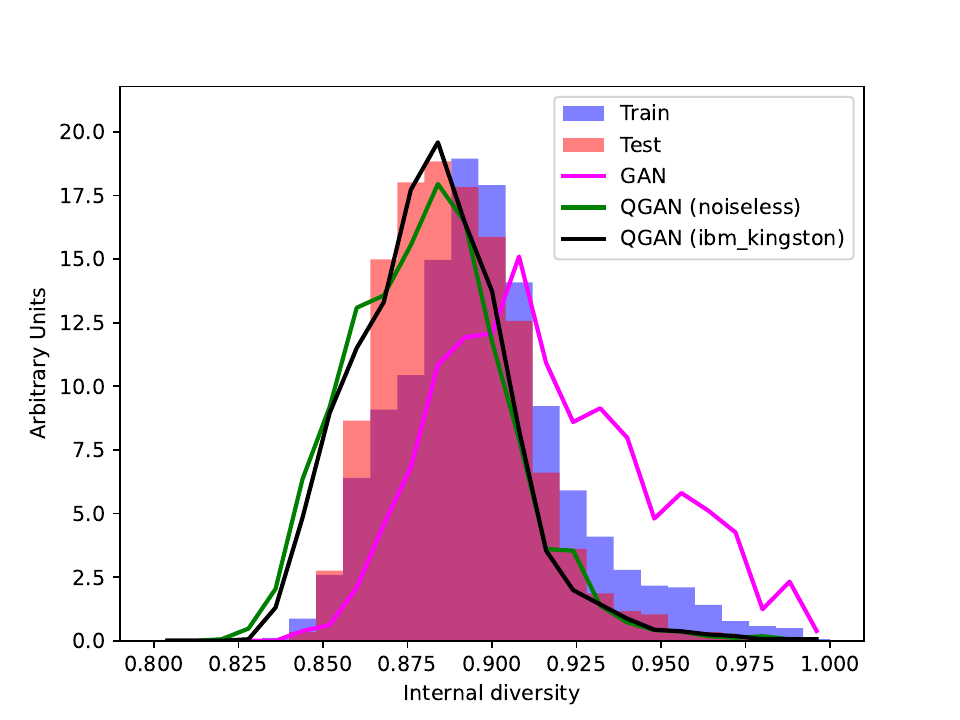}
    \includegraphics[width=0.32\linewidth]{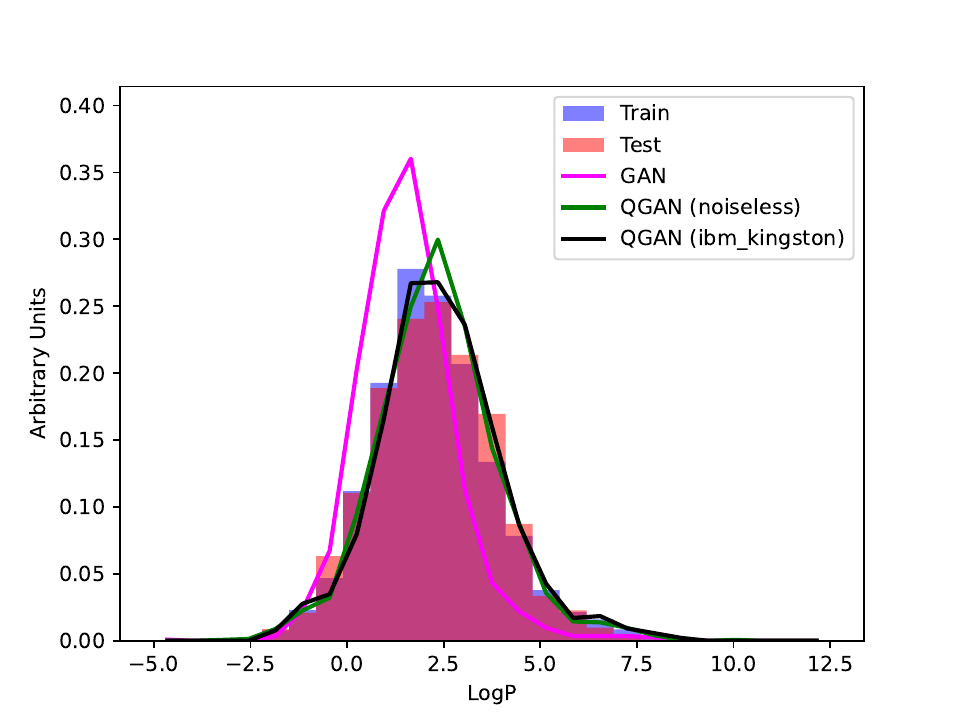}
    \includegraphics[width=0.32\linewidth]{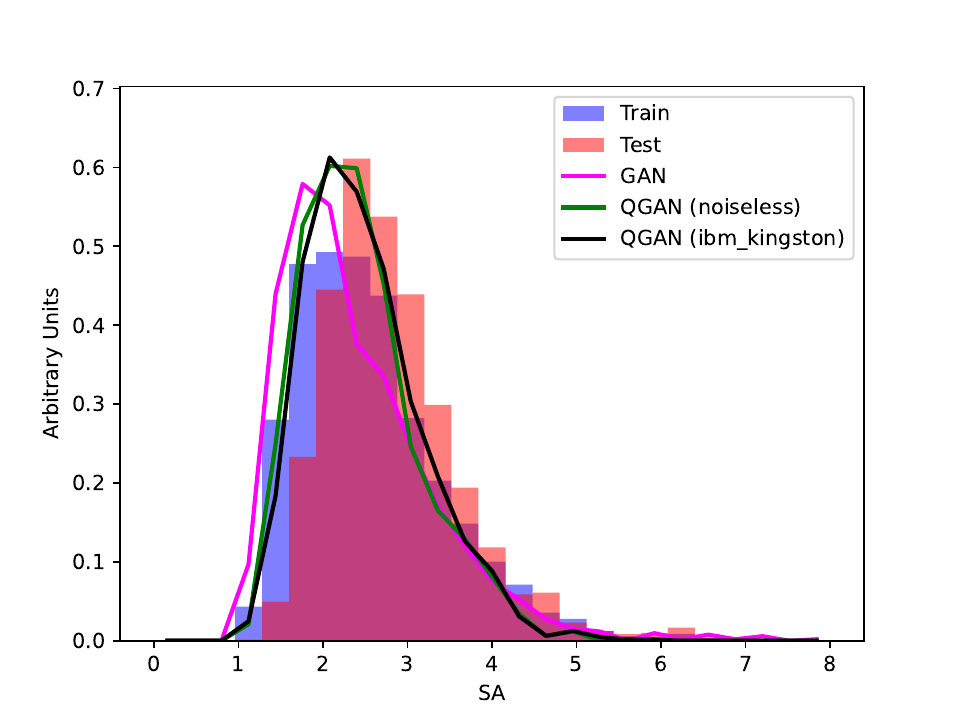}
    \includegraphics[width=0.32\linewidth]{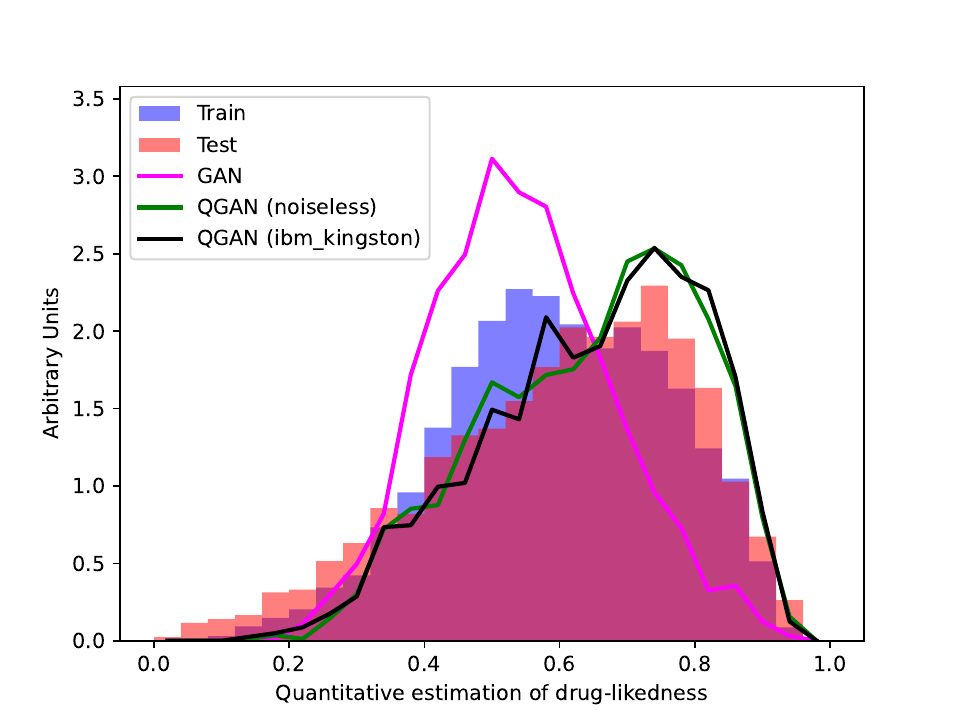}
    \includegraphics[width=0.32\linewidth]{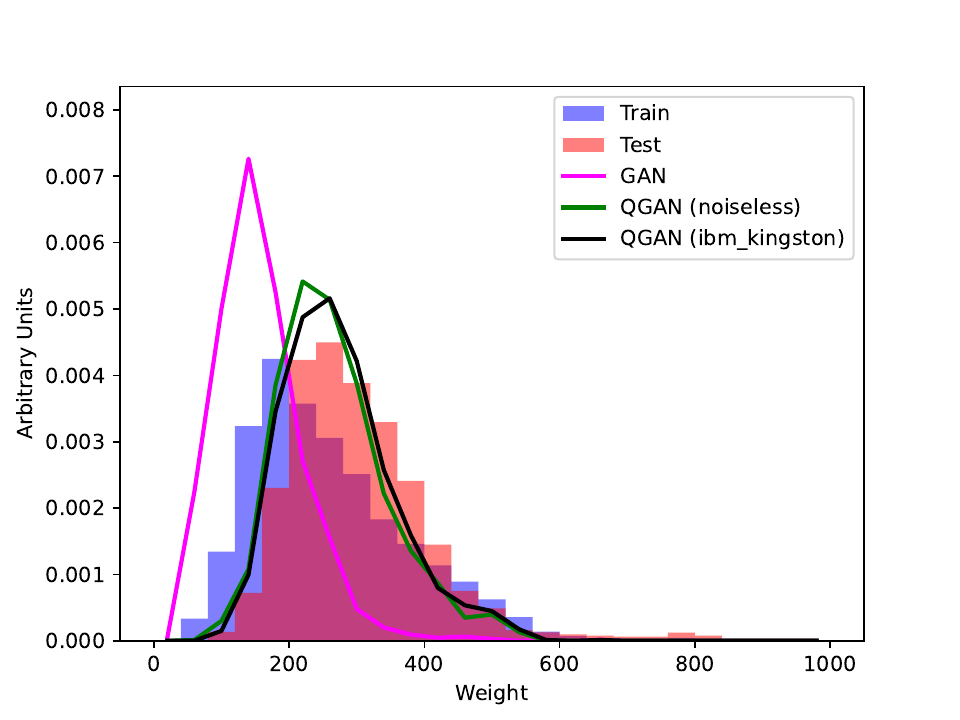}
    \caption{Distributions of LogP, SA, and QED metrics for the generation of 2,500 molecules using a classical GAN, the style \qgan\ run on noiseless simulator, and the style \qgan\ run on \ibm. The quantum generator uses $n_{qb} = 5$, a number of layers $n_{l} = 2$, and dual readout.}
    \label{fig:ibm_comp}
\end{figure}

Four of the seven reported metrics satisfy $Z_0^m \geq -1$, i.e.\ show no significant degradation at this bootstrap precision: novelty ($Z_0=+0.18$), internal diversity ($+0.85$), filters ($+0.49$), and $\langle\text{QED}\rangle$ ($+0.28$).
In particular, the hardware $\langle\text{QED}\rangle=0.649$ matches the noiseless-simulator value $0.647$ within the quoted intervals. Three metrics show a resolved shift relative to the simulator: $\epsilon_{LogP}$ ($Z_0=-1.04$), $\langle\text{SA}\rangle$ ($Z_0=-2.81$), and $\langle\text{Weight}\rangle$ ($Z_0=-3.43$). As shown in Supporting Information Table~\ref{tab:app:hw:noisy}, a matched \texttt{FakeKingston} noisy simulation reproduces the noiseless value for $\langle\text{QED}\rangle$ within bootstrap precision at $0.642$, and the hardware result $0.649$ is likewise compatible with both the noiseless and noisy-emulator values. The aggregate $\langle Z_0\rangle=-0.78$ indicates no broad degradation pattern dominated by a single observable. Overall, hardware noise does not induce a broad degradation of the pipeline, beyond the mean SA and Weight, which are still very much reasonable.
Comparing the \qgan\ results to the comparable classical GAN pipeline with an identical number of generated samples, we also observe reasonable consistency between the set of results. This statement is also supported by Figure~\ref{fig:ibm_comp}, which displays the distribution of metrics, comparing the three GAN scenarios: classical GAN in purple, noiseless \qgan\ simulation in red, and \qgan\ run on \ibm\ in black.

% Requires: \usepackage{booktabs}
\begin{table}[h!]
    \centering
    \begin{tabular}{lccc}
        \toprule
        Metrics & Classical GAN & \qgan\ (noiseless) & \qgan\ (\ibm)\\
        \midrule
        $\epsilon_d$ & 0.689 & 0.892 & 0.891 \\
        $\epsilon_v$ & 0.947 & 0.934 & 0.904 \\
        $\epsilon_u$ & 0.990 & 1.000 & 0.999 \\
        %$N_{Canonical}$ & 1,614 & 2,082 & 2,128 \\
        Novelty & 0.310 [0.287, 0.333] & 0.338 [0.318, 0.358] & 0.341 [0.320, 0.362] \\
        IntDiv & 0.915 [0.912, 0.918] & 0.881 [0.879, 0.883] & 0.882 [0.880, 0.884] \\
        Filters & 0.719 [0.698, 0.741] & 0.706 [0.687, 0.726] & 0.713 [0.693, 0.733] \\
        $\epsilon_{LogP}$  & 0.907 [0.893, 0.921] & 0.899 [0.886, 0.913] & 0.889 [0.875, 0.902] \\
        $\langle \mathrm{SA} \rangle$ & 2.388 [2.343, 2.432] & 2.424 [2.395, 2.453] & 2.485 [2.455, 2.515] \\
        $\langle \mathrm{QED} \rangle$ & 0.543 [0.537, 0.549] & 0.647 [0.641, 0.654] & 0.649 [0.642, 0.656] \\
        $\langle \text{Weight} \rangle$ & 156.2 [153.0, 159.4] & 264.9 [261.4, 268.3] & 273.6 [270.0, 277.2] \\
        \bottomrule
    \end{tabular}
        \caption{Metrics comparison calculated with 2,500 generated molecules, between the classical GAN, the style \qgan\ on a noiseless simulator, and the style \qgan\ inference on \ibm. The quantum generator uses $n_{qb} = 5$, a number of layers $n_{l} = 2$, and dual readout. For all metrics except the distinctness, validity, and uniqueness fractions ($\epsilon_d$, $\epsilon_v$, $\epsilon_u$, quoted as point estimates), the values in brackets give the $95\%$ confidence interval obtained by nonparametric bootstrap resampling of the generated molecules ($B=10,000$ resamples; $2.5$th--$97.5$th percentiles).}
    \label{tab:ibm_comp}
\end{table}

\begin{figure}[p]
    \centering
    \includegraphics[width=\linewidth]{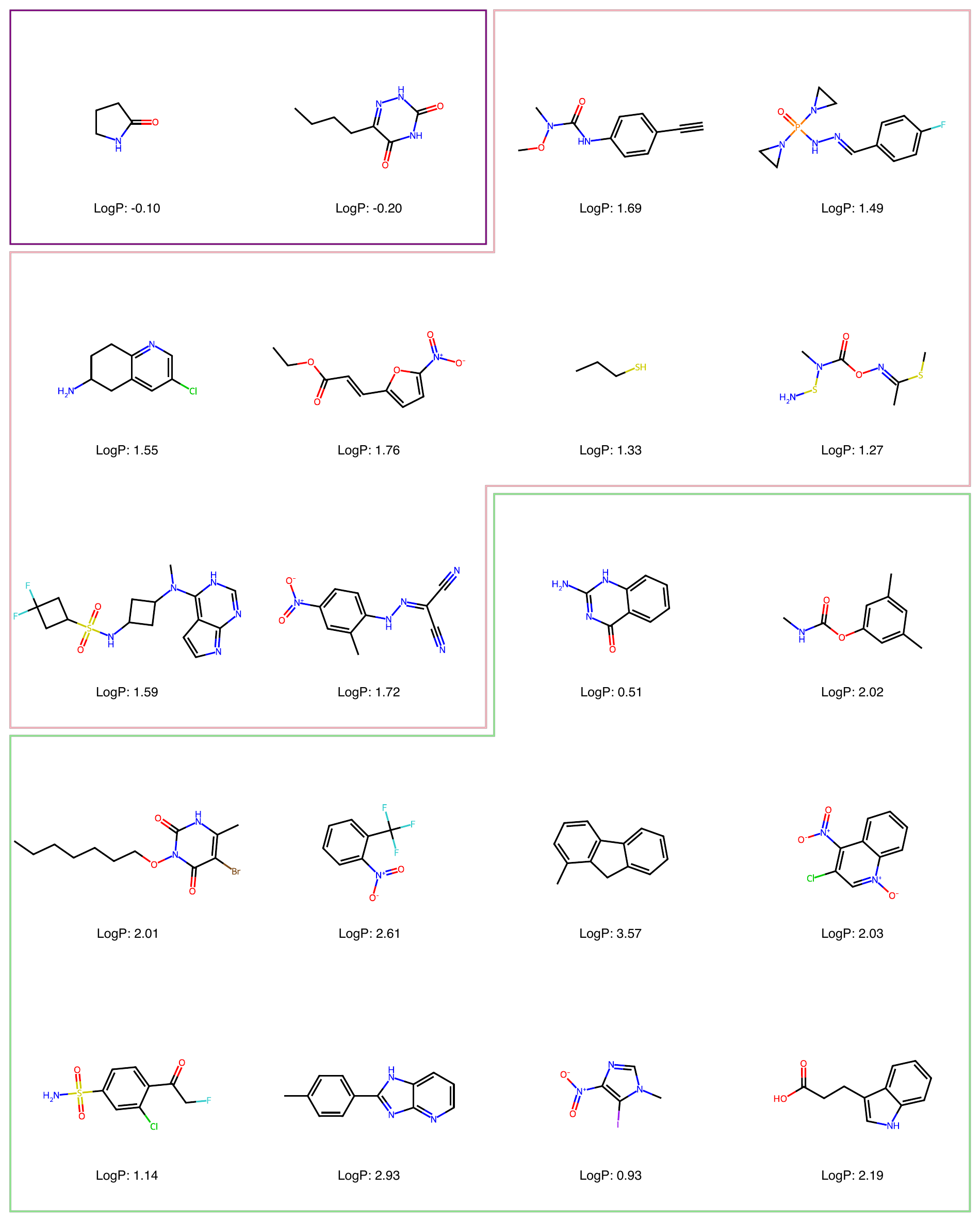}
    \caption{Random selection of the 2,500 generated molecules sampled on \ibm, passing the RDKit filters and all new compared to both the training and test dataset. The purple cluster contains molecules with LogP values compatible with injectable drugs; the pink cluster contains molecules with LogP values ideal for oral ingestion; the green cluster contains all other molecules with LogP not in the previous categories and still viable drug candidates with $|\text{LogP}|<5$.}
    \label{fig:ibm_mols}
\end{figure}

For illustration, Figure~\ref{fig:ibm_mols} presents a subset of molecules generated on the \ibm\ quantum computer. These molecules were not present in the dataset (training and validation),  demonstrating the ability of the \qgan\ to create new molecules on the quantum hardware in the inference step.

%%%%%%%%%%%%%%%%%%%%%%%%%%%%%%%%%%%%%%%%%%%%

\section{Discussion and outlook}
\label{sec:discussion}

In this work, we have adapted the MOSES framework for quantum-assisted drug design using a latent style-based Wasserstein GAN where the generator is replaced by a quantum network. The underlying VAE, which encodes the molecules in the latent space where the \qgan\ operates, has been optimized for the dataset we used to benchmark our proposed pipeline against existing standard classical latent GAN pipelines implemented in the MOSES benchmark suite. We have also tuned the classical GAN parameters.

We have presented several innovations compared to the existing pipelines: 1) We have adapted for the first time the style-based approach into a computer-assisted drug design pipeline, leveraging data reuploading to increase expressivity; 2) We have also used a novel nonlinear expression of the rotation angle in the various quantum gates of the two different quantum circuits we have investigated, in order to further prevent quantum mode collapse; 3) We have used a subset of MOSES dataset, which is an industry-relevant dataset for machine-learning-assisted biomolecular studies. After training both classical and quantum GANs on classical computing resources with GPUs, we have performed a performance comparison using various standard metrics relevant for drug design, such as LogP, molecular weight, quantitative estimation of drug-likeness (QED), novelty, and uniqueness. By calculating an average $\langle Z_0\rangle$ score, we have quantified the statistical performance of the \qgan\ and found that our pipeline delivers statistically compatible, competitive performance. For some metrics, the use of the \qgan\ yielded statistically significant improvements over the classical baseline. Examples include the average QED and the fraction of unique molecules produced by the pipeline. These conclusions hold as well when comparing to classical latent GANs with matched generator capacity (94/110/115/346/682 parameters), under identical VAE, discriminator, and training protocol. The \qgan\ pipeline offers a significant improvement over the MOSES classical GAN baseline: our default \qgans\ have up to 110 trainable parameters, compared to the 705,162 trainable parameters of the nominal MOSES classical GAN pipeline. This is a massive reduction in the neural network capacity, by a factor of 6,400. It should be noted that, with a classical generator at matched capacity, the results do not drastically change, signaling an overparameterization of the nominal MOSES baseline; however, even for the matched-capacity architecture, the training robustness is not as good as that of the \qgan, which is hence more efficient. Compared to previous studies of quantum generative modeling pipelines for drug design, we have been able to obtain competitive results beyond the small-molecule regime. We also note the scope of our comparison: this work proposes a new latent style-based \qgan\ for de novo drug design and benchmarks it in a controlled comparison between classical and quantum generators embedded in an otherwise identical GAN-based latent pipeline. We do not compare against diffusion or flow-matching models~\autocite{Alakhdar2024,Dunn2026}, which typically use 3D or graph molecular representations, and we make no claim of superiority over such architectures. The MOSES GAN baseline is used so that only the generator is changed between the classical and quantum pipelines.

When building the architecture of the \qgan\ pipeline, two architectural choices have been fixed from the outset: the gradient-penalized Wasserstein objective (WGAN-GP) and the style-based reuploading of the latent noise at every parametrized layer of the circuit. This follows previous studies of both classical GANs (Wasserstein distance and gradient penalty) and quantum GAN architectures (style-based reuploading). 
The advantage of the Wasserstein objective over the original Kullback--Leibler/Jensen--Shannon-divergence-based GAN loss, in terms of training stability and the mitigation of mode collapse, was demonstrated in the original Wasserstein-GAN work~\autocite{Arjovsky2017}, and the gradient-penalty term was subsequently shown to further improve training over the weight-clipping scheme~\autocite{gulrajani2017improvedtrainingwassersteingans}. 
Likewise, the benefit of style-based data reuploading, in order to increase the expressivity of the quantum generator, was established in the work introducing the style-based QGAN architecture~\autocite{BravoPrieto2022}. 
A dedicated, drug-design-specific ablation of these two components, as well as a study of the on/off effect of the $\tanh$ angle rescaling, would further quantify their individual contributions and is a natural direction for future work.

It is also worth comparing our results with the closest prior \qgan\ studies for drug design~\autocite{Li2021,Kao2023,Anoshin2024}. A direct numerical comparison is, however, not meaningful: these works use the small-molecule QM9~\autocite{Ramakrishnan2014} and PC9~\autocite{Glavatskikh2019} datasets within a graph-based MolGAN framework, and report metrics (e.g.\ normalized QED, synthesizability, and solubility/LogP scores rescaled to $[0,1]$, with uniqueness evaluated over a few thousand generated graphs) whose definitions differ from the SMILES-based MOSES metrics used here. With this in mind, Kao et al.~\autocite{Kao2023} (QuMolGAN, a variational-quantum-circuit noise generator with only tens of trainable parameters) report a validity in the $\sim$70--80\% range on QM9 together with improved drug-likeness at small noise dimension, but a markedly reduced uniqueness, falling to the single-digit to $\sim$20\% range for the hybrid quantum variants. 
Anoshin et al.~\autocite{Anoshin2024} (hybrid quantum cycle MolGAN) report improvements of up to $30\%$ in QED, representative normalized scores such as $\text{QED}\simeq 0.62$, $\text{SA}\simeq 0.84$, and $\text{LogP}\simeq 0.84$ on PC9, an enhanced uniqueness brought by the cycle component, and noise resilience on IBM simulators. For comparison, our latent style-based \qgan\ on the industry-relevant MOSES dataset reaches, for the BEL circuit, a validity of $0.875$, a uniqueness of $0.993$, a novelty of $0.548$, and a mean QED of $0.643$ (Table~\ref{tab:cgan_qgan_nom}). In contrast with the uniqueness degradation reported for the hybrid models of~\autocite{Kao2023}, our pipeline maintains a very high uniqueness ($\simeq 0.99$). These statements are indicative given the differences in datasets, molecular representations, and metric definitions.

As a complementary hardware check to estimate any potential degradation of the results when running on real, noisy quantum hardware, 2,500 samples have been generated on the \ibm\ 156-qubit Heron device for the BEL \qgan\ circuit. The resulting metrics are in reasonable agreement with the results from the noiseless quantum simulator, suggesting that the pipeline is not strongly degraded at this illustrative level of scrutiny and is a rather robust pipeline with respect to expected quantum noise. Unlike the noisy-simulator assessments of~\autocite{Anoshin2024}, our results are obtained directly on a physical IBM Heron processor. In addition, our MOSES inference run is also the largest molecular \qgan\ hardware execution reported to date, to our knowledge, with 2,500 molecules, compared to the 1,000-sample QM9 hardware inference in~\autocite{Li2021}. We do not claim a detailed, gate-resolved hardware noise characterization beyond this single inference session, as this would require repeated controlled executions, which is beyond the scope of this work. Execution metadata and execution-date device calibration for \ibm\ are summarized in the Supporting Information (Tables~\ref{tab:app:hw:config} and~\ref{tab:app:hw:calib}).

There are various directions for future work, in addition to the ones mentioned above. The first, immediate extension requires including conditioning in the \qgan\ pipeline~\autocite{Zoufal:2019rxf}. This would restrict the generation of molecules to specific criteria tailored to, e.g., the targeted disease, or to defined toxicity thresholds. The generation mechanism is then enhanced by allowing the (quantum) generator to sample only specific molecules of interest. This can be viewed as a type of fine-tuning of the generative pipeline. The simplest way to perform this in the current setup is to add ancilla qubits and then entangle them with the measured qubits with trainable controlled rotations.
A second direction of high interest is to improve the efficiency of the minimax game between the discriminator and the generator. We have performed several preliminary studies by varying the ratio between discriminator and generator steps in a given training epoch, potentially leading to a higher discriminator-to-generator training frequency ratio for \qgan, which is different from the tuned choice for classical GANs.
Finally, a very exciting direction of research is investigating hybrid training schemes, where both classical resources (GPUs) and quantum resources are used in the training phase. So far, we have performed the training entirely on a quantum-state-vector simulator running on a GPU. Performing the training in a hybrid mode, with some epochs trained directly on the quantum hardware, would capture the quantum noise characteristics in the model itself, which, in turn, is expected to enhance the quality of the final sampling on the quantum hardware. Along this direction, better training methods or more hardware-training-efficient circuits~\autocite{jcdk-q3xc} to perform the entire training efficiently on quantum hardware should be explored to increase the power of the pipeline and allow for training much wider quantum networks with hundreds of qubits, exploring molecular datasets of high complexity requiring a latent space of dimension 100 or more.

\section*{Data and Software Availability}
The underlying dataset used in this study is the MOSES benchmark dataset, publicly available from its official Github repository: \url{https://github.com/molecularsets/moses}. The specific data points behind all the figures of the study are publicly available on a Zenodo repository at \url{https://doi.org/10.5281/zenodo.19682669}.
The source code used for the training and inference of the pipeline is available on the following Github repository: \url{https://github.com/Quantum-Drug-Discovery/DrugQGAN}.

\section*{Acknowledgments}
The work from R. P. has been supported by PricewaterhouseCoopers (PwC). The authors acknowledge the use of IBM Quantum services for this work. The views expressed are those of the authors and do not reflect the official policy or position of IBM or the IBM Quantum team. Y. H. is part of the CERN Quantum Technology Initiative (QTI) and of the Open Quantum Institute (OQI); the OQI is an initiative hosted by CERN, born at GESDA, and supported by UBS.

%
% ---- Bibliography ----
\printbibliography

%%%%%%%%%%%%% Supplementary material %%%%%%%%%%%%

\pagebreak
\appendix

\section{Supplementary material: Input distributions}
\label{sec:app:input}

\begin{figure}[h!]
    \centering
    \includegraphics[width=0.32\linewidth]{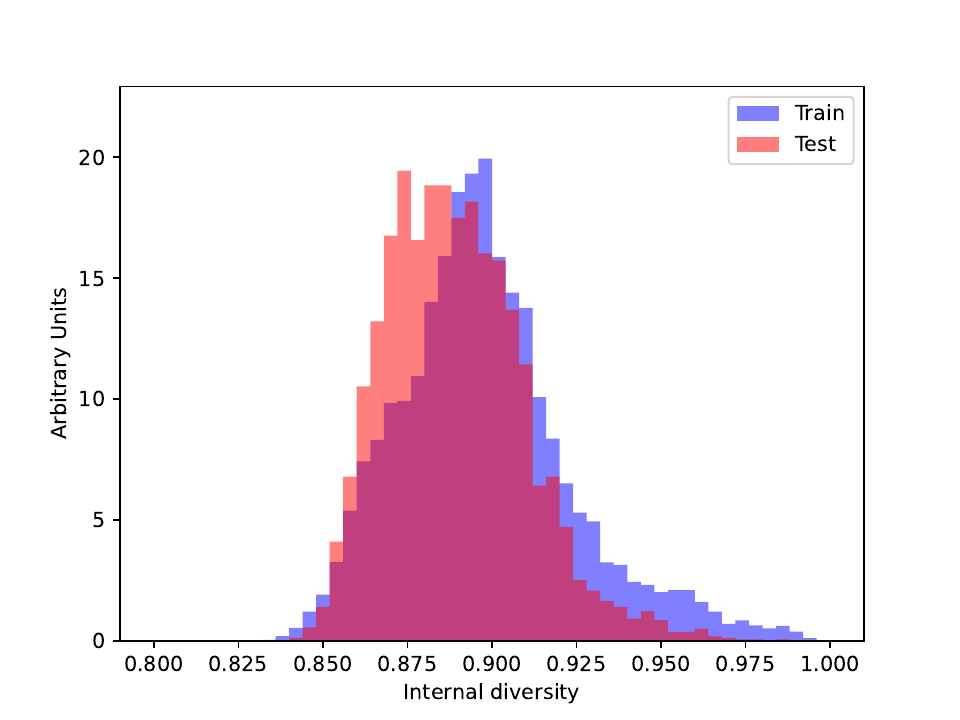}
    \includegraphics[width=0.32\linewidth]{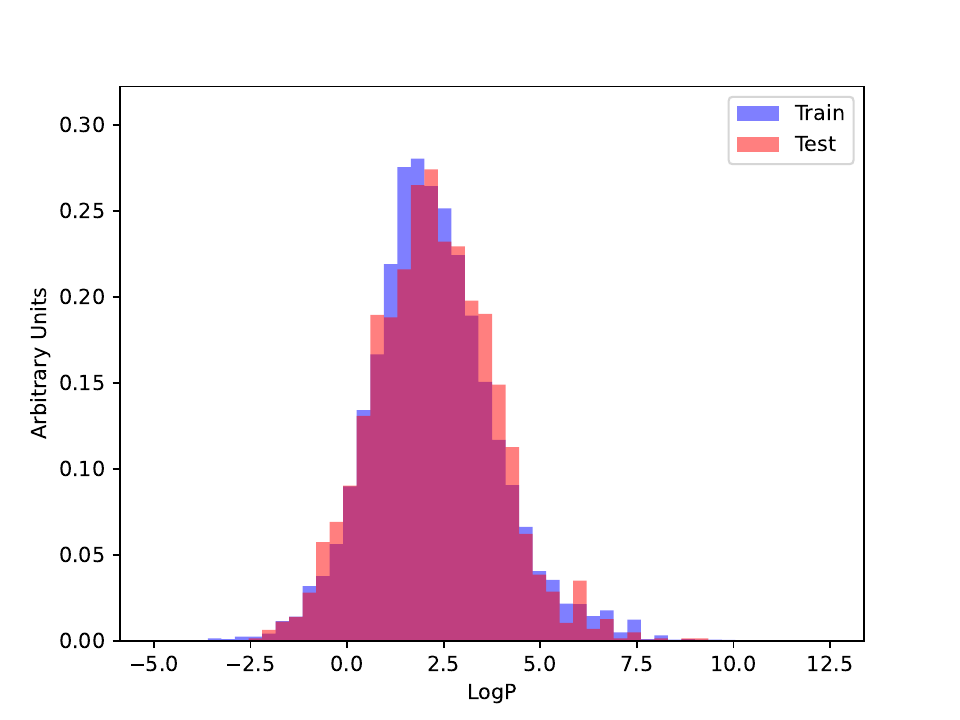}
    \includegraphics[width=0.32\linewidth]{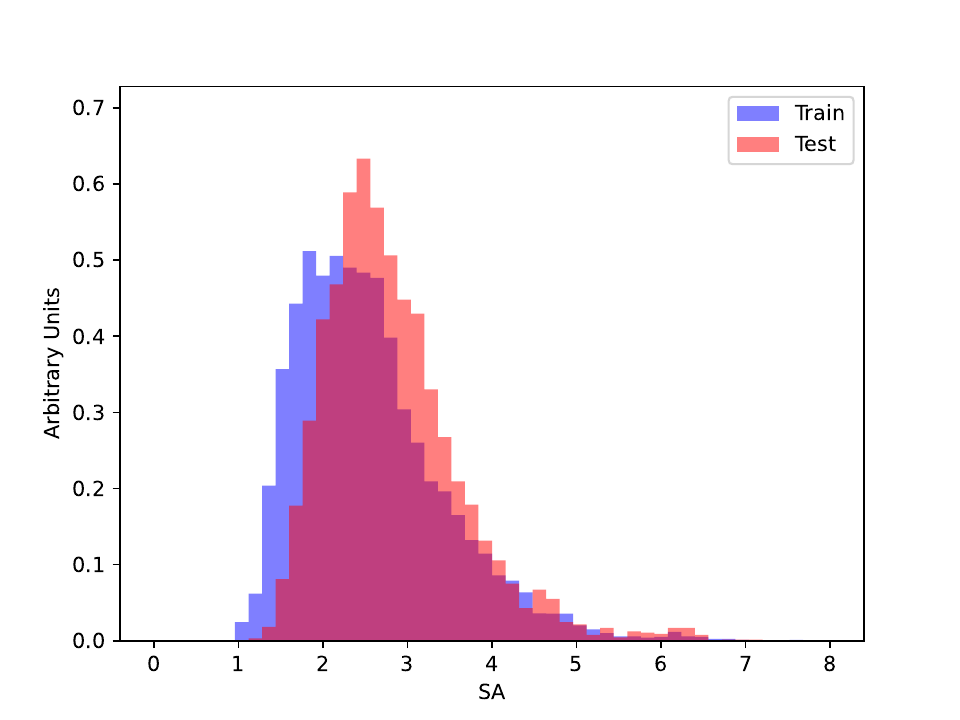}
    \includegraphics[width=0.32\linewidth]{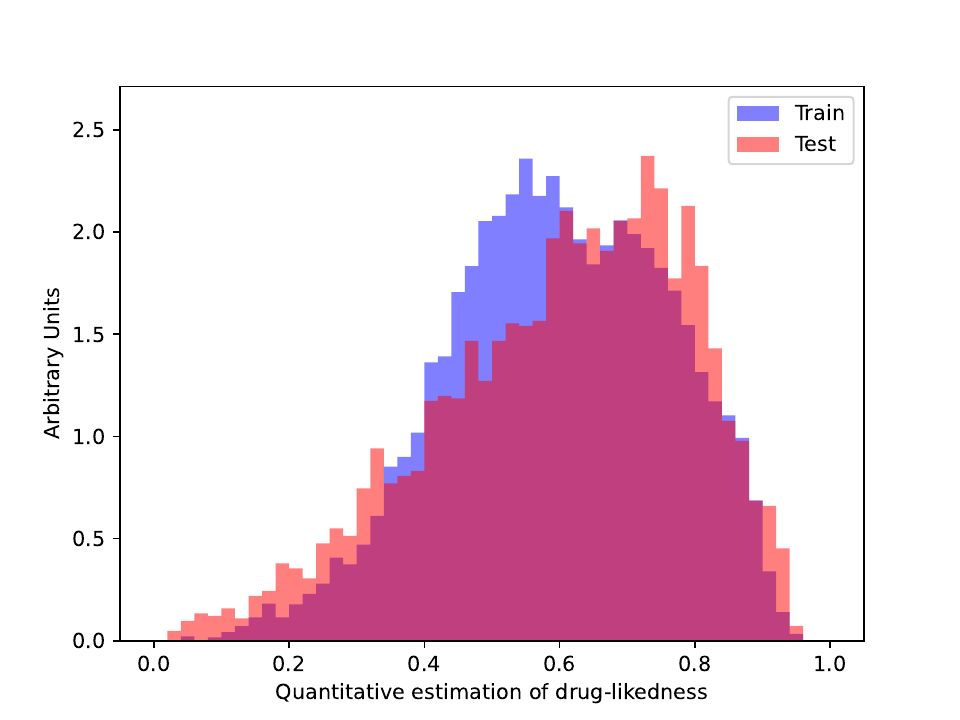}
    \includegraphics[width=0.32\linewidth]{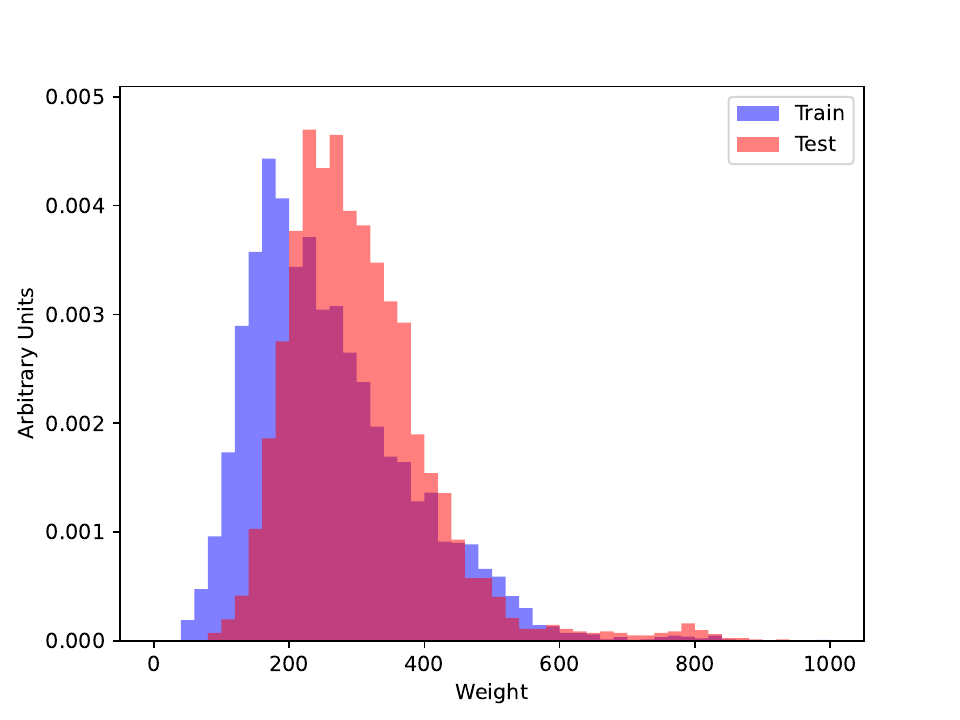}
    \caption{Comparison of metrics distributions between the input train and test sets.}
    \label{fig:sec:app:input}
\end{figure}

In this section, the five main metrics are presented, calculated from the input training and test datasets. The plots are in Figure~\ref{fig:sec:app:input} and show certain differences between the two distributions, except for the LogP metric.

Table~\ref{tab:sec:app:input} then shows the mean internal diversity and the Wasserstein distance from the test sample.

\begin{table}[h!]
    \centering
    \begin{tabular}{lcc}
        \toprule
        \textbf{Metrics} &\textbf{Train set} & \textbf{Test set} \\
        \midrule
        IntDiv            & 0.898   & 0.889   \\
        $\mathcal{W}(\text{LogP})$    & 0.091   & --      \\
        $\mathcal{W}(\text{SA})$       & 0.268   & --      \\
        $\mathcal{W}(\text{QED})$      & 0.027   & --     \\
        $\mathcal{W}(\text{Weight})$   & 42.8  & --    \\
        \bottomrule
    \end{tabular}
    \caption{Comparison of metrics between the input train and test sets.\label{tab:sec:app:input}}
\end{table}

\section{Supplementary material: VAE study of $N_{ep}$}
\label{sec:app:vae_nep}

To ensure stable model training, the sensitivity of the loss function and metrics to the number of training epochs has to be studied. The general rule is that the loss function around the final chosen point of $N_{ep}$ should be stable for the training as well as the test dataset.

\begin{figure}[h!]
    \centering
    \includegraphics[width=0.75\linewidth]{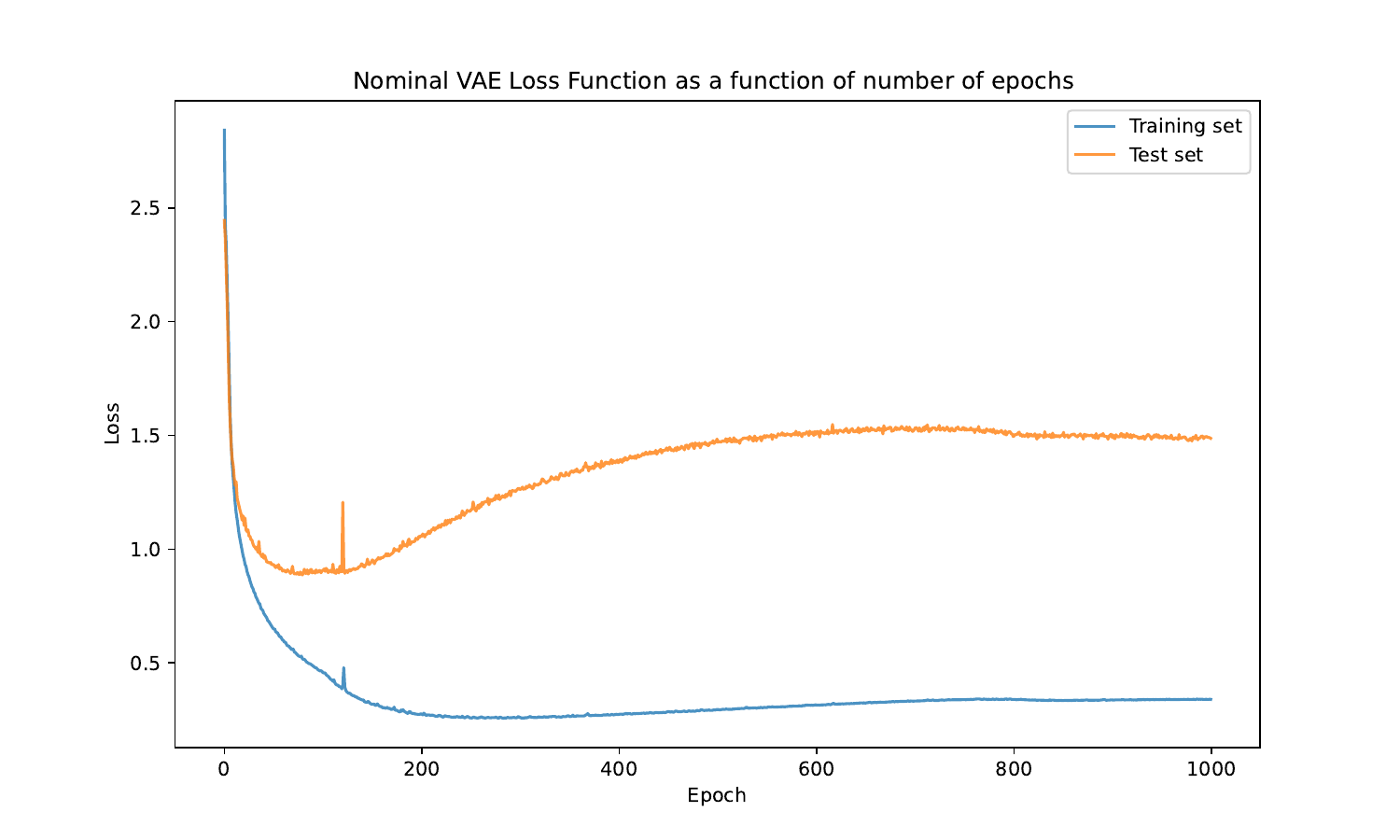}
    \caption{VAE loss function as a function of number of training epochs for the nominal VAE.}
    \label{fig:app:vae_nep}
\end{figure}

Figure~\ref{fig:app:vae_nep} presents the dependency of the loss function for both sets as a function of the number of training epochs. From the requirement of stability, it follows that anything below $N_{ep} = 800$ might not be stable enough. Four working points have been chosen to test the hypothesis: 

\begin{enumerate}
    \item $N_{ep} = 100$: default value from the MOSES library;
    \item $N_{ep} = 250$: value where the training set seems to be stable but the test set loss function is still increasing;
    \item $N_{ep} = 800$: value where both loss functions seem to start to be stable;
    \item $N_{ep} = 1000$: value chosen to be far from the beginning of the stable period of both loss functions.
\end{enumerate}

\begin{figure}[h!]
    \centering
    \includegraphics[width=0.32\linewidth]{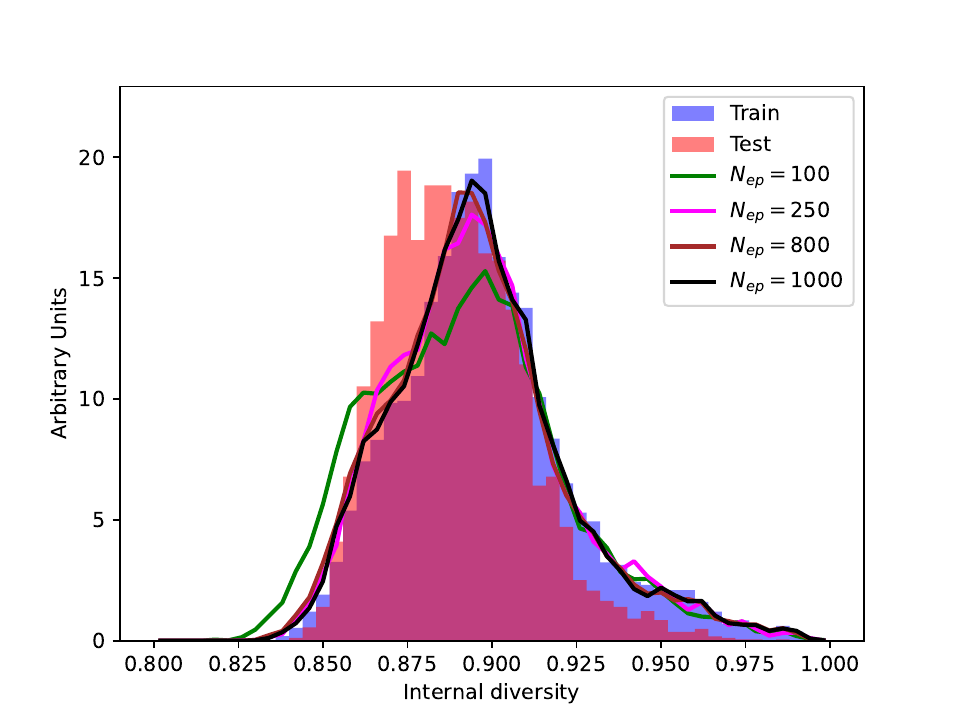}
    \includegraphics[width=0.32\linewidth]{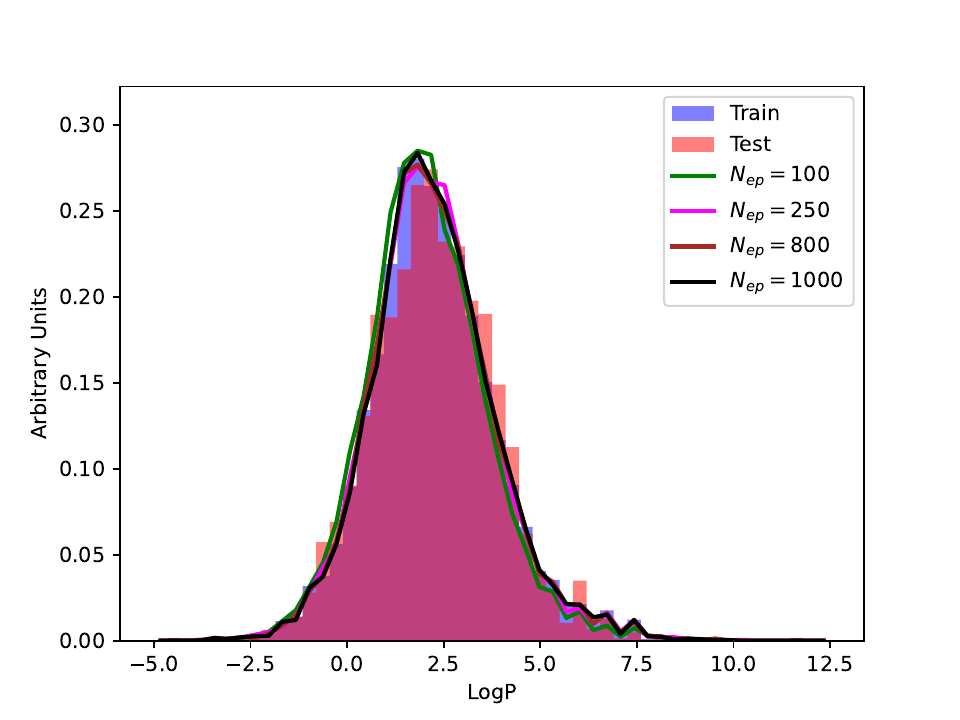}
    \includegraphics[width=0.32\linewidth]{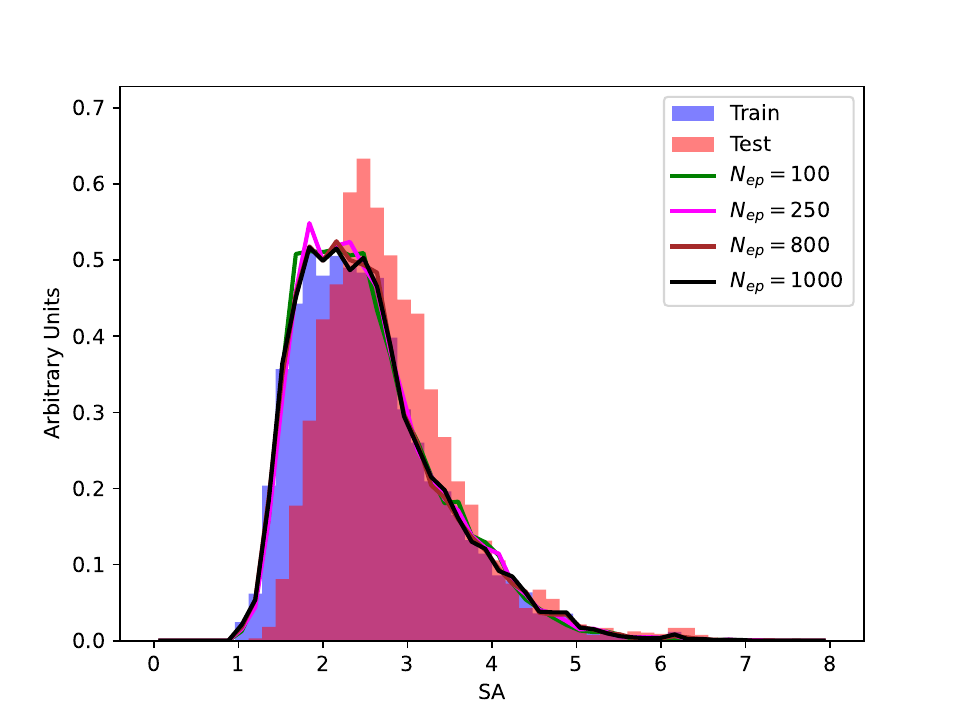}
    \includegraphics[width=0.32\linewidth]{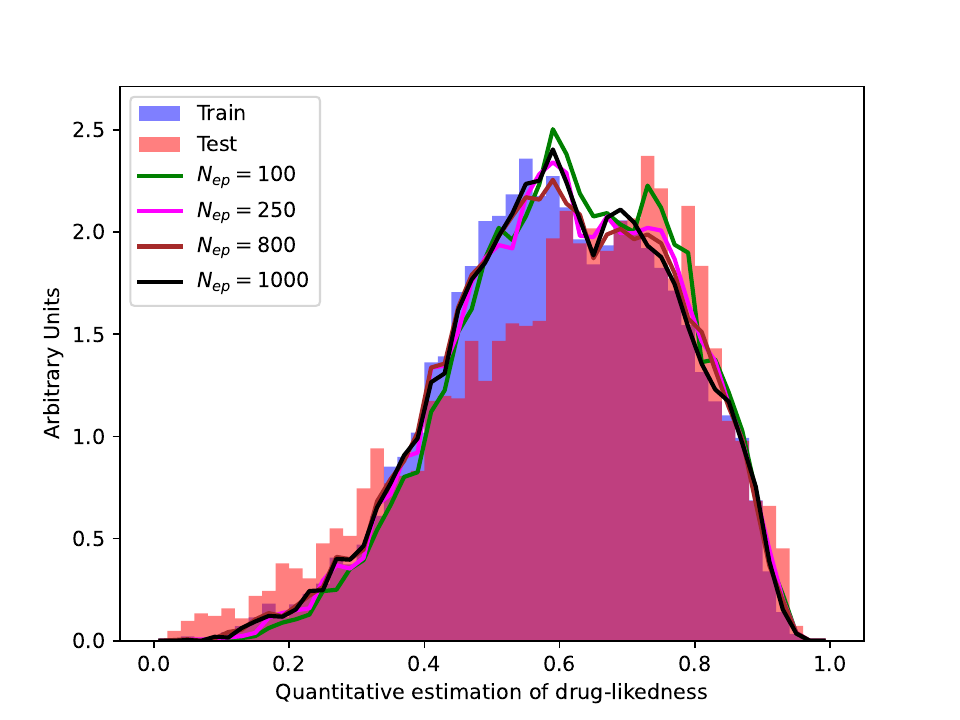}
    \includegraphics[width=0.32\linewidth]{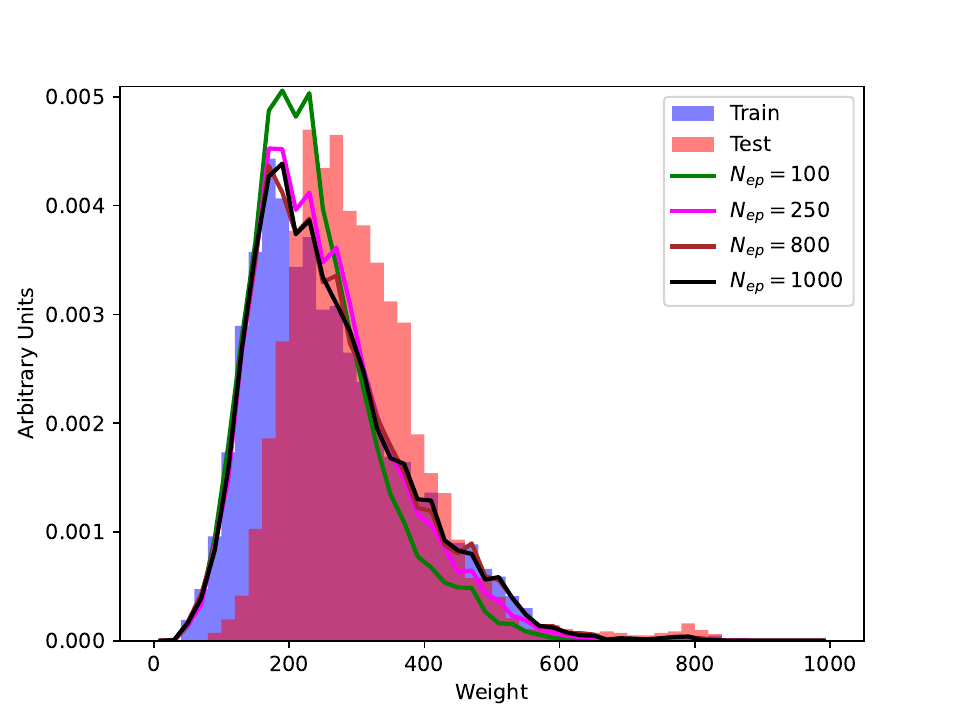}
    \caption{Metrics distributions for models trained with different number of epochs. Stabilization of the results starts with 800 epochs.}
    \label{tab:app:vae_nep_met}
\end{figure}

The trainings have been performed for the four amounts of epochs, and the results are summarized in Table~\ref{tab:app:vae_nep}. As expected, the values for the working point of 800 and 1,000 are consistent, hence supporting the theory that stable loss functions mean stable metric results and justifying the final choice of $N_{ep} = 1,000$ for the VAE training. The same conclusion can be drawn from Figure~\ref{tab:app:vae_nep_met} where an agreement is observed for the 800 and 1,000 working points while (in particular for internal diversity and molecular weight) the results for the 100 and 250 working points diverge.

% Requires: \usepackage{booktabs}
\begin{table}[h!]
    \centering
    \begin{tabular}{lcccc}
        \toprule
        %\textbf{Metrics} &\textbf{$N_{ep} = 100$} & \textbf{$N_{ep} = 250$} & \textbf{$N_{ep} = 800$} & \textbf{$N_{ep} = 1000$} \\
        \midrule
        $N_{ep}$ & 100 & 250 & 800 & 1,000 \\
        \midrule
                GPU runtime & 9m 10s & 17m 24s & 48m 49s  & 52m 35s  \\
        \midrule

        IntDiv            & 0.892   & 0.896   & 0.895   & 0.896   \\
        $\mathcal{W}(\text{LogP})$    & 0.190   & 0.086   & 0.088   & 0.087   \\
        $\mathcal{W}(\text{SA})$       & 0.285   & 0.265   & 0.280   & 0.275   \\
        $\mathcal{W}(\text{QED})$      & 0.025   & 0.023   & 0.023   & 0.025   \\
        $\mathcal{W}(\text{Weight})$   & 67.6  & 51.2  & 43.8  & 43.9  \\
        \bottomrule
    \end{tabular}
    \caption{VAE metrics comparison for the four $N_{ep}$ points. The Wasserstein distance is measured between the test and generated samples. While the internal diversity is maximized, all other metrics are minimized. We also report the GPU runtime for comparison.\label{tab:app:vae_nep}}
\end{table}

\section{Supplementary material: Dependency on $N_{Gen}$}
\label{sec:app:ngen}

This section investigates the question of whether 30,000 samples generated for the purpose of validating the molecular sets are sufficient.

Once the sampler is called and generates the requested amount, a three-step efficiency estimation (as described in Section~\ref{sec:setup:metric}) is performed. The number of relevant molecules that are used to derive all the metrics is defined in Equation~\ref{eq:ncanon} as:

\begin{equation}
    N_{Canonical} = N_{Generated} \times \epsilon_{d} \times \epsilon_{v} \times \epsilon_{u}
    \label{eq:ncanon}
\end{equation}

where $\epsilon_{d}$ quantifies the amount of identically duplicated strings, $\epsilon_{v}$ the amount of chemically valid molecules, and $\epsilon_{u}$ the amount of canonically distinct molecules.

Table~\ref{tab:app:ngen} presents the results for the nominal VAE settings and three different sizes of the generated sample. The trend of increasing inefficiency is clearly visible: while the amount of identical strings for the sample with 30,000 molecules (the 30k sample) is almost 50\%, for the 1M set it is almost 90\%. While the fraction of unique molecules stays almost constant, the fraction of valid molecules also drops significantly with increasing number of generated samples. The combined effect is then visible in the number of canonical generated molecules, which, for a 33-fold increase in initial statistics, grew only by a factor of six.

Worth mentioning is also the observed trend in the novelty, which increases dramatically. However, this increase does not immediately mean an increased capacity to generate new molecules. The novelty is defined as the number of generated molecules that were not present in the training sample. Although this holds well for the original MOSES analysis with more than 1M molecules used for training, it is important to keep in mind that this analysis used only 12,000 molecules for training. That means that even if all training samples would be present in the 1M set, the novelty would be 83\%. Same calculation for the 30k sample yields such minimal novelty of 6\%. Comparing with the results from Table~\ref{tab:app:ngen}, we see that the 30k sample actually generates more new molecules than the 1M sample. The 30k scenario, resulting in about 13,000 valid canonical molecules, is therefore a good candidate for a reliable novelty calculation in an analysis where only 12,000 training molecules are used.

All these arguments, together with minimizing the sampling time, lead to the decision of using 30,000 generated samples for subsequent result estimation.

% Requires: \usepackage{booktabs}
\begin{table}[h!]
    \centering
    \begin{tabular}{lccc}
        \toprule
        \textbf{$N_{generated}$} & \textbf{30,000} & \textbf{300,000} & \textbf{1,000,000} \\
        \midrule
        GPU runtime       & 52m  & 1h 37m  & 2h 12m  \\
        Sampling runtime       & 10m  & 55m  & 1h 30m  \\
        \midrule
        $\epsilon_d$    & 0.479   & 0.168  & 0.120  \\
        $\epsilon_v$     & 0.889    & 0.704  & 0.609  \\
        $\epsilon_u$    & 0.996   & 0.982  & 0.977  \\
             \midrule
   $N_{Canonical}$          & 12,725    & 34,763   & 71,118   \\
           \midrule
     Novelty           & 0.237   & 0.662  & 0.835  \\
        IntDiv            & 0.896   & 0.896  & 0.895  \\
        $\mathcal{W}(\text{LogP})$     & 0.087   & 0.066  & 0.094  \\
        $\mathcal{W}(\text{SA})$       & 0.275   & 0.188  & 0.127  \\
        $\mathcal{W}(\text{QED})$      & 0.025   & 0.020  & 0.017  \\
        $\mathcal{W}(\text{Weight})$   & 43.9  & 38.4 & 32.6 \\
        \bottomrule
    \end{tabular}
    \caption{Metrics comparison for the nominal VAE training for three different sizes of sets of generated molecules. We also report runtime on the GPU.}
        \label{tab:app:ngen}
\end{table}

\section{Supplementary material: VAE hyperparameter tuning}
\label{sec:app:vae_opt}
The tuning of the VAE was performed for three distinct domains: the learning rate, the encoder parameters, and the decoder parameters. Tables~\ref{tab:app:vae_opt:lr}, \ref{tab:app:vae_opt:enc} and \ref{tab:app:vae_opt:dec} present the parameter variations as well as the metrics comparison.

\begin{table}[h!]
    \centering

    \begin{tabular}{lcccccc}
        \toprule
        lr max & $3\times 10^{-4}$ & $1\times 10^{-4}$ & $1\times 10^{-2}$ & $5\times 10^{-5}$ & $5\times 10^{-3}$ & $1\times 10^{-3}$ \\
        lr min & $3\times 10^{-4}$ & $1\times 10^{-6}$ & $1\times 10^{-4}$ & $5\times 10^{-5}$ & $5\times 10^{-3}$ & $1\times 10^{-5}$ \\
        \midrule
        $\epsilon_d$ & 0.479 & 0.805 & 0.938 & 0.842 & 0.936 & 0.417 \\
        $\epsilon_v$ & 0.889 & 0.660 & 0.018 & 0.626 & 0.041 & 0.957 \\
        $\epsilon_u$ & 0.995 & 0.986 & 0.956 & 0.985 & 0.940 & 0.997 \\
        Novelty & 0.237 & 0.702 & 0.925 & 0.768 & 0.856 & 0.111 \\
        IntDiv & 0.896 & 0.893 & 0.887 & 0.891 & 0.929 & 0.898 \\
        $\mathcal{W}(\text{LogP})$ & 0.087 & 0.104 & 13.789 & 0.133 & 0.823 & 0.083 \\
        $\mathcal{W}(\text{SA})$ & 0.275 & 0.292 & 1.731 & 0.305 & 0.398 & 0.265 \\
        $\mathcal{W}(\text{QED})$ & 0.025 & 0.026 & 0.394 & 0.025 & 0.131 & 0.026 \\
        $\mathcal{W}(\text{Weight})$ & 43.9 & 57.3 & 1261.2 & 62.3 & 169.9 & 43.1 \\
        \bottomrule
    \end{tabular}
        \caption{Comparisons of the results of VAE learning rate tuning, the first column corresponds to the nominal scenario. Individual scenarios are defined by the learning rate minima/maxima at the top of the Table.}
    \label{tab:app:vae_opt:lr}
\end{table}

\begin{figure}[h!]
    \centering
    \includegraphics[width=0.32\linewidth]{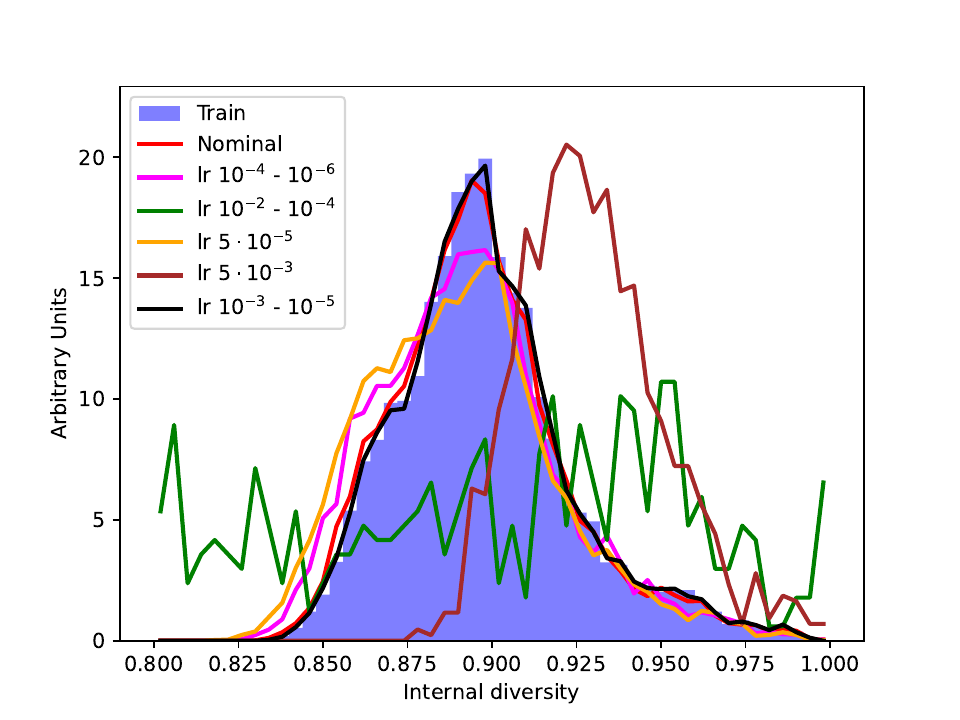}
    \includegraphics[width=0.32\linewidth]{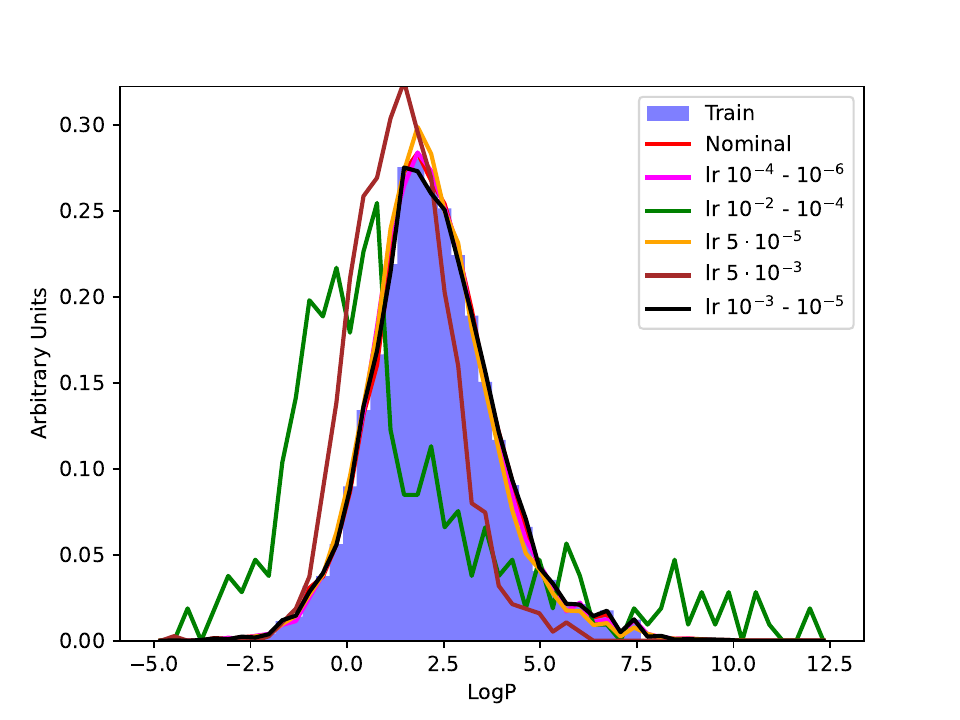}
    \includegraphics[width=0.32\linewidth]{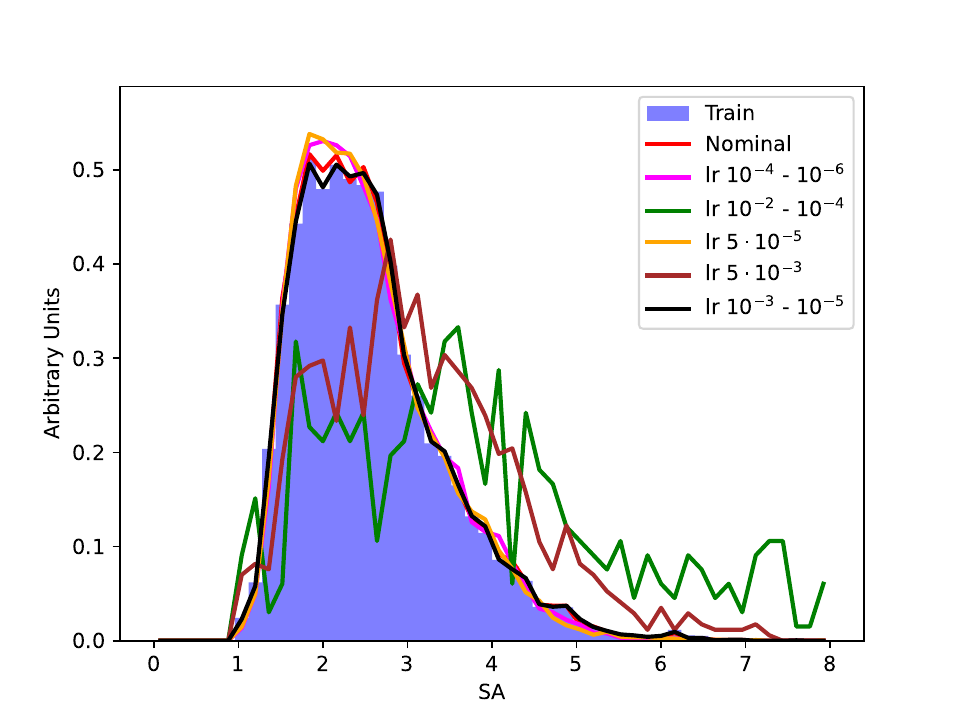}
    \includegraphics[width=0.32\linewidth]{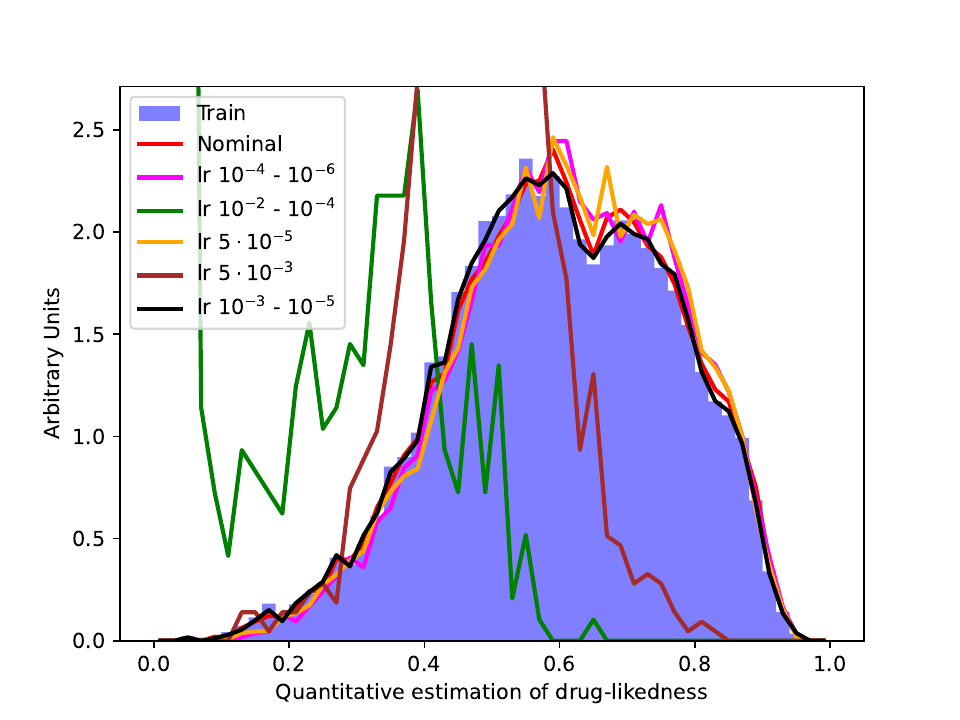}
    \includegraphics[width=0.32\linewidth]{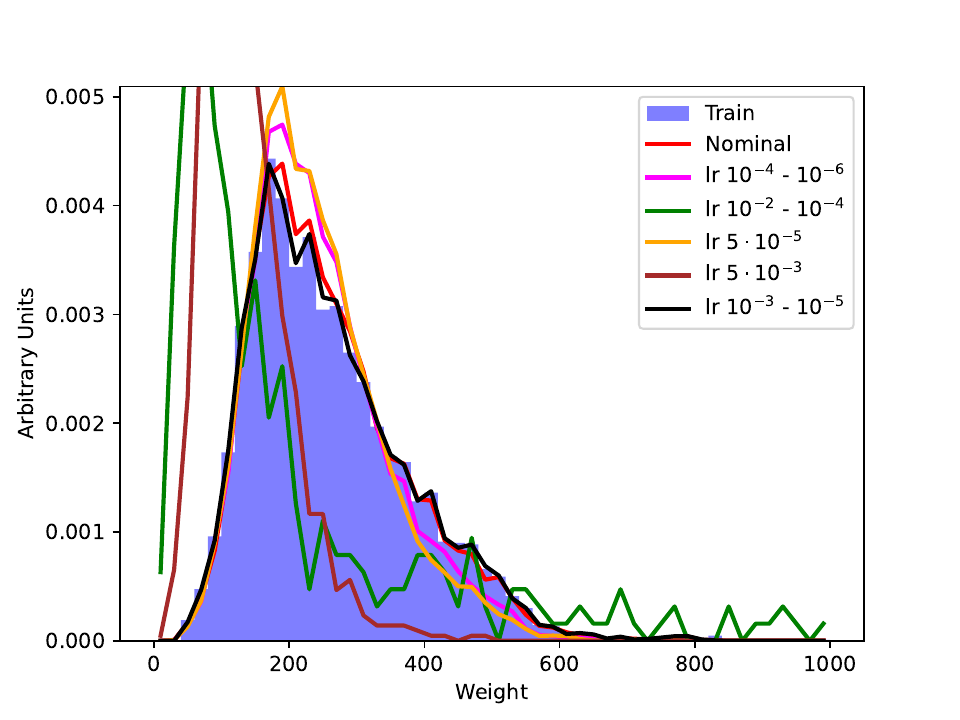}
    \caption{Metrics distributions for models trained with different learning rates. Clear breakdown for the 0.01-0.0001 and 0.005 scenarios is visible.}
    \label{fig:app:vae_opt:lr}
\end{figure}

Table~\ref{tab:app:vae_opt:lr} and Figure~\ref{fig:app:vae_opt:lr} present the results of the learning rate optimization. It is obvious from the plots that, when the learning rate is too large (0.01-0.0001 and 0.005 scenarios), the results collapse, and the metrics become meaningless, as reflected in Table~\ref{tab:app:vae_opt:lr}. Comparison of the remaining scenarios prefers the 0.001-0.00001 tuning.

% Requires: \usepackage{booktabs}
\begin{table}[h!]
    \centering
    \begin{tabular}{lc|c|cc}
        \toprule
         $n_{l}$ & 1 & \textbf{2} & 1 & 1  \\
         $D_h$ & 256 & 256 & \textbf{128} & \textbf{512}  \\
         Dropout & 0.0 & \textbf{0.5} & 0.0 & 0.0  \\
                \midrule
        $\epsilon_d$ & 0.479 & 0.548 & 0.475 & 0.5321  \\
        $\epsilon_v$ & 0.889 & 0.838 & 0.888 & 0.8552  \\
        $\epsilon_u$ & 0.996 & 0.994 & 0.995 & 0.9933  \\
%        $N_{canonical}$ & 12,725 & 13,696 & 12,587 & 13,560 & 12,725 & 12,725 \\
        Novelty & 0.237 & 0.349 & 0.238 & 0.327  \\
        IntDiv & 0.896 & 0.895 & 0.896 & 0.896  \\
        $\mathcal{W}(\text{LogP})$ & 0.087 & 0.094 & 0.075 & 0.087 \\
        $\mathcal{W}(\text{SA})$ & 0.275 & 0.273 & 0.271 & 0.269  \\
       $\mathcal{W}(\text{QED})$ & 0.025 & 0.024 & 0.026 & 0.025 \\
        $\mathcal{W}(\text{Weight})$ & 43.9 & 44.3 & 43.6 & 45.3  \\
        \bottomrule
    \end{tabular}
        \caption{Comparisons of the results of VAE encoder parameter tuning, the first column corresponds to the nominal scenario. The scenario variation is highlighted at the top of the table. The dropout is active only for a scenario with more than one encoder layer, in such case value of 0.5 was used.}
    \label{tab:app:vae_opt:enc}
\end{table}

\begin{figure}[h!]
    \centering
    \includegraphics[width=0.32\linewidth]{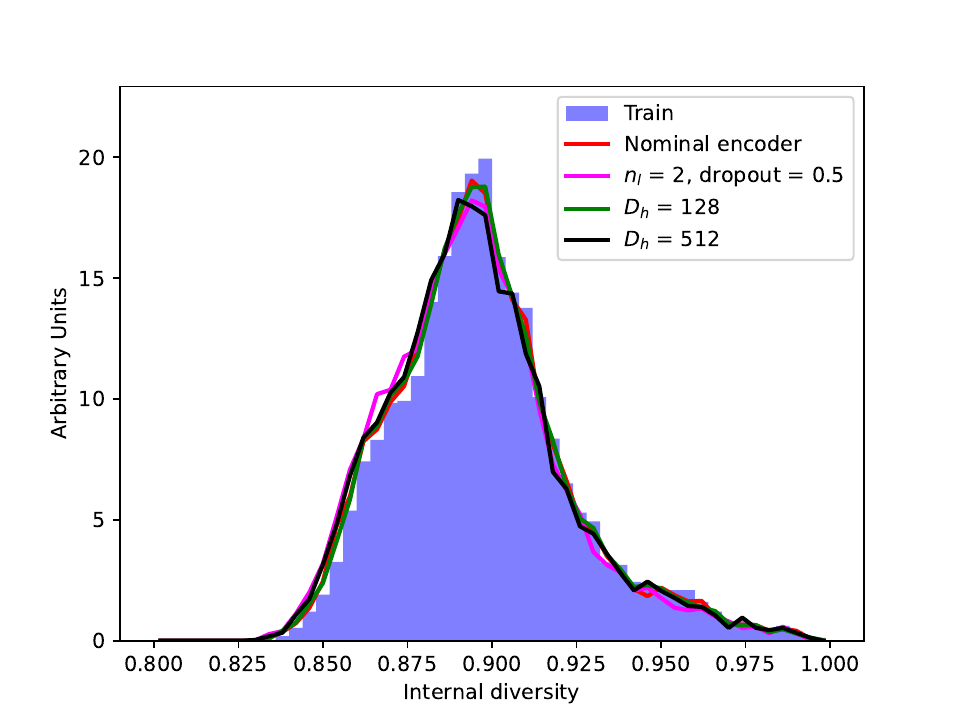}
    \includegraphics[width=0.32\linewidth]{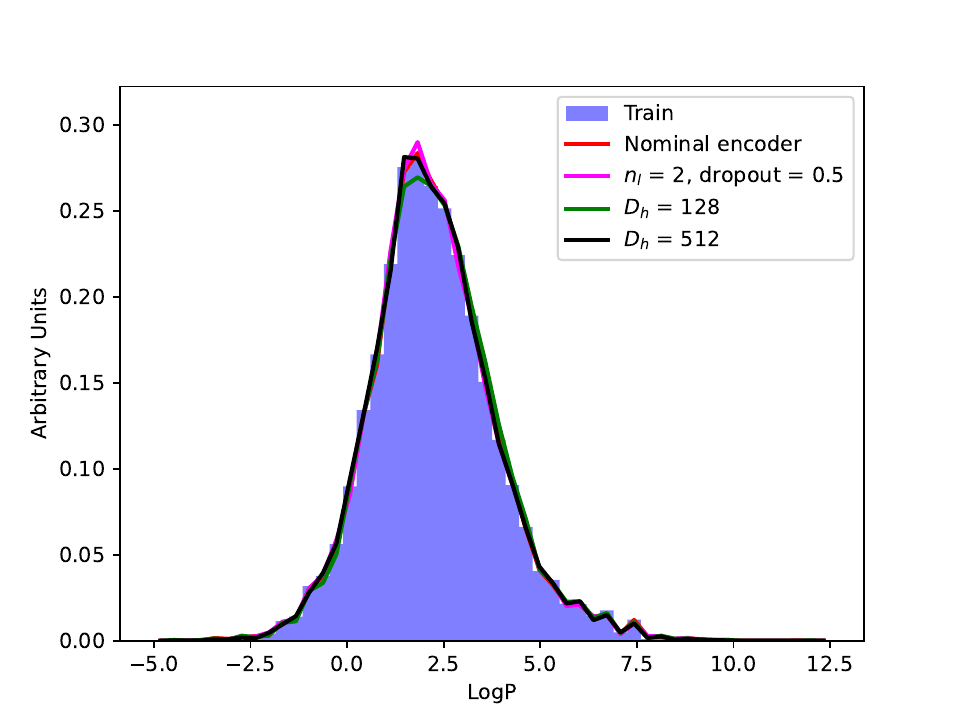}
    \includegraphics[width=0.32\linewidth]{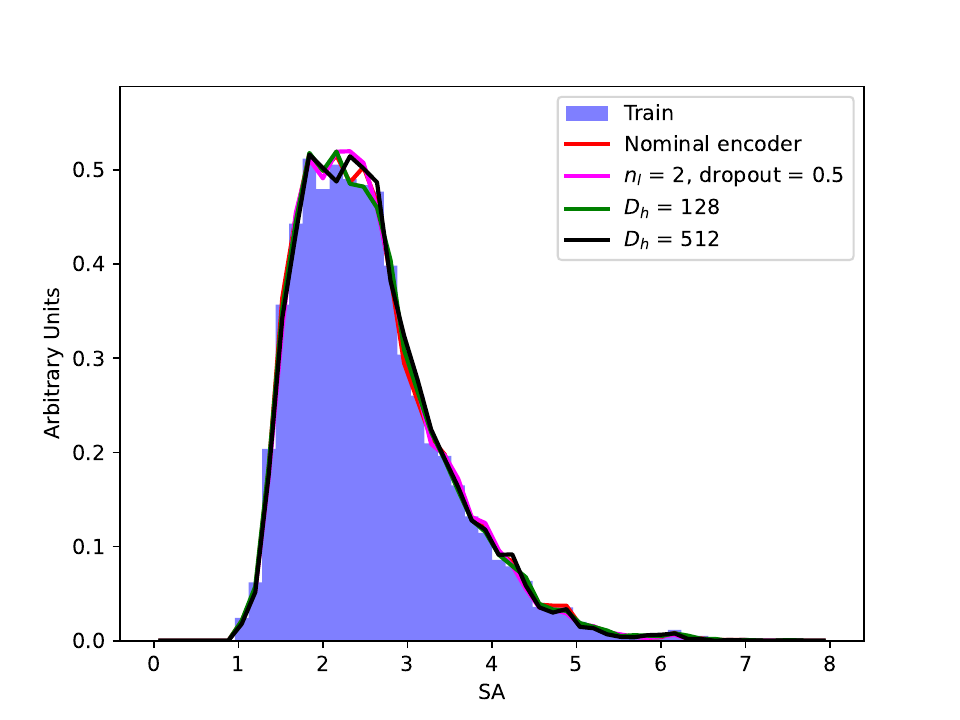}
    \includegraphics[width=0.32\linewidth]{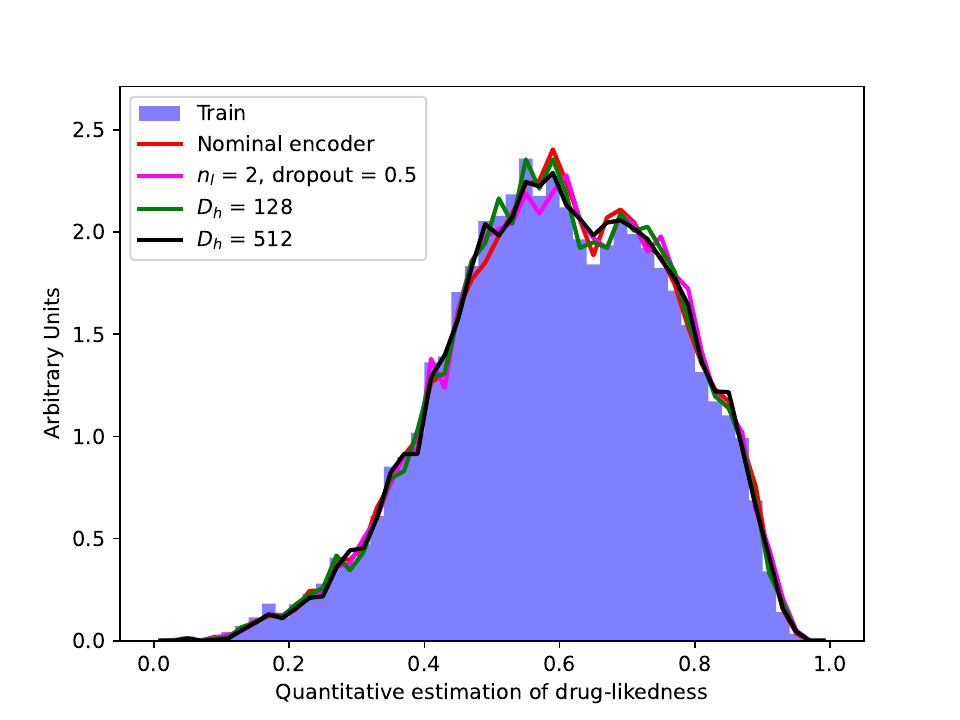}
    \includegraphics[width=0.32\linewidth]{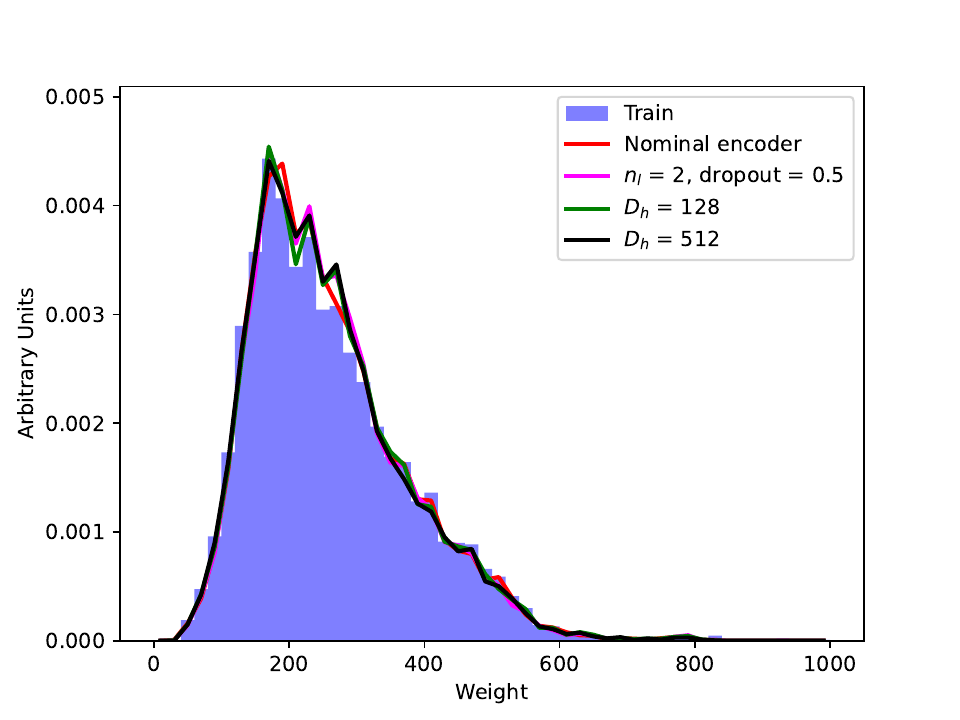}
    \caption{Metrics distributions for models trained with different encoder settings. Results are very similar across different scenarios.}
    \label{fig:app:vae_opt:enc}
\end{figure}

Table~\ref{tab:app:vae_opt:enc} and Figure~\ref{fig:app:vae_opt:enc} present the results of the encoder optimization. Although there seems to be a slight inclination to increase the hidden dimension to 512, changing the number of layers to two does not provide any improvement.%, and varying dropout has no effect at all.

% Requires: \usepackage{booktabs}
\begin{table}[h!]
    \centering
    \begin{tabular}{lc|cc|cc|cc}
        \toprule
        $n_{l}$      & 3      & \textbf{2}      & \textbf{4}      & 3      & 3      & 3      & 3      \\
        $D_h$     & 512    & 512    & 512    & \textbf{256}    & \textbf{1024}   & 512    & 512    \\
        Dropout        & 0.0    & 0.0    & 0.0    & 0.0    & 0.0    & \textbf{0.25}   & \textbf{0.75}   \\
        \midrule
        $\epsilon_d$       & 0.479 & 0.590   & 0.487 & 0.718 & 0.416 & 0.480 & 0.556 \\
        $\epsilon_v$       & 0.889  & 0.818 & 0.869 & 0.718 & 0.944 & 0.911 & 0.875 \\
        $\epsilon_v$       & 0.996 & 0.990 & 0.996 & 0.988 & 0.997 & 0.996 & 0.994 \\
%        $N_{canonical}$    & 12,725  & 14,333  & 12,637  & 15,270  & 11,767  & 13,066  & 14,501  \\
        Novelty               & 0.237 & 0.429  & 0.240 & 0.573 & 0.126 & 0.248 & 0.378 \\
        IntDiv                & 0.896 & 0.896  & 0.897 & 0.894 & 0.898 & 0.897 & 0.896 \\
        $\mathcal{W}(\text{LogP})$         & 0.087 & 0.098  & 0.088 & 0.096 & 0.087 & 0.084 & 0.081 \\
        $\mathcal{W}(\text{SA})$           & 0.275 & 0.244 & 0.252 & 0.289  & 0.268 & 0.250 & 0.270 \\
        $\mathcal{W}(\text{QED})$          & 0.025 & 0.025 & 0.025 & 0.023  & 0.027  & 0.025 & 0.023  \\
        $\mathcal{W}(\text{Weight})$       & 43.9 & 48.4 & 41.7 & 51.3 & 45.3 & 42.6 & 43.6 \\
        \bottomrule
    \end{tabular}
            \caption{Comparisons of the results of VAE decoder parameter tuning, the first column corresponds to the nominal scenario. The scenario variation is highlighted at the top of the table. The default decoder dropout is zero regardless the number of layers.}
    \label{tab:app:vae_opt:dec}
\end{table}

\begin{figure}[h!]
    \centering
    \includegraphics[width=0.32\linewidth]{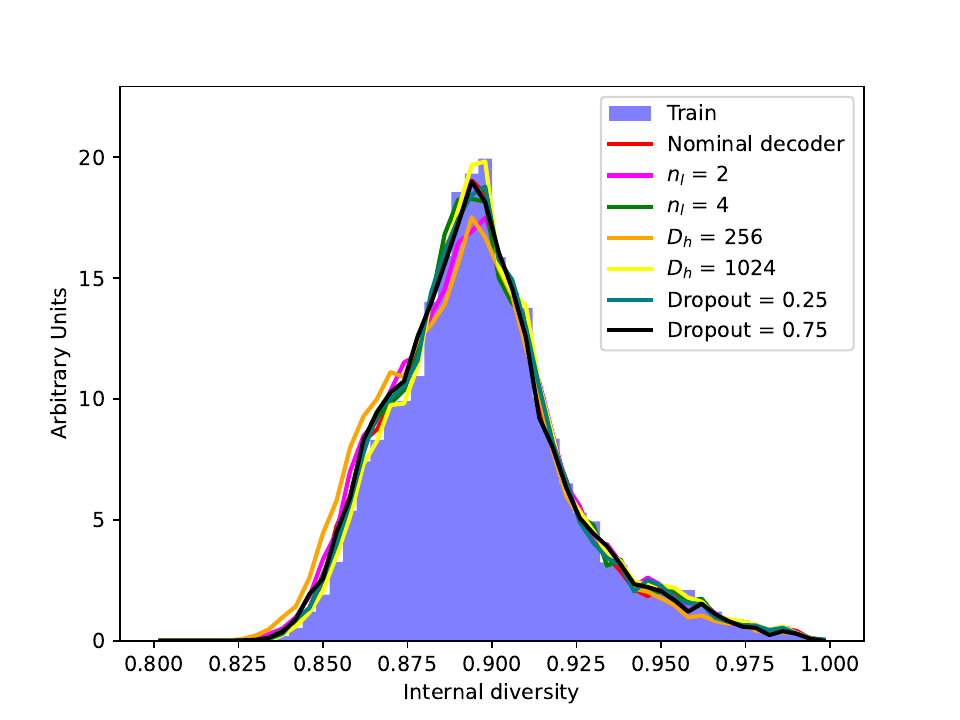}
    \includegraphics[width=0.32\linewidth]{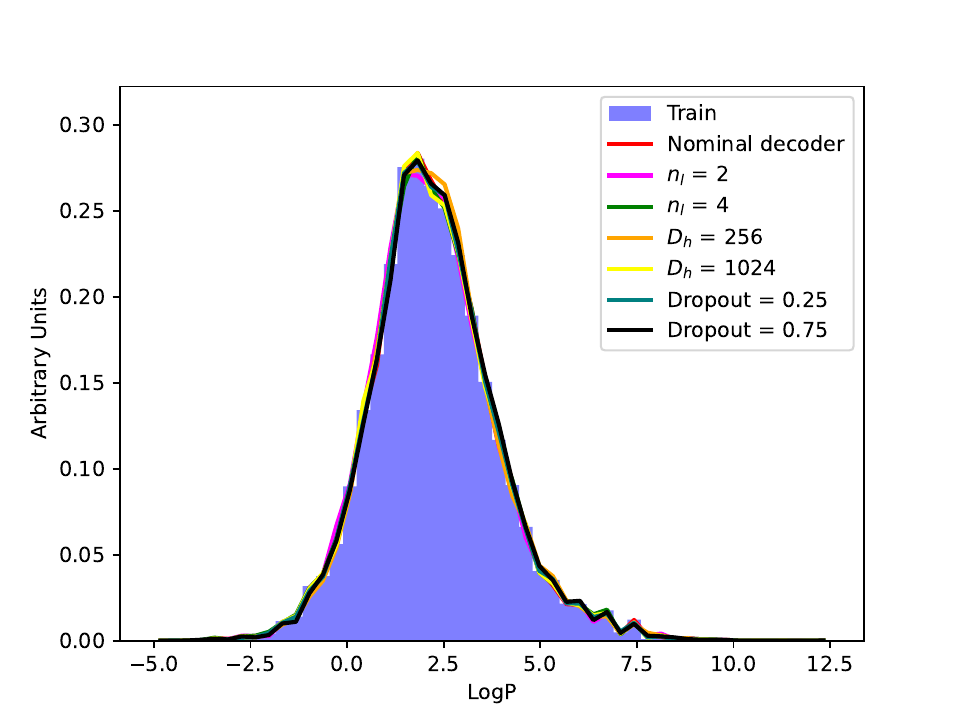}
    \includegraphics[width=0.32\linewidth]{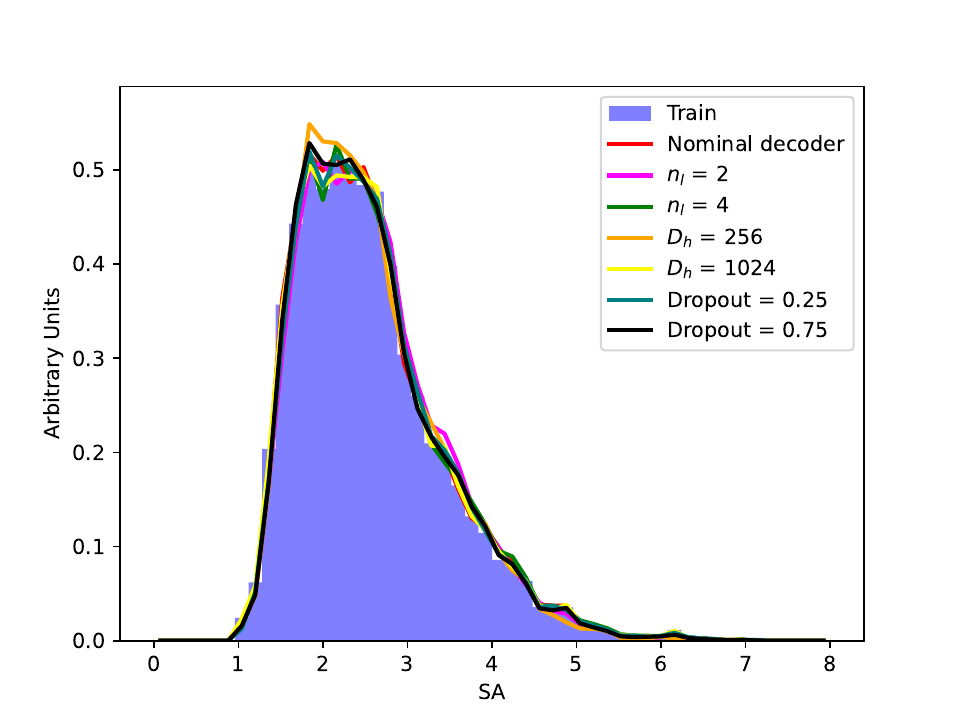}
    \includegraphics[width=0.32\linewidth]{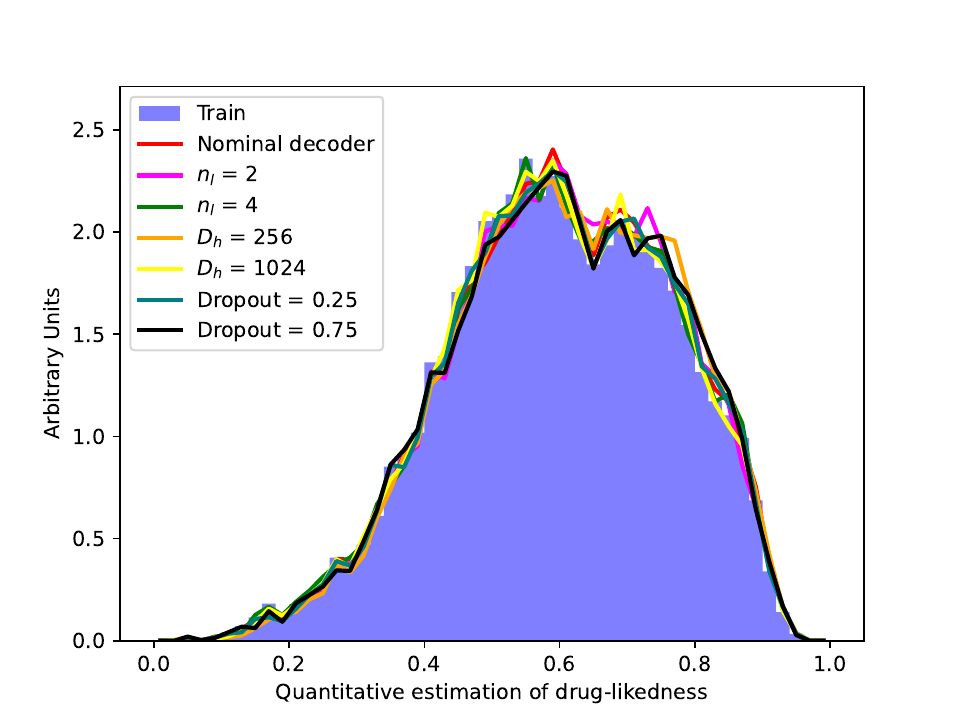}
    \includegraphics[width=0.32\linewidth]{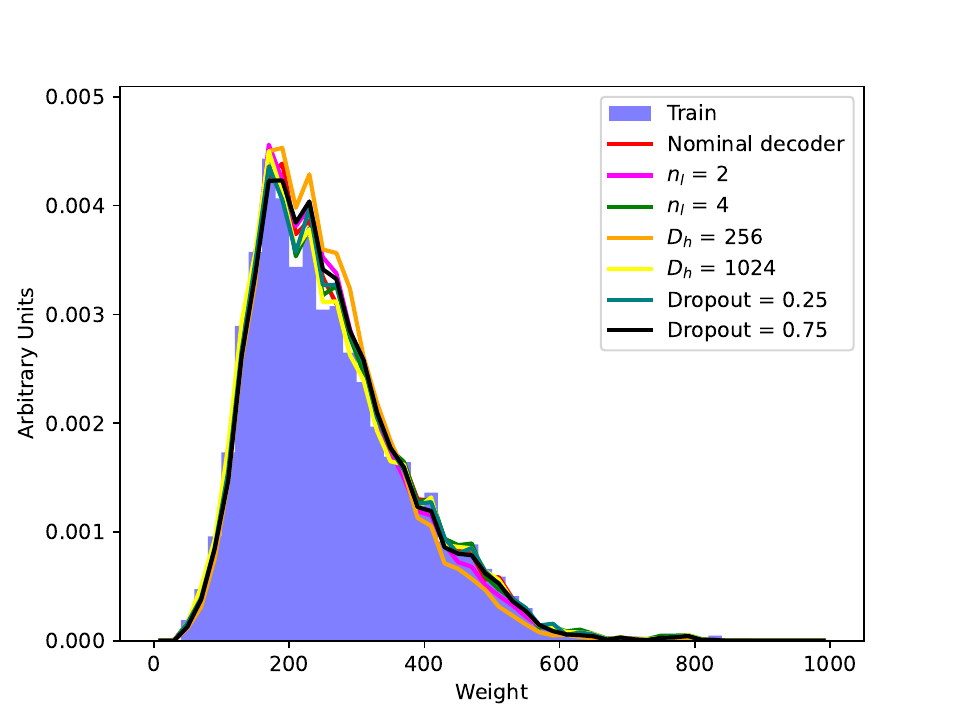}
    \caption{Metrics distributions for models trained with different decoder settings. Results are very similar across different scenarios.}
    \label{fig:app:vae_opt:dec}
\end{figure}

Table~\ref{tab:app:vae_opt:dec} and Figure~\ref{fig:app:vae_opt:dec} present the results of the decoder optimization. Unlike the encoder, there seems to be a preference for increasing the dropout to 0.75 and the number of layers to four, while the hidden layer dimension stays the same.

\FloatBarrier

\section{Supplementary material: Test of the VAE decoder}
\label{sec:app:vae_dec}

In order to be able to claim that the VAE is functioning properly, it is important to test whether its decoding part is able to distinguish between different inputs. In other words, whether the decoder is sensitive to the shape of the distributions of the latent vectors or whether any distribution collapses to the same output.

The base latent space of dimension 10 was used, and numbers drawn from the following distributions were decoded and processed using the standard metric evaluation:

\begin{enumerate}
\item The analysis training sample of 12,000 SMILES strings encoded by the VAE;
\item The analysis test sample of \ntest\ SMILES strings encoded by the VAE;
\item 10,000 samples drawn from a uniform distribution between 0.0 and 1.0;
\item 10,000 samples drawn from a normal distribution with $\mu=$ 0.0 and $\sigma=$ 1.0;
\item 10,000 samples drawn from a log-normal distribution with $\mu=$ 0.0 and $\sigma=$ 0.558 shifted by two units to the negative values;
\item 10,000 samples drawn from a $sin$ distribution with three peaks in the range $[-3;3]$.
\end{enumerate}

\begin{figure}[h!]
    \centering
    \includegraphics[width=1\linewidth]{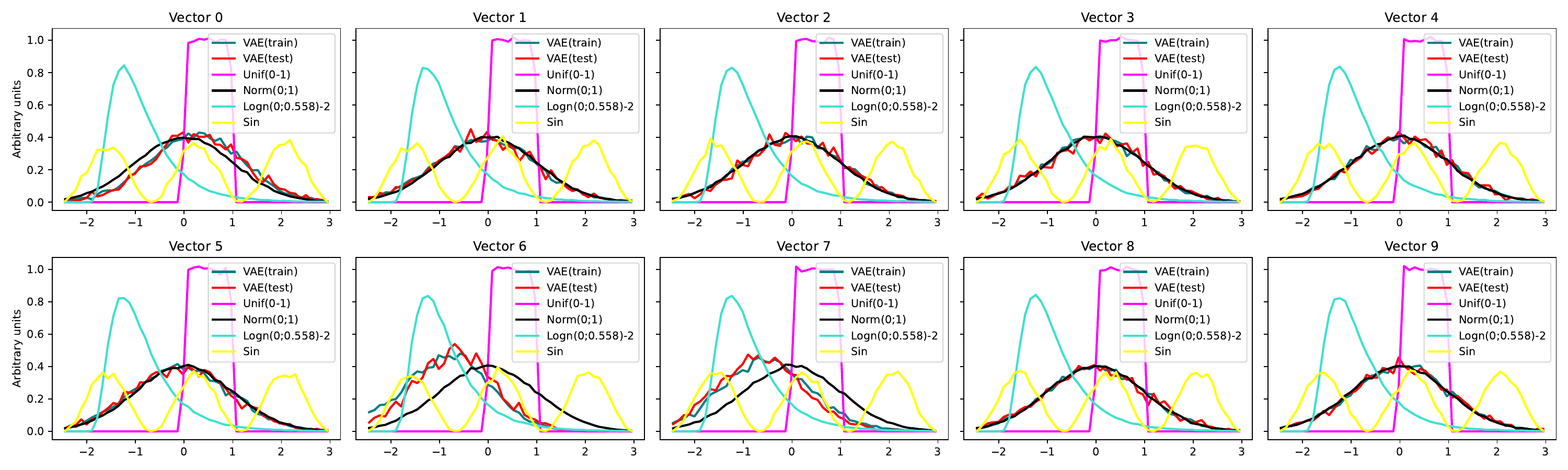}
    \caption{Distributions of 10 latent vectors used as inputs to the VAE decoder.}
    \label{sec:app:vae_dec_latent}
\end{figure}

The distributions of the 10 vectors are presented in Figure~\ref{sec:app:vae_dec_latent}, clearly showing the different shapes. The first two samples corresponding to the train and test encoded datasets exhibit similar behavior consistent with the normal(0;1) distribution, except for vectors number six and seven.

\begin{figure}[h!]
    \centering
    \includegraphics[width=0.32\linewidth]{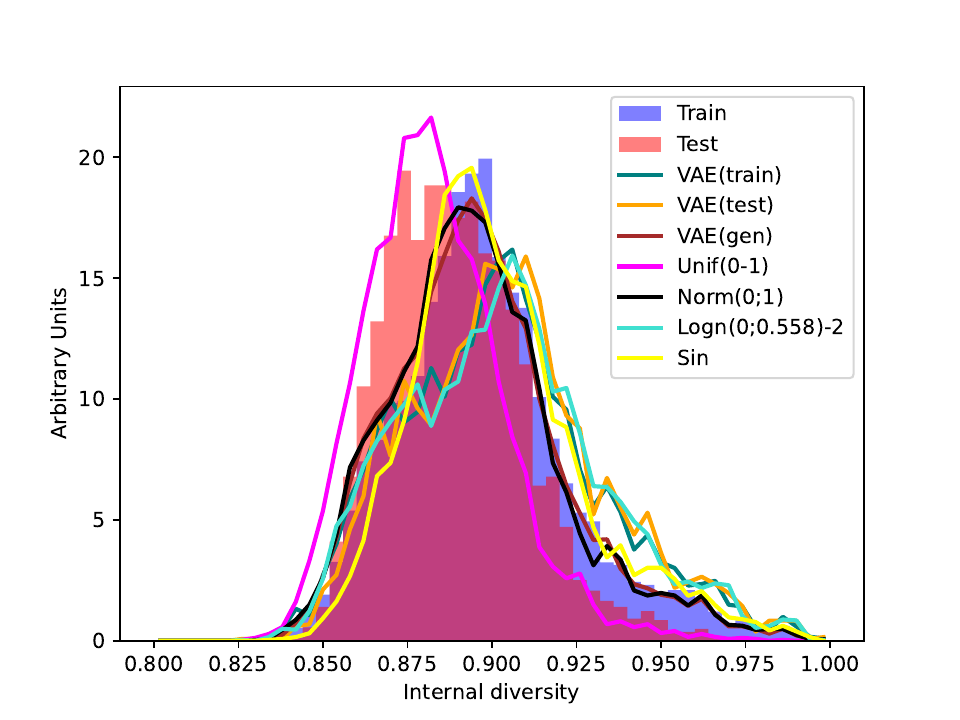}
    \includegraphics[width=0.32\linewidth]{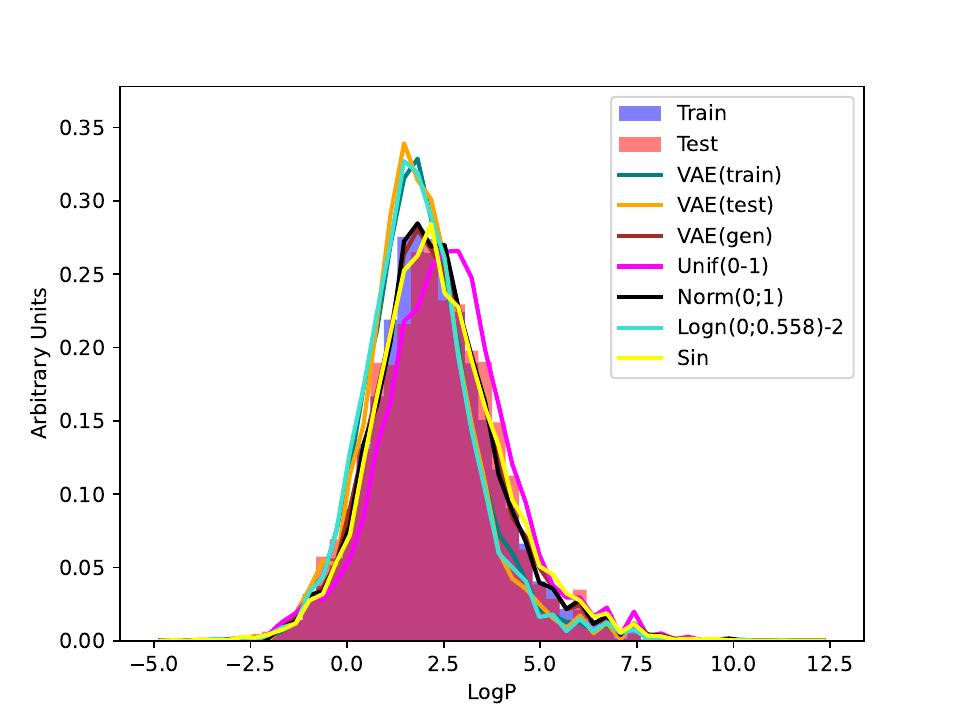}
    \includegraphics[width=0.32\linewidth]{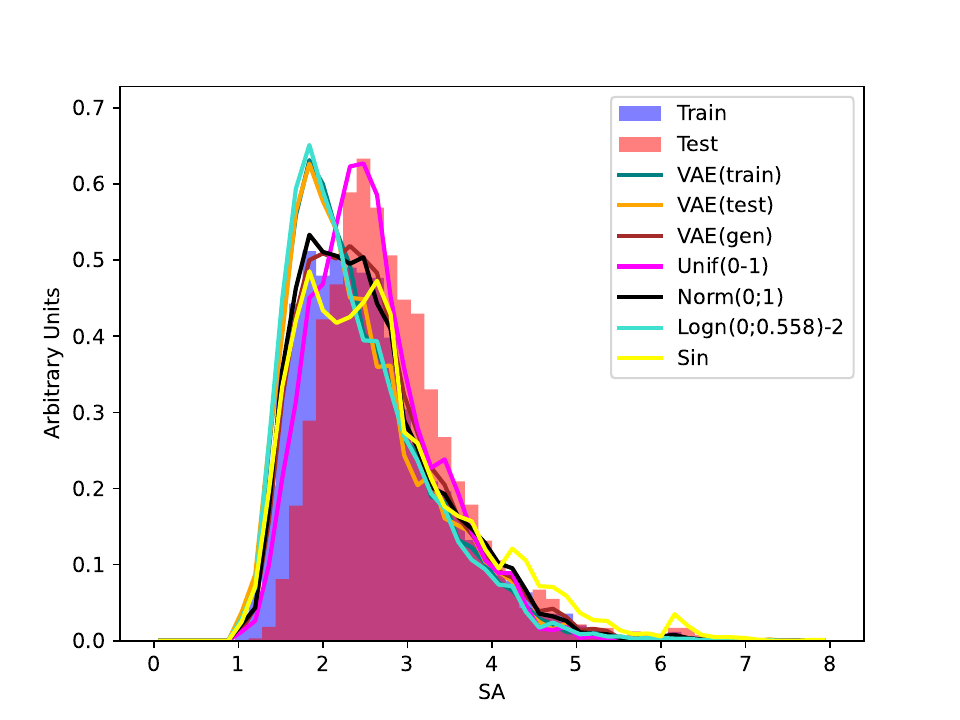}
    \includegraphics[width=0.32\linewidth]{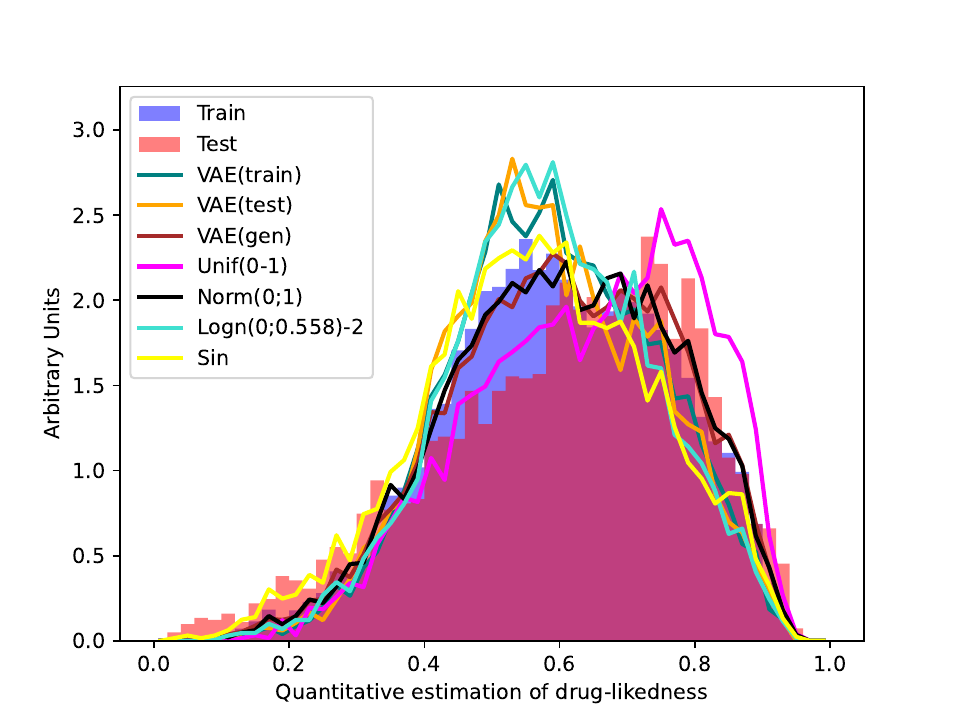}
    \includegraphics[width=0.32\linewidth]{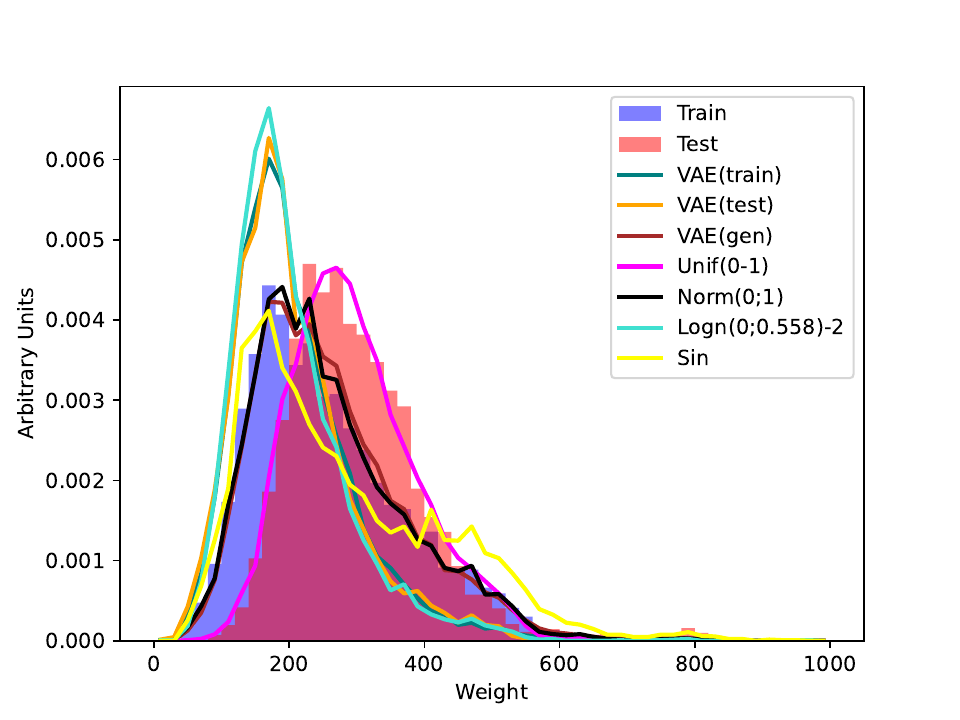}
    \caption{Metric distribution obtained from SMILES sets which were generated by the VAE decoder from various latent vector distributions. Additional scenario -- VAE(gen) -- corresponds to a set which is generated by the standard MOSES sampling procedure. The distributions are normalized to unit area.}
    \label{sec:app:vae_dec_metric}
\end{figure}

The sets of SMILES strings produced by the VAE decoder were subsequently evaluated in order to obtain the metrics. The comparison is in Figure~\ref{sec:app:vae_dec_metric}.

The two shaded histograms correspond to the initial train and test samples without any use of VAE and clearly exhibit distinct behavior. However, the same samples, when encoded and decoded by the VAE (red and green), end up very consistent with each other but not with either the original train or test samples. This might hint at some level of information loss in the process of encoding and decoding the sample through a latent space of dimension only 10.

The VAE(gen) is very similar to the distribution originating from the uniform sampling in the latent space (brown and black). This is not very surprising because the internal MOSES VAE routine uses a uniform distribution as input for the sampler.

The uniform distribution has the most extreme shape difference with respect to the others (flat vs peak-y structure) and also shows the largest deviations in the metric distributions. The VAE decoder seems to treat all distributions that exhibit a peak structure in a similar way.

Surprisingly, the metrics distribution originating in the log-normal scenario seems to be the most similar to the VAE(test) and VAE(gen) samples. That might be explained by the fact that the input distributions for vectors six and seven seem to be more consistent for the VAE(train) and VAE(test) with the log-normal rather than normal distributions. That would also hint that these two vectors are the ones with the largest discriminating power. 

However, one can conclude that the VAE decoder is working as intended and is capable of producing SMILES sets with different feature characteristics depending on the shape of the latent vectors.

\section{Supplementary material: Correlations between latent vectors}
\label{app:corr_mat}

To complement the study of the latent space of dimension 10, this section presents the correlation matrices for the train, test, classical GAN, and \qgan\ datasets. While the train and test latent samples are simply a transformation of the SMILES sets through the VAE encoder, the samples for the GANs stem from the respective generators.

\begin{figure}[h!]
    \centering
    \includegraphics[width=0.48\linewidth]{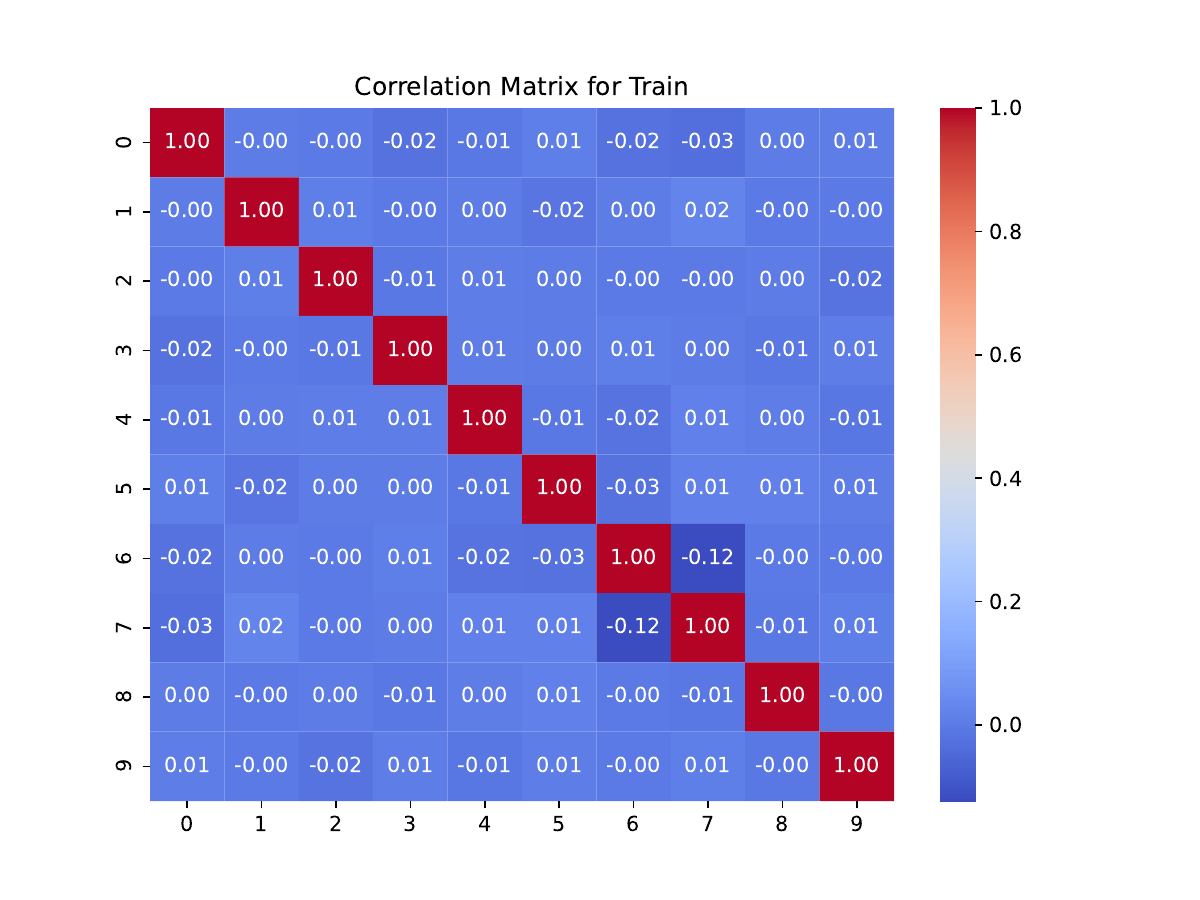}
    \includegraphics[width=0.48\linewidth]{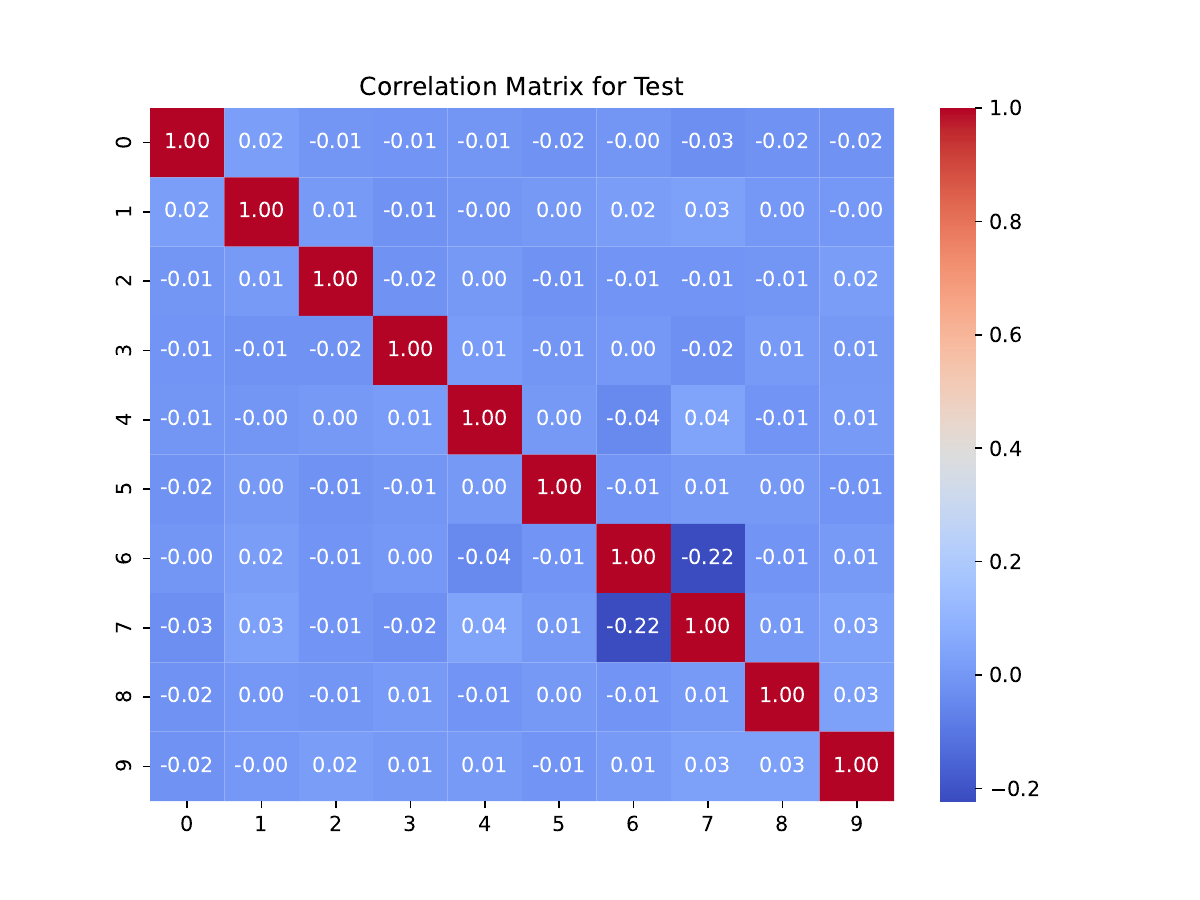}
    \includegraphics[width=0.48\linewidth]{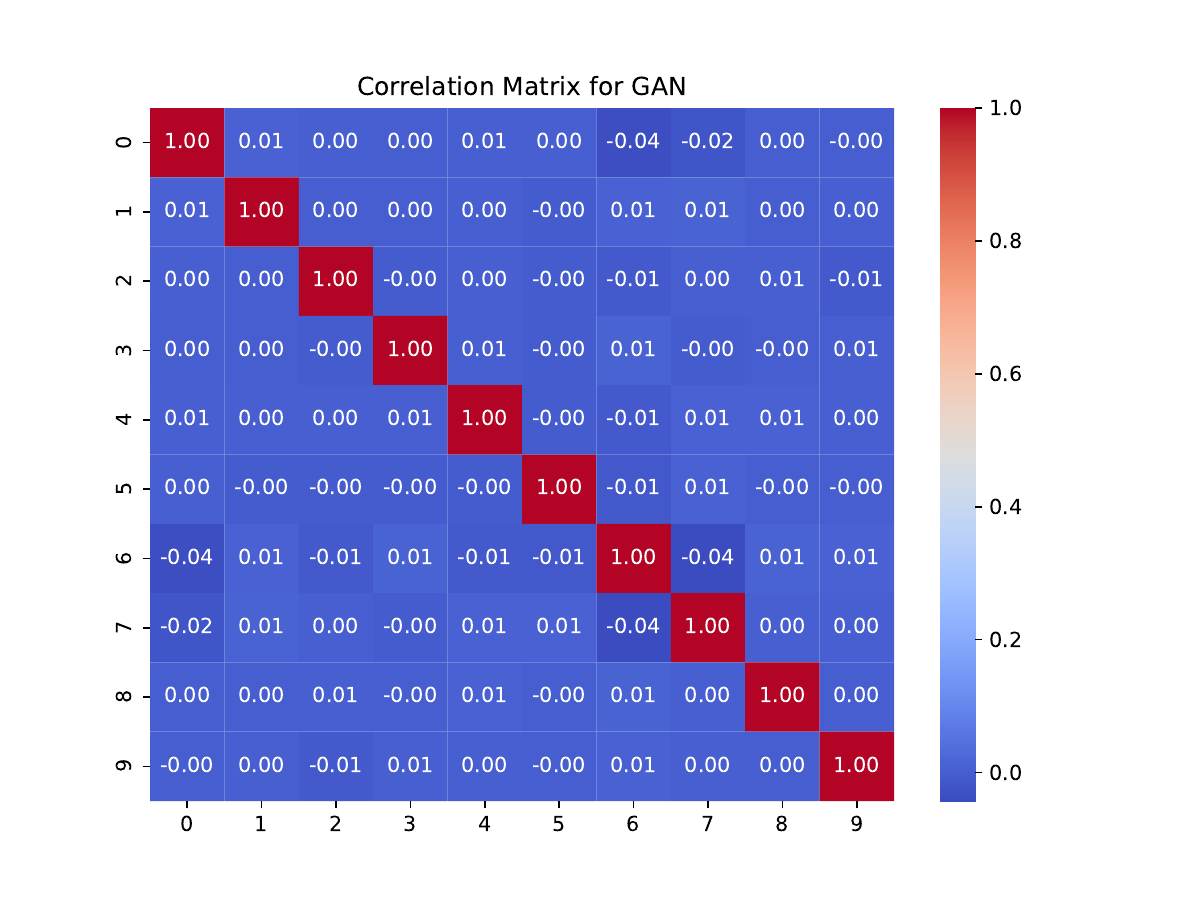}
    \includegraphics[width=0.48\linewidth]{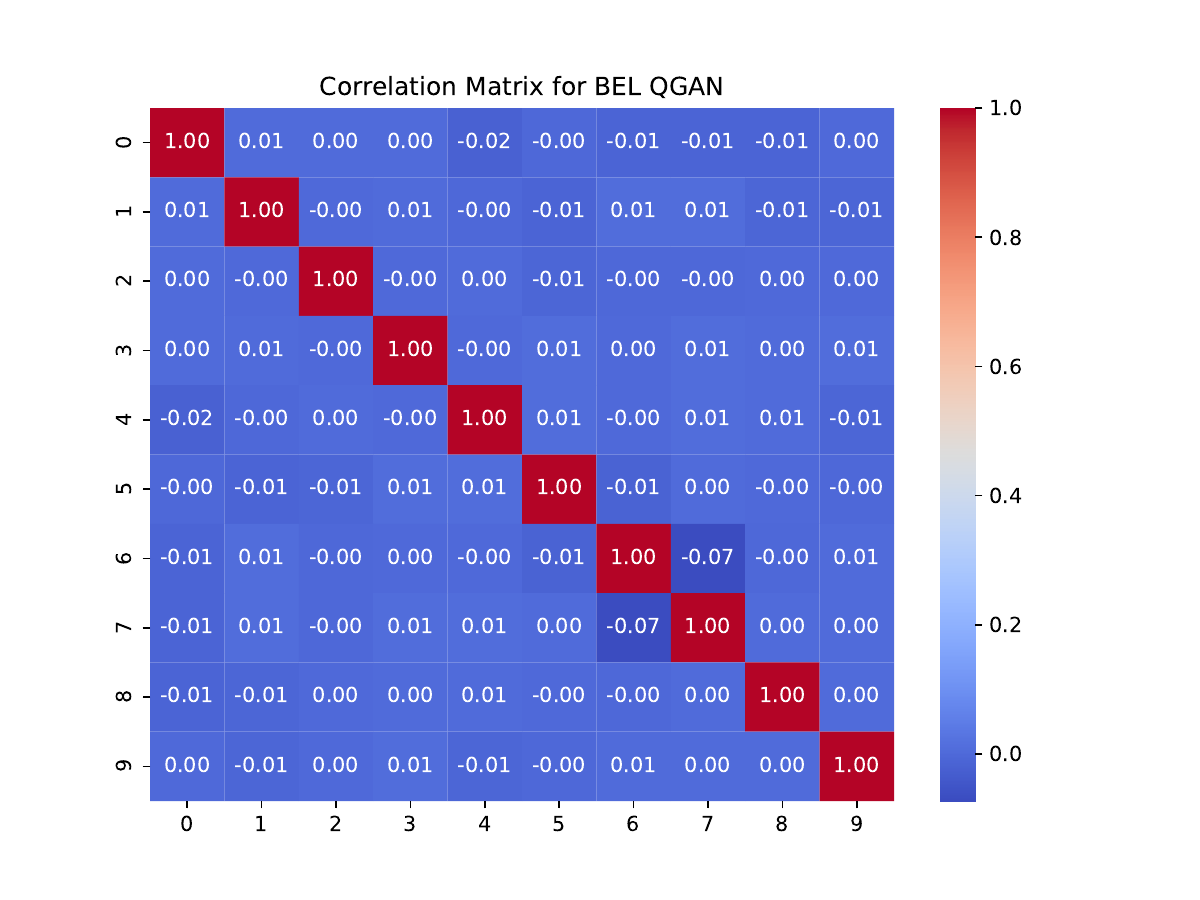}

    \caption{Correlations between individual latent vectors for the training set, test set, classical GAN training output, and BEL \qgan\ training output, all for a latent dimension of 10.}
    \label{fig:app:corr_mat}
\end{figure}

Figure~\ref{fig:app:corr_mat} presents the correlation matrices. All vectors apart from numbers six and seven are almost perfectly uncorrelated, while the numbers six and seven show a small level of anti-correlation. This effect has different sizes in the four samples and is manifested by differing distributions for the respective vectors as presented in Figure~\ref{fig:cgan_qgan_lat}.

\section{Supplementary material: Supporting material for \qgan\ scenarios}
\label{sec:app:scen}

In this section, all detailed results supporting summary Table~\ref{tab:scen} are presented. Each of the following scenarios is represented by the corresponding metrics table and set of plots, for easy orientation, see Table~\ref{tab:app:scen:overview}.

\begin{table}[h!]
    \centering
    \begin{tabular}{c|ccc|cc}
        \toprule
    \multirow{2}{*}{Latent dimension} & \multicolumn{3}{c|}{\qgan} & \multirow{2}{*}{Table} & \multirow{2}{*}{Figures} \\ 
  & $n_{qb}$ & $n_{l}$ & Readout & & \\
         \midrule
10 & 5 & 2 & Dual & \ref{tab:app:scen:cqgan_5qb2lrs} & \ref{fig:cgan_qgan_nom} \\
10 & 5 & 4 & Dual & \ref{tab:app:scen:cqgan_5qb4lrs} & \ref{fig:app:scen:cqgan_5qb4lrs} \\
10 & 10 & 2 & Single & \ref{tab:app:scen:cqgan_10qb2lrs} & \ref{fig:app:scen:cqgan_10qb2lrs} \\
10 & 10 & 4, 6 & Single & \ref{tab:app:scen:cqgan_10qb46lrs} & \ref{fig:app:scen:cqgan_10qb46lrs} \\
20 & 10 & 2 & Dual & \ref{tab:app:scen:cqgan_latdim20} & \ref{fig:app:scen:cqgan_latdim20} \\
30 & 15 & 2 & Dual & \ref{tab:app:scen:cqgan_latdim30} & \ref{fig:app:scen:cqgan_latdim30} \\
10, 20, 30 & -- & -- & -- & \ref{tab:app:scen:cgan_latdim} & \ref{fig:app:scen:cgan_latdim} \\
        \bottomrule

    \end{tabular}
        \caption{Summary of the different scenarios and links to the corresponding individual tables and figures.}
    \label{tab:app:scen:overview}
\end{table}

All detailed results support the conclusion from Section~\ref{sec:results:scen}. Tables \ref{tab:app:scen:cqgan_latdim20} and \ref{tab:app:scen:cqgan_latdim30} use the GAN with the corresponding latent dimension for comparison with the QGANs. The average statistical significance is well below one standard deviation and hence statistically compatible with the results from the scenario comparison Table \ref{tab:scen}.

\begin{table}[h!]
    \centering
\begin{tabular}{lc|cc|cc}
\toprule
Metrics & Classical tuned & Styled simple \qgan & $Z_0$ & styled BEL \qgan & $Z_0$\\
\midrule
$N_{params}$  & 705,162  & 20 &  & 110 &  \\
\midrule
    $\epsilon_d$ & $0.481 \pm 0.031$ & $0.529 \pm 0.003$ & $+1.56$ & $0.523 \pm 0.003$ & $+1.36$ \\
    $\epsilon_v$ & $0.917 \pm 0.012$ & $0.831 \pm 0.002$ & $-7.21$ & $0.875 \pm 0.003$ & $-3.43$ \\
    $\epsilon_u$ & $0.986 \pm 0.003$ & $0.994 \pm 0.000$ & $+3.50$ & $0.993 \pm 0.001$ & $+2.90$ \\
    Novelty & $0.536 \pm 0.015$ & $0.572 \pm 0.003$ & $+2.40$ & $0.548 \pm 0.003$ & $+0.82$ \\
    IntDiv & $0.898 \pm 0.004$ & $0.882 \pm 0.000$ & $-4.60$ & $0.883 \pm 0.000$ & $-4.21$ \\
    Filters & $0.721 \pm 0.011$ & $0.737 \pm 0.003$ & $+1.43$ & $0.719 \pm 0.002$ & $-0.17$ \\
    $\epsilon_{LogP}$ & $0.898 \pm 0.007$ & $0.883 \pm 0.001$ & $-2.35$ & $0.896 \pm 0.002$ & $-0.23$ \\
    $\langle \mathrm{SA} \rangle$ & $2.391 \pm 0.080$ & $2.575 \pm 0.007$ & $-2.31$ & $2.448 \pm 0.007$ & $-0.71$ \\
    $\langle \mathrm{QED} \rangle$ & $0.599 \pm 0.007$ & $0.641 \pm 0.001$ & $+6.17$ & $0.643 \pm 0.001$ & $+6.55$ \\
    $\langle \mathrm{Weight} \rangle$ & $203.9 \pm 10.6$ & $294.8 \pm 0.8$ & $-8.56$ & $261.4 \pm 0.8$ & $-5.42$ \\
        \midrule
$\langle Z_0 \rangle$  & Reference & -- & -1.00 & -- & -0.26 \\
\bottomrule
\end{tabular}
        \caption{Tuned classical GAN vs \qgans\ with $n_{qb} = 5$, $n_{l} = 2$, and dual readout.}
    \label{tab:app:scen:cqgan_5qb2lrs}
\end{table}

% Requires: \usepackage{booktabs}
\begin{table}[h!]
    \centering
    \begin{tabular}{lc|cc|cc}
        \toprule
        Metrics & Classical tuned & Styled simple \qgan & $Z_0$ & Styled BEL \qgan & $Z_0$ \\
        \midrule
        $N_{params}$ & 705,162 & 40 & & 210 & \\
        \midrule
    $\epsilon_d$ & $0.481 \pm 0.031$ & $0.526 \pm 0.003$ & $+1.46$ & $0.522 \pm 0.004$ & $+1.32$ \\
    $\epsilon_v$ & $0.917 \pm 0.012$ & $0.830 \pm 0.002$ & $-7.24$ & $0.876 \pm 0.004$ & $-3.36$ \\
    $\epsilon_u$ & $0.986 \pm 0.003$ & $0.994 \pm 0.000$ & $+3.42$ & $0.993 \pm 0.001$ & $+2.80$ \\
    Novelty      & $0.536 \pm 0.015$ & $0.577 \pm 0.003$ & $+2.72$ & $0.548 \pm 0.004$ & $+0.76$ \\
    IntDiv       & $0.898 \pm 0.004$ & $0.882 \pm 0.000$ & $-4.66$ & $0.883 \pm 0.000$ & $-4.21$ \\
    Filters      & $0.721 \pm 0.011$ & $0.740 \pm 0.003$ & $+1.64$ & $0.719 \pm 0.003$ & $-0.12$ \\
    $\epsilon_{LogP}$ & $0.898 \pm 0.007$ & $0.882 \pm 0.003$ & $-2.29$ & $0.896 \pm 0.002$ & $-0.36$ \\
    $\langle \mathrm{SA} \rangle$ & $2.391 \pm 0.080$ & $2.587 \pm 0.007$ & $-2.45$ & $2.450 \pm 0.006$ & $-0.74$ \\
    $\langle \mathrm{QED} \rangle$ & $0.599 \pm 0.007$ & $0.641 \pm 0.001$ & $+6.11$ & $0.644 \pm 0.001$ & $+6.61$ \\
        $\langle \mathrm{Weight} \rangle$ & $203.9 \pm 10.6$ & $297.4 \pm 0.6$ & $-8.82$ & $261.2 \pm 0.5$ & $-5.41$ \\
        \midrule
        $\langle Z_0 \rangle$  & Reference & -- & $-1.01$ & -- & $-0.27$ \\
        \bottomrule
    \end{tabular}
        \caption{Tuned classical GAN vs \qgans\ with $n_{qb} = 5$, $n_{l} = 4$, and dual readout.}
    \label{tab:app:scen:cqgan_5qb4lrs}
\end{table}

\begin{figure}[h!]
    \centering
    \includegraphics[width=0.32\linewidth]{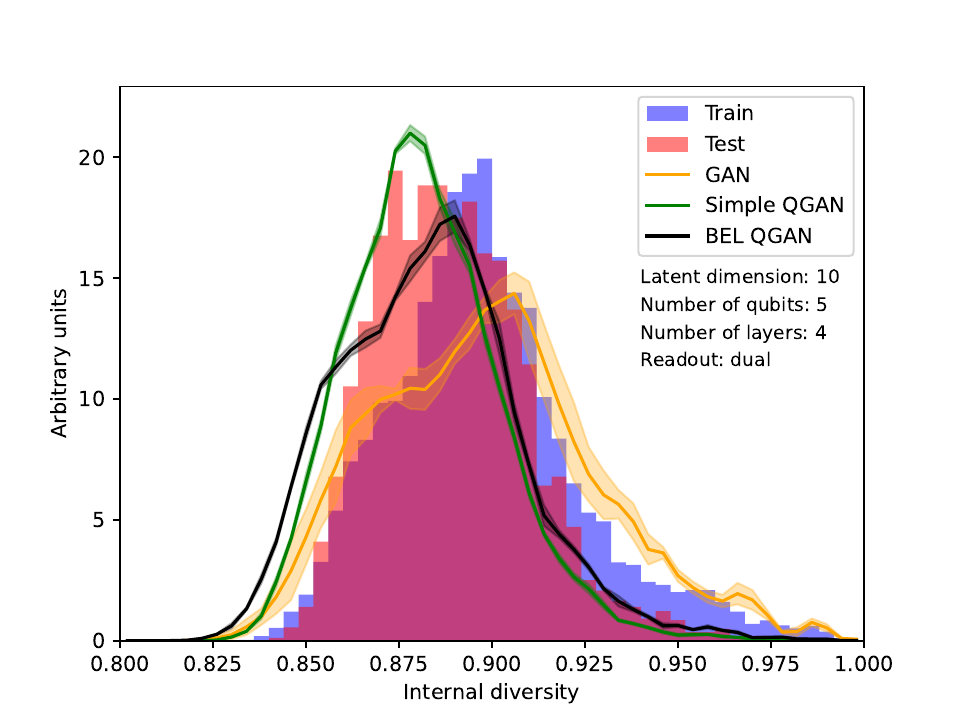}
    \includegraphics[width=0.32\linewidth]{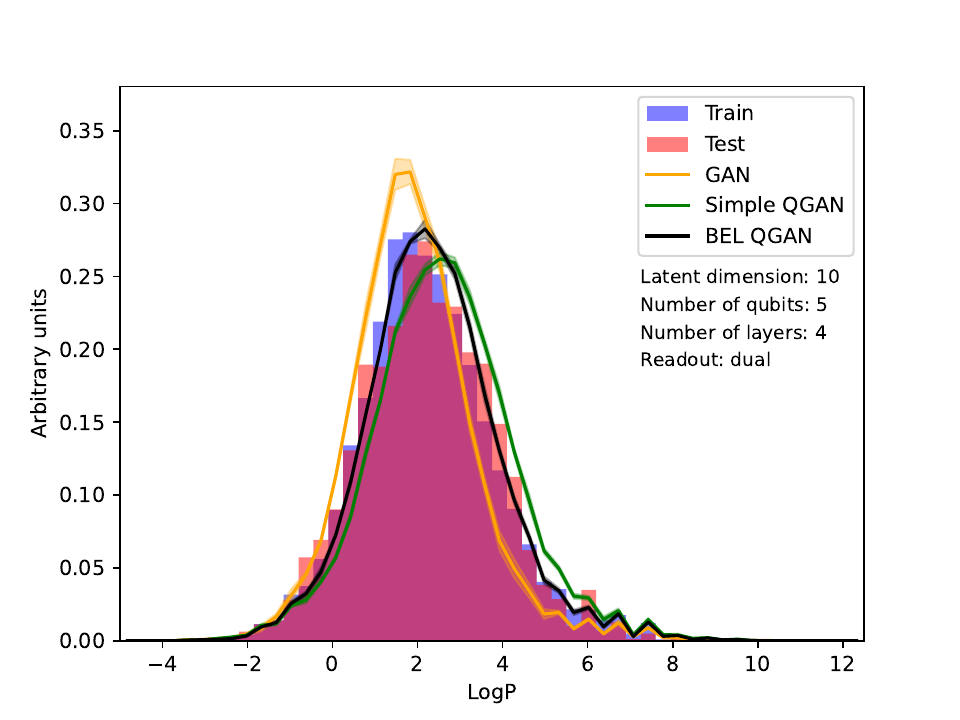}
    \includegraphics[width=0.32\linewidth]{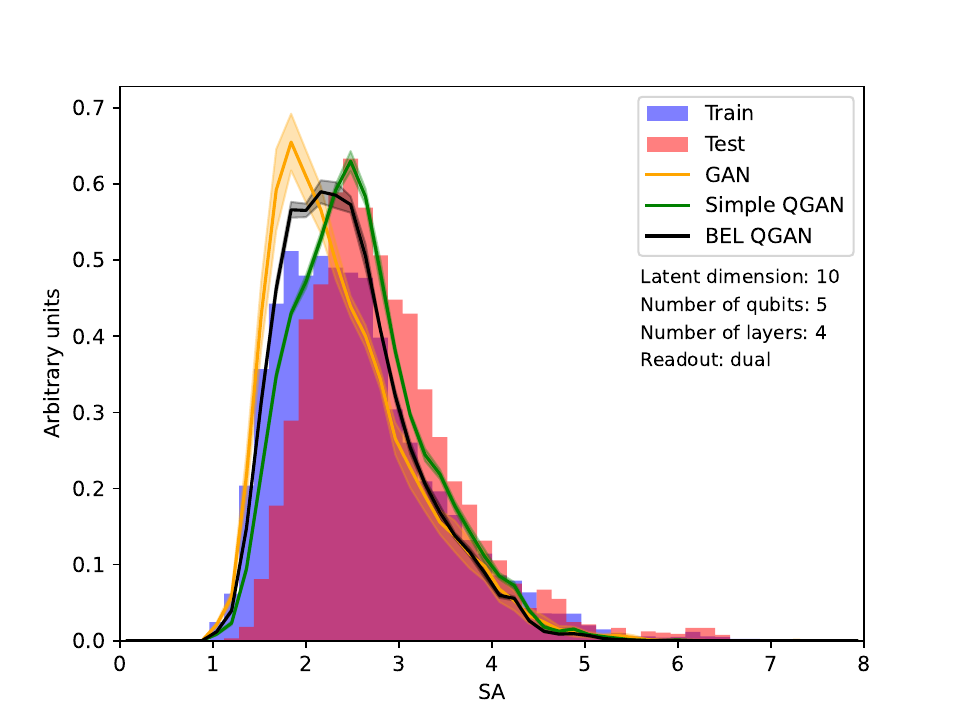}
    \includegraphics[width=0.32\linewidth]{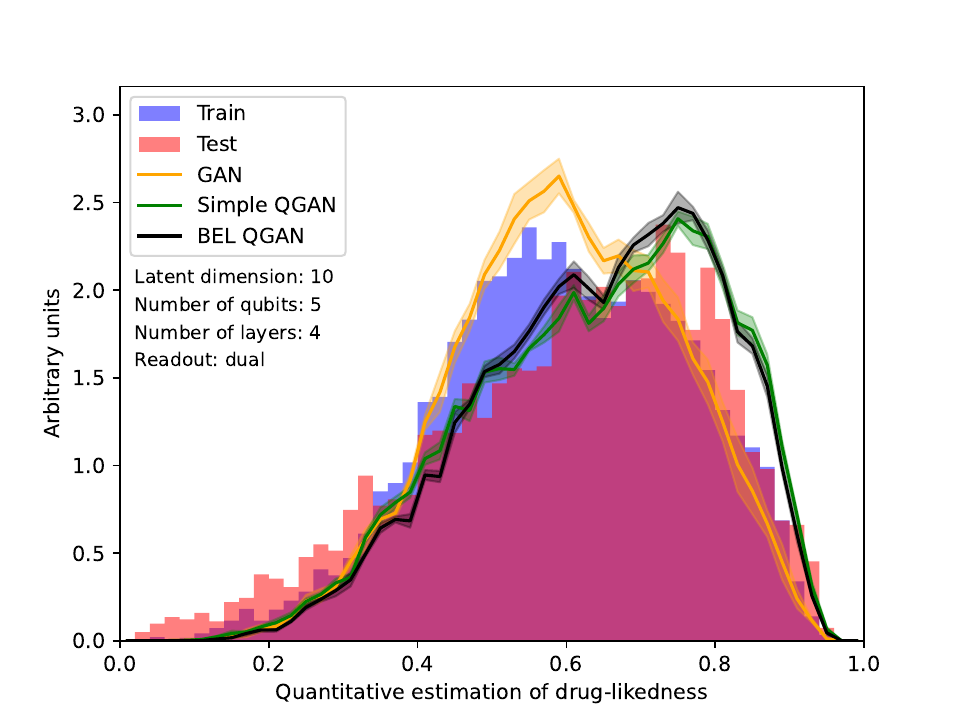}
    \includegraphics[width=0.32\linewidth]{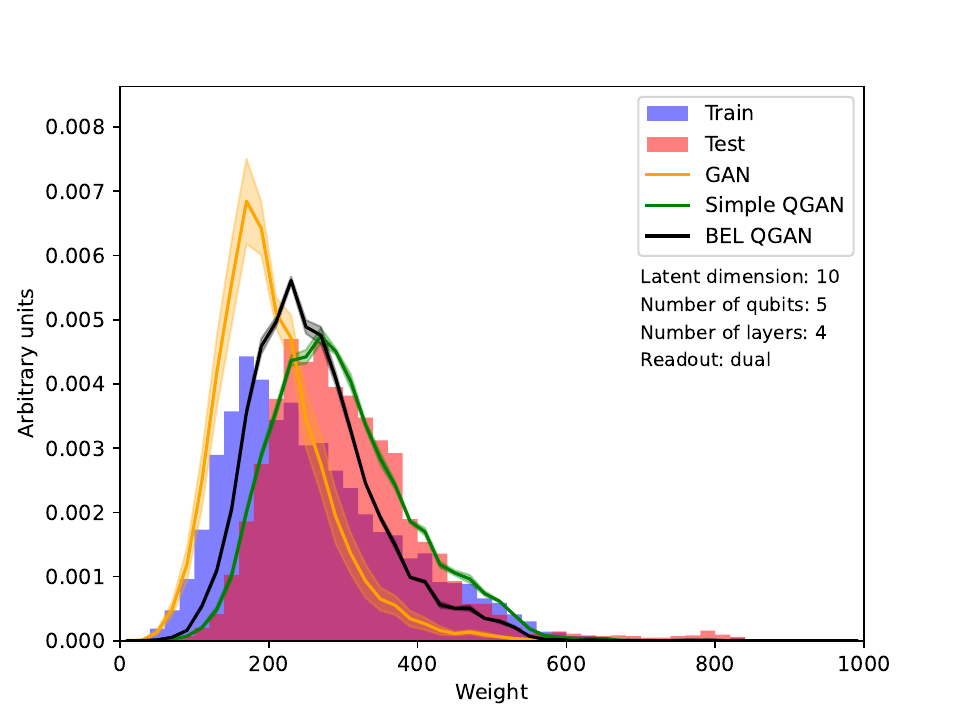}
    \caption{Metrics distributions for classical and quantum models. All GANs are trained with a latent dimension of 10. The quantum GANs use $n_{qb}=5$, $n_{l}=4$, and dual readout.}
    \label{fig:app:scen:cqgan_5qb4lrs}
\end{figure}

% Requires: \usepackage{booktabs}
\begin{table}[h!]
    \centering
    \begin{tabular}{lc|cc|cc}
        \toprule
        Metrics & Classical tuned & Styled simple \qgan & $Z_0$ & Styled BEL \qgan & $Z_0$ \\
        \midrule
        $N_{params}$ & 716,692 & 40 &  & 220 &  \\
        \midrule
    $\epsilon_d$ & $0.481 \pm 0.031$ & $0.514 \pm 0.004$ & $+1.06$ & $0.514 \pm 0.002$ & $+1.06$ \\
    $\epsilon_v$ & $0.917 \pm 0.012$ & $0.833 \pm 0.003$ & $-6.99$ & $0.887 \pm 0.003$ & $-2.45$ \\
    $\epsilon_u$ & $0.986 \pm 0.003$ & $0.994 \pm 0.000$ & $+3.27$ & $0.992 \pm 0.001$ & $+2.45$ \\
    Novelty & $0.536 \pm 0.015$ & $0.586 \pm 0.003$ & $+3.34$ & $0.548 \pm 0.002$ & $+0.82$ \\
    IntDiv & $0.898 \pm 0.004$ & $0.882 \pm 0.000$ & $-4.68$ & $0.884 \pm 0.000$ & $-3.99$ \\
    Filters & $0.721 \pm 0.011$ & $0.737 \pm 0.003$ & $+1.40$ & $0.712 \pm 0.004$ & $-0.70$ \\
    $\epsilon_{LogP}$  & $0.898 \pm 0.007$ & $0.885 \pm 0.002$ & $-1.91$ & $0.900 \pm 0.001$ & $+0.30$ \\
    $\langle \text{SA} \rangle$ & $2.391 \pm 0.080$ & $2.569 \pm 0.005$ & $-2.23$ & $2.415 \pm 0.007$ & $-0.30$ \\
    $\langle \text{QED} \rangle$ & $0.599 \pm 0.007$ & $0.645 \pm 0.001$ & $+6.81$ & $0.641 \pm 0.001$ & $+6.19$ \\
    $\langle \text{Weight} \rangle$ & $203.9 \pm 10.6$ & $293.6 \pm 1.1$ & $-8.43$ & $250.1 \pm 0.9$ & $-4.36$ \\
        \midrule
        $\langle Z_0 \rangle$ & Reference & -- & $-0.84$ & -- & $-0.10$ \\
        \bottomrule
    \end{tabular}
        \caption{Tuned classical GAN vs \qgans\ with $n_{qb} = 10$, $n_{l} = 2$, and single readout.}
    \label{tab:app:scen:cqgan_10qb2lrs}
\end{table}

\begin{figure}[h!]
    \centering
    \includegraphics[width=0.32\linewidth]{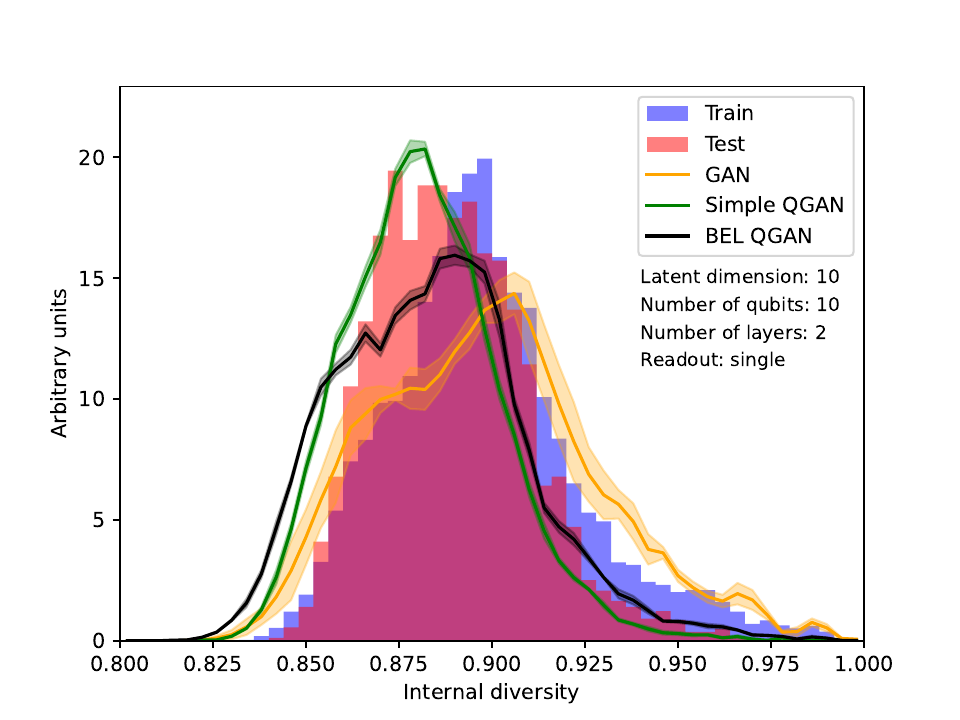}
    \includegraphics[width=0.32\linewidth]{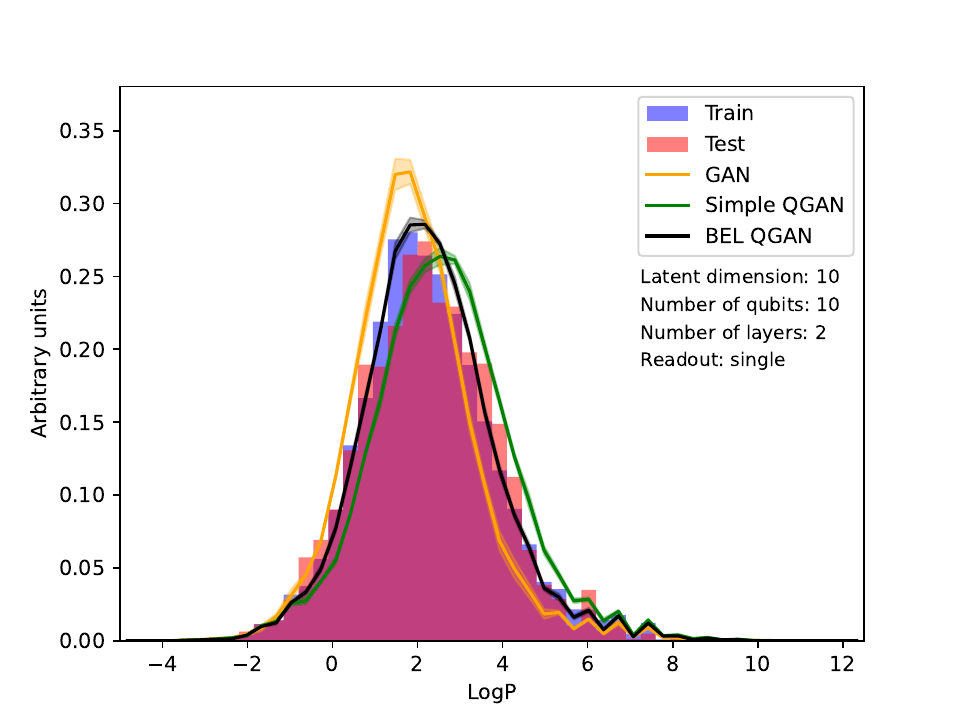}
    \includegraphics[width=0.32\linewidth]{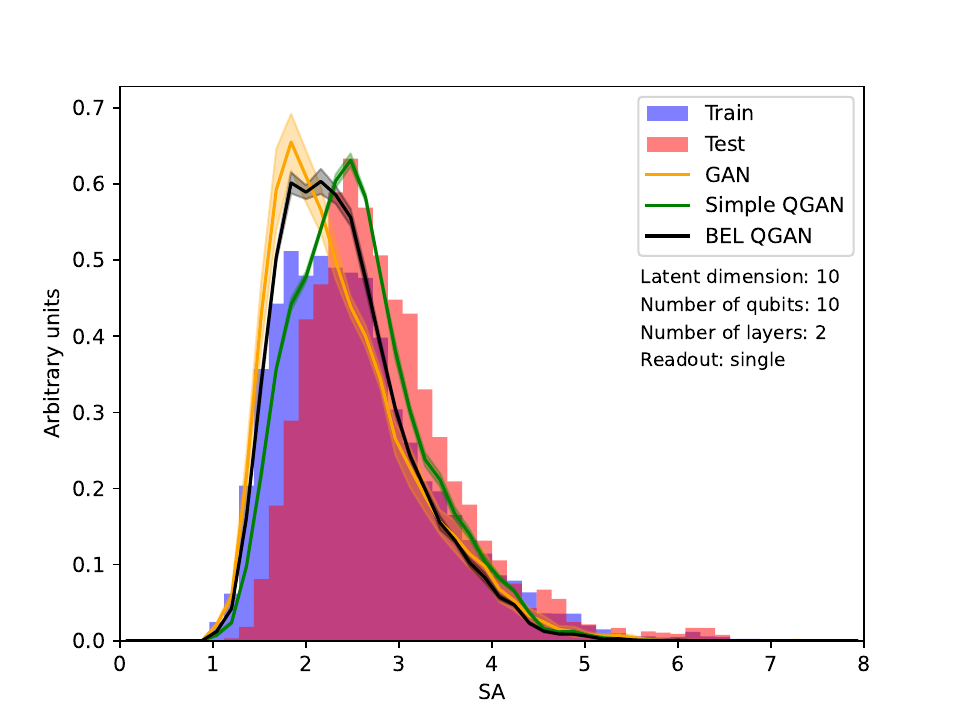}
    \includegraphics[width=0.32\linewidth]{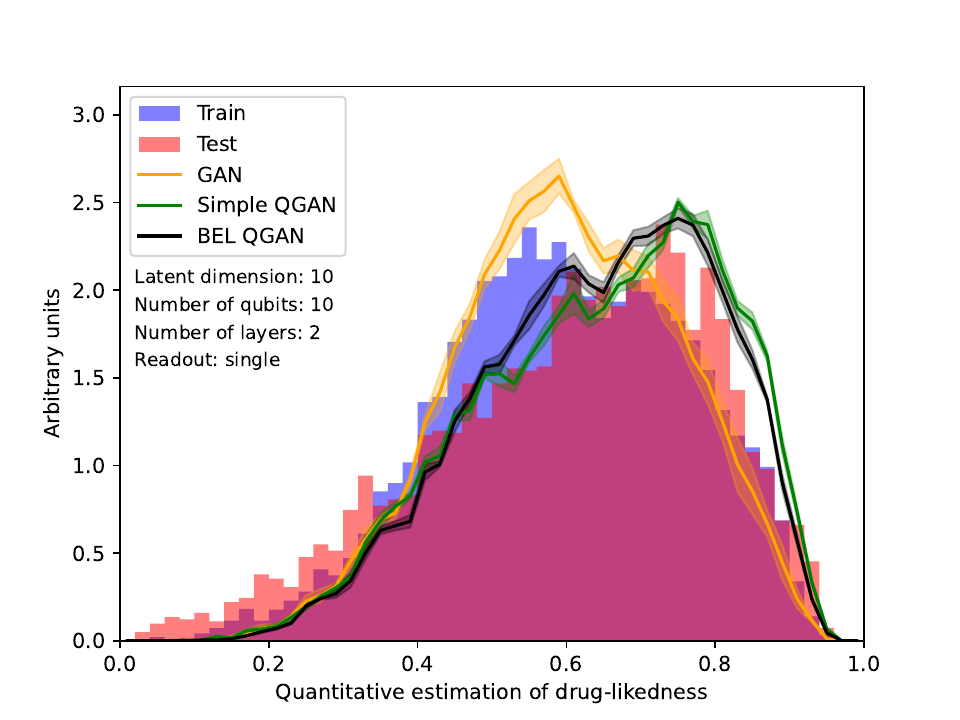}
    \includegraphics[width=0.32\linewidth]{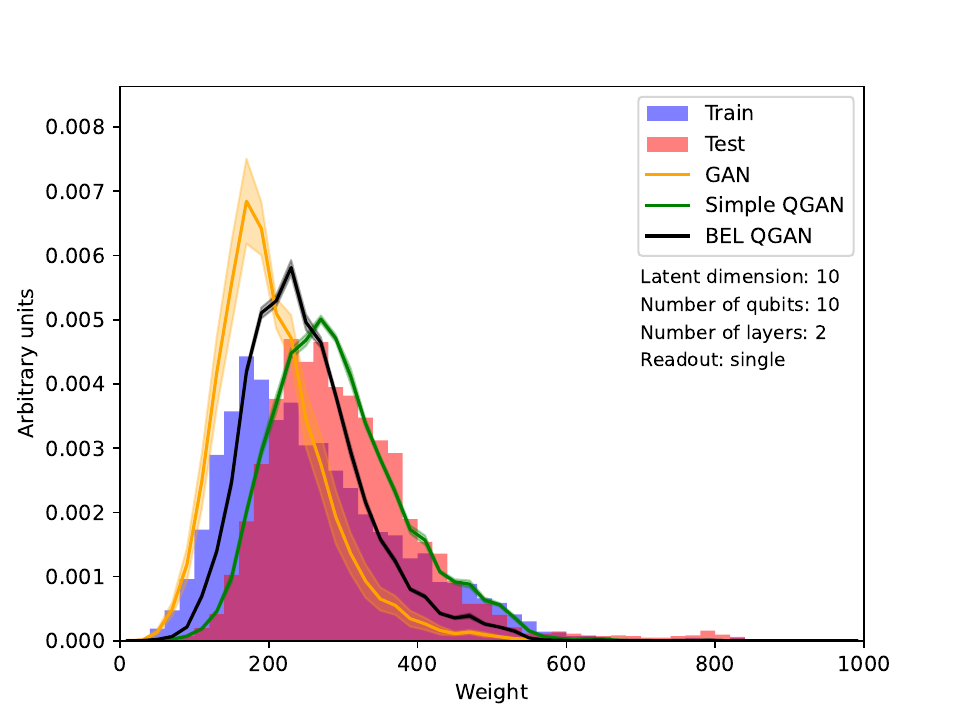}
    \caption{Metrics distributions for classical and quantum models. All GANs are trained with a latent dimension of 10. The quantum GANs use $n_{qb}=10$, $n_{l}=2$, and single readout.}
    \label{fig:app:scen:cqgan_10qb2lrs}
\end{figure}

% Requires: \usepackage{booktabs}
\begin{table}[h!]
    \centering
    \resizebox{\columnwidth}{!}{
    \begin{tabular}{lc|cc|cc|cc}
        \toprule
        Metrics & Classical tuned & Styled simple \qgan & $Z_0$ & Styled BEL \qgan  & $Z_0$ & styled BEL & $Z_0$\\
        \midrule
        $n_{l}$ & - & 4 & & 4 & & 6 & \\
         \midrule
        $N_{params}$ & 705,162 & 80 & & 420 & & 620 &       \\
         \midrule
    $\epsilon_d$ & $0.481 \pm 0.031$ & $0.515 \pm 0.005$ & $+1.08$ & $0.516 \pm 0.005$ & $+1.13$ & $0.515 \pm 0.003$ & $+1.10$ \\
    $\epsilon_v$ & $0.917 \pm 0.012$ & $0.833 \pm 0.002$ & $-6.95$ & $0.887 \pm 0.003$ & $-2.42$ & $0.888 \pm 0.002$ & $-2.41$ \\
    $\epsilon_u$ & $0.986 \pm 0.003$ & $0.994 \pm 0.000$ & $+3.26$ & $0.992 \pm 0.001$ & $+2.48$ & $0.992 \pm 0.001$ & $+2.62$ \\
    Novelty                                 & $0.536 \pm 0.015$ & $0.586 \pm 0.002$ & $+3.37$ & $0.549 \pm 0.002$ & $+0.87$ & $0.550 \pm 0.001$ & $+0.93$ \\
    IntDiv                                  & $0.898 \pm 0.004$ & $0.882 \pm 0.000$ & $-4.69$ & $0.884 \pm 0.000$ & $-4.01$ & $0.884 \pm 0.000$ & $-3.99$ \\
    Filters                                 & $0.721 \pm 0.011$ & $0.737 \pm 0.003$ & $+1.41$ & $0.714 \pm 0.002$ & $-0.56$ & $0.712 \pm 0.003$ & $-0.77$ \\
    $\epsilon_{LogP}$                   & $0.898 \pm 0.007$ & $0.885 \pm 0.003$ & $-1.87$ & $0.900 \pm 0.002$ & $+0.25$ & $0.897 \pm 0.002$ & $-0.11$ \\
    $\langle \text{SA} \rangle$             & $2.391 \pm 0.080$ & $2.571 \pm 0.007$ & $-2.25$ & $2.417 \pm 0.011$ & $-0.32$ & $2.421 \pm 0.004$ & $-0.37$ \\
    $\langle \text{QED} \rangle$            & $0.599 \pm 0.007$ & $0.645 \pm 0.001$ & $+6.83$ & $0.641 \pm 0.001$ & $+6.15$ & $0.641 \pm 0.001$ & $+6.23$ \\
    $\langle \text{Weight} \rangle$         & $203.9 \pm 10.6$ & $293.7 \pm 1.2$ & $-8.44$ & $251.1 \pm 1.9$ & $-4.39$ & $249.9 \pm 0.9$ & $-4.33$ \\
        \midrule
        $\langle Z_0 \rangle$  & Reference  & -- & -0.83 & --  & -0.08 & - & -0.11 \\
        \bottomrule
    \end{tabular}}
        \caption{Tuned classical GAN vs \qgans\ with $n_{qb} = 10$, $n_{l} = 4, 6$, and single readout.}
    \label{tab:app:scen:cqgan_10qb46lrs}
\end{table}

\begin{figure}[h!]
    \centering
    \includegraphics[width=0.32\linewidth]{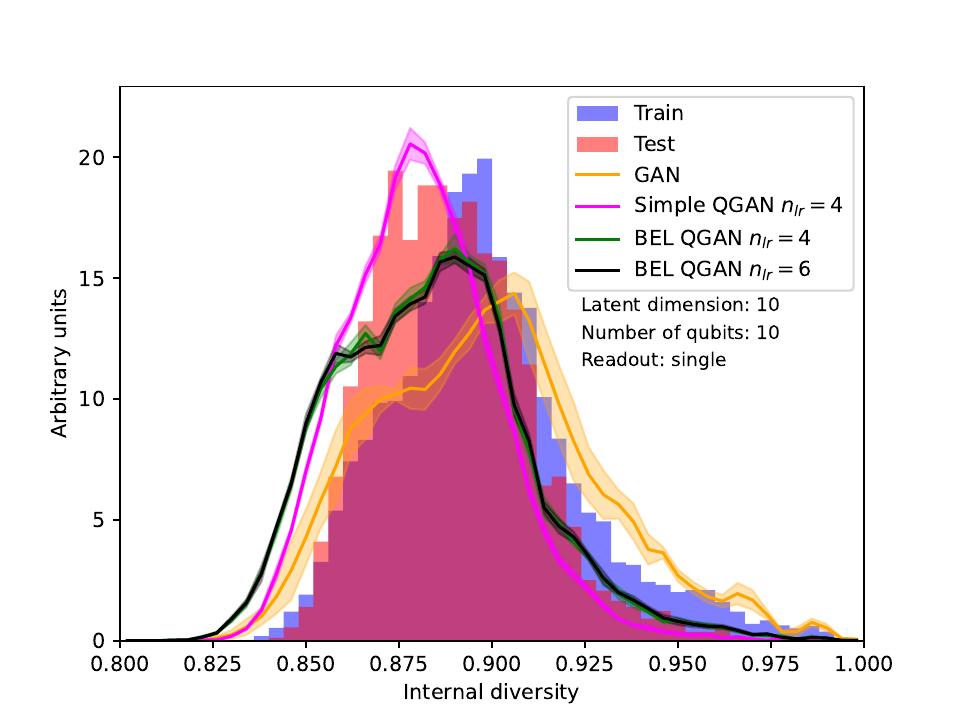}
    \includegraphics[width=0.32\linewidth]{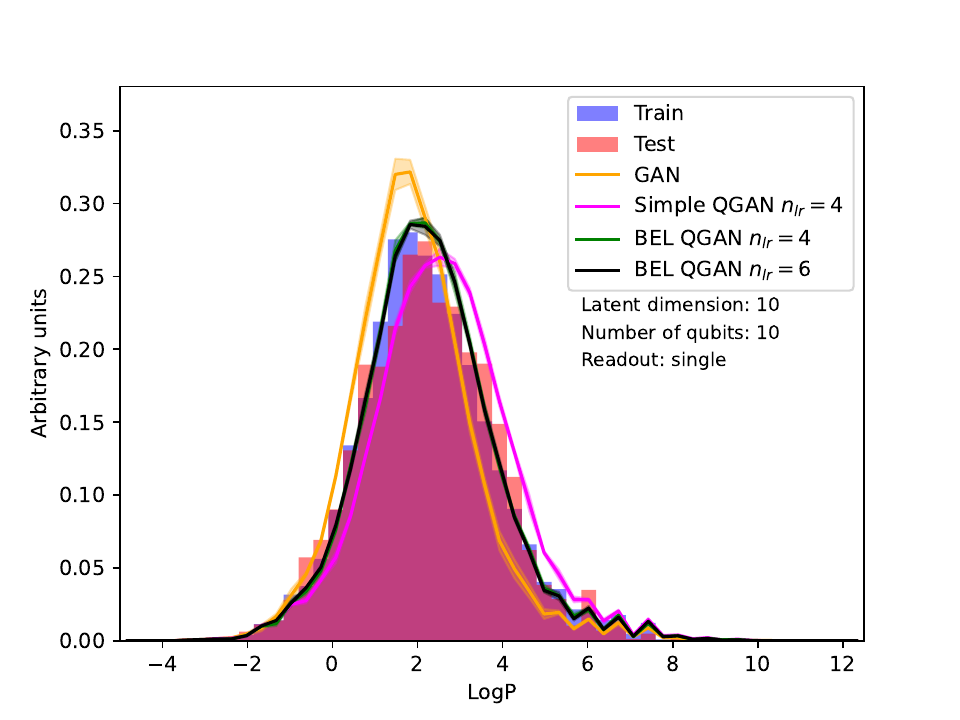}
    \includegraphics[width=0.32\linewidth]{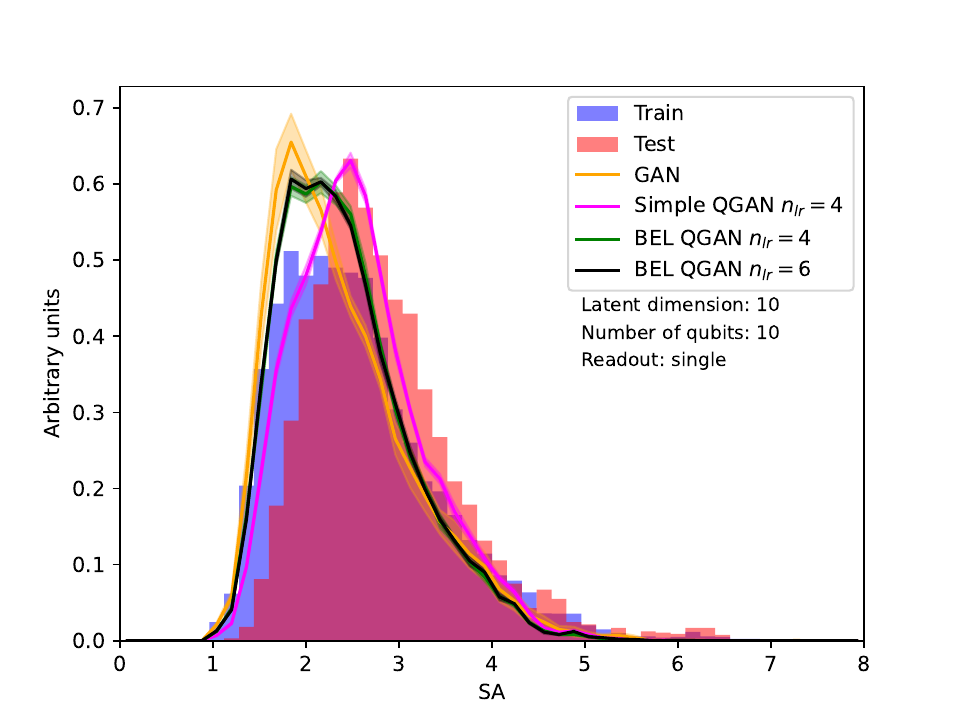}
    \includegraphics[width=0.32\linewidth]{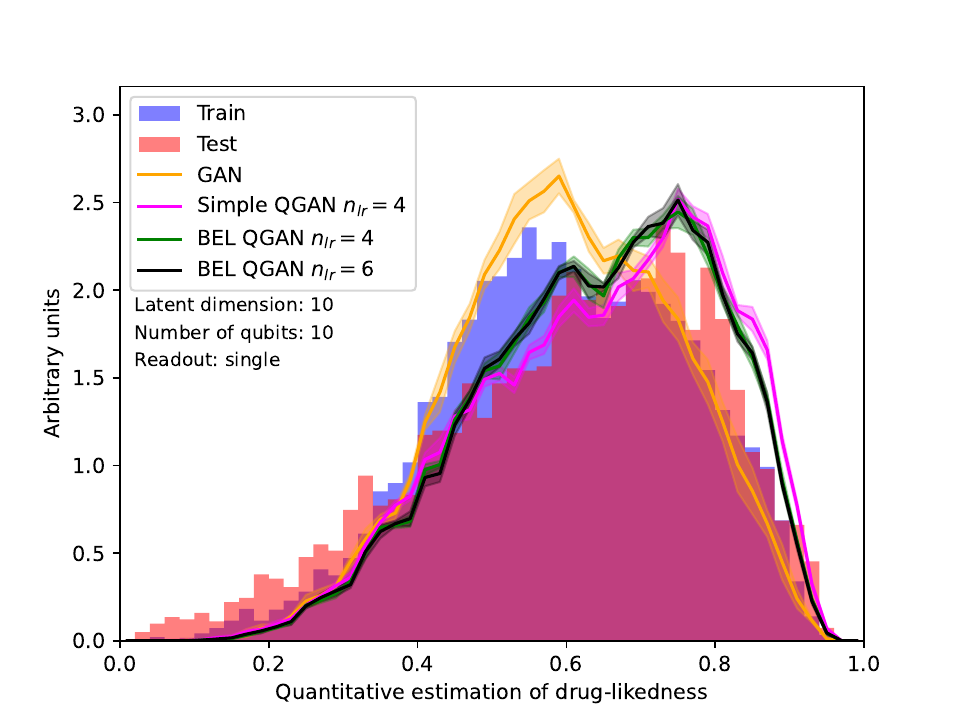}
    \includegraphics[width=0.32\linewidth]{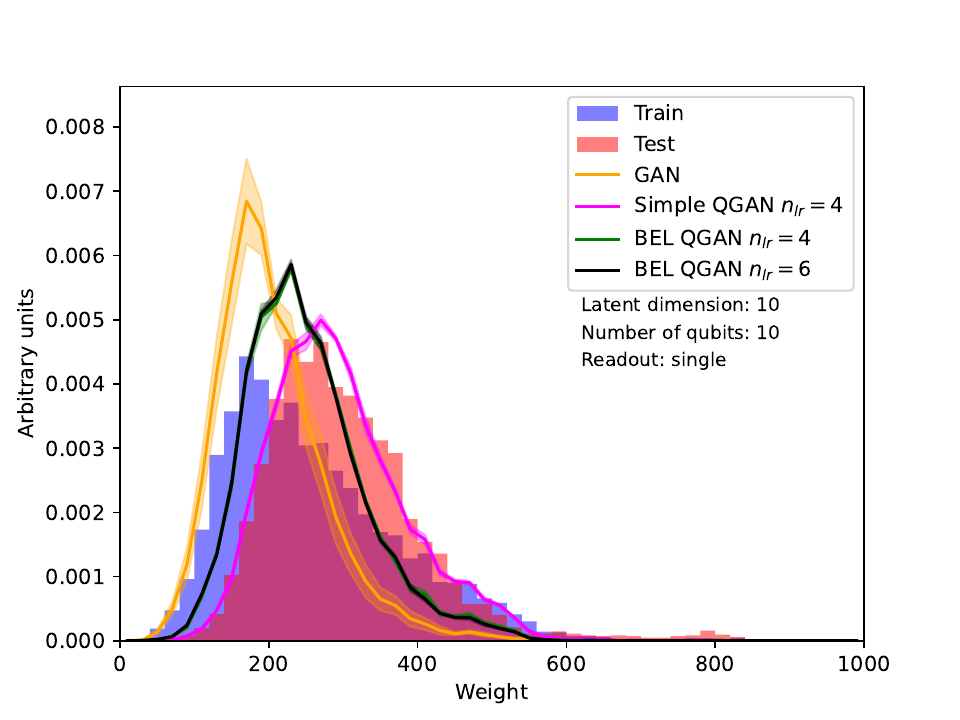}
    \caption{Metrics distributions for classical and quantum models. All GANs are trained with a latent dimension of 10. The quantum GANs use $n_{qb}=10$, $n_{l}=4$~and 6, and single readout.}
    \label{fig:app:scen:cqgan_10qb46lrs}
\end{figure}

% Requires: \usepackage{booktabs}
\begin{table}[h!]
    \centering
    \begin{tabular}{lc|cc|cc}
        \toprule
        Metrics & Classical tuned & Styled simple \qgan & $Z_0$ & Styled BEL \qgan & $Z_0$ \\
        \midrule
        \textbf{Latent dimension} & 20 & 20 &  & 20 &  \\

        \midrule
        $N_{params}$ & 716,692 & 40 &  & 220 &  \\
        \midrule
$\epsilon_d$ & 0.526 $\pm$ 0.022 & 0.507 $\pm$ 0.003 & $-0.85$ & 0.519 $\pm$ 0.005 & $-0.32$ \\
$\epsilon_v$ & 0.901 $\pm$ 0.008 & 0.911 $\pm$ 0.004 & $+1.23$ & 0.910 $\pm$ 0.002 & $+1.22$ \\
$\epsilon_u$ & 0.990 $\pm$ 0.002 & 0.991 $\pm$ 0.001 & $+0.25$ & 0.991 $\pm$ 0.001 & $+0.05$ \\
Novelty & 0.488 $\pm$ 0.018 & 0.484 $\pm$ 0.001 & $-0.22$ & 0.484 $\pm$ 0.002 & $-0.25$ \\
IntDiv & 0.893 $\pm$ 0.002 & 0.890 $\pm$ 0.000 & $-1.42$ & 0.892 $\pm$ 0.000 & $-0.64$ \\
Filters & 0.724 $\pm$ 0.004 & 0.720 $\pm$ 0.002 & $-0.88$ & 0.722 $\pm$ 0.003 & $-0.56$ \\
$\epsilon_{LogP}$ & 0.897 $\pm$ 0.004 & 0.900 $\pm$ 0.002 & $+0.69$ & 0.897 $\pm$ 0.003 & $+0.11$ \\
$\langle \text{SA} \rangle$ & 2.446 $\pm$ 0.034 & 2.393 $\pm$ 0.004 & $+1.56$ & 2.421 $\pm$ 0.003 & $+0.75$ \\
$\langle \text{QED} \rangle$ & 0.613 $\pm$ 0.006 & 0.620 $\pm$ 0.001 & $+1.24$ & 0.618 $\pm$ 0.001 & $+0.93$ \\
$\langle \text{Weight} \rangle$ & 234.4 $\pm$ 11.7 & 231.7 $\pm$ 0.9 & $+0.23$ & 233.4 $\pm$ 0.7 & $+0.09$ \\
        \midrule
        $\langle Z_0 \rangle$ & Reference & -- & +0.18 & -- & +0.14 \\
        \bottomrule
    \end{tabular}
        \caption{Quantum GANs with latent dimension of 20, $n_{qb} = 10$, $n_{l} = 2$, and dual readout.}
    \label{tab:app:scen:cqgan_latdim20}
\end{table}

\begin{figure}[h!]
    \centering
    \includegraphics[width=0.32\linewidth]{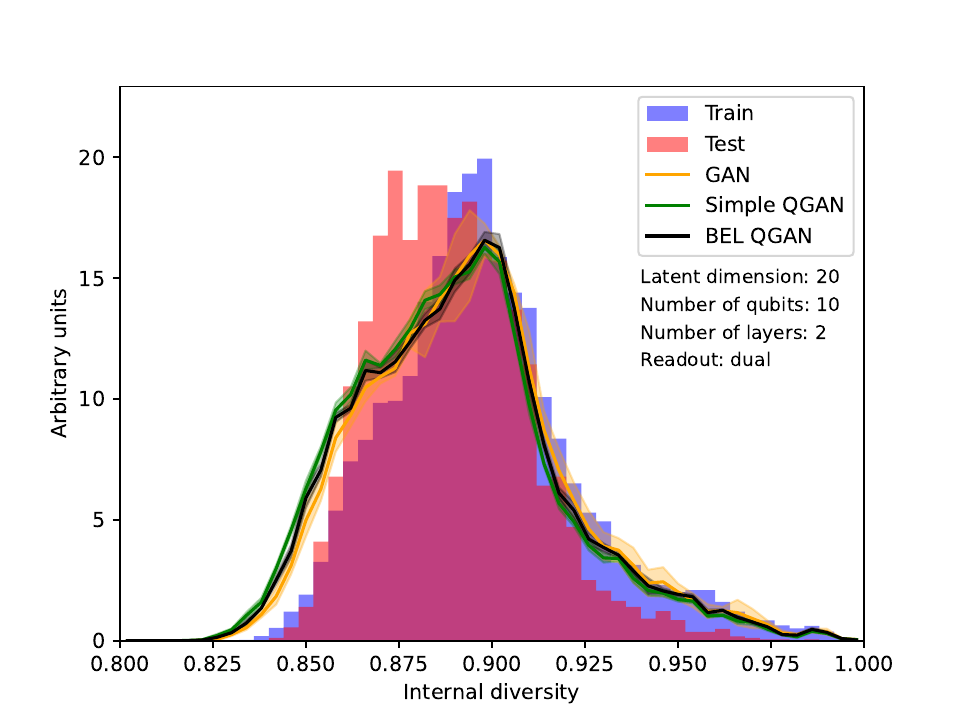}
    \includegraphics[width=0.32\linewidth]{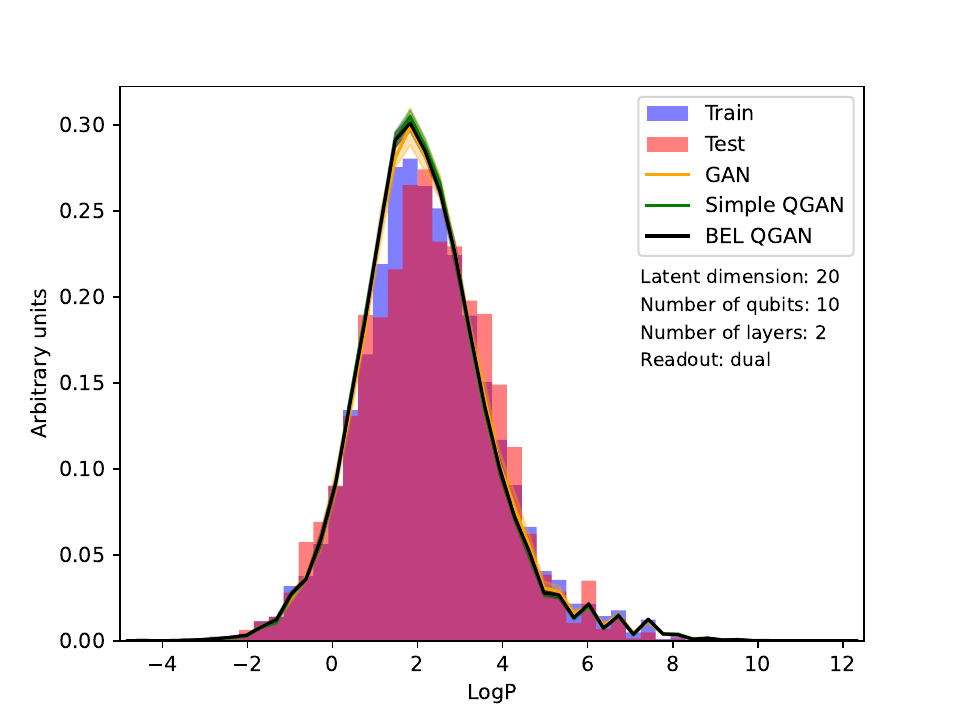}
    \includegraphics[width=0.32\linewidth]{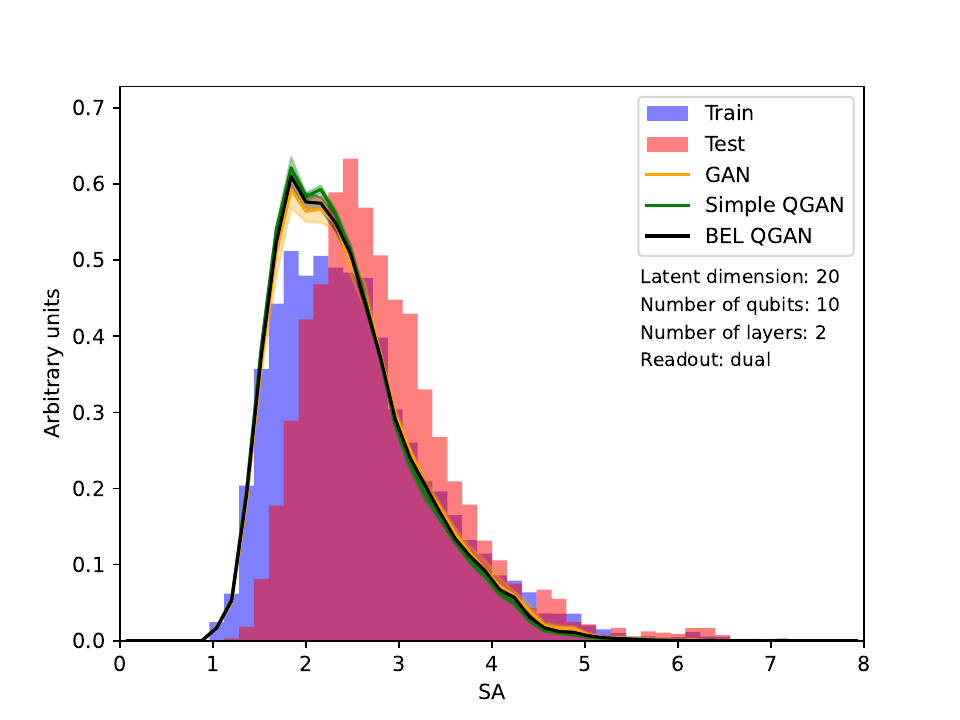}
    \includegraphics[width=0.32\linewidth]{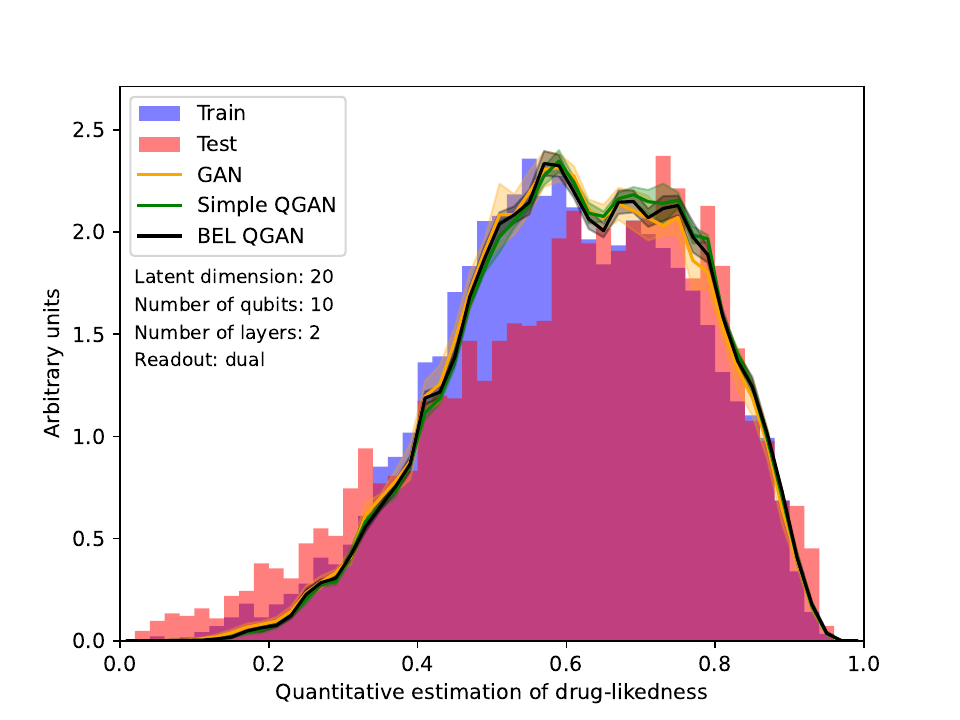}
    \includegraphics[width=0.32\linewidth]{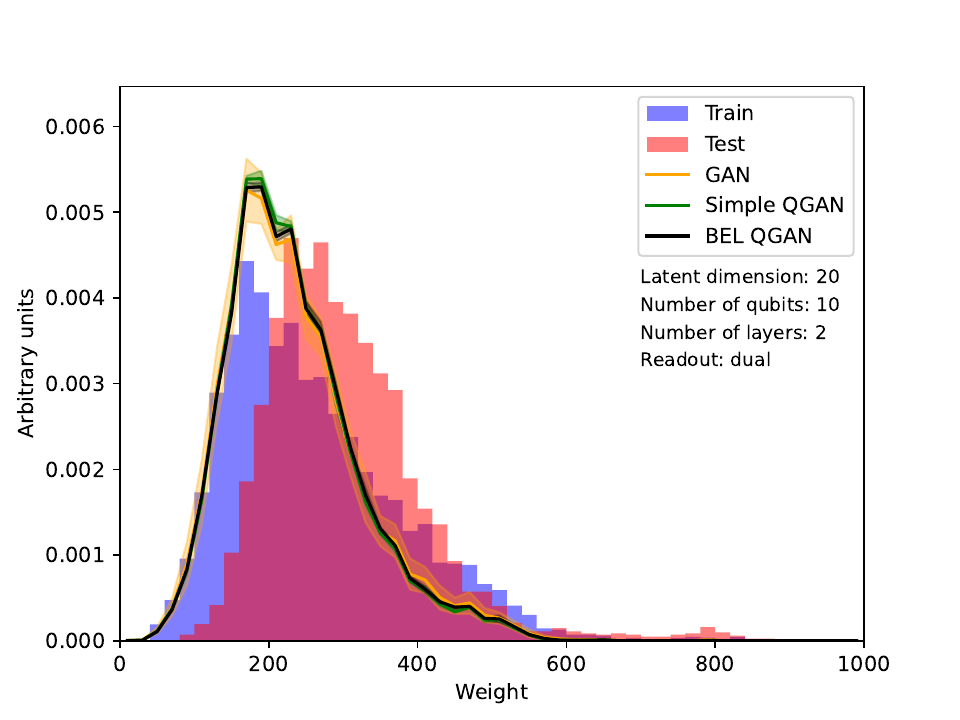}
    \caption{Metrics distributions for classical and quantum models. All GANs are trained with a latent dimension of 20. The quantum GANs use $n_{qb}=10$, $n_{l}=2$, and dual readout.}
    \label{fig:app:scen:cqgan_latdim20}
\end{figure}

\begin{table}[h!]
    \centering
    \begin{tabular}{lc|cc|cc}
        \toprule
        Metrics & Classical tuned & Styled simple \qgan & $Z_0$ & Styled BEL \qgan & $Z_0$ \\
        \midrule
        \textbf{Latent dimension} & 30 & 30 &  & 30 &  \\
        \midrule
        $N_{params}$ & 728,222 & 60 &  & 330 &  \\
        \midrule
$\epsilon_d$ & 0.500 $\pm$ 0.066 & 0.496 $\pm$ 0.003 & $-0.06$ & 0.504 $\pm$ 0.002 & $+0.07$ \\
$\epsilon_v$ & 0.884 $\pm$ 0.032 & 0.895 $\pm$ 0.002 & $+0.32$ & 0.903 $\pm$ 0.003 & $+0.57$ \\
$\epsilon_u$ & 0.990 $\pm$ 0.004 & 0.992 $\pm$ 0.001 & $+0.42$ & 0.991 $\pm$ 0.001 & $+0.39$ \\
Novelty & 0.504 $\pm$ 0.018 & 0.476 $\pm$ 0.003 & $-1.55$ & 0.471 $\pm$ 0.003 & $-1.81$ \\
IntDiv & 0.896 $\pm$ 0.003 & 0.888 $\pm$ 0.000 & $-2.57$ & 0.891 $\pm$ 0.000 & $-1.75$ \\
Filters & 0.719 $\pm$ 0.015 & 0.725 $\pm$ 0.002 & $+0.38$ & 0.723 $\pm$ 0.003 & $+0.25$ \\
$\epsilon_{LogP}$ & 0.894 $\pm$ 0.012 & 0.901 $\pm$ 0.002 & $+0.55$ & 0.900 $\pm$ 0.002 & $+0.44$ \\
$\langle \text{SA} \rangle$ & 2.451 $\pm$ 0.144 & 2.349 $\pm$ 0.002 & $+0.71$ & 2.367 $\pm$ 0.005 & $+0.59$ \\
$\langle \text{QED} \rangle$ & 0.604 $\pm$ 0.006 & 0.624 $\pm$ 0.001 & $+3.23$ & 0.619 $\pm$ 0.001 & $+2.34$ \\
$\langle \text{Weight} \rangle$ & 225.8 $\pm$ 30.1 & 232.8 $\pm$ 0.5 & $-0.23$ & 226.8 $\pm$ 0.9 & $-0.03$ \\
        \midrule
        $\langle Z_0 \rangle$ & Reference & -- & +0.12 & -- & +0.11 \\
        \bottomrule
    \end{tabular}
        \caption{Quantum GANs with a latent dimension of 30, $n_{qb} = 15$, $n_{l} = 2$, and dual readout.}
    \label{tab:app:scen:cqgan_latdim30}
\end{table}

\begin{figure}[h!]
    \centering
    \includegraphics[width=0.32\linewidth]{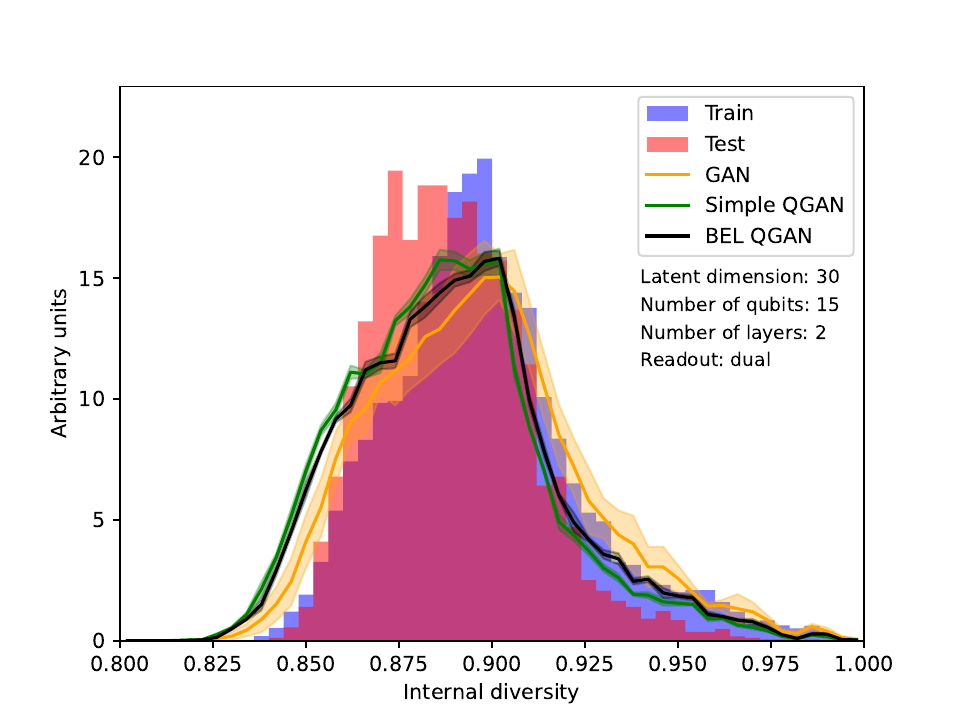}
    \includegraphics[width=0.32\linewidth]{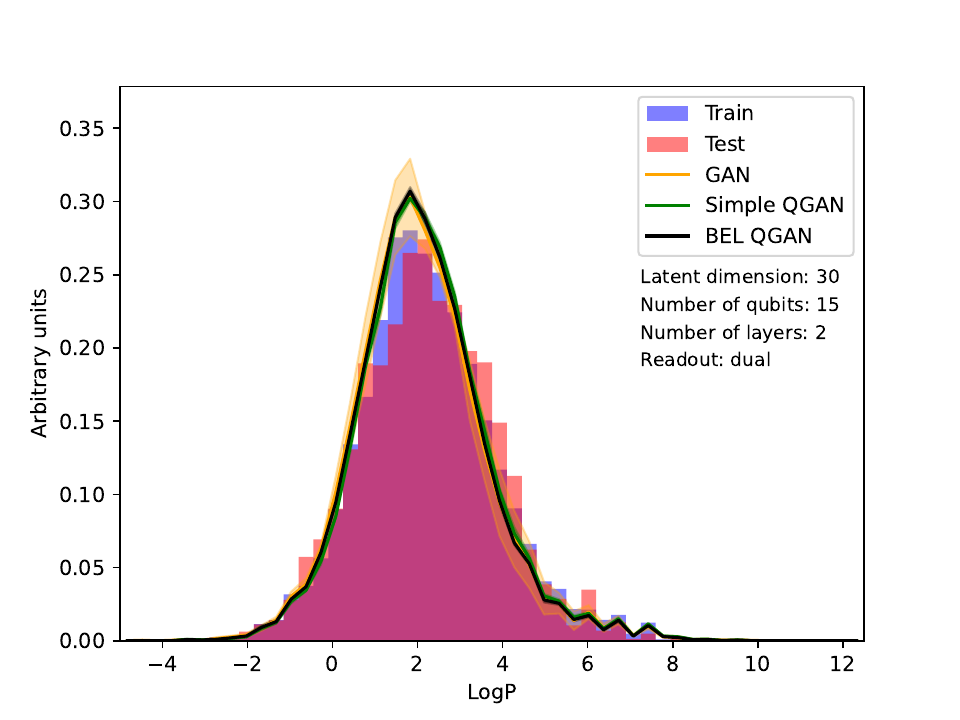}
    \includegraphics[width=0.32\linewidth]{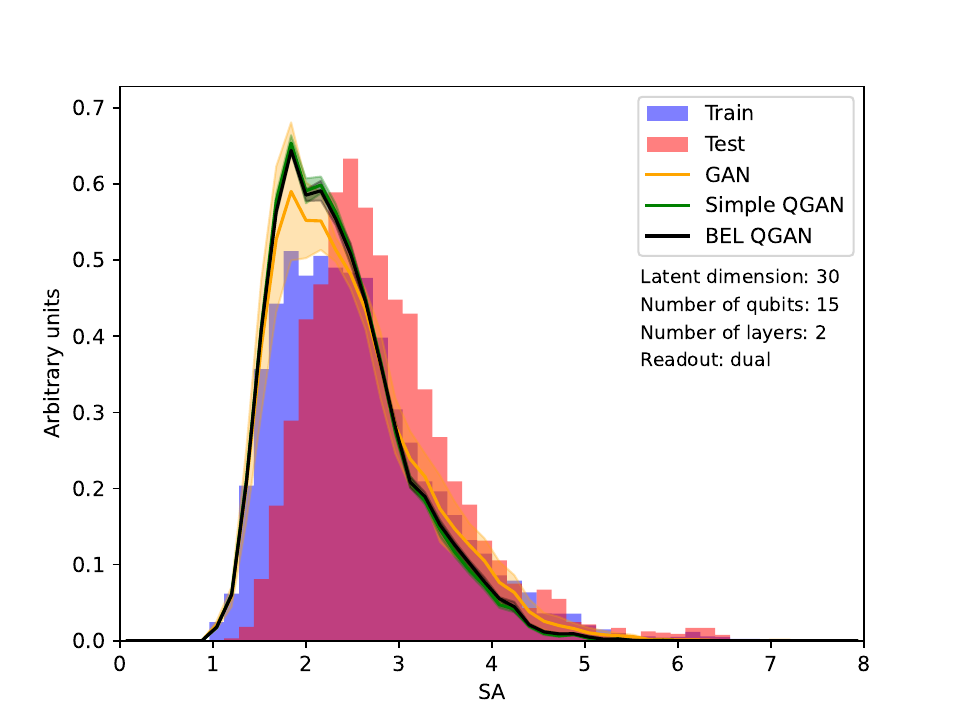}
    \includegraphics[width=0.32\linewidth]{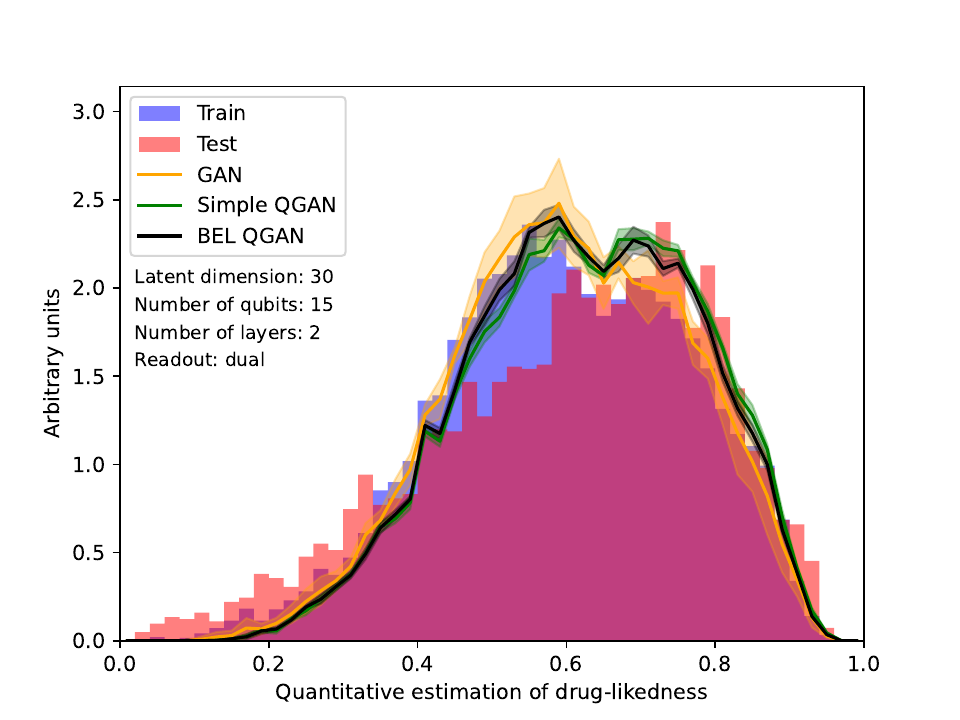}
    \includegraphics[width=0.32\linewidth]{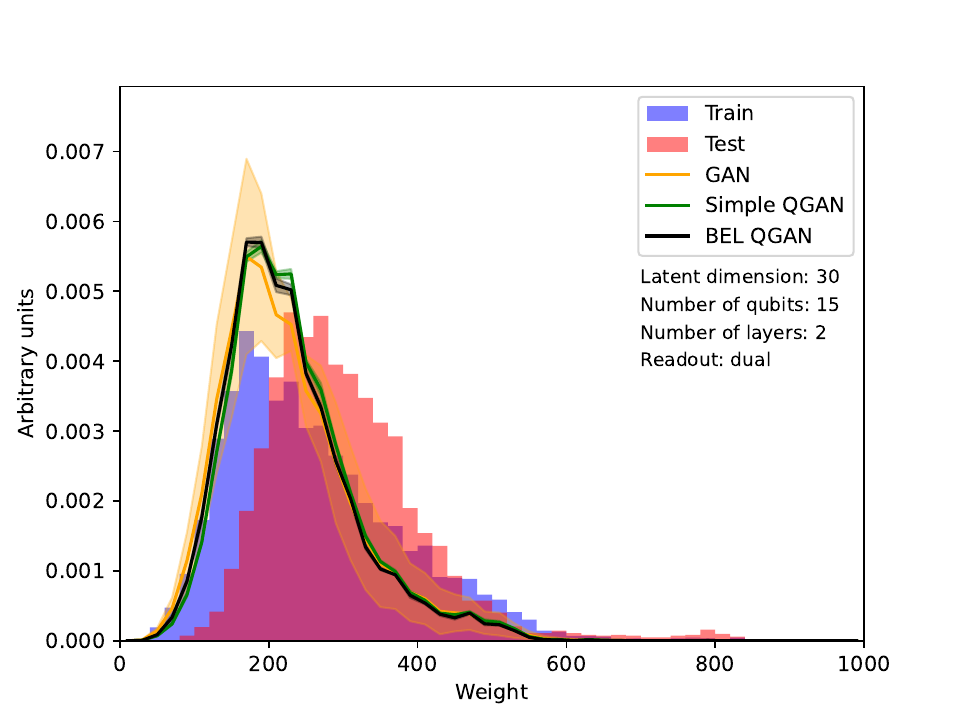}
    \caption{Metrics distributions for classical and quantum models. All GANs are trained with a latent dimension of 30. The quantum GANs use $n_{qb}=15$, $n_{l}=2$, and dual readout.}
    \label{fig:app:scen:cqgan_latdim30}
\end{figure}

% Requires: \usepackage{booktabs}
\begin{table}[h!]
    \centering
    \begin{tabular}{lc|cc|cc}
        \toprule
        \textbf{Latent Dimension} & 10 & 20 & $Z_0$ & 30 & $Z_0$ \\
        \midrule
        $N_{params}$ & $705,162$ & $716,692$ &  & $728,222$ &  \\
        \midrule
$\epsilon_d$ & $0.481 \pm 0.031$ & $0.526 \pm 0.022$ & $+1.19$ & $0.500 \pm 0.066$ & $+0.25$ \\
    $\epsilon_v$ & $0.917 \pm 0.012$ & $0.901 \pm 0.008$ & $-1.16$ & $0.884 \pm 0.032$ & $-0.96$ \\
    $\epsilon_u$ & $0.986 \pm 0.003$ & $0.990 \pm 0.002$ & $+1.59$ & $0.990 \pm 0.004$ & $+0.84$ \\
    Novelty & $0.536 \pm 0.015$ & $0.488 \pm 0.018$ & $-2.04$ & $0.504 \pm 0.018$ & $-1.39$ \\
    IntDiv & $0.898 \pm 0.004$ & $0.893 \pm 0.002$ & $-1.19$ & $0.896 \pm 0.003$ & $-0.43$ \\
    Filters & $0.721 \pm 0.011$ & $0.724 \pm 0.004$ & $+0.28$ & $0.719 \pm 0.015$ & $-0.07$ \\
    $\epsilon_{LogP}$ & $0.898 \pm 0.007$ & $0.897 \pm 0.004$ & $-0.14$ & $0.894 \pm 0.012$ & $-0.26$ \\
    $\langle \text{SA} \rangle$ & $2.391 \pm 0.080$ & $2.446 \pm 0.034$ & $-0.64$ & $2.451 \pm 0.144$ & $-0.36$ \\
    $\langle \text{QED} \rangle$ & $0.599 \pm 0.007$ & $0.613 \pm 0.006$ & $+1.54$ & $0.604 \pm 0.006$ & $+0.49$ \\
    $\langle \text{Weight} \rangle$ & $203.9 \pm 10.6$ & $234.4 \pm 11.7$ & $-1.94$ & $225.8 \pm 30.1$ & $-0.69$ \\
        \midrule
        $\langle Z_0 \rangle$ & Reference & -- & $-0.25$ & -- & $-0.26$ \\
        \bottomrule
    \end{tabular}
        \caption{Classical GAN comparison of different latent dimensions.}
    \label{tab:app:scen:cgan_latdim}
\end{table}

\begin{figure}[h!]
    \centering
    \includegraphics[width=0.32\linewidth]{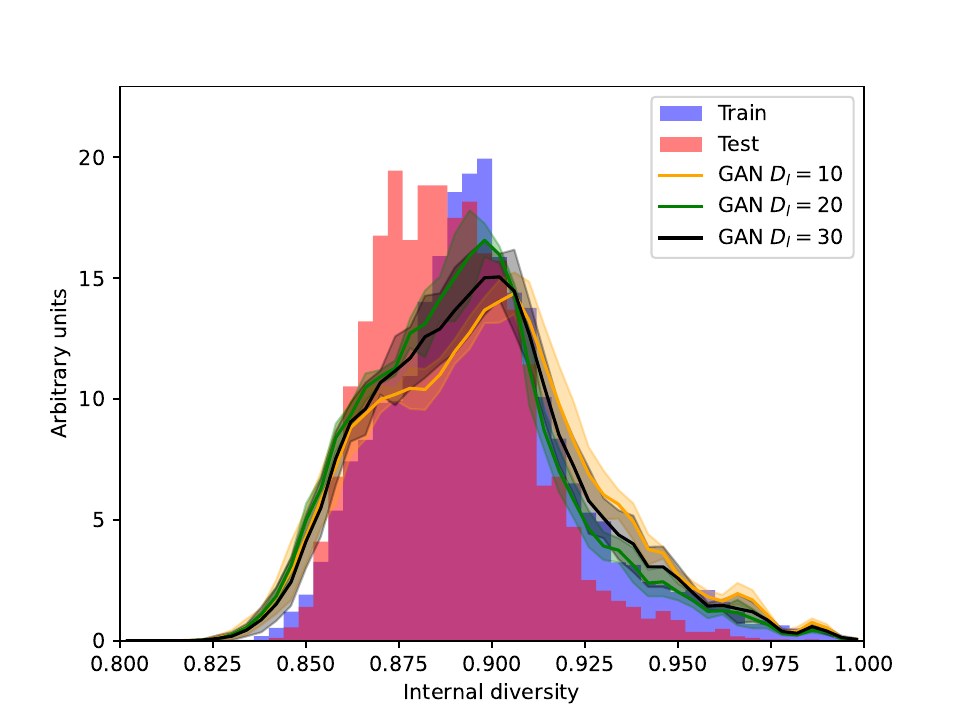}
    \includegraphics[width=0.32\linewidth]{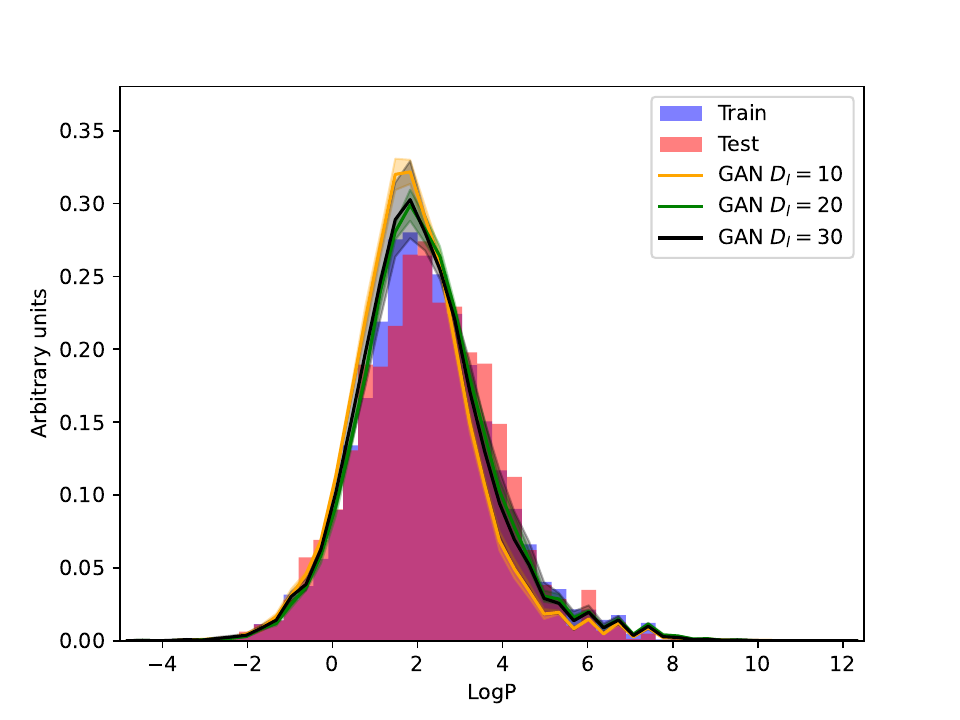}
    \includegraphics[width=0.32\linewidth]{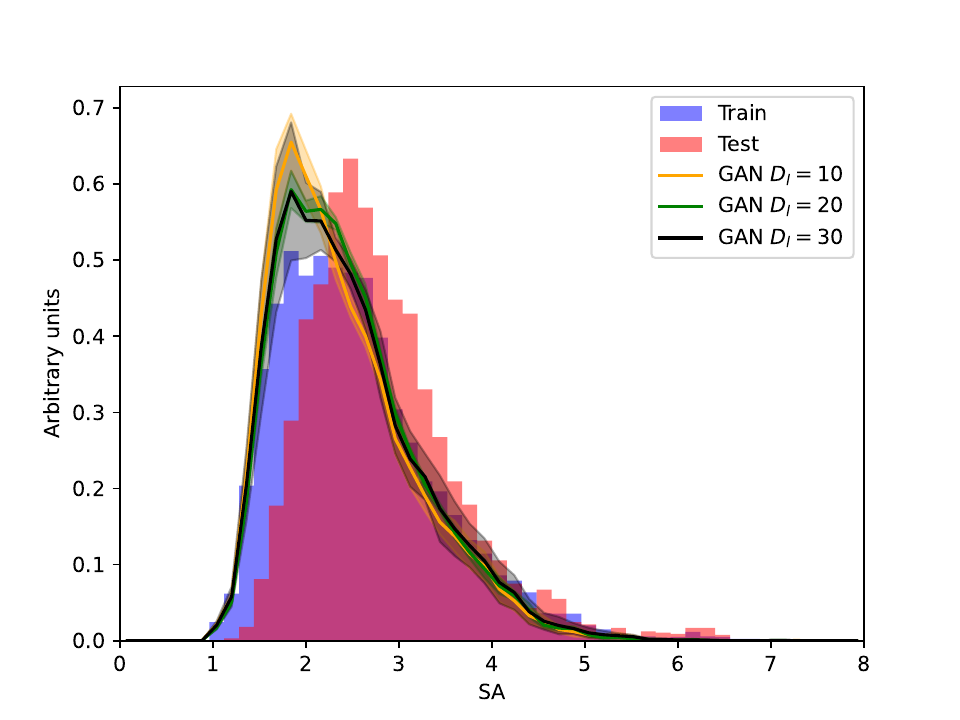}
    \includegraphics[width=0.32\linewidth]{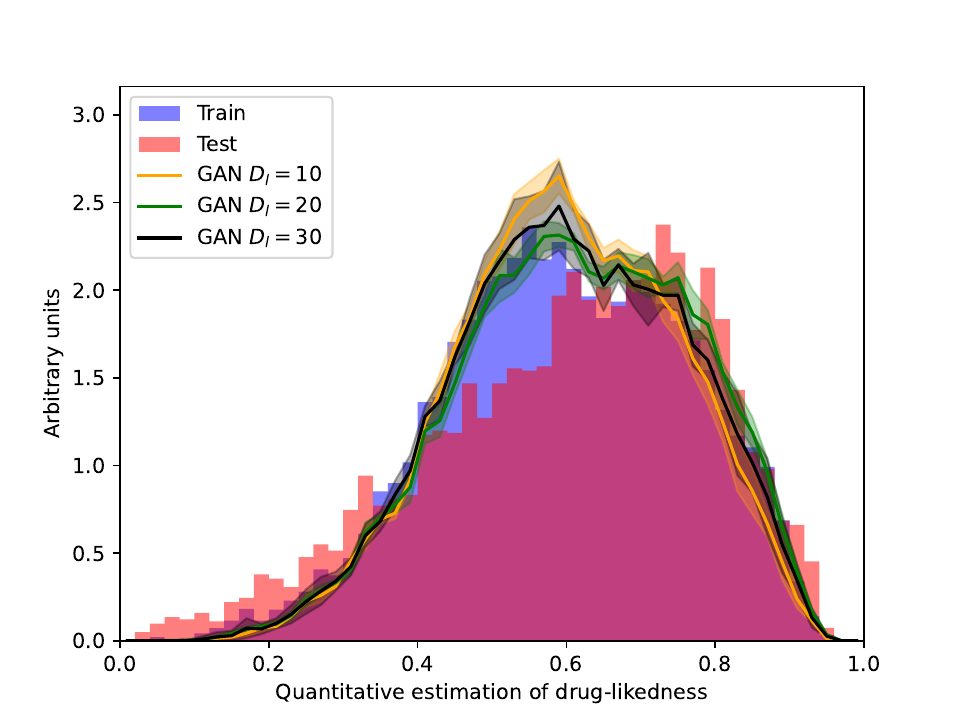}
    \includegraphics[width=0.32\linewidth]{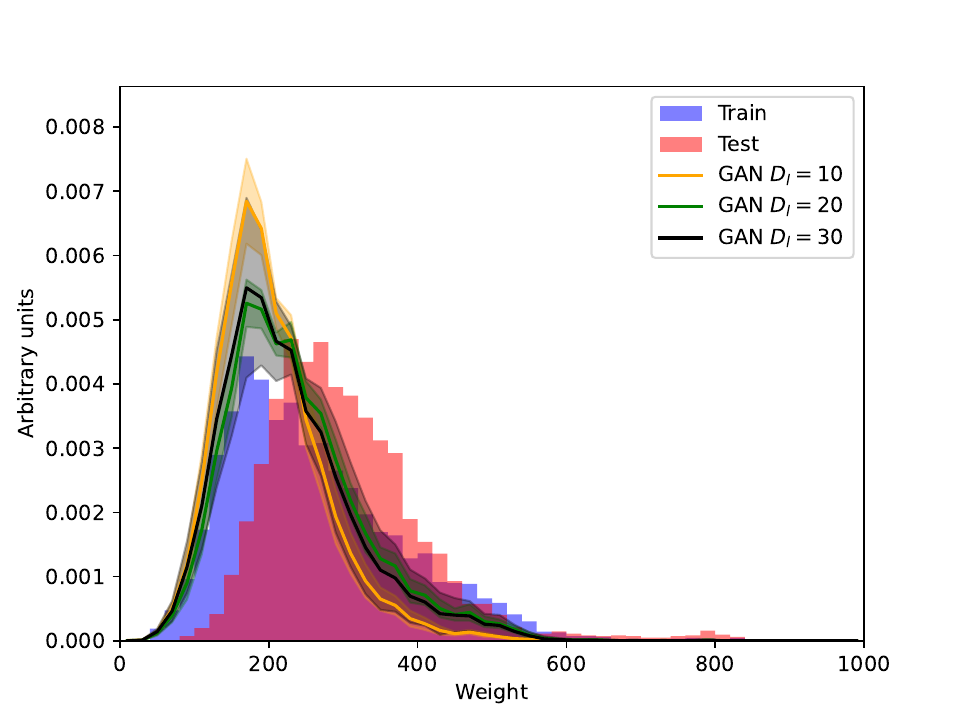}
    \caption{Metrics distributions for classical models trained with different latent dimensions.}
    \label{fig:app:scen:cgan_latdim}
\end{figure}

\FloatBarrier
\section{Supplementary material: Classical GAN study with same-order capacity as QGAN architecture}
\label{sec:app:cgan_lowparam}

The nominal architecture and parameter settings in the main text have been taken from the MOSES package, resulting in 705,162 trainable parameters. Given that training was performed on a subset of the entire MOSES dataset, it is important to check whether a comparably smaller capacity for the classical GAN could outperform the nominal classical GAN architecture, such that more robust training would be observed and the corresponding metrics would improve significantly. We have included in the main text results from a classical latent GAN with exact matched capacity for the generator; we extend this analysis with other capacities in this section.

We have performed a supplementary control study with a downgraded architecture, such that the number of trainable parameters is roughly the same order of magnitude as the number of trainable parameters of the BEL QGAN (110 trainable parameters). As described in Section~\ref{sec:setup:sqGAN} in the main text, the classical generator consists of five linear layers with increasing number of nodes. Three low-parameter scenarios were defined by replacing the five-layer MOSES generator architecture $[128,256,512,1024,D_l]$ by a single hidden layer followed by the output layer, $[D, D_l]$ with 1) $D=32$ resulting in 682 parameters; 2) $D=16$ resulting in 346 parameters; 3) $D=10$ resulting in the exact matched capacity, 110 parameters. Two extra scenarios used a two-hidden-layer architecture, $[10,D,D_l]$, with $D=5$ resulting in 115 parameters and $D=4$ resulting in 94 parameters. We use the same VAE, the same discriminator, noise distribution, and tuned hyperparameters as in the main text.

\begin{figure}
    \centering
    \includegraphics[width=0.49\linewidth]{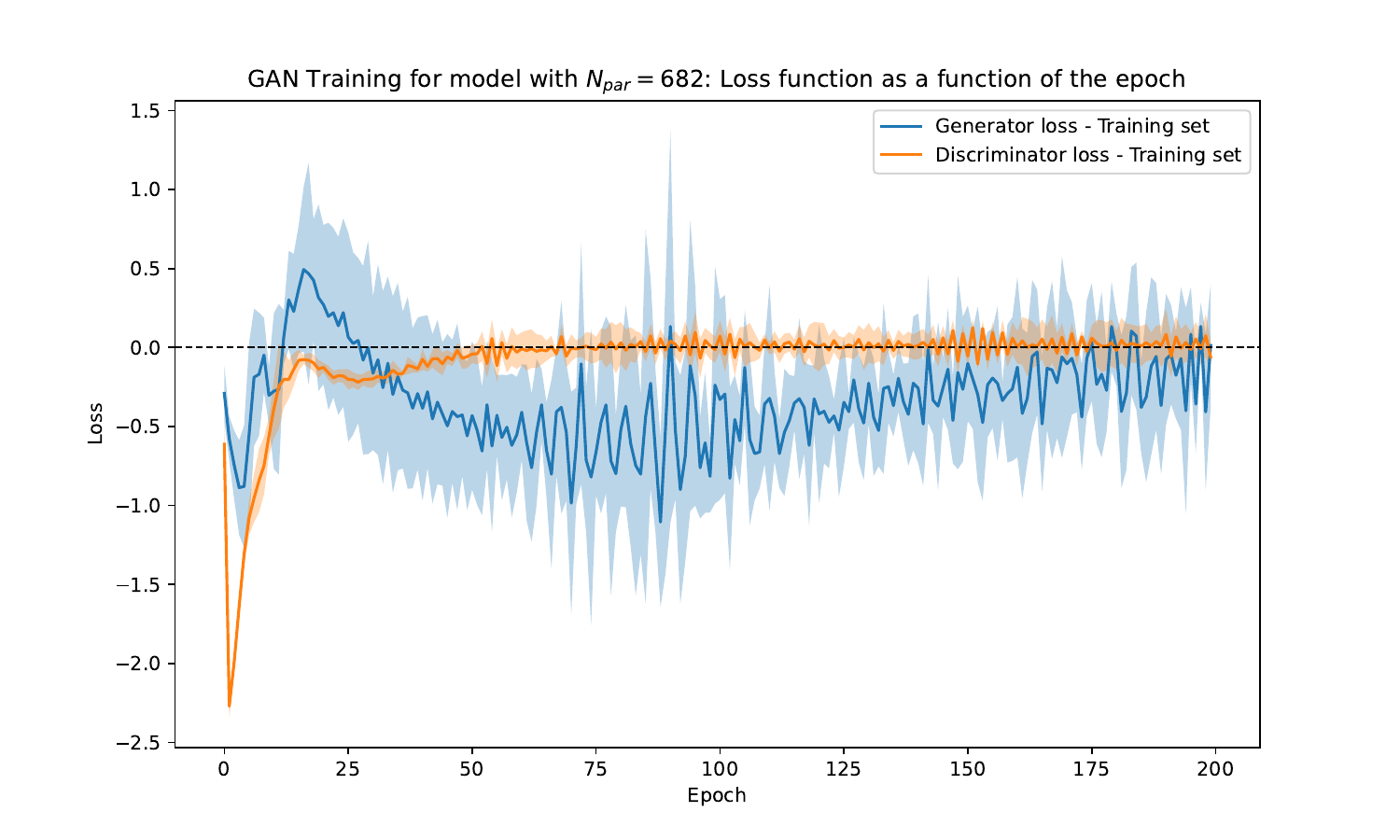}
    \includegraphics[width=0.49\linewidth]{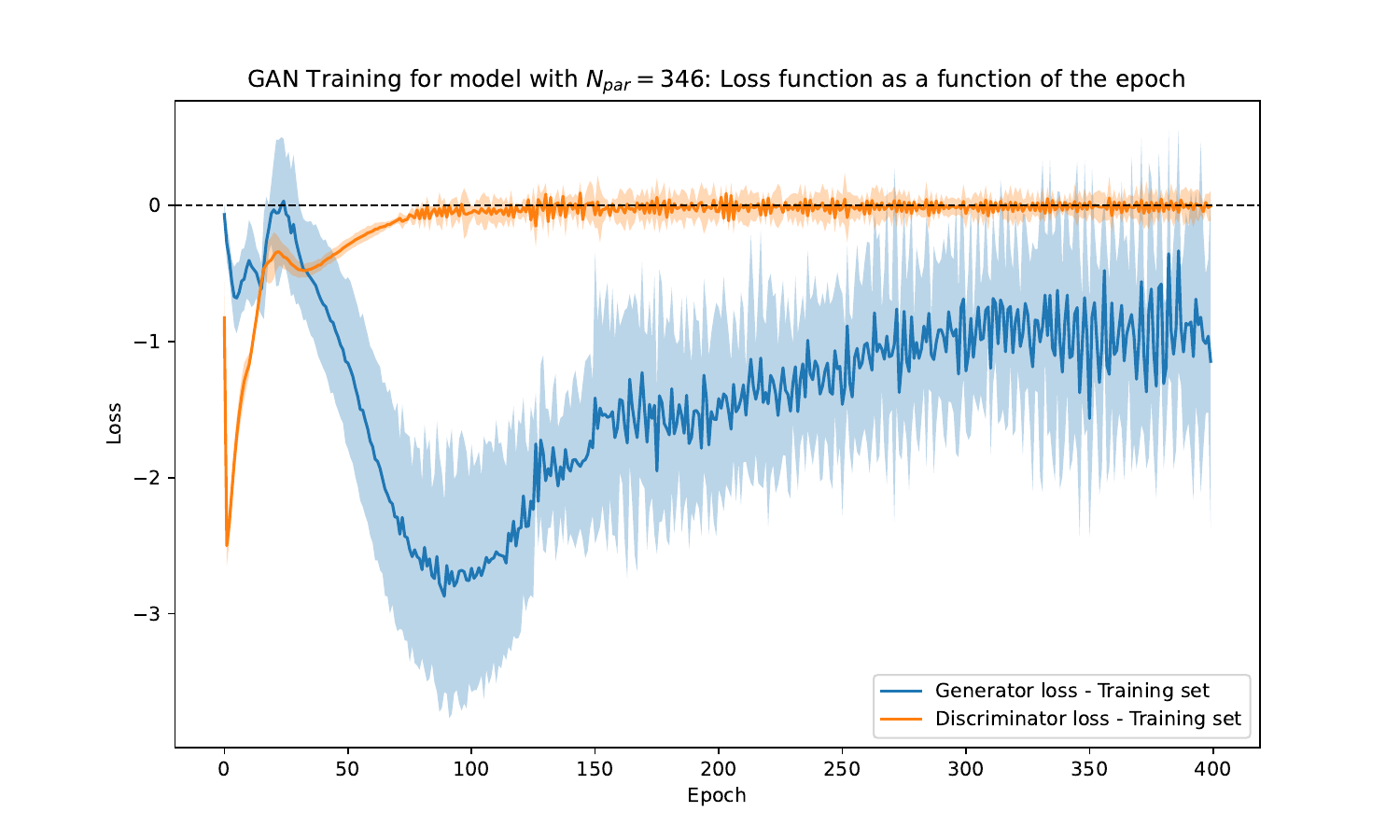}\\
    \includegraphics[width=0.49\linewidth]{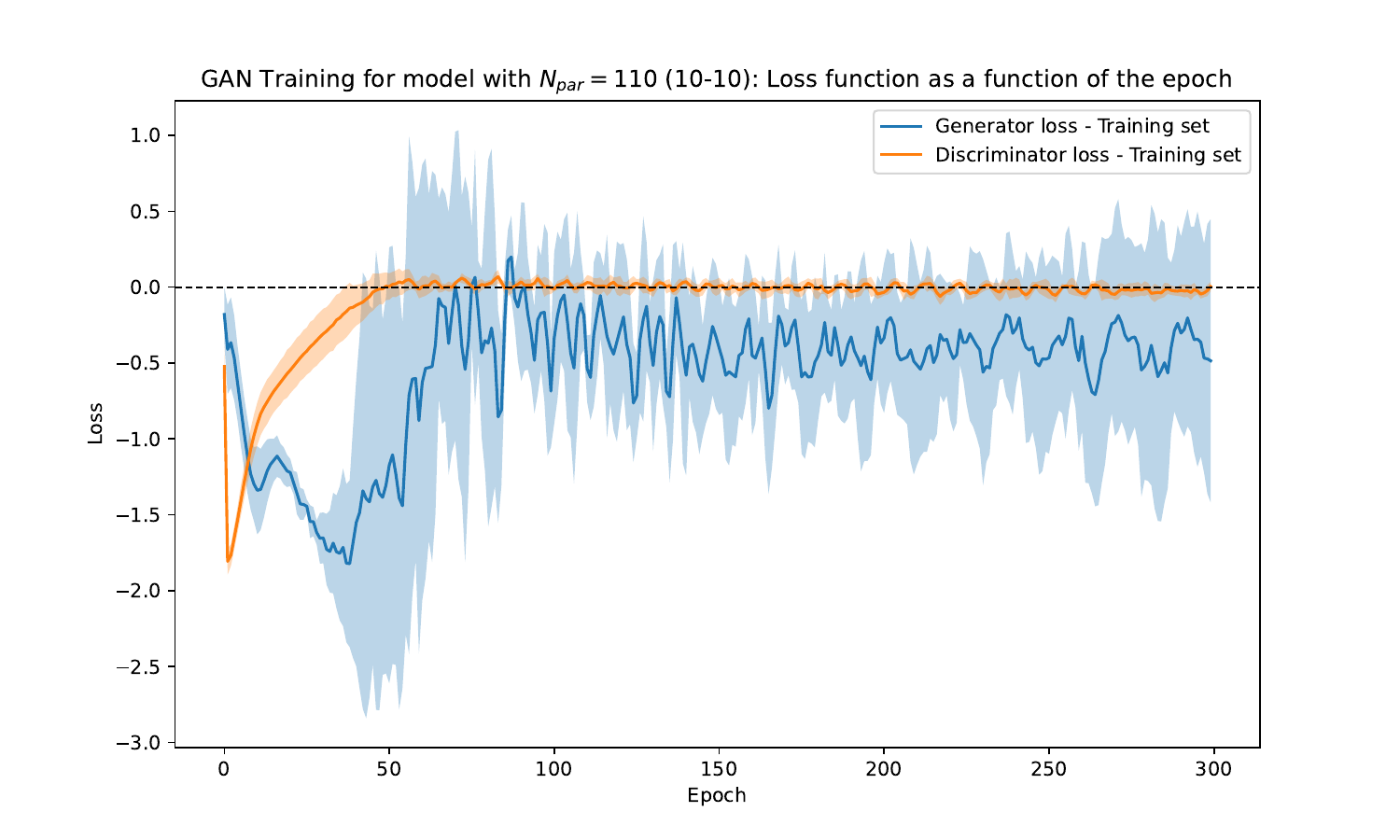}
    \caption{Loss functions for the low parametric GANs with 682 (upper left), 346 (upper right), and 110 (lower row) trainable parameters. The number of epochs is chosen so that the three loss functions are stable. Each model was run with five different seeds, and the shaded band is defined as $\pm 1\sigma$ deviation.}
    \label{fig:app:robustness:cgan_loss}
\end{figure}

\begin{figure}
    \centering
    \includegraphics[width=0.49\linewidth]{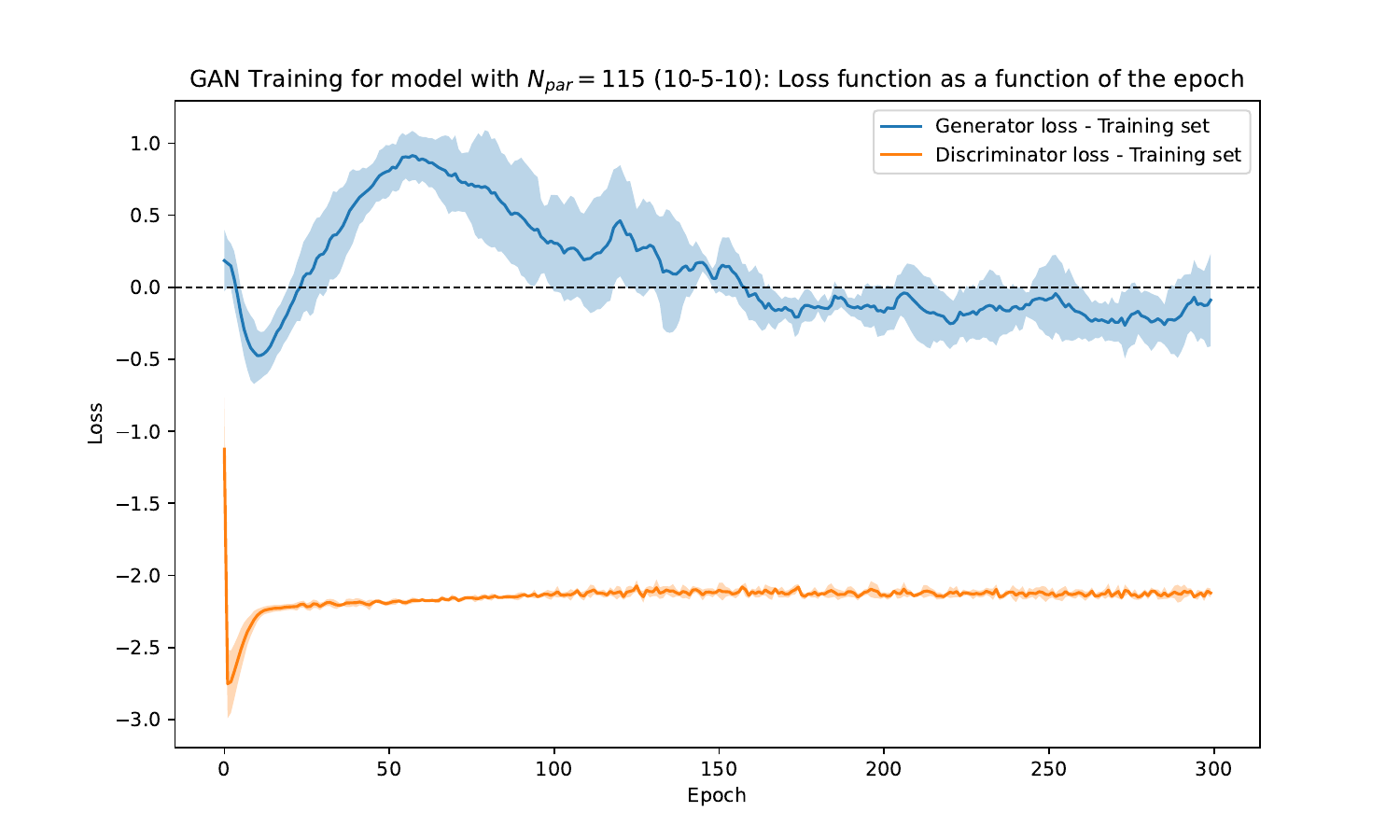}
    \includegraphics[width=0.49\linewidth]{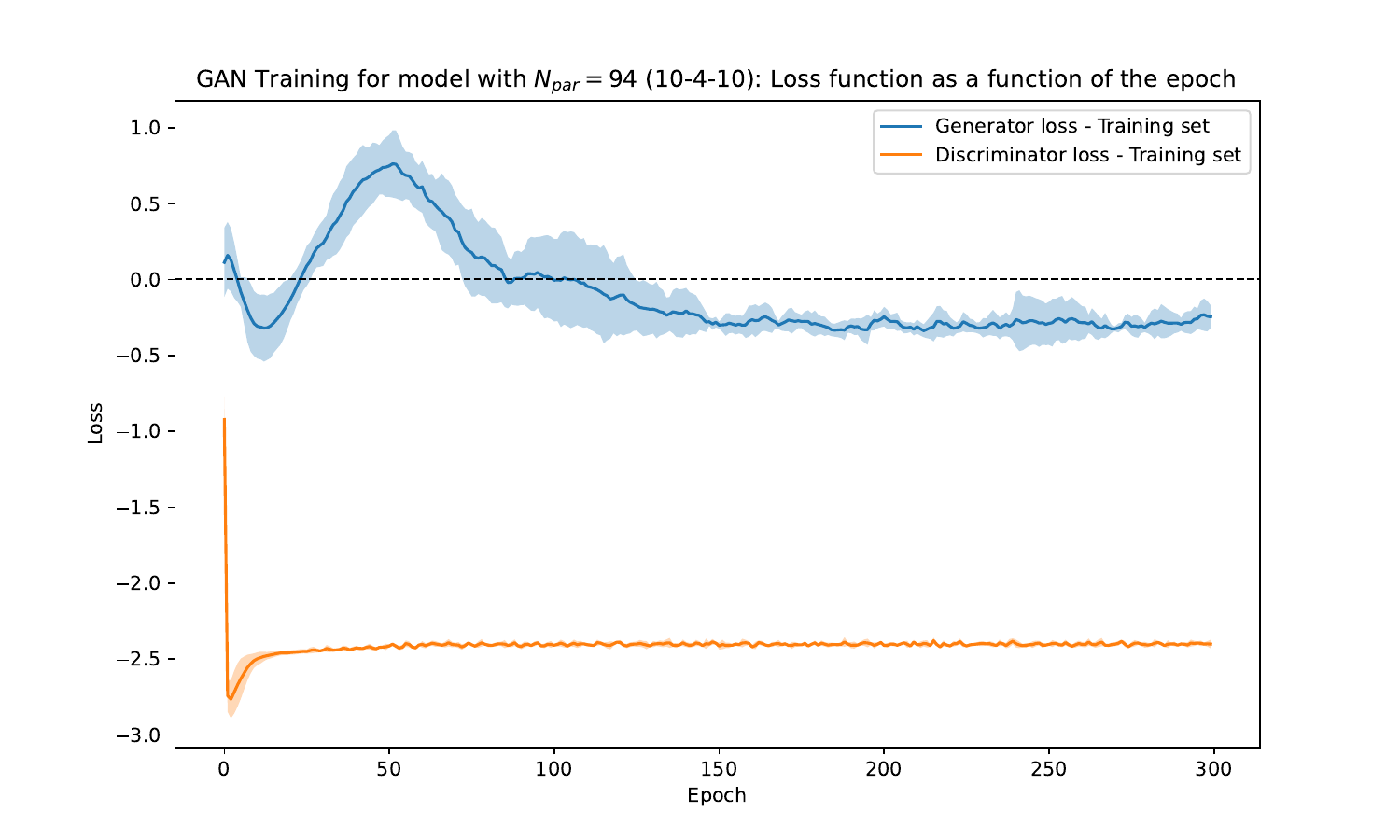}
    \caption{Same as Figure~\ref{fig:app:robustness:cgan_loss} but for two-hidden-layer scenarios with 115 (left panel) and 94 (right panel) trainable parameters.}
    \label{fig:app:robustness:cgan_loss2}
  \end{figure}

We present in Figure~\ref{fig:app:robustness:cgan_loss} the mean values of the loss function with the uncertainty band defined as the $1\sigma$ standard deviation calculated with five runs with different seeds, as in the main text, for the single-hidden-layer scenarios with 682, 346, and 110 trainable parameters. The number of epochs where the loss function shows stability has been estimated to be $N_{ep} = 200$ for the 682-parameter scenario, $N_{ep} = 400$ for the 346-parameter scenario, and $N_{ep}=300$ for the 110-parameter scenario and the two-hidden-layer scenarios with 115 and 94 trainable parameters. When compared with the loss function of the reference classical GAN in Figure~\ref{fig:gan_loss} in the main text, similar behavior is obtained, and the robustness of the training is not significantly enhanced, which would have corresponded to a smoothing of the generator loss function and a much smaller band. This indicates that overfitting of the large number of trainable parameters is likely not the reason we observe a less robust training with the reference classical GAN compared to our styled BEL QGAN. We also present in Figure~\ref{fig:app:robustness:cgan_loss2} the same analysis, but for the two-hidden-layer scenario with 115 and 94 trainable parameters. The loss curves are definitively smoother than for the nominal GAN and for the one-hidden-layer loss functions, including the matched-capacity scenario. However, they are still not as smooth as the \qgan\ loss functions, indicating that the \qgan\ architecture is more robust than the classical generators investigated in this work.

Table~\ref{tab:app:robustness:metrics} presents the comparison and per-metric statistical compatibility between the reference GAN and the one-hidden-layer low-parameter scenarios presented in this section. While the difference is within one standard deviation, with $|\langle Z_0\rangle| \leq 1$, we observe a consistent trend toward negative $Z_0$ across all drug-relevant metrics in the three scenarios, including the exact matched capacity scenario compared to BEL QGAN: this indicates a consistent slight degradation of performance. This contrasts with the results we have obtained with the styled QGAN in the main text, which displayed statistically significant improvement in some of the individual metrics. We display in Table~\ref{tab:app:robustness:metrics2} the corresponding comparison, but against the two-hidden-layer scenarios with 115 and 94 trainable parameters. There is still an overall statistical compatibility with the nominal reference scenario (within one standard deviation), but the situation slightly improves overall for the 94-parameter scenario with $\langle Z_0\rangle = +0.16$. Interestingly, the drug-efficient metrics are now slightly better than for the nominal scenario, in contrast with the previous single-hidden-layer scenarios. However, not a single individual metric yields a significant improvement over the reference MOSES baseline, contrary to what we observe with the BEL QGAN.

% Requires: \usepackage{booktabs}
\begin{table}[h!]
    \centering
    \begin{tabular}{lc|cc|cc|cc}
        \toprule
        \textbf{$N_{params}$} & $705,162$ & $682$ &  $Z_0$ & $346$ & $Z_0$ & $110$ & $Z_0$ \\
        \midrule
$\epsilon_d$ & 0.481 $\pm$ 0.031 & 0.519 $\pm$ 0.006 & +1.21 & 0.522 $\pm$ 0.005 & +1.33 & 0.523 $\pm$ 0.001 & +1.36 \\
$\epsilon_v$ & 0.917 $\pm$ 0.012 & 0.910 $\pm$ 0.003 & -0.57 & 0.912 $\pm$ 0.003 & -0.41 & 0.910 $\pm$ 0.001 & -0.56 \\
$\epsilon_u$ & 0.986 $\pm$ 0.003 & 0.985 $\pm$ 0.001 & -0.27 & 0.985 $\pm$ 0.001 & -0.04 & 0.986 $\pm$ 0.001 & +0.00 \\
Novelty & 0.536 $\pm$ 0.015 & 0.496 $\pm$ 0.004 & -2.67 & 0.498 $\pm$ 0.004 & -2.52 & 0.495 $\pm$ 0.003 & -2.79 \\
IntDiv & 0.898 $\pm$ 0.004 & 0.902 $\pm$ 0.000 & +1.05 & 0.902 $\pm$ 0.000 & +0.91 & 0.902 $\pm$ 0.000 & +0.97 \\
Filters & 0.721 $\pm$ 0.011 & 0.721 $\pm$ 0.002 & +0.03 & 0.722 $\pm$ 0.002 & +0.11 & 0.722 $\pm$ 0.003 & +0.09 \\
$\epsilon_{LogP}$ & 0.898 $\pm$ 0.007 & 0.894 $\pm$ 0.002 & -0.57 & 0.895 $\pm$ 0.001 & -0.50 & 0.895 $\pm$ 0.002 & -0.37 \\
$\langle \text{SA} \rangle$ & 2.391 $\pm$ 0.080 & 2.473 $\pm$ 0.006 & -1.03 & 2.462 $\pm$ 0.007 & -0.89 & 2.472 $\pm$ 0.005 & -1.02 \\
$\langle \text{QED} \rangle$ & 0.599 $\pm$ 0.007 & 0.594 $\pm$ 0.001 & -0.74 & 0.595 $\pm$ 0.002 & -0.59 & 0.594 $\pm$ 0.001 & -0.73 \\
$\langle \text{Weight} \rangle$ & 203.9 $\pm$ 10.6 & 214.2 $\pm$ 2.1 & -0.96 & 213.9 $\pm$ 1.1 & -0.94 & 215.2 $\pm$ 1.0 & -1.07 \\
        \midrule
        $\langle Z_0 \rangle$ & Reference & -- & $-0.45$ & -- & $-0.35$ & --  & -0.41 \\
        \bottomrule
    \end{tabular}
        \caption{Metrics comparison for GANs trained with different number of generator trainable parameters. Each model was run with five different seeds to calculate the $\pm 1\sigma$ deviation on the metrics.}
    \label{tab:app:robustness:metrics}
  \end{table}

% Requires: \usepackage{booktabs}
\begin{table}[h!]
    \centering
    \begin{tabular}{lc|cc|cc}
        \toprule
        \textbf{$N_{params}$} & $705,162$ & $115$ &  $Z_0$ & $94$ & $Z_0$ \\
        \midrule
$\epsilon_d$ & 0.481 $\pm$ 0.031 & 0.491 $\pm$ 0.025 & +0.23 & 0.473 $\pm$ 0.013 & -0.26 \\
$\epsilon_v$ & 0.917 $\pm$ 0.012 & 0.920 $\pm$ 0.009 & +0.17 & 0.924 $\pm$ 0.002 & +0.55 \\
$\epsilon_u$ & 0.986 $\pm$ 0.003 & 0.986 $\pm$ 0.002 & +0.27 & 0.986 $\pm$ 0.002 & +0.26 \\
Novelty & 0.536 $\pm$ 0.015 & 0.518 $\pm$ 0.013 & -0.90 & 0.529 $\pm$ 0.003 & -0.47 \\
IntDiv & 0.898 $\pm$ 0.004 & 0.898 $\pm$ 0.002 & -0.19 & 0.898 $\pm$ 0.002 & -0.19 \\
Filters & 0.721 $\pm$ 0.011 & 0.723 $\pm$ 0.003 & +0.23 & 0.725 $\pm$ 0.004 & +0.38 \\
$\epsilon_{LogP}$ & 0.898 $\pm$ 0.007 & 0.898 $\pm$ 0.004 & +0.00 & 0.901 $\pm$ 0.004 & +0.47 \\
$\langle \text{SA} \rangle$ & 2.391 $\pm$ 0.080 & 2.389 $\pm$ 0.046 & +0.02 & 2.359 $\pm$ 0.025 & +0.39 \\
$\langle \text{QED} \rangle$ & 0.599 $\pm$ 0.007 & 0.601 $\pm$ 0.005 & +0.23 & 0.600 $\pm$ 0.005 & +0.08 \\
$\langle \text{Weight} \rangle$ & 203.9 $\pm$ 10.6 & 206.5 $\pm$ 10.5 & -0.17 & 199.7 $\pm$ 3.5 & +0.38 \\
        \midrule
        $\langle Z_0 \rangle$ & Reference & -- & $-0.01$ & -- & $+0.16$ \\
        \bottomrule
    \end{tabular}
        \caption{Same as in Table~\ref{tab:app:robustness:metrics} for the two-hidden-layer scenarios with 115 and 94 trainable parameters.}
    \label{tab:app:robustness:metrics2}
  \end{table}

\begin{figure}[h!]
    \centering
    \includegraphics[width=0.32\linewidth]{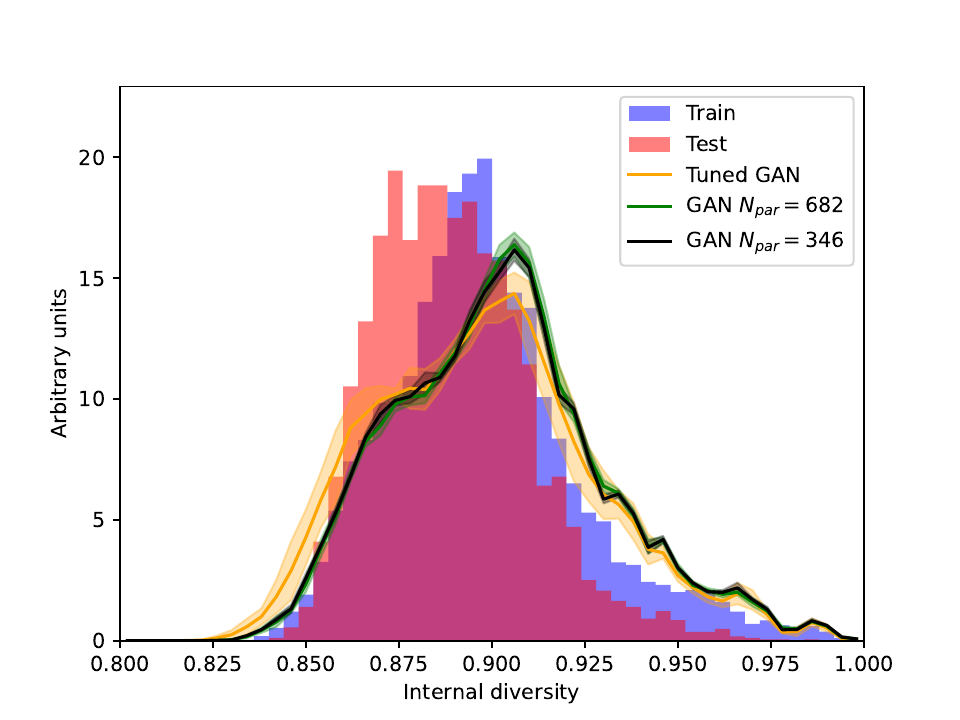}
    \includegraphics[width=0.32\linewidth]{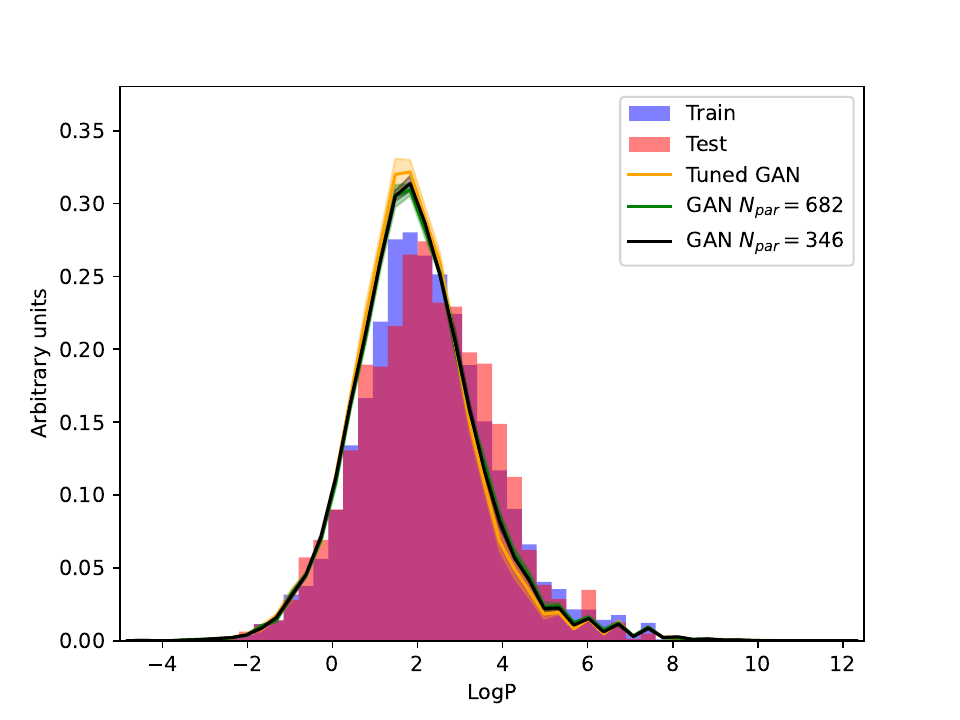}
    \includegraphics[width=0.32\linewidth]{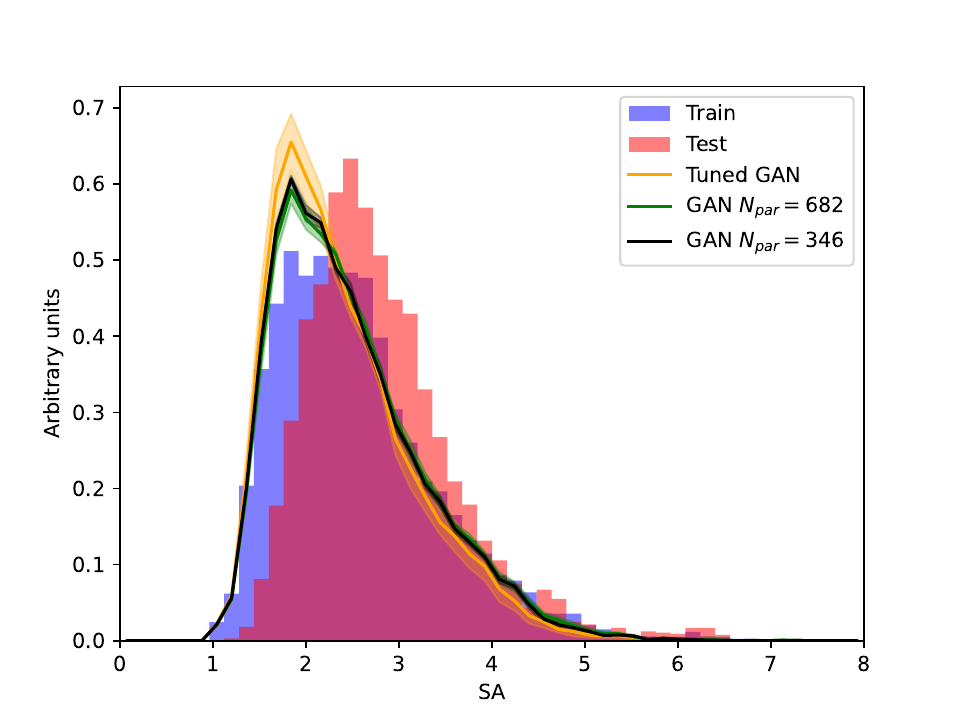}
    \includegraphics[width=0.32\linewidth]{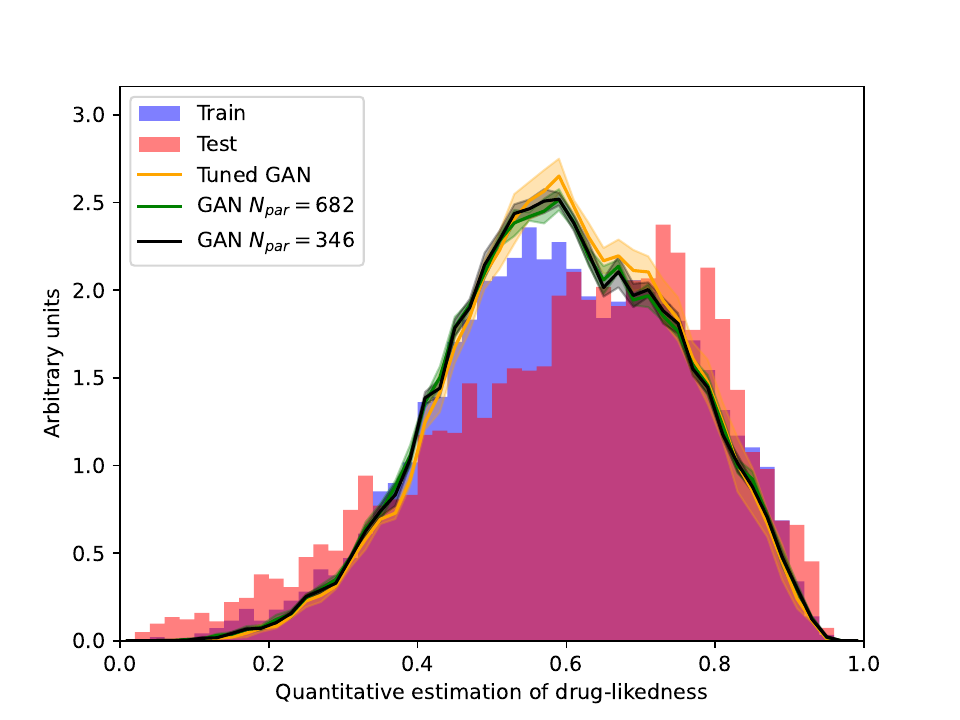}
    \includegraphics[width=0.32\linewidth]{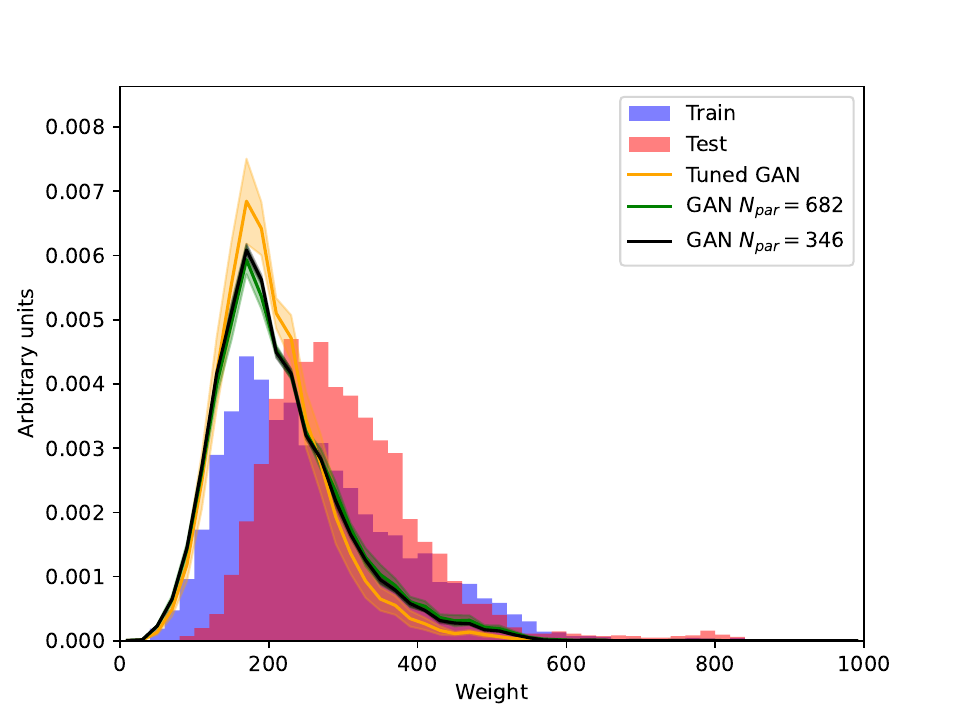}
    \caption{Metrics distributions for GANs trained with different number of generator trainable parameters (scenarios with 682 and 346 trainable parameters). Each model was run with five different seeds, and the shaded band is defined as the $\pm 1\sigma$ deviation.}
    \label{fig:app:robustness:metrics}
\end{figure}

\begin{figure}[h!]
    \centering
    \includegraphics[width=0.32\linewidth]{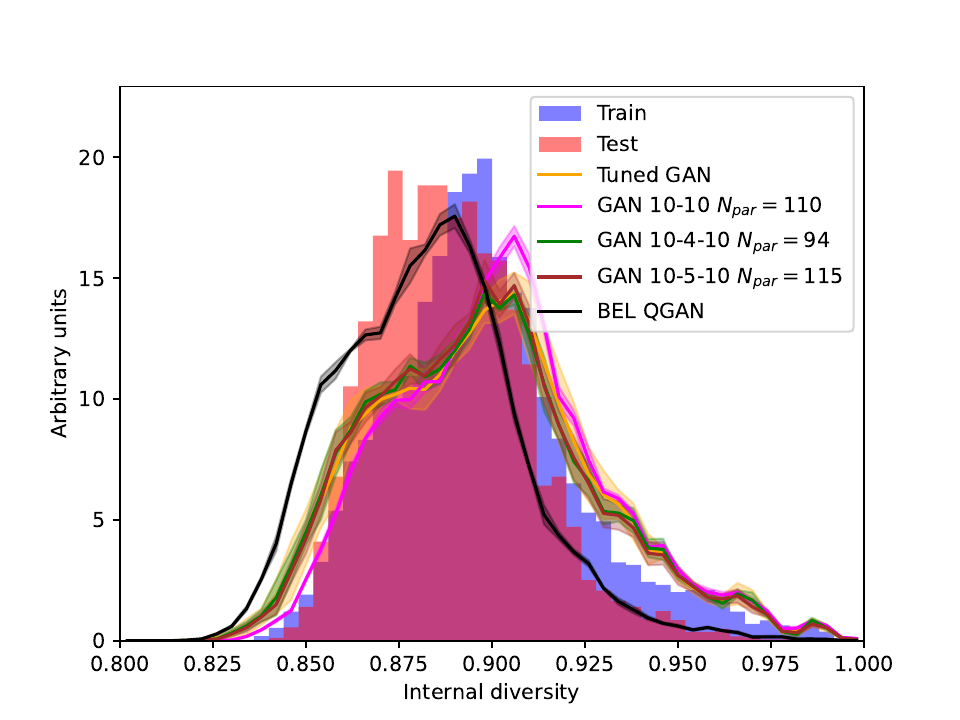}
    \includegraphics[width=0.32\linewidth]{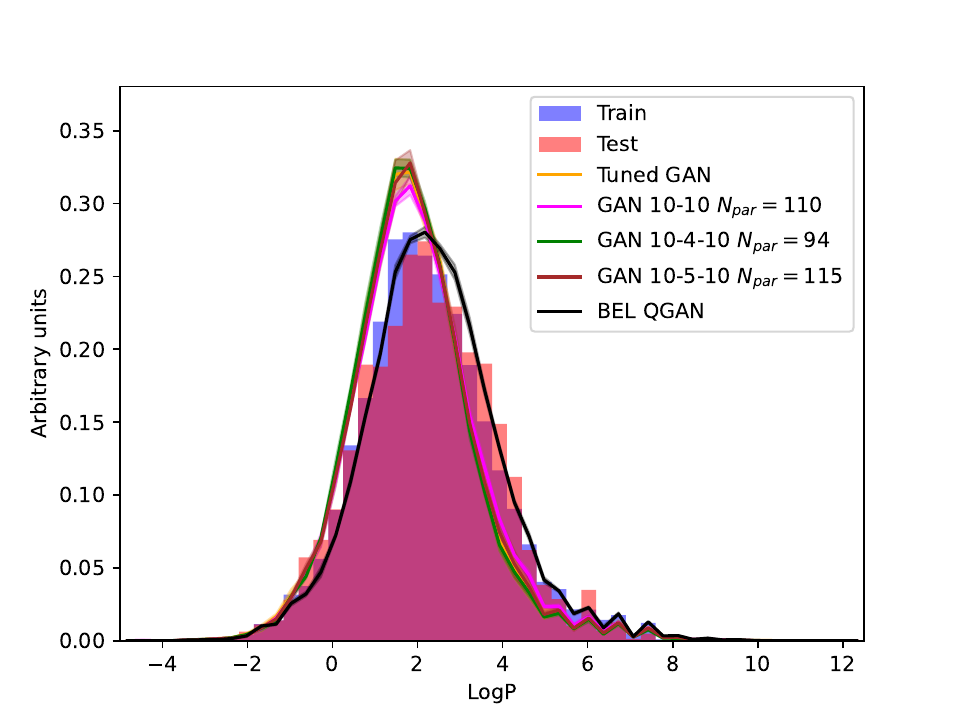}
    \includegraphics[width=0.32\linewidth]{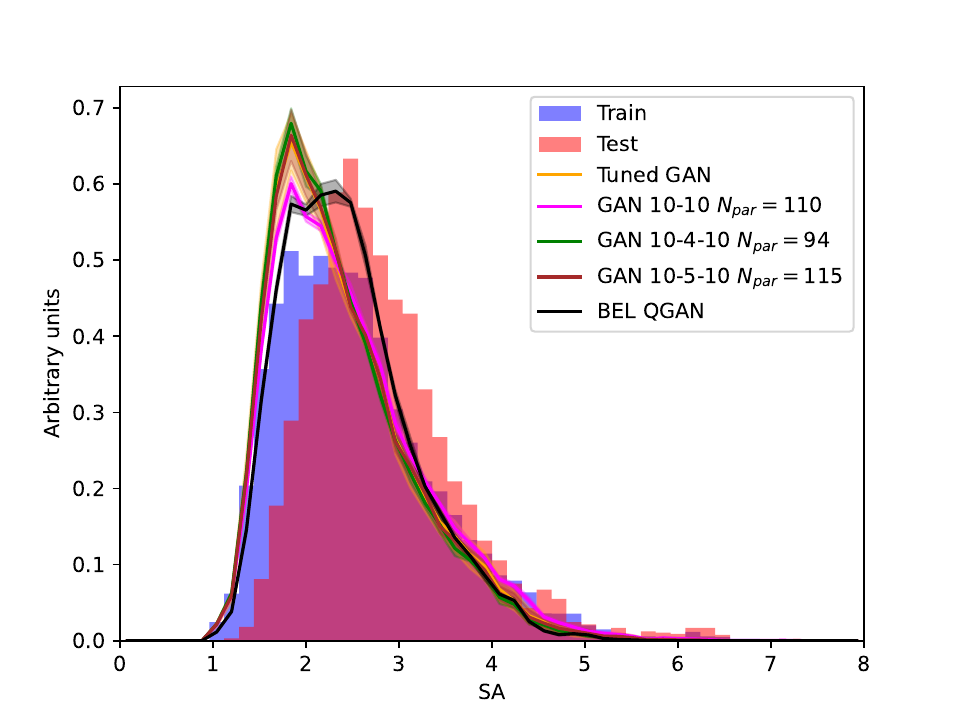}
    \includegraphics[width=0.32\linewidth]{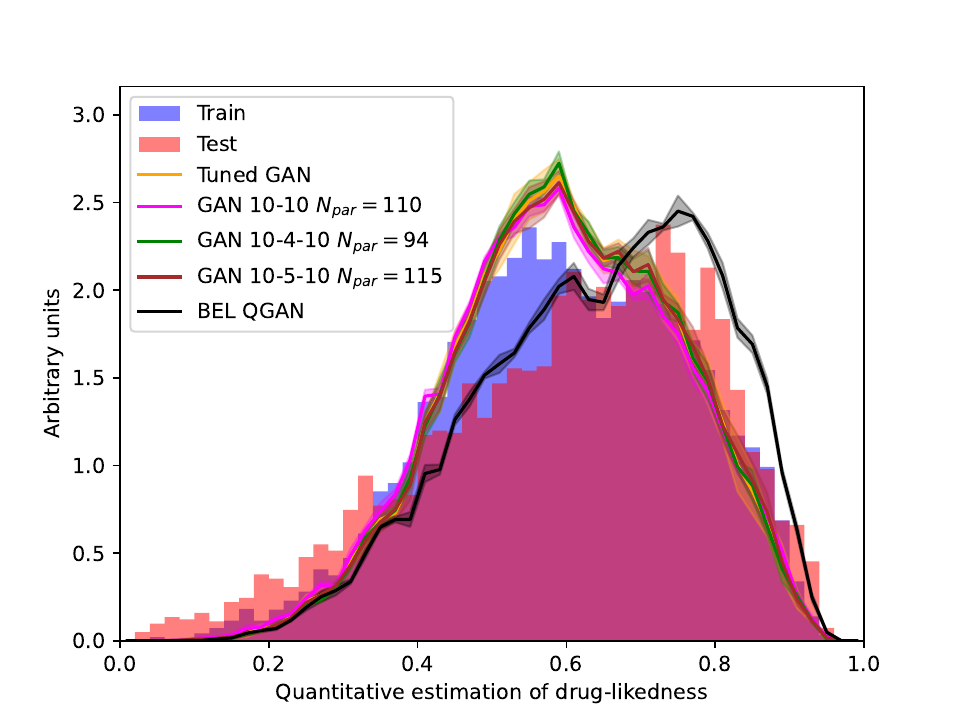}
    \includegraphics[width=0.32\linewidth]{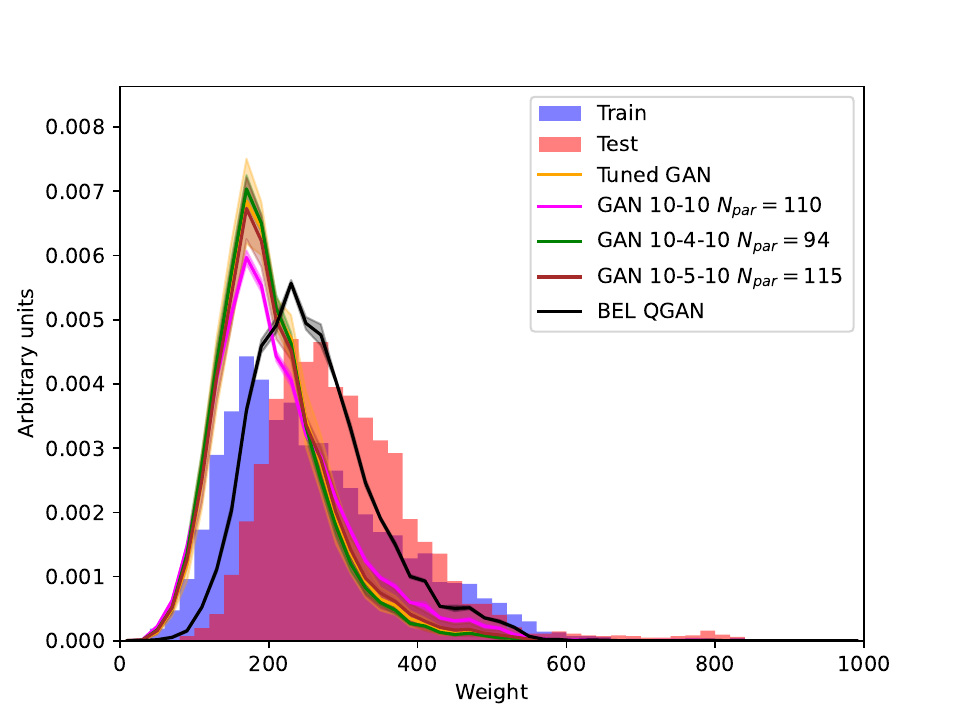}
    \caption{Same as in Figure~\ref{fig:app:robustness:metrics} for the low-parameter scenarios with 115, 110, and 94 trainable parameters.}
    \label{fig:app:robustness:metrics2}
  \end{figure}

We present in Figure~\ref{fig:app:robustness:metrics} and Figure~\ref{fig:app:robustness:metrics2} the distributions of the metrics themselves for the samples generated by the five classical GANs. While the results are statistically consistent, there is a slight observable difference between the reference and the low-parameter scenarios.

In conclusion, this test shows that the reference GAN with architecture taken from MOSES does not suffer from performance degradation due to the large number of parameters. In contrast, single-hidden-layer classical GANs with a number of trainable parameters roughly matching the number of trainable parameters of the styled BEL QGAN slightly underperform for most of the metrics compared to the reference classical GAN in the main text. Only the 94-parameter scenario with two hidden layers slightly overperforms, but it still does not provide with any individual metric significantly improving over the nominal MOSES baseline, contrary to our proposed BEL QGAN.

\section{Supplementary material: Quantum hardware characterization, noisy simulation, and latent-vector distributions}
\label{sec:app:hardware}

\begin{figure}[h!]
    \centering
    \includegraphics[width=\linewidth]{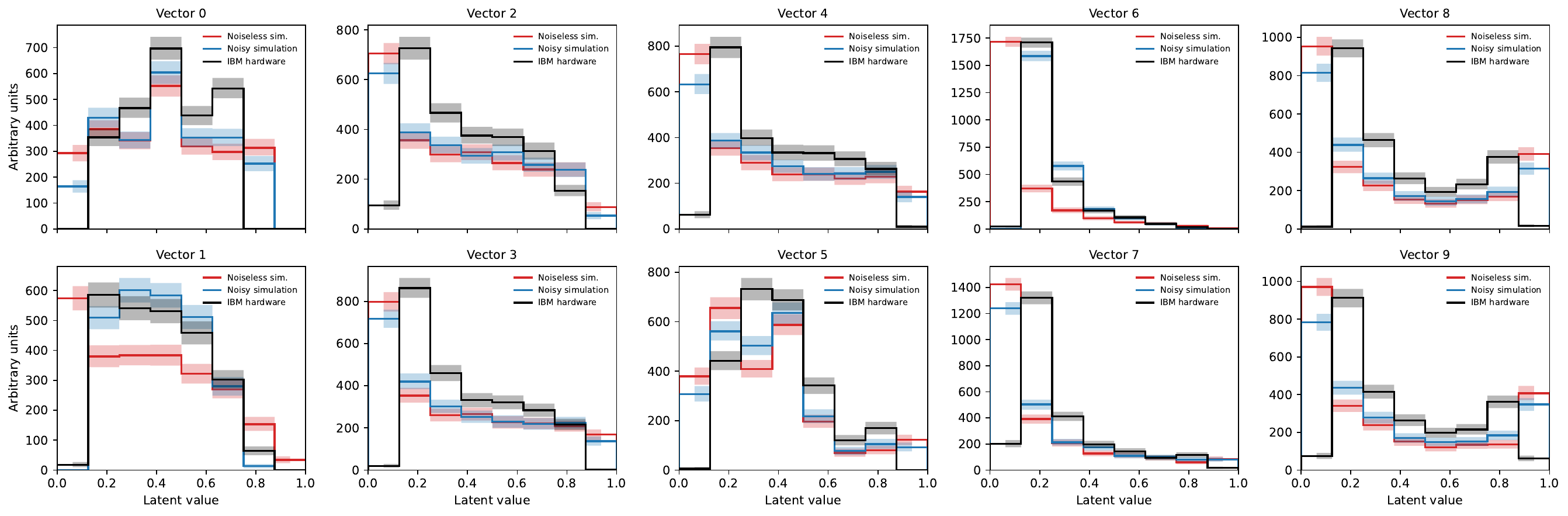}
    \caption{Pre-VAE latent-vector distributions on an inference with 2,500 samples from the styled-BEL \qgan\ ($n_{qb}=5$, $n_l=2$, dual readout) with a noiseless state-vector simulator (red), a \texttt{FakeKingston} Aer noisy simulation (blue), and the \ibm\ hardware run (black). The step count histograms use eight bins, and the shaded regions display the $95\%$ bootstrap percentile bands ($B=2,000$ resamples per scenario).}
    \label{fig:app:hw:latent}
\end{figure}

This section documents the quantum hardware and the execution settings used for the inference-only run reported in Section~\ref{sec:results:inference} of the main text, in which 2,500 molecules have been sampled from the simulator-trained \qgan\ on the \ibm\ device. No training was performed on hardware. Table~\ref{tab:app:hw:config} summarizes the model and execution configuration, Table~\ref{tab:app:hw:calib} reports representative device calibration properties, and Table~\ref{tab:app:hw:zscore} gives the per-metric $Z_0$ significance of the hardware run relative to the noiseless simulation, see Equation~\ref{eq:z0def}, calculated with bootstrap standard errors.
We also report in Table~\ref{tab:app:hw:noisy} a matched noise-model simulation on the same trained generator, and in Figure~\ref{fig:app:hw:latent} and Table~\ref{tab:app:hw:latent} a comparison of the latent-vector distributions for the matched noiseless simulation, noisy simulation, and \ibm\ hardware run, before VAE decoding. Transpiled circuit depth, two-qubit-gate count, and physical-qubit layout have been read from the executed circuits stored in the IBM Runtime job records for all 5,000 Estimator jobs (2,500 $X$/ $Z$ pairs).
We observe 13 distinct logical-to-physical layout mappings and nine distinct five-qubit physical subsets across the run (the mode layout maps logical qubits 0--4 to physical indices 96, 83, 82, 76, 81 in 34\% of jobs). Device-wide calibration medians in Table~\ref{tab:app:hw:calib} are taken from actual calibration data for \ibm\ (using \texttt{backend.properties(datetime=...)}) on 2025-10-04~UTC, contemporaneous with the inference session.

\begin{table}[h!]
    \centering
    \begin{tabular}{ll}
        \toprule
        Property & Value \\
        \midrule
        Quantum processor & \ibm\ (IBM Heron r2) \\
        Qubit topology & heavy-hex \\
        Total physical qubits on device & 156 \\
        Circuit & styled BEL \\
        Qubits / readout & 5 / dual \\
        Number of layers $n_l$ & 2 \\
        Latent-vector dimension & 10 \\
        Trainable parameters & 110 \\
        Number of generated molecules & 2,500 \\
        Shots per circuit & 1,000 \\
        Transpilation optimization level & 2 (Qiskit default) \\
        Resilience level & 0 (no error mitigation) \\
        Execution mode & batch (IBM Cloud) \\
        Calibration snapshot date & 2025-10-04, 17:00~UTC \\
        Physical qubit indices (mode layout) & 96, 83, 82, 76, 81 \\
        Distinct layouts / physical-qubit sets & 13 / 9 (over 5,000 jobs) \\
        X/Z readout scheme & Separate Estimator jobs (alternating $X$ then $Z$) \\
        Qiskit IBM Runtime primitive & Estimator \\
        Transpiled circuit depth (median) & 171 ($X$: 171; $Z$: 171; range 147--190) \\
        Transpiled two-qubit-gate count (median) & 75 CZ ($X$: 75; $Z$: 75; range 71--88) \\
        Number of jobs & 5,000 \\
        Execution date(s) & 2025-10-04, 14:42--22:32~UTC \\
        Total QPU time & 10,377~s ($\approx 2.9$~h) \\
        \bottomrule
    \end{tabular}%
    
    \caption{Model and execution configuration for the illustrative inference-only run on \ibm. The quantum generator uses $n_{qb}=5$, $n_l=2$, and dual readout to produce a latent vector of dimension~10. Transpiled depth, CZ counts, and physical-qubit layout statistics are computed over all 5,000 Estimator jobs from the transpiled circuits stored in the IBM Runtime job records. The listed physical indices are the mode logical-to-physical mapping for qubits~0--4 (34\% of jobs); see text for layout variation across the run.}
    \label{tab:app:hw:config}
\end{table}

\begin{table}[h!]
    \centering
    \begin{tabular}{lll}
        \toprule
        Calibration metric & Value & Unit \\
        \midrule
        Basis gates & \texttt{cz}, \texttt{id}, \texttt{rz}, \texttt{sx}, \texttt{x} & -- \\
        Median $T_1$ & 249 & $\mu$s \\
        Median $T_2$ & 142 & $\mu$s \\
        Median single-qubit ($\sqrt{X}$) gate error & $2.4 \times 10^{-4}$ & -- \\
        Median two-qubit (CZ) gate error & $2.0 \times 10^{-3}$ & -- \\
        Median readout assignment error & $8.5 \times 10^{-3}$ & -- \\
        Median single-qubit gate time & 32 & ns \\
        Median two-qubit gate time & 68 & ns \\
        \bottomrule
    \end{tabular}%
    \caption{Representative device-wide calibration properties for the IBM Heron~r2 \ibm\ device on the execution date. Medians are computed over all 156 device qubits from \texttt{backend.properties(datetime=2025-10-04)}, with snapshot taken at 17:00~UTC on 2025-10-04. The five physical qubits listed in Table~\ref{tab:app:hw:config} are a subset of the device qubits. Note that only the default basis gates are reported here to match the transpilation setup of the hardware run.}
    \label{tab:app:hw:calib}
\end{table}

\begin{table}[h!]
    \centering
    \begin{tabular}{lcc|c}
        \toprule
        Metrics & \qgan\ (noiseless) & \qgan\ (\ibm) & $Z_0$ \\
        \midrule
        Novelty & $0.338 \pm 0.010$ & $0.341 \pm 0.011$ & $+0.18$ \\
        IntDiv & $0.881 \pm 0.001$ & $0.882 \pm 0.001$ & $+0.85$ \\
        Filters & $0.706 \pm 0.010$ & $0.713 \pm 0.010$ & $+0.49$ \\
        $\epsilon_{LogP}$ & $0.899 \pm 0.007$ & $0.889 \pm 0.007$ & $-1.04$ \\
        $\langle \mathrm{SA} \rangle$ & $2.424 \pm 0.015$ & $2.485 \pm 0.016$ & $-2.81$ \\
        $\langle \mathrm{QED} \rangle$ & $0.647 \pm 0.003$ & $0.649 \pm 0.004$ & $+0.28$ \\
        $\langle \mathrm{Weight} \rangle$ & $264.8 \pm 1.8$ & $273.6 \pm 1.8$ & $-3.43$ \\
        \midrule
        $\langle Z_0 \rangle$ & Reference & -- & $-0.78$ \\
        \bottomrule
    \end{tabular}%
    \caption{Per-metric $Z_0$ significance (Equation~\ref{eq:z0def}) for the inference-only hardware run versus the noiseless simulation, using the bootstrap standard errors of both runs as $\sigma_1^m$ and $\sigma_2^m$ ($B=10,000$ resamples). The noiseless simulation is the reference; a negative $Z_0^m$ for a maximized metric indicates degradation of the hardware result at this bootstrap precision. These uncertainties quantify finite-sample precision of each single run and do not reflect run-to-run initialization spread.}
    \label{tab:app:hw:zscore}
\end{table}

We have also performed a matched noisy simulation of the hardware run based on the \texttt{FakeKingston} noise model provided by IBM. We have generated 2,500 latent vectors from the same simulator-trained styled-BEL generator ($n_{qb}=5$, $n_l=2$, dual readout, 110 trainable parameters) using the Aer simulator backend, initialized with the \texttt{FakeKingston} snapshot, using per-sample execution and 1,000 shots per circuit.
The resulting latent vectors were decoded with the identical tuned VAE used in our work for the classical GAN, the noiseless-simulation QGAN run, and the actual hardware runs. We then evaluated the metrics with the same MOSES pipeline as in Table~\ref{tab:ibm_comp} of the main text, using again nonparametric bootstrap resampling, with $B=10,000$ bootstrap resamples, to calculate the confidence intervals in Table~\ref{tab:app:hw:noisy}. This offline simulation isolates the effect of gate, readout, and decoherence errors encoded in the device noise model, without transpilation onto a specific physical layout or the separate alternating $X$/ $Z$ Estimator job structure used on hardware.

The results in Table~\ref{tab:app:hw:noisy} show that the noisy simulation is in good agreement with the noiseless simulation on most chemically relevant metrics, for example with $\langle\text{QED}\rangle = 0.642$ (0.647 in the noiseless simulation) or Novelty = $0.337$ (0.338 in the noiseless simulation). Some metrics do differ beyond the bootstrap precision, for example $\langle\text{Weight}\rangle = 280.1$ (264.9 in the noiseless simulation) or $\langle\text{SA}\rangle = 2.522$ (2.424 in the noiseless simulation), with bootstrap confidence intervals that do not overlap, but the difference is still limited.
These results indicate that the static device noise model alone does not predict a large degradation relative to ideal simulation, meaning that e.g. the gate errors, readout errors, and decoherence errors captured in the noise model are not causing a significant degradation of the pipeline. Comparing the noisy simulation with the real hardware run, we find an overall good agreement on the central drug-design observables. In particular, $\langle\text{QED}\rangle = 0.649$ on hardware agrees with $0.647$ on the noiseless simulator and $0.642$ on the noisy emulator within bootstrap precision. Modest resolved shifts appear for $\epsilon_{LogP}$, $\langle\text{SA}\rangle$, and $\langle\text{Weight}\rangle$ (Supplementary Information Table~\ref{tab:app:hw:zscore}), without indicating a broad failure of simulator-trained parameters on current hardware. This suggests that effects beyond the snapshot noise model, such as layout-dependent transpilation or the separate-basis readout protocol, contribute to the residual noisy-simulation--hardware gap on these metrics. We do not expect the noisy simulation to bracket every metric exactly, consistent with the simplified execution model.

Taken together, gate, readout, and decoherence noise as modeled by \texttt{FakeKingston} does not lead to a collapse of the pipeline relative to the noiseless simulation, and the hardware decoded metrics remain broadly compatible with both the noiseless and noisy-emulator results at this bootstrap precision. We hence do not observe a broad failure of simulator-trained parameters on current quantum hardware.

\begin{table}[h!]
    \centering
    \begin{tabular}{lccc}
        \toprule
        Metrics & \qgan\ (noiseless) & \qgan\ (noisy simulation) & \qgan\ (\ibm) \\
        \midrule
        $\epsilon_d$ & 0.892 & 0.893 & 0.891 \\
        $\epsilon_v$ & 0.934 & 0.914 & 0.904 \\
        $\epsilon_u$ & 1.000 & 0.999 & 0.999 \\
        Novelty & $0.338$ [$0.318$, $0.358$] & $0.337$ [$0.316$, $0.357$] & $0.341$ [$0.320$, $0.362$] \\
        IntDiv & $0.881$ [$0.879$, $0.883$] & $0.883$ [$0.881$, $0.885$] & $0.882$ [$0.880$, $0.884$] \\
        Filters & $0.706$ [$0.687$, $0.726$] & $0.705$ [$0.685$, $0.724$] & $0.713$ [$0.693$, $0.733$] \\
        $\epsilon_{LogP}$ & $0.899$ [$0.886$, $0.913$] & $0.882$ [$0.868$, $0.896$] & $0.889$ [$0.875$, $0.902$] \\
        $\langle \mathrm{SA} \rangle$ & $2.424$ [$2.395$, $2.453$] & $2.522$ [$2.492$, $2.554$] & $2.485$ [$2.455$, $2.515$] \\
        $\langle \mathrm{QED} \rangle$ & $0.647$ [$0.641$, $0.654$] & $0.642$ [$0.635$, $0.649$] & $0.649$ [$0.642$, $0.656$] \\
        $\langle \text{Weight} \rangle$ & $264.9$ [$261.4$, $268.3$] & $280.1$ [$276.4$, $283.9$] & $273.6$ [$270.0$, $277.2$] \\
        \bottomrule
    \end{tabular}%
    \caption{Metrics comparison for 2,500 molecules generated with the same simulator-trained styled-BEL \qgan\ ($n_{qb}=5$, $n_l=2$, dual readout) on the noiseless state-vector simulator, on a \texttt{FakeKingston} Aer noise-model simulator (1,000 shots), and on \ibm\ hardware. The noisy-simulation run uses the publicly available Heron~r2 device snapshot; it does not reproduce the hardware transpilation layout or the separate $X$/ $Z$ Estimator job structure. The metrics $\epsilon_d$, $\epsilon_v$, and $\epsilon_u$ are point estimates; for all other metrics, values in brackets give the $95\%$ bootstrap percentile intervals ($B=10,000$ resamples).}
    \label{tab:app:hw:noisy}
\end{table}

We have also performed a direct comparison of the latent-vector distributions for the noiseless simulation, the noisy simulation with the \texttt{FakeKingston} noise model, and the \ibm\ hardware run, \emph{before} VAE decoding.
We compare the 2,500 predecoder latent vectors from the three inference settings in Figure~\ref{fig:app:hw:latent}. For this comparison, we have generated the noiseless-simulator latent vectors by running the trained styled-BEL generator on the noiseless state-vector simulator to emit 2,500 latent vectors without outputting the corresponding molecules with the VAE decoder, using the same trained styled-BEL checkpoint as in the main text to obtain Table~\ref{tab:ibm_comp}.
We have rescaled the latent vectors to the $[0,1]$ range using the $(1+\langle\hat{O}\rangle)/2$ formula, to match the hardware latent vectors, which are the expectation values saved from the IBM Runtime jobs and rescaled identically. The noisy-simulation latent vectors are those produced by the matched \texttt{FakeKingston} run of Table~\ref{tab:app:hw:noisy}. Note that we have checked that the noiseless-simulator latent vectors we use for this comparison do reproduce the intended noiseless inference distribution within bootstrap precision.

We present our results in Figure~\ref{fig:app:hw:latent} and Table~\ref{tab:app:hw:latent}. Similarly to Figure~\ref{fig:cgan_qgan_lat} of the main text, we display the per-component count histograms, including shaded uncertainty bands. In Figure~\ref{fig:app:hw:latent}, they correspond to the nonparametric bootstrap $95\%$ percentile interval with $B=2,000$ resamples with replacement from the 2,500 latent values for each corresponding scenario: red lines for the noiseless simulated inference, blue lines for the noisy simulated inference using the \texttt{FakeKingston} noise model, and black lines for the hardware inference on \ibm. Because the hardware inference is only performed for 2,500 samples, the binning is coarser than in Figure~\ref{fig:cgan_qgan_lat} of the main text: eight bins in Figure~\ref{fig:app:hw:latent} (roughly 310 counts per bin) instead of 50 bins in Figure~\ref{fig:cgan_qgan_lat} (roughly 600 counts per bin). Note that we compare here the direct generator output rescaled in $[0,1]$, while in the main text the comparison is in the VAE latent coordinate system, roughly $[-3,3]$. Most components show broad overlap between the three inference scenarios (within bootstrap uncertainty bands), except for large shifts in the sixth and seventh components, as well as in the very first bin. The observed shifts for the sixth and seventh components are consistent with the anti-correlated pair discussed in Section~\ref{app:corr_mat} and reflect genuine hardware and shot-noise effects rather than sample-size artifact alone.

We quantify our observations in Table~\ref{tab:app:hw:latent}, which summarizes two-sample Kolmogorov--Smirnov (KS) statistics and one-dimensional Wasserstein ($W_1$) distances aggregated over the ten latent dimensions. All pairwise comparisons reject the null of identical marginal distributions at $p<0.05$ for every component, reflecting finite-sample differences in means and variances. Hardware latent vectors are generally shifted toward higher values.
We note that, with $N=2,500$ samples per arm, the KS test is sensitive to the modest mean and variance shifts expected under finite-shot readout and device noise; therefore, the statistical rejection of exact equality does not by itself imply a wholesale replacement of the generator output distribution or a breakdown of the inference pipeline.
Importantly, the mean $W_1$ distances remain small, with $\langle W_1\rangle \lesssim 0.09$ for the noiseless--hardware comparison, which is consistent with the close agreement of the post-VAE drug-design metrics in Tables~\ref{tab:ibm_comp} and~\ref{tab:app:hw:noisy}, notably $\langle\text{QED}\rangle = 0.649$ on hardware versus $0.647$ on the noiseless simulator. The noisy simulation lies between the noiseless simulation and the hardware result on several components (notably $z_6$--$z_7$), but does not uniformly bracket hardware, in line with the simplified execution model of Table~\ref{tab:app:hw:noisy}.

These results support the conclusion that simulator-trained parameters transfer well to real hardware in this inference run. The latent-vector histograms display a broad overlap between the noiseless simulation, noisy simulation, and hardware inference: noise affects some components, in some bins and notably for $z_6$ and $z_7$, but not enough to collapse the distribution. The mean Wasserstein distance between noiseless simulation and hardware, $\langle W_1\rangle \simeq 0.086$, is consistent with the post-decoder drug-design metrics, for which $\langle\text{QED}\rangle$ matches the noiseless simulator within bootstrap precision and four of seven reported observables show no significant degradation ($Z_0^m \geq -1$; Table~\ref{tab:app:hw:zscore}).

\begin{table}[h!]
    \centering
    \begin{tabular}{lcccc}
        \toprule
        Comparison & Mean KS & Max KS & Mean $W_1$ & Max $W_1$ \\
        \midrule
        Noiseless sim.\ vs.\ hardware & 0.343 & 0.681 & 0.086 & 0.122 \\
        Noiseless sim.\ vs.\ noisy simulation & 0.227 & 0.739 & 0.044 & 0.143 \\
        Hardware vs.\ noisy simulation & 0.251 & 0.434 & 0.052 & 0.081 \\
        \bottomrule
    \end{tabular}%
    \caption{Summary of pre-VAE latent-distribution distances for the 2,500-sample inference runs ($d_{\text{latent}}=10$). Mean and maximum KS statistics and one-dimensional Wasserstein distances ($W_1$) are computed per latent component and aggregated over the ten latent vectors $z_0,\ldots,z_9$. The $p$-values are $<0.05$ for all ten components in each pairwise comparison.}
    \label{tab:app:hw:latent}
\end{table}

\end{document}